\documentclass[10pt,b5paper,twoside]{UGthesis}

\usepackage{aas_macros}
\usepackage{amsmath, amssymb}
\usepackage{arcs}
\usepackage{array}
\usepackage[icelandic,english]{babel}
\usepackage{booktabs}
\usepackage[small,bf]{caption}
\usepackage[usenames,dvipsnames]{color}
\usepackage{colortbl}
\usepackage[b5]{crop}
\usepackage{dcolumn}
\usepackage{enumerate}
\usepackage{enumitem}
\usepackage[dvips]{epsfig}
\usepackage{fancyhdr}
\usepackage[T1]{fontenc}
\usepackage[margin=2.3cm,b5paper,includehead]{geometry}
\usepackage{graphicx}
\usepackage[utf8]{inputenc}
\usepackage{lscape}
\usepackage{listings}
\usepackage{lmodern}
\usepackage{longtable}
\usepackage{multirow}
\usepackage{multicol}
\usepackage{natbib}
\usepackage{pdflscape}
\usepackage{psfrag}
\usepackage{relsize}
\usepackage{setspace}
\usepackage{sidecap}
\usepackage{subfig}
\usepackage{url}
\usepackage{verbatim}
\usepackage{wrapfig}
\usepackage[table]{xcolor}
\usepackage[linktoc=none]{hyperref}
\hypersetup{citecolor=blue, colorlinks=true, pagecolor=OliveGreen, linkcolor=BrickRed, urlcolor=BrickRed}
\definecolor{grey}{gray}{0.90}

\newcolumntype{g}{>{\columncolor{grey}}l}
\newcolumntype{G}{>{\columncolor{grey}}r}
\newcolumntype{H}{>{\columncolor{grey}}c}

\newcommand{\parcsec}{\mbox{$\stackrel{\prime\prime}{\textstyle .}$}}
\newcommand{\parcmin}{\mbox{$\stackrel{\prime}{\textstyle .}$}}
\newcommand{\nitrogen}{[\mbox{N\,{\sc ii}}]}
\makeatletter

\newcommand{\Rmnum}[1]{\expandafter\@slowromancap\romannumeral #1@}
\makeatother
\newcommand{\dtm}{$\mathcal{{DT\!\!M}}$}

\newcommand\titill{Exploring Gamma-Ray Bursts, Their Immediate Environment and Host Galaxies}

\begin{document}

\pagenumbering{roman}
\begin{titlepage}
~\\[-3cm] 

\hfill{\sffamily RH-10-2015} 
\vspace{1cm}
\begin{center}
{\sffamily\bfseries Dissertation for the degree of doctor of philosophy}
 
\vspace{1.0cm}
 
\upshape\huge\sffamily\bfseries \titill  \\[0.5cm]

\vspace{0.5cm}
 
\rule{11cm}{2pt}
 	
\vspace{1cm}
 
{\Large\sffamily\bfseries Mette Friis}
 
\vspace{3.cm}
\begin{figure}[!ht]
\centering
\includegraphics[width=6cm]{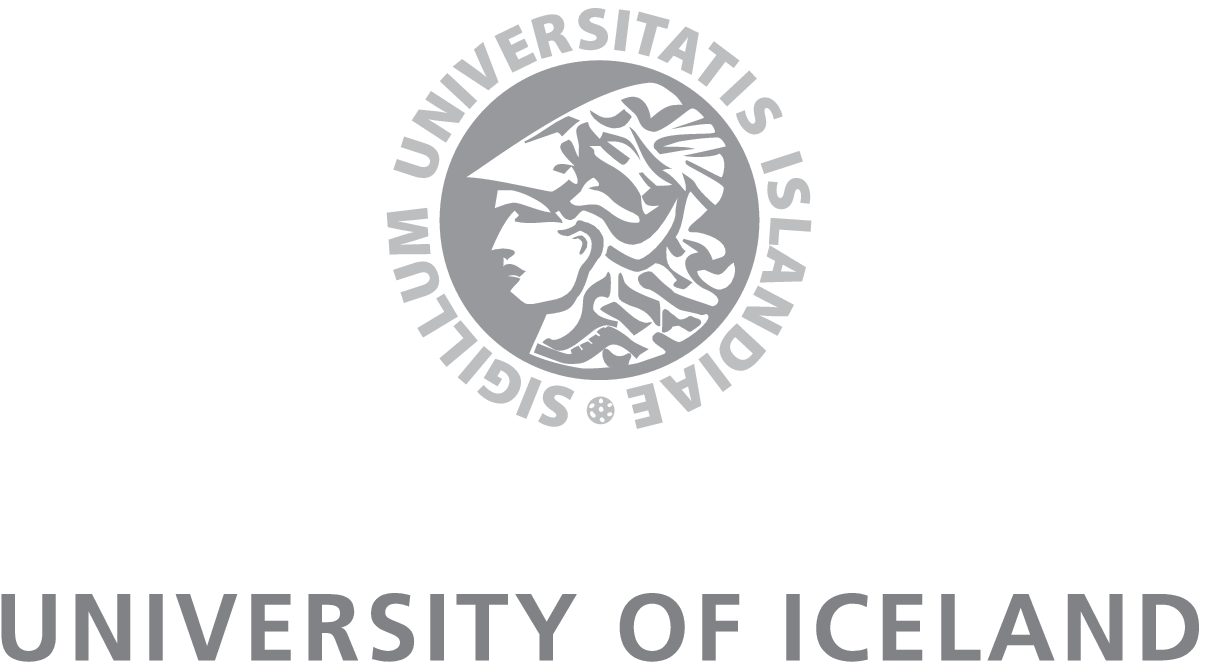}
\end{figure}
 
\vspace{2.cm}
\small\sffamily
School of Engineering and Natural Sciences \\
Faculty of Physical Sciences \\ 
Reykjav\'ik, October 2015
\end{center}
\end{titlepage}

\thispagestyle{empty}
\noindent
A dissertation presented to the University of Iceland, School of Engineering and \\ Natural Sciences, in candidacy for the degree of doctor of philosophy

\vspace{1cm}
\noindent
{\bf Doctoral committee} \\
Gunnlaugur Bj\"ornsson, Research Professor\\
Science Institute, University of Iceland\\
\\
P\'all Jakobsson, Professor\\
University of Iceland\\
\\
Einar H. Gu\dh{}mundsson, Professor\\
University of Iceland

\vspace{1cm}
\noindent
{\bf Opponents} \\
Professor Sandra Savaglio \\
Physics Department \\
University of Calabria \\
Via P. Bucci \\
I-87036 Arcavacata di Rende \\
Italy \\
\\
Dr. Stefano Covino \\
INAF / Brera Astronomical Observatory \\
Via Bianchi 46 \\
23807 Merate (LC) \\
Italy \\

\vspace{1cm}

\noindent	
\titill\\
\noindent
\copyright\  2015 Mette Friis\\
\noindent
Printed in Denmark by Reklamernes F\ae llestryk \\
\noindent
Science Institute Report RH-10-2015 \\
\noindent
ISBN 978-9935-9263-6-4

\null\vfill

\begin{flushright}
\textit{Dedicated to You. \\ Thank you for reading.}
\end{flushright}

\vspace{14cm}


\cleardoublepage

\pagestyle{fancy}
\renewcommand{\chaptermark}[1]{\markboth{Chapter \thechapter\ \ \ #1}{#1}}
\renewcommand{\sectionmark}[1]{\markright{\thesection\ \ #1}}
\lhead[\fancyplain{}{\sffamily\thepage}]%
  {\fancyplain{}{\sffamily\rightmark}}
\rhead[\fancyplain{}{\sffamily\leftmark}]%
  {\fancyplain{}{\sffamily\thepage}}
\cfoot{}
\setlength\headheight{14pt}

\clearpage
\tableofcontents

\cleardoublepage

\chapter*{Abstract}
\markboth{Abstract}
TThe Universe has always been the source of wonder and the home to many mysteries. What goes on in distant parts, great beyond our own solar system, is difficult to decipher, even with the enormous technological progress of the last hundred years. One opportunity to study the distant, young Universe, is presented through observations of Gamma-Ray Bursts (GRBs). GRBs are extremely bright transient objects, in principle observable out to cosmological redshifts or $z>10$. Not only are they extremely bright in $\gamma$-rays, GRBs also have powerful afterglows, with bursts such as GRB\,080319B, dubbed the 'naked-eye' burst, as it was bright enough to be visible with the human eye, despite a redshift of $z=0.9$. Much work has been done on these objects, but there are still open questions to explore. 

In the first paper of this thesis I present work on the radiative mechanisms of GRBs. In particular, the work concerns the interpretation of the soft X-ray components observed in the early afterglow. To examine this component, we selected a sample of the X-ray brightest bursts observed by the \emph{Swift} telescope XRT, on which we performed a systematic search for a thermal component. The results show that this emission is more ubiquitous than previously thought. Most of the afterglows show signs of a thermal component, with about one fourth having a statistical significant detection. Previously, these components have often been interpreted as the supernova (SN) shock break-out. Model calculations show that the total energy from the break-out will never be more than $10^{47}$\,ergs. This result fits poorly with the already discovered thermal components, and is even more difficult to reconcile with the new-found additions from our search. The energetics are simply too large. The blackbody temperatures and luminosities are so high that a physical origin is difficult to construct. On that background, and because thermal emission is frequently observed in the prompt ($\gamma$- and hard X-ray) emission of the GRBs, we propose the model of late photospheric emission. In this model, the emission originates within the jet, from material moving at relativistic speed. This is supported by the evolution of the simple blackbody radius, seen to evolve with an apparent velocity faster than the speed of light. Compared with the results from the prompt phase, the Lorentz-factors are of similar or smaller values here, and the photospheric radii about two magnitudes larger, as expected at these later times. In the prompt phase, the blackbody temperature and luminosity is seen to decay with power-law indices that all fall within a narrow range. The indices we find in the later soft X-ray regime, are seen to be consistent with this range. The proposed model of late photospheric emission, is hence a physically well-motivated theory, that has the added advantage of allowing the determination of the jet Lorentz factor and the photospheric radius. Furthermore, as this component is caused by the GRB inner engine, it allows us to probe the prompt emission of GRBs at late times, with the more sensitive soft X-ray telescopes, making it a possibly important step towards determining the physics behind the GRB itself.

The brightness of the optical afterglows makes GRBs excellent light houses, used to study the host galaxy and environment. Using an extensive dataset of the afterglow and host galaxy of GRB\,121024A, this thesis includes a study of a $z=2.3$ galaxy. One of several results to come out of this study, is a comparison between different metallicity estimates, namely from absorption line fitting and strong-line diagnostics (emission lines), as well as a comparison to the mass-metallicity relation (all three methods agree on a metallicity of [$M$/H]\,=\,$\sim-0.4$). This is the first time for a GRB-damped Ly$\alpha$ absorber that the metallicity has been determined both in absorption and emission. We also observe molecular hydrogen, the fourth such detection in a GRB host, as well as determine the star-formation rate, the stellar mass of the host, and dust absorption properties. This last point is the subject of the last part of the thesis, as we found that the dust in the line-of-sight to GRB\,121024A contains a large component of 'grey' dust. Grey dust refers to the lack of colour dependance in the dust absorption, which is physically caused by the dust grain size distribution being skewed towards larger grains, but otherwise similar to dust that is observed in the Milky Way, as seen by our fits. This is not the first detection of grey dust towards GRBs, and in fact we find that up to $30$\% of observed GRB lines-of-sight could contain such a dust component. Lastly, the consequences of ignoring the possibility of grey dust in the high-$z$ Universe is explored. We find that, realistically, the star formation rate density at a given redshift, could be underestimated by as much as 30\%. I end the thesis with a short discussion of future prospects.
\addcontentsline{toc}{chapter}{\sffamily\bfseries Abstract}
\cleardoublepage

\pagestyle{plain}

\chapter*{Acknowledgements}

First of all people I want to acknowledge the huge help from Steve and Annalisa. I bet you did not know it at the time, but by accepting your positions to be students in Iceland, you also entered into what I fear for you might be a life-long commitment to help out with Palli's future students. You have both been invaluable in your knowledge and your (almost) endless patience with my stupid questions. Annalisa, thank you so much for taking all that time to teach me about Voigt-profile fitting, and for your extensive help in navigating the thought process of a referee. Steve, my faith in your technical skills seem to surpass your own, but in the end I feel I am always right. There are very few things you cannot fix.

The big help I have had in Annalisa and Steve is very illustrative of the brilliant way in which Palli supervises. One person cannot know everything there is to know, so knowing who to ask and where to get help, is a key aspect of getting ahead in life. Thank you so much to both Palli and Gulli for the best possible supervising. There are very few phd students, I think, who get to work so independently purely on their own projects, while still knowing that the door is always open should she or he need advise. Also a big thank you to Einar, for actively supporting gender equality in science. As the only girl at the astronomy centre, it is a subject close to my heart. While I am on the subject of work environment, I also want to thank Zach and Andreas for fruitful/entertaining discussions on science, and everything else that matters (or not) in life. It was always boring days when none of you were in the office. A further thanks to Andreas, for teaching me how to fix my bike.

I would also like to thank Johan, Daniele, and everyone else at Dark (especially Jens-Kristian for programming help in \emph{Python}) for helping me in my work, and for always giving me a warm welcome when I come for a visit. Also a big thank you to Christina and Antonio for hosting such awesome workshops in Spain, and for helping me with GTC data, and to Thomas for helping with just about everything else on my GRB\,121024A paper. Thank you to Rhaana Starling for teaching me to use isis. A special thanks to Darach and Anja, who are the most supportive people imaginable, in each their way, and without whom it is doubtful that I would have ventured into the research world. Your support has meant the world to me. The same goes for the other girls abroad; Maria, Bitten, Mathilde, and Eva who knows what it is like to be away from home, and who has provided instrumental support. Thanks to Maria, Bitten and Simone for physics discussions on Skype, and an extra thanks to Maria for lending a virtual ear to all my complaints about everything and everyone. When all your friends are far away, having almost non-stop communications with all you girls on Skype (including also Ann-Katrine) has been a life-saver. When mentioning friends on Skype, I would be amiss if I did not also thank Asbj\o rn. Thank you, for being the best friend a person could ever ask for, and for making me laugh so much.

I have not been completely alone and friendless in the cold north however. One person in particular, has been a little ray of constant sunshine. Emilia, thank you so much for every trip we have been on, for every party you have brought me to, for getting me involved with the volunteers, but most of all, thank you, for your eternal optimism and friendliness. Never have I seen you without a smile on your face. A further thanks to Bo, Ambi, Sebrina, Natalia, Steve \& Becci (and little Sasha), Mara, Ana, Gudjon, Andreas, Kyra, Hannah and Gregory and everyone else who has helped make Iceland feel like home. Thank you to Krissa, Gugga, Eyrun, Eydis, Sara, Gudbj\"org, Hofi, and everyone else at Fylkir for awesome volleyball tournaments, for the good mood on the team, for letting me kick your asses in badminton, and thank you so much to Krissa, Eydis, Birgitta, Eyrun and Gudjon for chauffeuring me around Iceland. Super many thank you to the whole team for welcoming my mother on our trip, and for continuously asking about her wellbeing. It means a lot to me! 

My mother.. How could I end this long babble of people for whom I am so grateful, without mentioning my parents. I know most people say that \emph{they} have the best parents in the world, but most people are wrong. I am not. My parents are the most supporting, most trusting and most loving parents in the whole world. I knew this, of course I did. But I did not understand it fully, until I got to travel the world, and see and hear how other people were brought up. I have had the best childhood one could possibly wish for, and you guys continue to be there for me. I love you, as I love my little sister, even when she is annoying, as is only fitting for the youngest sibling. My best of luck to Maria, I know life has not always treated you as fairly as it did me, but that makes it all the more impressive that you ended up the smart, mature and loving person that you are today. I am so proud that you have kept trying, and never compromised yourself.
\addcontentsline{toc}{chapter}{\sffamily\bfseries Acknowledgements}
\cleardoublepage
\allowdisplaybreaks
\pagenumbering{arabic}
\cleardoublepage
\pagestyle{fancy}	

\chapter{Gamma-Ray Bursts}\label{chapter:grb}
This chapter describes some of the general properties of gamma-ray bursts (GRBs) as well as the properties of the environment and galaxies in which we find these burst. The classical introduction to GRBs includes an account of the discovery of these curious phenomena, a story which I will briefly describe here as well, as it is quite unique. 

The story is set in the last years of the Cold War, after the nuclear test ban treaty was signed by the US and Moscow in 1963. The Americans, as part of an effort to check that Russia was upholding the treaty, launched the Vela satellites in 1967. The purpose of these were to monitor the earth's atmosphere for the characteristic $\gamma$-ray signature of nuclear bomb testing. While the majority of satellites orbit well below the $35,786$\,km for a geostationary orbit, the Vela satellites orbited at an altitude of $\sim100,000$\,km, to monitor tests in space (including possible explosions behind the Moon). To everyone's surprise the satellites did indeed detect a $\gamma$-ray signature. But the signal was not coming from below. It originated from space \citep{1973ApJ...182L..85K}.

\section{In Short: What is a Gamma-Ray Burst?}
Gamma-Ray Bursts are extremely bright flashes of $\gamma$-rays originating outside our own galaxy, the Milky Way. During a burst, the $\gamma$-ray sky (MeV) is completely dominated by photons from the GRB. The isotropic luminosities are $10^{50}$ -- $10^{53}$\,erg\,s$^{-1}$, comparable to that of the Sun during its entire lifetime. This makes GRBs the most energetic events in the Universe, since its creation (the Big Bang). These energies, however, are just the part that is emitted in form of electromagnetic radiation. Theory predicts that a large part is emitted as neutrinos and gravitational waves \citep{2010A&A...524A...4R,2011APh....35....1B,2012ApJ...752...31S}. We know that this enormous energy originates from a very small region in space, as we observe variations in the light curve on a millisecond scale \citep{1992Natur.359..217B,1997ApJ...485..270S}. These variations cannot be faster than the light-crossing time for the emitting volume, or they would have been smeared out. The small size and large energy output together indicate an origin related to compact objects, i.e. black holes or neutron stars. 

After the burst of $\gamma$-rays, long GRBs (see Section~\ref{sec:long}) are observed to have an afterglow in X-rays (usually observed), ranging through optical all the way down to radio (not always observed). These afterglows are thought to be largely due to interactions with the surrounding media, and provide an opportunity to study the birth environment of these bursts, which are seen to be actively star-forming regions (this differs from the bursts termed `short' GRBs). 

GRBs are among the furthest objects observed. Some of the highest spectroscopically confirmed redshifts belong to GRBs \citep[the highest being $z=8.3$ for GRB 090423,][]{2009Natur.461.1254T,2009Natur.461.1258S}. The typical observed redshift for GRBs is in the range $z\sim2$--$3$ \citep[the mean redshift of the TOUGH Survey, see Section~\ref{sec:hosts}, for instance, is $z=2.2$, see][]{2012ApJ...752...62J}, firmly putting GRBs in the class of cosmological phenomena.

\section{Classification}
GRBs are divided into subgroups depending on the length and spectral properties of the prompt (initial $\gamma$-ray) emission. The main classification used is short- and long-duration GRBs \citep[there have been suggestions of a third, intermediate class, e.g.][which I will not explore here]{2011A&A...525A.109D}. The physical mechanism is believed to be different for bursts in the different categories (I include here also X-ray flashes, which may not be a classification of its own, but rather a matter of viewing angle). 

\subsection{Short Gamma-Ray Bursts}
\begin{figure}[h]
\centering
\includegraphics[width=0.98\textwidth]{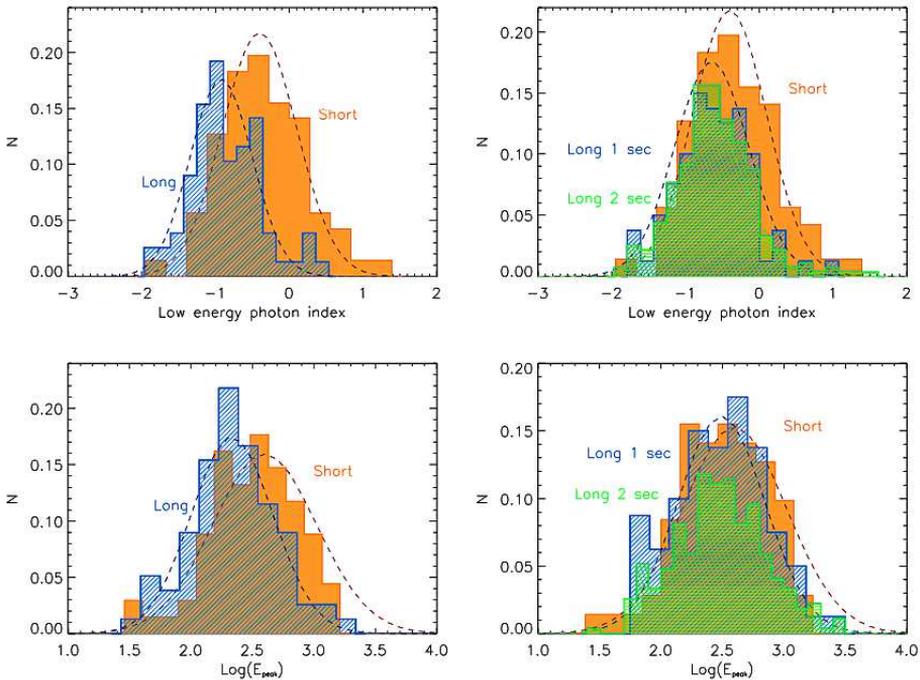}
    \caption{Distribution of spectral parameters (low energy photon index and $E_p$) for long vs. short GRBs for a sample of BATSE bursts reported in \cite{2009A&A...496..585G}. On the left side are shown parameters for the time integrated spectra, while the plots on the right side compares parameters from short GRB spectra with those from spectra comprised of data from only the first one (blue) and two (green) seconds of long GRBs. Dashed lines show a Gaussian fit to the distributions.}
    \label{fig:s_vs_l}
\end{figure}

Short GRBs are usually defined as GRBs with a prompt-emission duration $<2$\,s (traditionally the duration of a GRB is described as $T_{90}$, the time in which 90\% of the total fluence is recorded). They have been observed to have durations of only a few tens of milliseconds \citep[e.g.][it should be noted that these short time-scales are in the order of the detector's temporal resolution, which prevents a full sampling of duration]{1992Natur.359..217B}. The spectra of short GRBs are observed to be harder than their longer counterparts, with a spectral peak at higher energies $E_{p}\sim400$\,keV and a shallower spectral slope. Short GRBs are also observed to have a lower isotropic luminosity when compared to long GRBs that have similar spectral peak energies. These differences are less apparent though, when compared to only the first $\sim2$\,s of the long GRB spectra, see Fig.~\ref{fig:s_vs_l}. 

Fewer short GRBs have been observed in general \citep[today $\sim75$, see e.g.][for a review]{2014ARA&A..52...43B}, and there are especially few observations of short GRB afterglows. The latter was first reported in 2005 \citep{2005Natur.437..851G}, as opposed to 1997 for long GRBs \citep{1997Natur.387..783C,1997Natur.389..261F,1997Natur.386..686V}. This has led to a late start in modelling the origin of these bursts, and especially determining the characteristics of the birthplace, as the afterglow is needed to determine a redshift and precise location on the sky.

Some short GRBs ($15$--$30$\%) exhibit extended $\gamma$-ray emission lasting $\sim10$--$100$\,s with softer spectra than the prompt emission. There has been a variety of interpretations of this emission, e.g. stellar winds or outflows, in the frame-work of different progenitor models, see Section~\ref{sect:model} \citep{2008MNRAS.385.1455M,2014MNRAS.438..240G}.

\subsection{Long Gamma-Ray Bursts}\label{sec:long}
The majority of GRBs observed fall into the category of long gamma-ray bursts (LGRBs). These have a prompt-emission duration of $>2$\,s (mean duration of $\sim30$\,s observed with BATSE, see Section~\ref{sec:telescopes}), which is nearly always observed to be followed by an X-ray afterglow. About 60\% / 50\% of long GRBs have observed afterglows at optical/radio wavelengths, though the latter is not well sampled due to the sensitivity of radio telescopes. Bursts without an optical counterpart are sometimes classified as 'dark bursts', often defined as bursts with an optical--to--X--ray spectral index $\beta_{\text{OX}}<0.5$ \citep{2004ApJ...617L..21J}. The X-ray afterglow is composed of different phases (e.g. steep decays and plateau), as well as often observed flares (X-ray flares have been observed for short GRBs as well), which are believed to be associated with the prompt emission mechanism \citep[see e.g.][]{2007RSPTA.365.1213B,2010MNRAS.406.2113C}. 

Some LGRBs are observed to be associated with a supernova (SN) type Ic \citep{1998Natur.395..670G,2003Natur.423..847H,2003ApJ...591L..17S,2006ARA&A..44..507W}. The GRB and SN has been spectroscopically linked for an increasing number of bursts \citep[e.g.][]{2011MNRAS.413..669C,2013ApJ...776...98X,schulze14}, and a small 'bump' in the optical afterglow which matches expectations for the SN in colour, timing, and brightness has been observed for many GRBs \citep[e.g.][]{1999Natur.401..453B,2004ApJ...609..952Z}.

Long GRBs typically have a spectral peak energy of  $E_{p}\sim200$\,keV and an isotropic luminosity of $\sim10^{50}$\,erg\,s$^{-1}$. There has been several attempts to redefine the separation criteria between short and long GRBs, as the current criteria used are detector dependent, and there are high redshift bursts such as GRBs 080913 and 090423, which are observed to be long GRBs, but due to cosmological time dilation are short GRBs in the rest-frame.

\subsection{Ultra Long Gamma-Ray Bursts}
An additional class of GRBs has been proposed for long bursts with a duration of $\sim1000$--$10,000$\,s.  An ultra long GRB typically has a similar total energy to that of a normal burst. There are only 5--7 bursts observed with this length; GRBs\,091024A \citep{2013ApJ...778...54V}, 101225A \citep{2011Natur.480...72T}, 111209A \citep{2013ApJ...766...30G}, 121027A \citep{2014ApJ...781...13L}, 130925A \citep{2014MNRAS.444..250E,2014ApJ...790L..15P} and possibly GRBs 141121A \citep{2014GCN..17108...1G,2015arXiv151000996C} and 150126 \citep{2015GCN..17357...1G}. It is unclear whether these bursts are indeed a different class, and not just extreme versions of long GRBs. Emission in $\gamma$-rays lasting for $10,000$\,s is so extreme though, that these bursts at least are likely to have a different progenitor. This is vaguely supported by the afterglow (X-ray and optical), which in several cases does not fit the standard LGRB model, with especially the afterglow of GRB\,101225A showing distinct chromatic behaviour. Due to the low statistics it is too early to draw a conclusion on the origin of these bursts. The host galaxies of the ultra-long GRBs are all star forming galaxies at low redshift (as the peak fluxes are low, the low redshift observed is due to observational constraints), and the GRBs seem to coincide with the brightest regions of their host. There are a multitude of models proposed for the origin of ultra-long GRBs, including collapsars (see Section~\ref{sec:progenitor}), tidal disruption flares and the collapse of a giant/supergiant star. For a short review see e.g. \cite{2015arXiv150603960L}.

\subsection{X-Ray Flashes}
X-ray flashes is a subclass of long GRBs which are dominated by the X-ray counterpart with little, or no, $\gamma$-ray emission. Besides from the lack of $\gamma$-rays, they are observed to have very similar properties (e.g. distribution on the sky, duration and afterglows) to normal GRBs. At least one \citep[][]{2006Natur.442.1008C} X-ray flash has been observed to be associated with a SN. The abbreviation XRF is used as well as GRB for these bursts. One theory for the lower energies observed in XRFs is that they are standard GRBs, but we are viewing the burst off axis, and hence do not observe the beamed $\gamma$-rays \citep[e.g.][]{2004ApJ...609..962F}. An alternative theory is what is called the 'dirty fireball model' (see below for an explanation of the fireball model), where an excess of baryonic matter in the explosion is slowing down the fireball, preventing the boost of photons up into $\gamma$-ray energies \citep{2007A&A...466..839S}. \\
\\
\indent In what follows 'GRB' will refer to a long gamma-ray burst, unless otherwise specified.

\section{The Gamma-Ray Burst Model}\label{sect:model}
The physics behind GRBs is still a very debated topic, with several models existing. Any model though, must fit the general observational features. Particularly, it must solve the problem of compactness; as previously discussed, we know that the engine behind the burst must have a small radius, as we universally observe variations in the light curve on millisecond scale. On the other hand, we observe so high energies (typically isotropic equivalent to $\sim10^{52}$\,erg\,s$^{-1}$), that the energy density in the burst engine must be high enough that the photons can create electron-positron pairs, i.e. pair-production. But those particles being produces should make the engine optically thick to $\gamma$-rays, prohibiting their direct escape. What we observe should then be thermal photons from the surface. But we do not observe a thermal $\gamma$-ray spectrum, the prompt spectrum of GRBs is predominantly power-law shaped, see Chapter~\ref{chap:spec}.

The solution to the compactness problem is relativistic motion \citep[see][]{1992MNRAS.258P..41R,1994ApJ...430L..93R}. A particle moving at relativistic speed $v$ with a Lorentz factor of $\Gamma = 1/\sqrt{1-(v/c)}$ radiates spherically in its own rest frame, but for an observer the radiation is strongly beamed into a cone with half-opening angle $1/\Gamma$.

This means, firstly that the rest frame \emph{isotropic} equivalent energy is lower than inferred, as the observer sees radiation beamed into an angular region of $\sim\Gamma^{-1}$, and secondly that the observed photons are blueshifted, so in the rest frame the burst emission is in hard X-rays, rather than $\gamma$-rays, lowering the number of photons that can produce particle pairs. Together with the fact that we only observe $R/\Gamma$ of the engine (where $R$ is the radius), so the GRB engine in the rest-frame is less compact, this solves the compactness problem, if $\Gamma$ for a typical GRB is $\sim100$ \citep{2004RvMP...76.1143P}.

Below I will describe what can be called the 'standard model' of long GRBs, the fireball Model, and what is possibly wrong or missing from the picture.

\subsection{The Fireball Model}

\begin{figure}[h]
\centering
\includegraphics[width=0.95\textwidth]{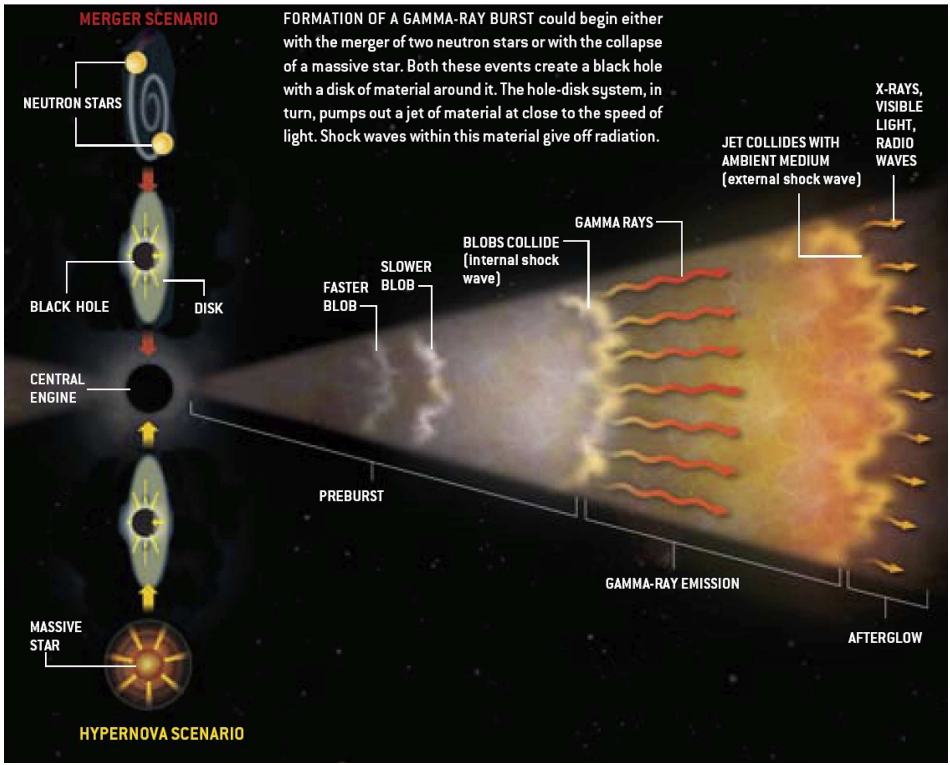}
    \caption{Illustration from \cite{2002SciAm.287f..52G} of the formation of a GRB.}
    \label{fig:fireball}
\end{figure}

In the fireball model an ultra-relativistic 'fireball' is emitted from an inner engine. This scenario is illustrated in Fig.~\ref{fig:fireball}. This energy flow is classically in the form of kinetic energy of particles, but is in some theories dominated by electromagnetic Poynting flux (energy of the magnetic field, see Section~\ref{sec:alt}). The energy flow is not constant, but rather made up of density shells. This leads to shock fronts in the flow, where particle number- and energy density jumps drastically. As these shells are travelling at slightly different velocities (close to the speed of light), they catch up with each other and interact, and the energy of the flow is dissipated. As the flow becomes optically thin, we can observe this dissipated energy in the form of $\gamma$-rays. This happens at a photospheric radius of 2$\Gamma^2ct\sim10^{11}$--$10^{13}$\,cm \citep{2005A&A...432..105R,2010ApJ...709L.172R}.

The $\gamma$-rays are radiated via synchrotron radiation (see Chapter~\ref{chap:spec} for more details on the radiation processes), which is the result of charged particles being accelerated by the magnetic field across the shock front. We also, in some bursts \citep[see e.g.][]{2013ApJ...776...95F}, observe a contribution from inverse Compton scattering, where photons scatter off the relativistic moving electrons so that energy is exchanged in favour of the photons, which gain $\gamma$-ray energies. The emission also include a low energy tail, from particles moving at an angle to the observer, which will have a smaller Lorentz factor in the frame of the observer, and hence a lower energy. This emission will be slightly delayed, again due to the lower $\Gamma$, and is observed as early X-ray emission (earlier than the X-ray afterglow, see below).

\begin{figure}
\centering
\includegraphics[width=0.45\textwidth]{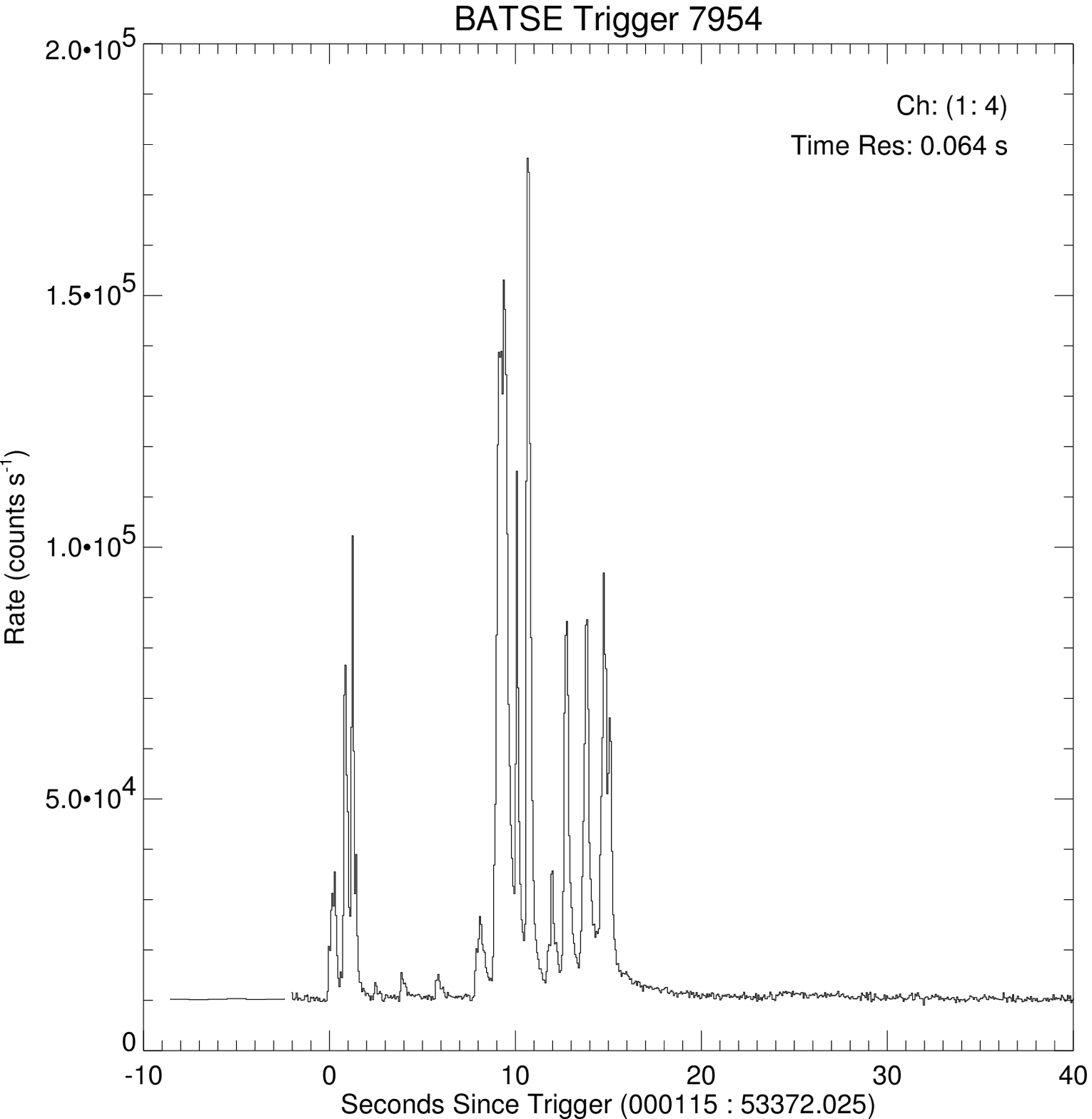}
\includegraphics[width=0.45\textwidth]{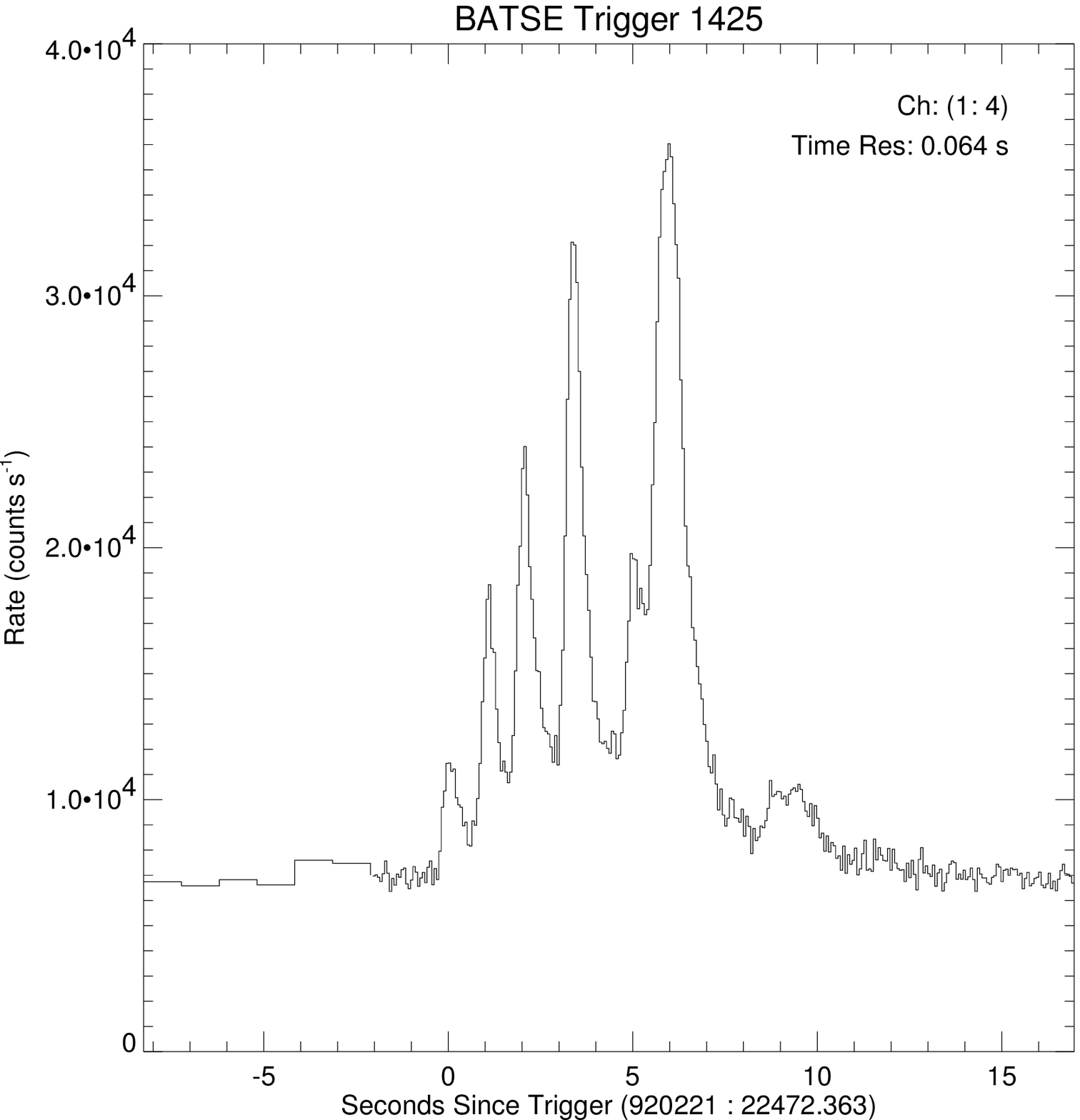}
\includegraphics[width=0.45\textwidth]{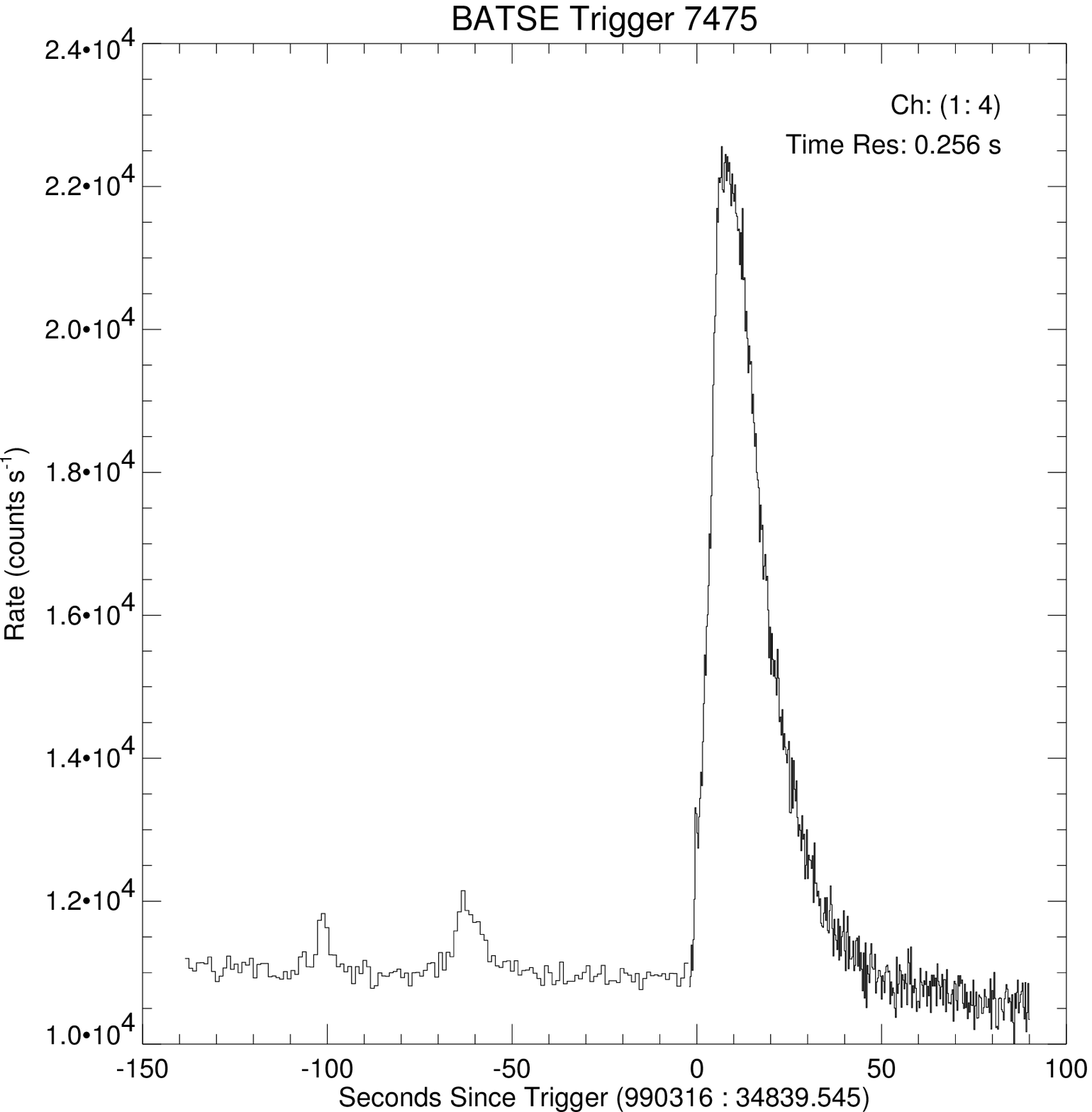}
\includegraphics[width=0.45\textwidth]{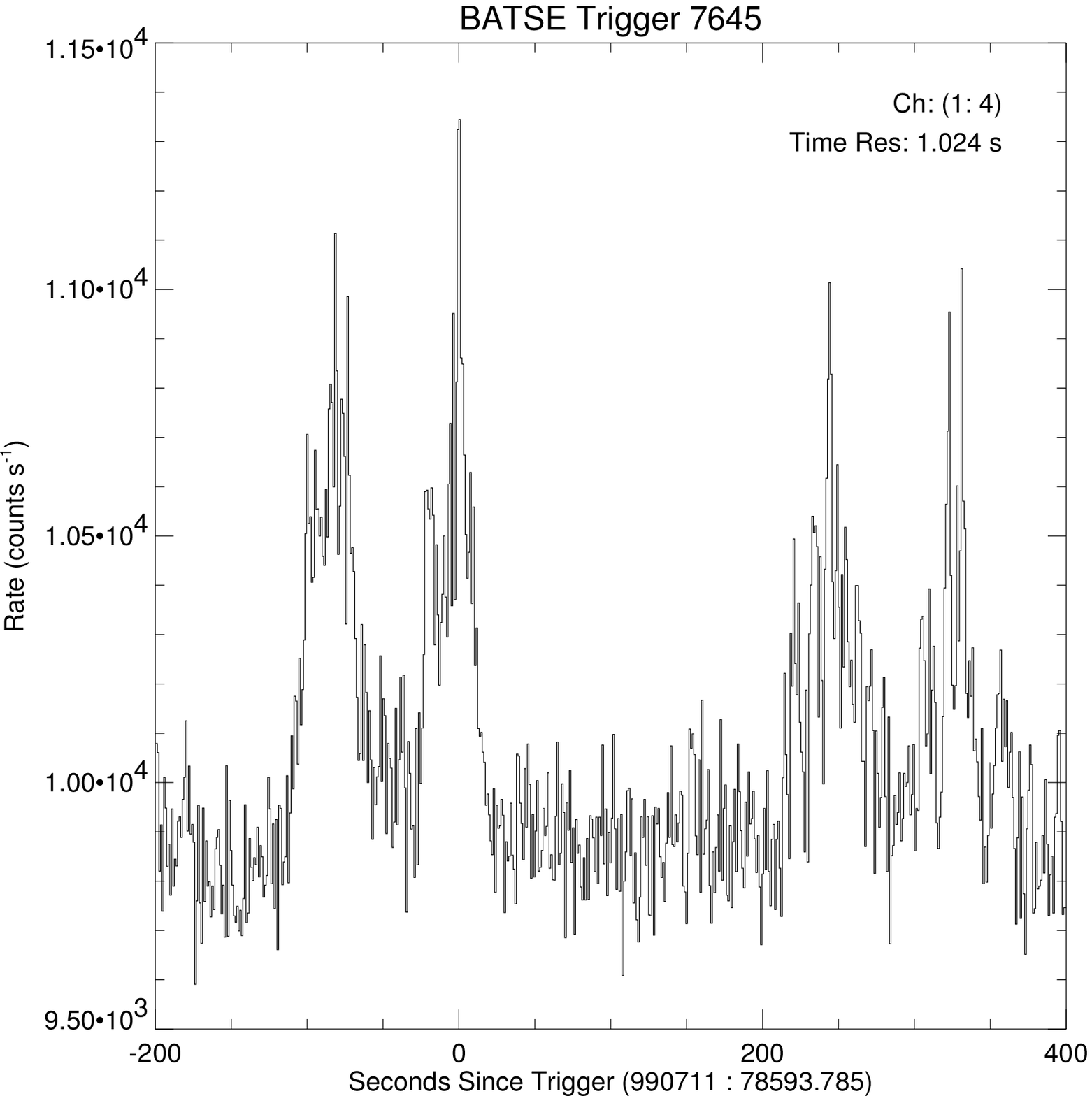}
    \caption{A sample of GRB light curves from the BATSE repository (from upper left corner: GRBs\,000115, 920221, 990316A, 990711).}
    \label{fig:lc_burst}
\end{figure}

The model of a multiple of internal (to the flow) shocks explains the large variety in observed number of peaks in the light curve of the GRB prompt emission, as seen in Fig.~\ref{fig:lc_burst}. One broad peak would be observed if the shells collide continuously compared to the time resolution of observations, while a longer time between individual or small groups of collisions would be observed as several thinner peaks (radiation from a given shell would always be observed with some broadening, as radiation from high-latitude reaches us at a later time).

The shocks themselves are only mildly relativistic with regards to the flow as a whole, so $\Gamma$ for the internal shock is of the order of a few. Assuming an adiabatic relativistic gas, it is possible to estimate the cooling frequency of the synchrotron spectrum, which indeed seems to agree with observations \citep{1999PhR...314..575P}.

Furthermore, both synchrotron and inverse Compton emission is polarised (though in a magnetic field with a low level of order, the polarisation can cancel out). Polarisation has probably been observed in GRB emission \citep[e.g.][]{2003Natur.423..415C,2013MNRAS.431.3550G}, supporting the fireball scenario. The observed polarisation is high, which requires a uniform, large-scale magnetic field over the entire emission region. This supports a theory of a GRB explosion driven by the magnetic field, and the formation of jets.

\subsection{The Jet Model}
In the basic fireball model, the energy flow is spherical in the rest frame. However, in most GRB progenitor models, the energy flow (kinetic or Poynting flux dominated) is driven in jets perpendicular to an accretion disc, as magnetic field lines fail to reconnect (see Section~\ref{sec:progenitor}). The beaming effect then entails that the emission is only observed by an observer sitting in the direction of the jet. 

Initially, it makes no difference whether we are observing down a jet, or a part of a spherical shell, as beaming restricts observations to an angle $\Gamma^{-1}$. But as the jet starts to decelerate, the Lorentz factor decreases to a point where the angular region we are observing from, becomes comparable to the opening angle of the jet, and we start observing the edge of the jet. This is called the jet-break, and has been observed to happen at $\sim1$\,day after the beginning of the burst, in observer's frame \citep[e.g.][]{1999ApJ...523L.121H,2006ApJ...637..889Z}. Observational evidence is in the form of a steepened decay of the light curve, because the radiating surface decreases in the observer's frame, and beaming effects decreases as the jet slows down and starts to expand sideways (note that these two effects may cause two breaks in the light curve, if the sideway expansion does not happen at the speed of light).

A consequence of the jet slowing down, is that suddenly off-axis observers can see the radiation, in the form of an 'orphan afterglow', that is, an afterglow without the observed prompt emission. An observation of such an afterglow would support the jet theory \citep[see e.g.][]{1997ApJ...487L...1R}, but so far none have convincingly been detected. This may just be due to the difficulty in searching for these, as there is no clear warning of when and where they will happen, and they may easily be confused with other transients \citep[though][report the detection of two GRB optical afterglows observed before an associated high-energy emission was found. In the second case, the source was subsequently observed in X-rays, but the first case could potentially be an orphan afterglow, though observations fit better with a pre jet-break decay of the afterglow]{2013ApJ...769..130C,2015ApJ...803L..24C}. 

If GRBs are indeed in the form of jets, then a large fraction of bursts going off in a direction pointing away from Earth, will never be observed. This indicates that the actual population of bursts is much larger, than would otherwise be inferred, a factor that needs to be accounted for before studying the statistical occurrence of GRBs.

\subsection{The Afterglow}\label{sec:afterglow}
In cases where the jet break has been observed, this is well after the prompt emission has faded, with the GRB light curve consisting of afterglow emission only. The afterglow is believed to be the consequence of an \emph{external} rather than internal shock. External shocks were originally suggested to be the cause of the prompt emission \citep[e.g.][]{1992MNRAS.258P..41R,1994ApJ...422..248K}, but it does not reproduce the light curve well, in particular it does not explain the multiple peaks often observed \citep[e.g.][]{1997ApJ...485..270S}. 

As the jet plows into the surrounding medium both a forward and a reverse shock is formed. The initial density contrast between the expanding shell and the interstellar medium (ISM) is large, and the energy conversion takes place mainly in the forward shock, which is responsible for the long-lasting bright GRB afterglows. The reverse shock is believed to produce prompt optical emission, known as the 'optical flash' \citep[e.g.][]{1999A&AS..138..537S}. As the shell expands, the density difference decreases and the energy converted into thermal energy in the two shocks becomes comparable. The peak emission from the early forward shock is in X-ray energies, while the reverse shock is traveling slower, with a smaller $\Gamma$, and hence the peak emission is at lower energies at optical/IR wavelengths. The afterglow radiation mechanism is due to similar physics as the prompt emission, and is predominantly synchrotron radiation, see Section~\ref{sec:agspec}. That the afterglow and prompt emission have different physical mechanisms (internal vs. external shocks) is supported by the fact that most observed afterglows are poorly back-interpolated to the prompt emission \citep[e.g.][]{1997ApJ...490..772K}.

\begin{figure}[h]
\centering
\includegraphics[width=0.98\textwidth]{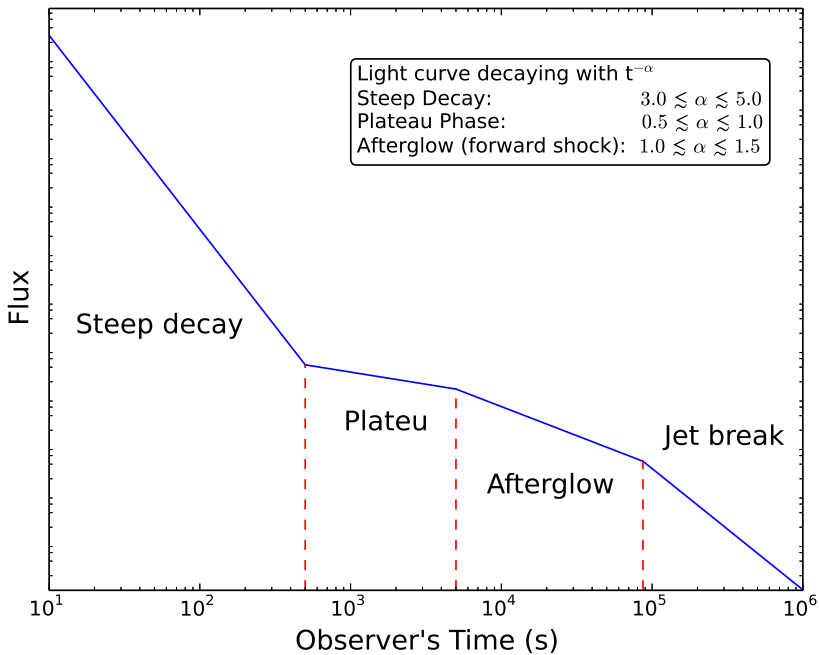}
\caption{Canonical light curve of the X-ray afterglow. Typical decay slopes are from the early \emph{Swift} sample, see \cite{2006ApJ...642..389N}.}
\label{fig:xray_lc}
\end{figure}

Observations of the X-ray afterglow have yielded a more complicated picture than can be explained purely by external shocks. The canonical X-ray light curve is made up of 4--5 segments with different decay slopes. Fig.~\ref{fig:xray_lc} displays the different segments using typical slopes from the \emph{Swift} sample \citep{2006ApJ...642..389N}.

The onset of the afterglow is observed as a smooth bump, expected to peak when about half of the fireball energy is transferred to the surrounding medium \citep{1999ApJ...520..641S}. Observations of the onset provides an opportunity to determine the Lorentz factor of the flow, as the Lorentz factor is proportional to the deceleration time, and the deceleration of the flow is signalled by the onset of the afterglow, see e.g. \cite{2010ApJ...725.2209L}. 

The early afterglow phase is characterised by a fast-decay corresponding to the high-latitude emission. This is part of the prompt emission, associated with the inner engine. This phase usually lasts about $10^2$--$10^3$\,s, and has a large evolution of spectral index $b$, $F_\nu\propto\nu^{-b}$, from hard to soft. The interpretation is supported by the fact that the emission joins smoothly with a back interpolation to the prompt emission, see e.g. \cite{2006NJPh....8..121O}.

After this, the light curve displays a very shallow decay phase. The decay rate is too small to fit with emission from the forward shock. The physical origin of this phase is somewhat debated. Some theories include (but are not limited to) energy injection from late-time activity of the inner engine \citep[e.g.][]{2014MNRAS.445.2414V}, a long-lived reverse shock \citep{2007MNRAS.381..732G,2007ApJ...665L..93U}, changes in the physical parameters of the surrounding ISM \citep{2006NCimB.121.1073G}, or an evolution of the microphysical parameters of the forward shock \citep{2006MNRAS.369.2059P}. There is no general consensus at this point. One observation that might support late-time engine activity, is that the light curve is often overlapped with X-ray flares \cite[e.g.][]{2007RSPTA.365.1213B}, as these are thought to be caused by the inner engine. However, the fluxes before and after a flare follows roughly the same power-law decay, which supports a scenario where the flares are superposed on the underlying power-law decay, so the two components likely have a different physical mechanism. The X-ray flares are observed to be most energetic at the earliest phases, but are observed at all times. They rise rapidly and decay again just as fast.

The next phase is what we refer to as the 'normal' X-ray afterglow. This is believed to be largely due to synchrotron emission from the forward shock interacting with the ISM. At this point, no more energy is injected into the flow by inner engine activity. As mentioned above, after about a day, some observations show a change to a steeper decay, caused by the edge of the jet becoming visible. This break is more often observed at optical wavelengths \citep[few observations of simultaneous X-ray and optical breaks exists, e.g. for GRB\,060614, see][]{2007A&A...470..105M}. Furthermore additional steepening is expected, simply because the external shock no longer has the energy required to accelerate electrons to radiate at X-ray energies.

Not all segments are observed for every bursts (in part due to observational constrains). Most commonly, the long lasting afterglow from the forward shock is observed. As the expanding shock-wave sweeps up more of the surrounding material, it slows down, resulting in a loss of energy. This naturally leads to a decay in peak energy. About 50\,\% of well-localised GRBs have an observed optical afterglow. The optical light curves have a greater variety than that of the X-rays. In a few cases an early optical flash has been observed, at times early enough to coincide with the GRB itself \citep[e.g.][]{2005Natur.435..178V,2006Natur.442..172V,2014Sci...343...38V}. One interpretation is that this flash is caused by the reverse shock. \cite{1999ApJ...520..641S} calculated that at its peak, the reverse shock has comparable energy to the GRB itself, but the temperature is significantly lower, so the wavelength of the emission would be longer, e.g. in the optical regime. The optical afterglow itself has, in some cases, been observed to show an initial increase in the light curve, as the synchrotron emission from the forward shock cools to optical frequencies \citep[e.g. for GRB\,091029,][]{2012A&A...546A.101F}. After its peak, some observations show a similar behaviour to the X-ray light curve, e.g. GRB\,050525A \citep{2006ApJ...637..901B} and GRB\,050801 \citep{2006ApJ...638L...5R}, while others show signs of chromatic breaks \citep[e.g. GRB\,060218, see][]{2006JCAP...09..013F}.

Radio afterglows are also frequently observed \citep[$\sim30$\,\% detection rate,][]{2012ApJ...746..156C}, typically at $\sim8$\,GHz. The radio emission peaks several days after the bursts itself \citep[the canonical radio light curve from the sample of][is found to peak three to six days after the burst itself]{2012ApJ...746..156C}, although earlier flares have been observed \citep{2015ApJ...806..179K}, believed to originate in the reverse shock. \cite{2012A&A...538A..44D} find a correlation between radio peak brightness and the X-ray brightness after 0.5 days. Unlike the optical afterglow, the radio afterglow is not affected by dust, which means that radio observations can be used to locate the host of dark bursts, such as GRB\,051022 \citep[see][]{2007A&A...475..101C}.

\subsection{Short Gamma-Ray Bursts}
Short GRBs are believed to have a different progenitor, but the radiative mechanisms for the two classes are thought to be similar, see Fig.~\ref{fig:fireball}, where the upper and lower formation scenarios lead to the same burst. \cite{2011MNRAS.418L.109G} find that they are indeed consistent with being the same by studying the prompt spectral evolution. This is also supported by the observations of jet-breaks \citep[e.g.][]{2006ApJ...653..468B,2006ApJ...653..462G}. 

This is not surprising since the progenitor models all share the physical condition leading to a relativistic outflow. Time scales and luminosities of individual flares are also similar, and the difference in spectral hardness could possibly be a selection effect \citep{2007PhR...442..166N}. The main difference then, is in the duration of inner engine activity. However, \cite{2009ApJ...701..824N} find that even the afterglows appear to have similar properties, only scaled with the fluence of the prompt emission, so that less-energetic bursts have a less energetic afterglow. It is perhaps surprising that there seem to be no other differences, as the ISM/stellar ejecta around the two types of bursts are thought to be different, see Section~\ref{sec:progenitor}.

\subsection{Alternative Models}\label{sec:alt}
The main 'branch' of alternative models to the fireball theory centres around magnetic dissipation. Rather than internal shocks in the jet, the radiation comes from the energy in the magnetic fields being dissipated due to magnetic hydrodynamic (MHD) instabilities. The milli-second variability in the light curve is then explained by relativistic magnetic turbulence and is not related to activity of the inner engine 

Evidence for these models include the lack of a photospheric emission component in at least some bursts \citep{2009ApJ...700L..65Z,2011ApJ...726...90Z}. Furthermore, \cite{2003astro.ph.12347L} argue that the high level of polarisation observed in some bursts rule out internal shocks. In the magnetic model, the GRB is still non-spherical, but emits over a large solid angle. This means that orphan afterglows are not expected, so an observation of this could rule out (at least some versions of) this model.

\section{Progenitor}\label{sec:progenitor}
We now turn towards the inner engine and the question of what creates the physical conditions necessary to set off the fireball. Besides explaining the emission, the model for the burst progenitor must also fit with observations of the typical GRB birthplace. We will first look at the 'classical' model for long GRBs, discuss potential issues and alternative models, and then end with the short GRB progenitor model.



\subsection{The Collapsar Model}\label{sec:collapsar}
Motivated by the large energies required in combination with the compactness of the source, and in line with later observations of GRBs in star-forming regions and the now observed association with SNe, \cite{1993ApJ...405..273W} suggested what is called the collapsar model for long GRBs. This, and all other models suggested for GRBs, starts off from the realisation that the only phenomenum we know of that can release the vast amount of energy needed, is a gravitational collapse into a compact object. 

In the collapsar model, the fireball is the result of the collapse of a fast rotating massive star into a black hole (BH). As matter falls onto the newly formed BH, an accretion disc is formed, due to the rotation, as matter along the rotational equator of the former star has too much angular momentum to fall straight onto the BH. The energy of the in-falling matter (potential energy converted into kinetic energy, or heat) has only one place to escape, which is along the rotation axis, where the stellar material has less rotational support. This leads to the creation of jets going out perpendicular to the rotating disc.

The jet first has to break through the envelope of the dying star. As it travels outwards, the density decreases, so the jet encounters less resistance. This leads to an acceleration, and by the time the jet breaks out of the stellar envelope, it has gained relativistic velocity. Accretion is not a steady process, but happens in parts, as clumps of matter falls onto the BH. This means that the energy deposited into the jet varies, leading to the different shells with varying Lorentz factors theorised in the fireball model. As the jet breaks out of the star a thermal pre-cursor to the burst is observed in some cases \citep[e.g.][]{2008ApJ...685L..19B}.

The progenitor star for a long GRB must satisfy a number of conditions to lead to the collapsar scenario. Firstly it must have a mass $M>30$\,M$_{\odot}$, to be able to form a central black hole \citep[though there are channels through which the mass could be lowered, see e.g.][]{2006ARA&A..44..507W,2006ApJ...637..914W}. Secondly it has to be rapidly rotating to have angular momentum enough to form an accretion disc and launch jets. Lastly, in order for the jet to be able to break through the star with enough energy left to power a GRB, the star must first have been stripped of its hydrogen/helium envelope, probably through stellar-winds. This is also supported by the fact the GRB-SNe are observed to be Type Ic \citep[e.g.][]{2011MNRAS.413..669C}, as these SNe have no hydrogen lines in their spectra, and are believed to have shed the outer hydrogen envelope. These conditions led to the suggestion that the progenitor star must be a Wolf-Rayet star, as these stars have gone through huge mass losses before collapsing \citep[e.g.][]{2006ARA&A..44..507W,2007ARA&A..45..177C}.

\subsection{Potential Problems}
The largest problem for the collapsar model lies in modelling a stellar progenitor with an angular momentum large enough to form the accretion disc needed to launch the jets. Theoretically, in order to retain a large angular momentum, the progenitor star cannot have too large a metallicity \citep[e.g.][]{2005A&A...443..643Y}. However, the progenitor is also expected to have gone through mass-loss, to get rid of the hydrogen envelope, in order for the jet to break out, and to coincide with the observed type of SNe (which show no hydrogen lines). This is problematic, as models of stellar mass-loss show a metal dependency \citep[e.g.][]{2001A&A...369..574V}, ruling out the low value needed for the angular momentum. As a solution, \cite{2005A&A...443..643Y} suggest that rapid rotation may induce chemically homogeneous evolution in the progenitor star. In this case, no hydrogen envelope is formed, so no mass-loss is needed, and the progenitor can have a low metallicity. However, near- or super-solar metallicity has been observed for some environments of long GRBs \citep[e.g. GRB\,130925A, see][]{2015A&A...579A.126S}. \cite{2015arXiv150402479P} examined whether there is a metallicity threshold for \emph{Swift} bursts. They did indeed find a sharp threshold, but this is near-solar. This is incompatible with all single-star progenitor models in the collapsar scheme, unless the progenitor somehow has a low metallicity in the middle of a high-metallicity environment. 


A binary origin of the collapsing star could potentially supply the angular momentum through tidal interactions spinning up the GRB progenitor, \citep[e.g.][]{1998ApJ...502L...9F}. This is supported by the fact that most massive stars are born in a binary system \citep{2012ASPC..465..363S}.  

Another problem is the so called ultra-long GRBs. Bursts such as GRB\,101225A \citep{2011Natur.480...72T,2014ApJ...781...13L} and GRB\,130925A \citep{2015A&A...579A.126S} have a duration much longer than what can be accounted for with a Wolf-Rayet like star as a progenitor. Instead, \cite{2011Natur.480...72T} suggest a merger between a helium- and a neutron star, where the jet then had to break through a common envelope of the pair, accounting for the unusual long burst. Alternative models are disruption events or magnetars \citep[e.g.][]{2014ApJ...781...13L}.

It has been suggested that rather than a BH being formed in the core straight away, the central object in GRBs is a millisecond magnetar (a neutron star with a very strong magnetic field), spinning down \citep[e.g.][]{2009MNRAS.396.2038B}. This can provide a longer lasting energy injection than accretion, and requires less angular momentum. Furthermore, the SN associated with the ultra-long GRB\,111209A has a high luminosity but low metal-line opacity, which fits best with a magnetar as the progenitor model, to avoid the blanketing expected from the high nickel mass in other SN models \citep{2012MNRAS.426L..76D,2015Natur.523..189G}. 

A possible solution is a mix of progenitors, in order to explain the large diversity of LGRBs.

\subsection{Short Gamma-Ray Bursts}
Short GRBs are not observed to be associated with the star-forming regions of their host galaxy \citep{2014ARA&A..52...43B}. We also do not observe any SN associated with the bursts, in fact in several cases, an SN connection is rules out \citep[see e.g.][]{2005Natur.437..845F,2005Natur.437..859H,2010MNRAS.408..383R}. This makes a massive (young) star a more unlikely progenitor for these bursts. 

The most commonly assumed model for short bursts, is the merger scenario, were two compact object, either a pair of NS, or a NS-BH pair, starts spiralling inwards due to energy-loss through gravitational radiation. In the NS-NS model, the two NSs merge and a BH is formed surrounded by an accretion disc, leading to a similar scenario as for the collapsar model, but on shorter time-scales. In the NS-BH model, the NS is tidally disrupted at the event horizon of the BH, which then leads to the formation of an accretion disc.

A way to test this theory is to look for short GRBs with a notable off-set from any possible host galaxy. The SN explosions leading to the formation of the progenitor pair could have resulted in a kick imparted to the system, sending it away from the birth sight \citep[see e.g.][finding no evidence for large off-sets]{2006ApJ...648.1110B}. Another prediction of this model is that short GRBs should be accompanied by a strong gravitational wave signal, as the compact objects spiral towards each other. Gravitational wave detectors such as Advanced LIGO \citep{2010CQGra..27h4006H} should be able to detect this signal in the future, for near GRBs \citep{2012ApJ...760...12A}. 

The resulting system after a merger will naturally have a high angular momentum, so rotation is not a problem in this model. However, the observation of X-ray flares pose a problem, as they are evidence of prolonged inner engine activity \citep[e.g.][]{2006A&A...454..113C,2011MNRAS.417.2144M}, which cannot be the case for a merger, where the accretion disc exists for only a very short time. The same is true of the observed extended emission \citep[e.g.][]{2006ApJ...643..266N}. A possible solution to this problem is if some mergers leads to a magnetar first, which then powers the short GRB \citep[e.g.][]{2015ApJ...805...89L}. 

\section{Host Galaxies}\label{sec:hosts}
The study of the environment and host galaxies of GRBs relies heavily on the detection of, first the X-ray afterglow to provide a more precise location on the sky, and then ground based telescopes to follow up with optical photometry and spectroscopy to determine a redshift.

The first host studies found that a typical GRB host have bluer colours and are fainter than the field galaxies at a comparable redshift \citep[e.g.][]{2003A&A...400..499L}, suggesting a young galaxy population. It was recognised from the beginning though, that the observations might be biased towards unobscured (and hence bluer) galaxies due to the dependance on optical afterglow detections. More recent studies have also showed, that with a more complete sample, GRB hosts are shown to span a wider range of properties. Host studies such as The Optically Unbiased GRB Host (TOUGH) sample \citep{2012ApJ...756..187H}, have defined samples such that observational biases, as for instance dust extinction, are non-existing, or at least minimised and understood. 

\begin{figure}
\includegraphics[width=1.0\textwidth]{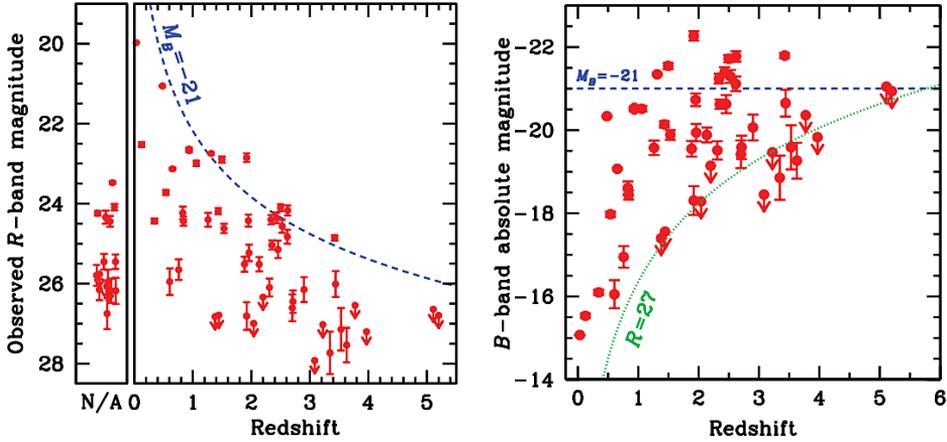}
\caption{Figure taken from \cite{2012IAUS..279..187J}. The left panel shows the R-band host magnitude as a function of redshift for all the bursts in the TOUGH sample. Upper limits are shown with arrows. Hosts without a reported redshift are plotted on the left side of the diagram. The dashed curve shows a galaxy with an absolute B-band magnitude of $-21$. The right panel shows the calculated absolute B-band magnitude (assuming $F_v\propto v^{-0.5}$), including only those hosts that have a detected redshift. The dotted curve shows a galaxy with an observed magnitude of R\,$=27$.}
\label{fig:mag}
\end{figure}

\begin{figure}
\includegraphics[width=1.0\textwidth]{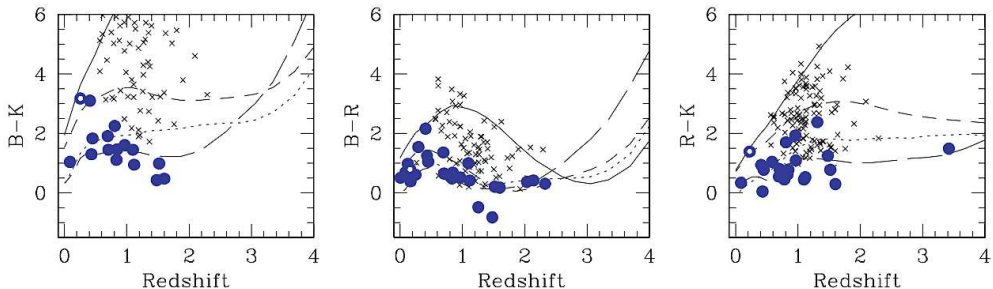}
\caption{Figure taken from \cite{2009ApJ...691..182S}. From left to right, B -- K, B -- R, and R -- K apparent colours (AB system) as a function of redshift for the GRB hosts (filled circles) and GDDS field galaxies (crosses). The curves are predicted colours of galaxies assuming different stellar populations: E (solid line), Sbc (short-dashed line), irregular (dotted), and starburst (long-dashed line).}
\label{fig:colour}
\end{figure}

Figure~\ref{fig:mag} shows host magnitude as a function of redshift for all the bursts in the TOUGH sample. The average GRB host is still seen to be relatively faint, with a median apparent magnitude R\,$\approx25$ (largely the same value that was found for earlier studies), with a distribution similar to the faint end of normal field galaxies. Similarly the average host is indeed blue, as seen in Figure~\ref{fig:colour} which shows a comparison to Gemini Deep Deep Survey (GDDS) field galaxies, as well as the predicted colours assuming different stellar populations (E/S0: pure elliptical/intermediate elliptical -- spiral galaxy, Sbc: spiral galaxy, irregular, and starburst). The figure indicates that GRB hosts are generally blue star-forming galaxies, also fitting well with irregular galaxies. While a few redder objects are observed, both \cite{2009ApJ...691..182S} and \cite{2012ApJ...756..187H} find clear evidence that GRBs are rarely formed in spiral galaxies. At low redshift, where we acquire imaging of the host morphology, we do often observe LGRBs in small irregular galaxies \citep[e.g.][]{2006A&A...447..891F,2007ApJ...657..367W,thomas2}

The dust properties of GRB hosts, or at least birth-regions within the hosts, are discussed in Chapter~\ref{chap:ext}.

\subsection{Galactic Environment}
The environment in which we find GRB host galaxies has not been studied much to date. \cite{2002A&A...388..425F} studied the fields of GRBs\,000301C and 000926 in Ly$\alpha$ emission. Similarly \cite{2005MNRAS.362..245J} report on the fields of GRB\,030226, 021004 and 020124. Several galaxies were observed in the fields, but it is difficult to form conclusions on whether the galaxy densities in these fields are particularly high, as no blank field studies have been carried out at similar depth and redshifts. \cite{2005MNRAS.362..245J} find a mean density of Ly$\alpha$ emitters in the field of the GRBs, similar to that in the field of the radio-galaxy PKS 1138--262. GRB\,000926 is shown to be located in a strong star-formation centre, while GRB\,000301C is in a faint galaxy far from any star-formation centre in its galactic environment. \cite{2003A&A...400..127G} derived photometric redshifts for all galaxies near the host of GRB\,000210, finding no near companion galaxy. Similarly \cite{2004ApJ...614...84B} searched the field of 6 GRBs, finding no indication of over-densities. 

Simulations of LGRB host galaxies, using the Millennium Simulation \citep{2005Natur.435..629S}, show that the hosts preferentially map different density environments at different redshifts, and that, at high redshifts, hosts are predicted to be in similar environments as the overall galaxy population, but have a slightly higher probability of a close companion \citep{2010MNRAS.408..647C}. The statistics are not good enough yet, that we can make any conclusions on the GRB host galaxy environment, but studies of the morphology of LGRB hosts show a high fraction of merging and interacting systems \citep{2007ApJ...657..367W}, so we might expect LGRB hosts to be found in high-density areas. Not much work has been done on short GRB host environments, but \cite{2006ApJ...642..989P} looked at the fields for short GRB host galaxies, finding that at least two out of four hosts reside in galaxy clusters. 

\subsection{Methods for Determining Redshift}
As mentioned above, ground-based optical telescope follow-up is needed to determine a redshift for the host galaxy (as well as a wealth of additional information, see Chapter~\ref{chapter:host}). I will now briefly describe the different methods in use to determine $z$.

\subsubsection{Line-of-sight absorption}\label{sec:redshift}
\begin{figure}
\centering
\includegraphics[width=0.9\textwidth]{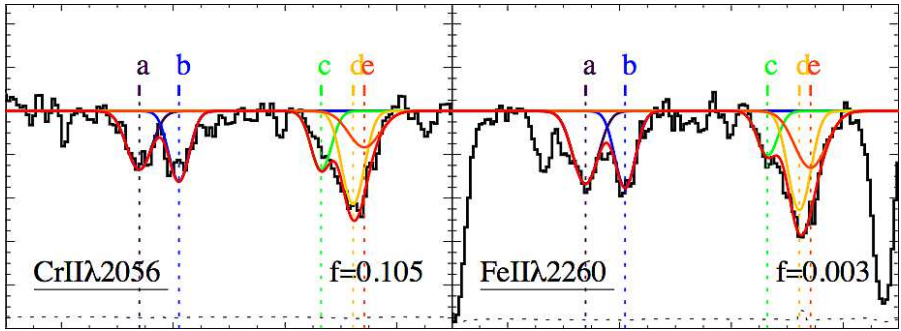}
\caption{The figure displays absorption lines from Fe and Cr towards the line of sight to GRB\,121024A using data from VLT/X-shooter, see Section~\ref{sec:xshooter}. The letters a-e mark the different absorption components. 'f' gives the oscillator strength for the displayed line, see Section~\ref{abs}. This figure is a small cut-out of Figure~\ref{fig:absorption}. The horizontal axis displays wavelength or velocity, while the vertical displays normalised flux.}
\label{fig:lines}
\end{figure}

The light from the optical afterglow passes through clouds of interstellar and intergalactic gas between the GRB and the observer. The different elements comprising the gas in the cloud (whether in neutral or ionised form) will absorb light at characteristic wavelengths. Very often it is possible to identify the redshift of these absorption lines by recognising doublets, and knowing which elements typically to expect. At $z>2$ the start of the Lyman-$\alpha$ forest will give the redshift of the GRB. At lower redshifts, if fine-structure lines are observed, we can associate the absorption system with the GRB, as the lines are likely excited by the GRB itself \citep[e.g.][]{2007A&A...468...83V}. Even if fine-structure lines are not observed, the GRB is expected to be found in molecular clouds in the hosts, so there is a large probability that the absorber system observed at the highest $z$ is from the host galaxy (gas clouds may have motion along the observed line of sight with respect to their host galaxy, resulting in a slightly different redshift than the GRB). Figure~\ref{fig:lines} shows a small cut-out of two such absorption lines from Cr and Fe towards the line of sight to GRB\,121024A (Chapter~\ref{sec:paper_121024A}). The letters a-e mark the different components corresponding to absorption from individual clouds. In reality the absorption is likely divided into several smaller components that we cannot resolve, so we have to give the integrated redshift (and column densities).

\subsubsection{Nebular-line emission}
Besides from line-of-sight absorption, it is sometimes possible to observe nebular lines in emission from the host galaxy. To observe these the GRB needs to be at low redshift, or be in a highly star-forming galaxy (as the nebular line flux is related to the star-formation rate, see Section~\ref{sec:starfr}). An advantage over absorption line analysis, is that we do not need to catch the GRB afterglow, but can use a host galaxy spectrum, as long as we are confident that we have identified the correct galaxy. Some of the lines typically observed are the Balmer lines H$\alpha$ and H$\beta$ and the oxygen lines [\mbox{O\,{\sc iii}}]  $\lambda$$\lambda$4959, 5007, as well as the [\mbox{O\,{\sc ii}}]  $\lambda$$\lambda$3726,3729 doublet. Since we always observe the same handful of nebular lines for GRB hosts, we only need two emission lines in our spectrum to correctly identify the lines from the relative position and strength, and hence determine the redshift. Even if we only observe one emission line we can make qualified guesses for $z$ (based on the non-detection of other lines), that then might be determined from e.g. fitting the stellar population, see Section~\ref{sec:spsm}. 

\subsubsection{Photometric redshift}\label{sec:balmer}
\begin{figure}
\centering
\includegraphics[width=0.8\textwidth]{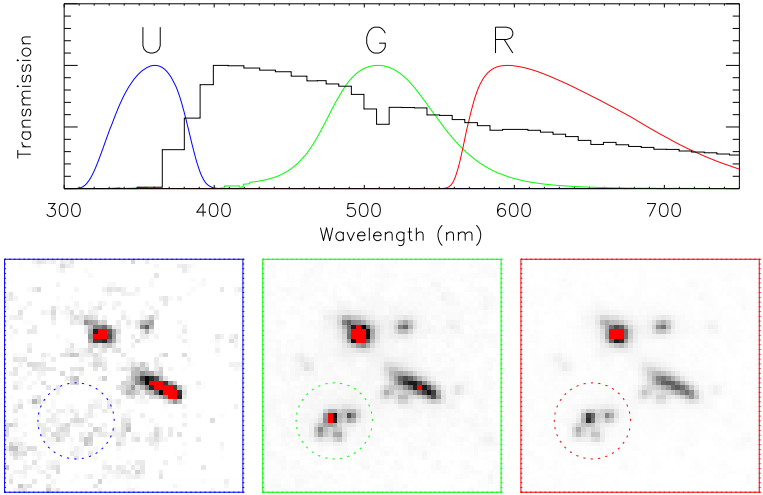}
\caption{Illustration of the Lyman-break technique (courtesy of Johan Fynbo). The lower panel shows Hubble images of a Lyman Break Galaxy in three photometric filters G, R and U, while the upper shows the wavelengths covered by these filters over the galaxy spectrum.}
\label{fig:lyman}
\end{figure}

In the cases where either no afterglow/host spectrum is taken, or the conditions did not allow for us to identify lines in the spectrum, it is still possible to determine the redshift if photometric measurements exist. This relies on features which are strong enough to be apparent in the crude wavelength bins of the photometric filters. One such feature is the Lyman limit. The Lyman limit is the high-energy end of the hydrogen Lyman series at 912\,\AA. At higher energies, photons can ionise hydrogen, and hence almost no light is observed below this wavelength, as neutral hydrogen along the line-of-sight absorbs it all. By comparing the magnitudes in the different imaging filters, we can locate the approximate position of the Lyman limit and hence determine a redshift, see Figure~\ref{fig:lyman} for an example of a Lyman Break Galaxy (LBG) detected in the photometric filters G and R, but not in the U-filter, due to the presence of the break. LBGs are high-redshift starburst galaxies, which are selected by having a strong Lyman break. This detection technique, is one of the more efficient ways to find galaxies at $z>2.5$ \citep{1995AJ....110.2519S,1996MNRAS.283.1388M,2001ApJ...554..981P}.

Another feature that is strong enough to be located through photometry is the Balmer break, analogous to the Lyman break, but for the Balmer series. The Balmer series is lines from the electron transitions from the principal quantum number n\,$\geq3$ to n\,$=2$, i.e. from a higher to the first ionised state. The Balmer break is located at 3650\,\AA.
 
\section{Prospects for Astronomy}

\subsection{Tracers of Star-Formation}\label{sec:tracesfr}
As already mentioned, long GRBs are observed in the star-forming regions of their host galaxy, and it has been suggested that they can be used to trace the cosmic star-formation (SF) history, as they are observed out to high redshifts. Unlike UV surveys, which are otherwise the most used probes at high-$z$, GRBs are, in principle, not affected by dust extinction, and can therefore locate star-forming galaxies that are otherwise too obscured to be detected (though the redshift determination often relies on the optical afterglow, which is obscured by dust), or which are at a sub-threshold UV luminosity.
One potential problem is a bias in metallicity. As noted in Sect~\ref{sec:collapsar}, the collapsar model predicts a cut-off in metallicity, due to the needs of a high angular momentum. There is a lesser restraint on the metallicity in the binary progenitor model, but observations do seem to indicate a bias \citep[e.g.][]{2015ApJ...801..102P}. 

Another problem is the comparison to core-collapse (CC) SNe. \cite{2006Natur.441..463F} reported that CC SNe and GRBs are found in different environments, with GRBs tracing the more brighter regions of their hosts. As described in Section~\ref{sec:hosts}, GRBs at low $z$ (close enough so that we may study the host) are often found in small irregular galaxies. This is not the case for CC SNe. On the other hand, \cite{2008ApJ...687.1201K} conducted a similar study comparing GRB environments with that of type Ic SNe, finding that these SNe \emph{does} seem to be located in similar regions with regards to the host luminosity distribution, while \cite{2008AJ....135.1136M} find that they are found in significantly more metal rich environments than GRBs, so it is far from a simple matter to determine the properties that sets the progenitor stars of the different types of explosions apart. 

\begin{figure}
\centering
\includegraphics[width=0.8\textwidth]{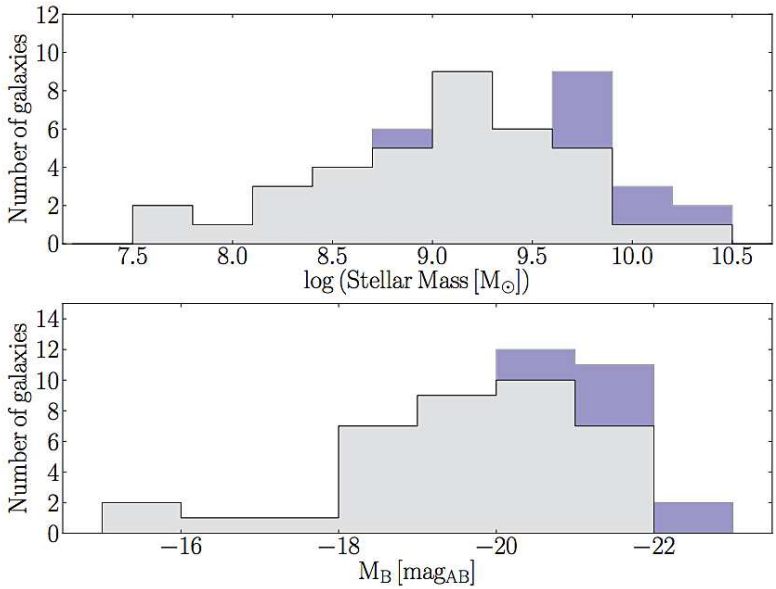}
    \caption{Figure taken from \cite{2011A&A...534A.108K}. Distribution of stellar masses and luminosities of the hosts of extinguished afterglows with $A_\text{V}\geq1$ (blue) and the host sample from \cite{2009ApJ...691..182S} (grey).}
    \label{fig:dustier}
\end{figure}

It has been suggested that the observed metallicity dependence is, at least in part, due to a bias against dustier host galaxies (as dust needs metals to form, high metallicity environments are assumed, and often observed, to be dusty). Figure~\ref{fig:dustier} illustrates the consequence of a dust bias. It shows host stellar mass and magnitude for the sample of \cite{2011A&A...534A.108K}, which is specifically chosen to have $A_\text{V}\geq1$, compared to the \cite{2009ApJ...691..182S} sample shown in Figure~\ref{fig:colour}. \cite{2011A&A...534A.108K} examined the possible observational differences between the two samples, such as a difference in median redshift, but concluded that their sample very likely probe a more luminous, massive, and chemically evolved population of GRB hosts, which hence might be underrepresented in other samples if these are generally more affected by extinction.

\begin{figure}
\includegraphics[width=0.65\textwidth]{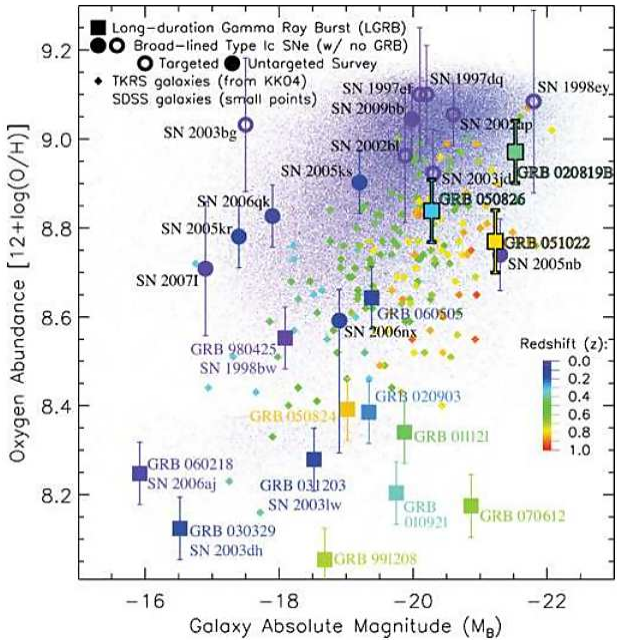}
  \begin{minipage}[b]{0.3\textwidth}
    \caption{Figure taken from \cite{2012IAUS..279..232G}. It shows galaxy central metallicity with the redshift indicated by colour. The figure shows host metallicity for local Type Ic SNe, compared to that of long GRB hosts. The small purple dots show SDSS galaxies at $z\sim0$. Note that the three high metallicity GRBs (highlighted with black edges) are at typical metallicity for galaxies of their luminosity and redshift. This is not consistent with the otherwise observed GRB metal aversion.}
    \label{fig:bias}
  \end{minipage}
\end{figure}

However, \cite{2012IAUS..279..232G} showed that while local Type Ic and Type II SNe track the star-formation weighted metallicity distribution of the Sloan Digital Sky Survey (SDSS) galaxies, GRBs are typically found at lower metallicities, as illustrated in Figure~\ref{fig:bias}. The figure shows the central metallicity of the hosts, rather than at the location of the burst/SN, in order to compare with field galaxies. They included dark bursts, and hence concluded that the metallicity bias is real. More recently, \cite{2013ApJ...772...42H} found no strong bias toward low metallicity in LGRB host galaxies, with a metallicity cut at Z\,$\geq0.6$Z$_{\odot}$. The overall conclusion from all these studies, must be that there is a low metallicity preference for GRBs, but the metallicity threshold cannot be much below solar. 

This restriction on the threshold means that while GRBs perhaps are poor tracers of star formation in the local Universe, the metallicity cut-off is unlikely to be a problem when using GRBs to trace the SF history at high redshift, as all galaxies were this poor in metal before a certain $z$ ($z\gtrsim2$--$3$). The difficulty then lies in better constraining this redshift, as direct metallicity measurements are generally difficult outside the local Universe. 

The SF history of the Universe is a very debated topic. It is generally agreed upon that the star-formation rate density (SFRD) increased with time up until a redshift of $z\sim2$, after which it has been in decline, but the precise slopes and turnover point is not clear (see also Section~\ref{sec:wrong} for a discussion of the subject). Compared with field-surveys of galaxies in UV or IR, GRB SFRDs appears to be higher at early times, as well as peaking earlier \citep[][]{2008ApJ...673L.119K,2011MNRAS.417.3025V,2012ApJ...744...95R,2012ApJ...752...62J}. It is possible that the field surveys systematically underestimate contributions from low-mass and high-$z$ galaxies, as these are below the observational threshold. Alternatively, this discrepancy is due to the metallicity bias.

\subsection{Probing the Epoch of Re-ionisation}
It has been suggested that GRBs may be used to probe the epoch of re-ionisation. This is the period in the Universe' history where the light from the very first stars (and possibly QSOs) began to ionise the surrounding neutral matter (mainly hydrogen) changing the Universe from being predominantly dark and neutral, to full of stars and galaxies and ionised matter (the Universe was previously entirely ionised, with all baryonic matter in the form of protons and electrons, until recombination when the Universe had cooled to form neutral gas). Since this is the period of the very first stars and galaxies, it is important to study to understand the evolution of the Universe. It is also a difficult period to study as the look-back time, and hence distance is so great (beginning as early as 500 million years after the Big Bang), that the electromagnetic radiation is extremely faint by the time it reaches us.

Simulations of the first stars show that these were likely massive, and inherently metal poor \citep[the formation process of these stars is still a debated topic though, see e.g.][]{2000ApJ...534..809O,2002Sci...295...93A,2003Natur.425..812B}. Since these are likely progenitors for (long) GRBs, it is reasonable to assume that the rate of GRBs was higher in the early Universe. The detection alone of GRBs at the highest redshift would help constrain the onset of star formation ($z\sim10$). \cite{2000ApJ...536....1L} calculated that \emph{Swift} would be able to detect the brightest GRBs out to $z\sim70$, which is higher than any model would predict that they are likely to occur. Furthermore, it should (and has proved to) be possible for us to observe the bright optical/near infrared (NIR) afterglow for very high-$z$ GRBs. Even though the flux should be lower for observations at higher $z$, the effect of time-dilation will counteract this. The afterglows flux decrease with time, so for a given observed time $\Delta t$, a larger time-dilation will mean that more photons are observed in $\Delta t$, and since $F=\frac{\text{number of photons}}{\Delta t \times \text{area}}$, the flux will increase \citep[see e.g.][]{2004A&A...427...87G}. Combined with the fact that at higher $z$ the distance-increase with $\Delta z$ becomes lower, there will be very little total decrease of the observed flux at high-$z$ for a typical GRB. 

Much of this afterglow light will be obscured underway by the Ly$\alpha$ forest. The forest refers to the multitude of Ly$\alpha$ absorption lines from neutral hydrogen in Lyman $\alpha$ clouds between galaxies. The strength of this will enable us to constrain the neutral hydrogen fraction of the intergalactic medium (IGM) at the burst's redshift. Until the end of re-ionisation, the fraction was so high, that the Universe was opaque to photons with energies high enough to ionise hydrogen, and the Ly$\alpha$ forest part of the spectrum is observed as completely absorbed, a feature that is known as the Gunn--Peterson trough. However given the larger densities of the early Universe, even a very small fraction of neutral hydrogen would result in complete absorption of the flux in the Ly$\alpha$ forest region, so this detection only provides a lower limit on the redshift of re-ionisation.

The Gunn--Peterson trough has been detected in a QSO at $z=6.28$. \cite{2001AJ....122.2850B} reported the observation of the spectrum along with spectra for two QSOs at $z$ just under $6$, where the trough is not observed. This is generally accepted as an indication that re-ionisation ended close to $z=6$. Other probes of the re-ionisation era include observations of the cosmic microwave background (CMB), which can be used to calculate the optical depth of electron-scattering through polarisation and temperature anisotropies. WMAP results indicated a high value, corresponding to an early onset of re-ionisation \citep{2003ApJS..148..161K,2003ApJ...596....9H}, but new results from Planck has lowered this value considerably, so that the current redshift of \emph{instantaneous} re-ionisation is $z\sim8.8$ \citep{2015arXiv150201589P}. 

\section{Gamma-Ray Bursts Observatories}\label{sec:telescopes}
As previously mentioned, GRBs were discovered by the Vela satellites, but the instrument that really kickstarted GRB science, was the Burst and Transient Source Experiment (BATSE) onboard the Compton Gamma Ray Observatory \citep[CGRO,][]{1994ApJS...92..351G}. CGRO was launched in 1991, and decommissioned in 2000. The main purpose of BATSE was to monitor the sky for GRBs (it was the first dedicated GRB instrument), as well as longer lived sources, in the energy range $\sim20$\,keV--$2$\,MeV. It had eight independent detectors, one at each corner of the satellite, providing a full cover of the sky, except the part blocked by Earth. 

The biggest results to come out of the first years of BATSE observations were that the bursts are located uniformly in the sky and were clearly divided into two classes, long and short GRBs \citep{1993ApJ...413L.101K}. BATSE also discovered the minute variations of the GRB light curves, being the first instrument to have such a good temporal resolution \citep{1998ApJ...496..849P}.

The next big step in GRB observations was made by the satellite BeppoSAX, which first detected the X-ray afterglows, providing better spacial resolution, allowing the redshifts of GRBs to be detected from ground-based follow-up observations, proving the extragalactic origin of the bursts. BeppoSAX was launched in 1996, and the mission ended in 2002. The energy range was $0.1$--$300$\,keV, and it was the first instrument able to detect a GRB with a high energy telescope, and then slew a lower energy telescope quickly to the burst position providing afterglow observations and a tighter constraint on the GRB location. Another telescope worth mentioning is The High Energy Transient Explorer 2 (HETE-2, the first HETE satellite was lost during launch), launched in 2000. HETE observed the first short GRB with an optical afterglow, and located GRB\,030329 which was the first cosmological burst to have a spectroscopically confirmed SN \citep{2003ApJ...591L..17S,2003Natur.423..847H}. 

The current GRB science is being led by two satellites, the \emph{Swift} and  \emph{Fermi}  missions. The two have different capabilities and compliment each other well. \emph{Swift} was launched in 2004, and is comprised of three telescopes; the Burst Alert Telescope (BAT) monitoring the sky in hard X-rays ($15$--$150$\,keV), detecting about 100 bursts per year, the X-Ray Telescope (XRT), following up on BAT triggers, observing in the energy range $0.2$--$10$\,keV, and the UltraViolet/Optical Telescope (UVOT) observing at $170$--$650$\,nm.

After a BAT trigger, the XRT and UVOT start observing within $\sim100$\,s, and immediately distribute an updated location of the burst to the ground, where ground-based optical telescopes can perform quick follow-up observations. In this way, \emph{Swift} has, by far, observed the largest number of well-localised bursts with observed afterglows and redshift determinations. This has led to a larger understanding of the afterglow, by characterising the different phases seen in Fig.~\ref{fig:xray_lc}. It also enabled the study of the GRB environment and host, leading to potential uses in cosmology, such as studying the global chemical evolution and the star-formation history, as highlighted above.

While \emph{Swift} has revolutionised our understanding of afterglows and hosts, \emph{Fermi} is specialised in the bursts themselves.
 \emph{Fermi}  was launched in 2008, and has onboard the Gamma-ray Burst Monitor (GBM), with an energy range of $8$\,keV--$40$\,MeV, and detection rate of $\sim300$ bursts per year, and the Large Area Telescope (LAT) which detects energies in the range $20$\,MeV-- $\sim300$\,GeV. The presence of the LAT enables the characterisation of the high energy part of the GRB, and has led to the discovery that the high-energy component is delayed with respect to the lower-energy bands, but seems to last for a longer time period. The LAT detects only $\sim10$\% of the bursts detected by GBM, giving us a handle of how common this high energy component is in GRBs \citep{2012Sci...337..932G}.
 
Both \emph{Fermi} and \emph{Swift} operations are planned to continue throughout at least 2016, and probably longer. After this, the mission planned to take over GRB detections is the space-based multi-band astronomical variable object monitor (\emph{SVOM}), which is currently to be launched in 2021. \emph{SVOM} is, like \emph{Swift}, planned to carry both a $\gamma$-ray, X-ray (MXT, 0.3--10\,keV) and optical telescope (VT, 400--950\,nm), but is also to have two dedicated ground-based optical/NIR telescopes which will automatically follow up on the $\gamma$-ray trigger \citep{2010arXiv1005.5008S}. The high energy telescope actually consists of two instruments, one for imaging (ECLAIRs, $4$--$120$\,keV) and a spectrometer (GRM) observing the sky in wide field. On the ground, a wide-field optical camera (GWAC, $450$--$900$\,nm) will observe the same part of the sky as ECLAIRs, in real time, enabling a better statistical study of early optical emission. It is expected that ECLAIRs will detect $70$--$80$ GRBs per year, while GRM will detect $>90$ \citep{2014styd.confE...5C}. It is the hope that with rapid optical follow-up, the GRB sample with detected redshifts will be more complete, with fewer bias than the \emph{Swift} sample. As the \emph{SVOM} instruments are still under construction, the exact specifications may change.

Other possible future GRB missions include the joint astrophysics nascent Universe satellite (\emph{JANUS}), though the funding is not currently secured. \emph{JANUS} is designed to observe GRBs and quasars in the high-redshift Universe, in order to study re-ionisation and the formation of the first galaxies. Since \emph{JANUS} is only meant to detect high-redshift GRBs (out to $z\sim12$), it has no need for a $\gamma$-ray telescope, as the peak energy of the prompt emission will be redshifted to a few tens of keV. Instead \emph{JANUS} is planned to have an X-ray and a NIR telescope onboard. The challenge will then be to make an X-ray telescope with a wide field of view, but which can localise the burst to great precision \citep{2012MSAIS..21...59B}. Similar missions proposed include the energetic X-ray imaging survey telescope \citep[\emph{EXIST},][this telescope is also currently without funding]{2010AIPC.1279..212G}, which is also designed to be sensitive to high-redshift bursts.

\cleardoublepage
\chapter{The Gamma-Ray Burst Spectrum}\label{chap:spec}
Having now gone through the general properties of GRBs and their birth environment, we will take a deeper look at the GRB prompt and afterglow spectrum, and the radiation processes responsible for its shape.

\section{The Prompt Spectrum}

\begin{figure}[h]
\centering
\includegraphics[width=0.8\textwidth]{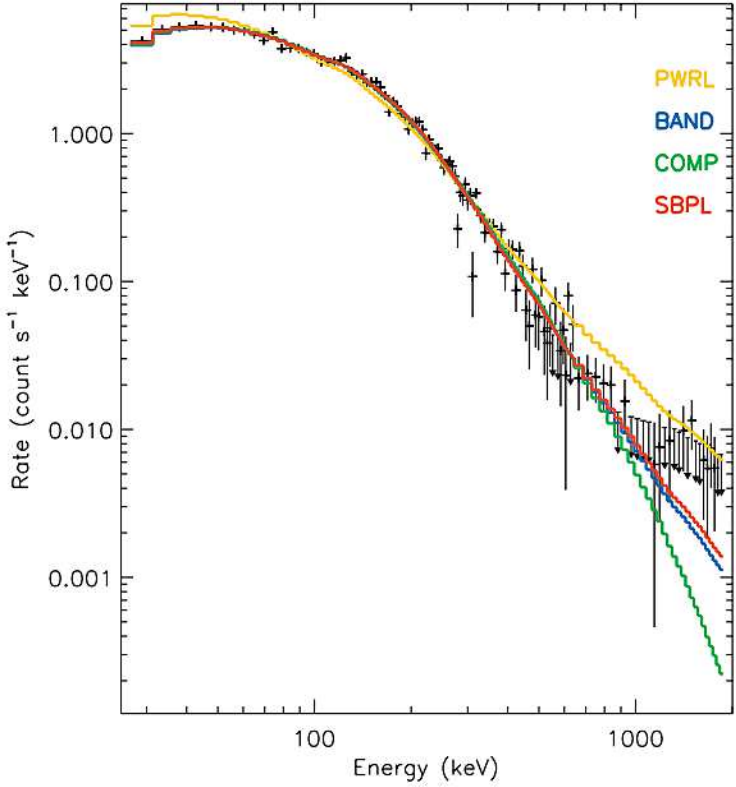}
\caption{Plot of the spectrum of GRB\,000429 from the BATSE catalogue \citep{2006ApJS..166..298K}. On top of the data (black crosses), is shown, in colours, fits to the most commonly used models, given in Equations~\ref{eq:band},~\ref{eq:PL},~\ref{eq:comp} and~\ref{eq:SBPL} (convolved with the BATSE detector response function).}
\label{fig:spec}
\end{figure}

Unlike the huge variability observed for GRB light curves, as seen in Fig.~\ref{fig:lc_burst}, the spectra of GRBs are relatively similar from burst to burst. Fig.~\ref{fig:spec} shows a typical example of a prompt spectrum. The spectrum is made up of power-law segments. The most commonly used model to fit the spectrum is the model suggested by \cite{1993ApJ...413..281B}; a broken power law with a smooth roll-over, usually referred to as the Band model: 
\begin{equation}
   f_{\text{Band}}(E)=%
   \begin{cases}
   A(\frac{E}{100\,\text{keV}})^\alpha \ \text{exp}(-\frac{E}{E_0})  & (\alpha-\beta)E_0\geq E \\
   A[\frac{(\alpha-\beta)E_0}{100\,\text{keV}}]^{\alpha-\beta} (\frac{E}{100\,\text{keV}})^\beta \ \text{exp}(\beta-\alpha) & (\alpha-\beta)E_0\leq E.
   \end{cases}
   \label{eq:band}
\end{equation}

$A$ is a normalisation constant, and $E_0$ is a characteristic or break energy. This function is purely mathematical, not physical, but for certain values of the low and high energy spectral indices $\alpha$ and $\beta$, it can represent physical models such as bremsstrahlung or synchrotron radiation, see below. Other mathematical models which are often used to fit the spectrum are a simple single power law (SPL or PWRL):
\begin{equation}
   f_{\text{SPL}}(E) = A \left(\frac{E}{100\,\text{keV}}\right)^{\psi};
   \label{eq:PL}
\end{equation}

\noindent an exponentially cutoff power-law (a Comptonised model, which represents the Comptonised spectrum from a thermal source in the special case where $\alpha=-1$), which is essentially a Band function where $\beta\rightarrow-\infty$:
\begin{equation}
   f_{\text{comp}}(E) = A \left(\frac{E}{100\,\text{keV}}\right)^\alpha \ \text{exp}\left[-\frac{(\alpha+2)E}{E_0}\right];
   \label{eq:comp}
\end{equation}

\noindent and a smoothly broken power law: 
\begin{equation}
   f_{\text{SBPL}}(E) = A \left(\frac{E}{100\,\text{keV}}\right)^{b} \ 10^a,
   \label{eq:SBPL}
\end{equation}

\noindent \citep[where $a$ is given in][]{2006ApJS..166..298K}. The difference between this and the Band function, is that the break is not coupled to the power laws, and the low energy part is closer to a true power law, where for the Band model the $\alpha$ index describes an asymptotic power law. It allows for the possibility of a sharper break than can be fit with the Band model.

\begin{figure}[h]
\centering
\includegraphics[height=6.67cm,width=0.51\textwidth]{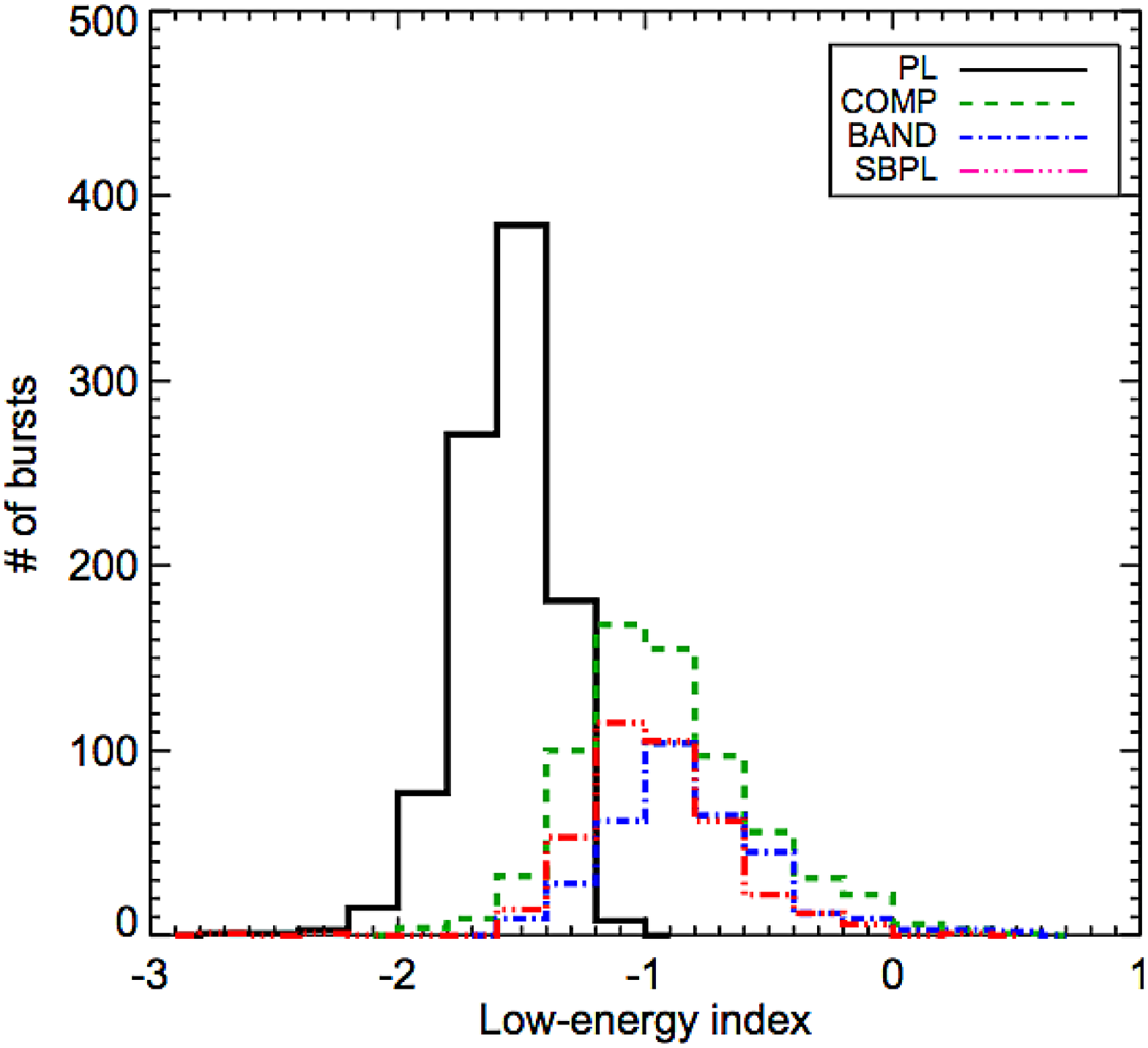}
\includegraphics[height=6.5cm,width=0.46\textwidth]{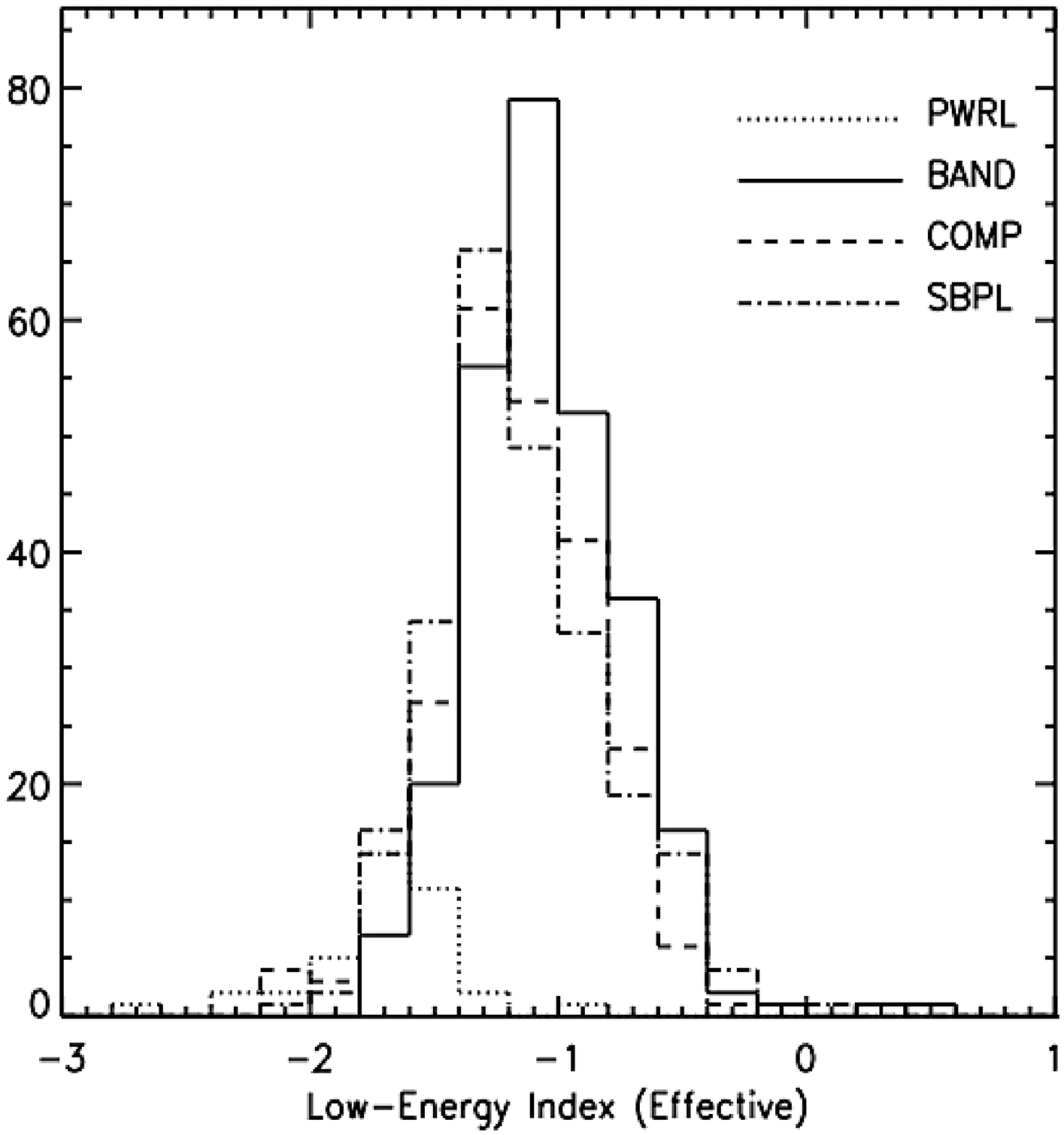}
\caption{Distribution of the low energy index $\alpha$ for the best fit model in the Fermi (left) and BATSE (right) burst catalogues. PWRL and PL is a simple power-law fit (Equation~\ref{eq:PL}), SBPL is a smoothly broken power-law fit (Equation~\ref{eq:SBPL}), Comp is the Comptonised model given in Equation~\ref{eq:comp}, and Band is the \cite{1993ApJ...413..281B} model given in Equation~\ref{eq:band}. See \cite{1998ApJ...506L..23P} for conversions between the effective and fitted low-energy index.}
\label{fig:alpha}
\end{figure}

\begin{figure}[h]
\centering
\includegraphics[height=6.65cm,width=0.51\textwidth]{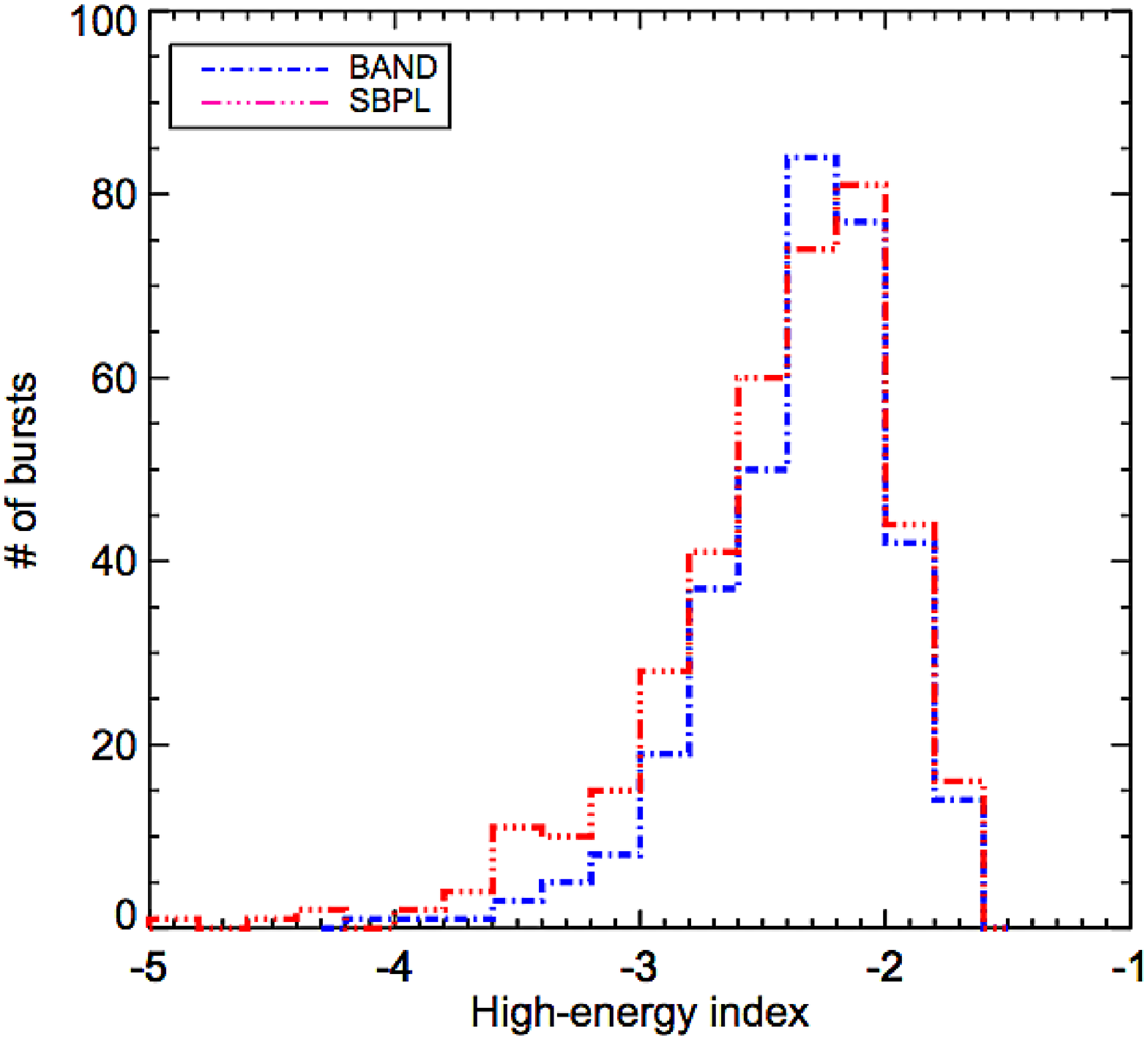}
\includegraphics[height=6.5cm,width=0.46\textwidth]{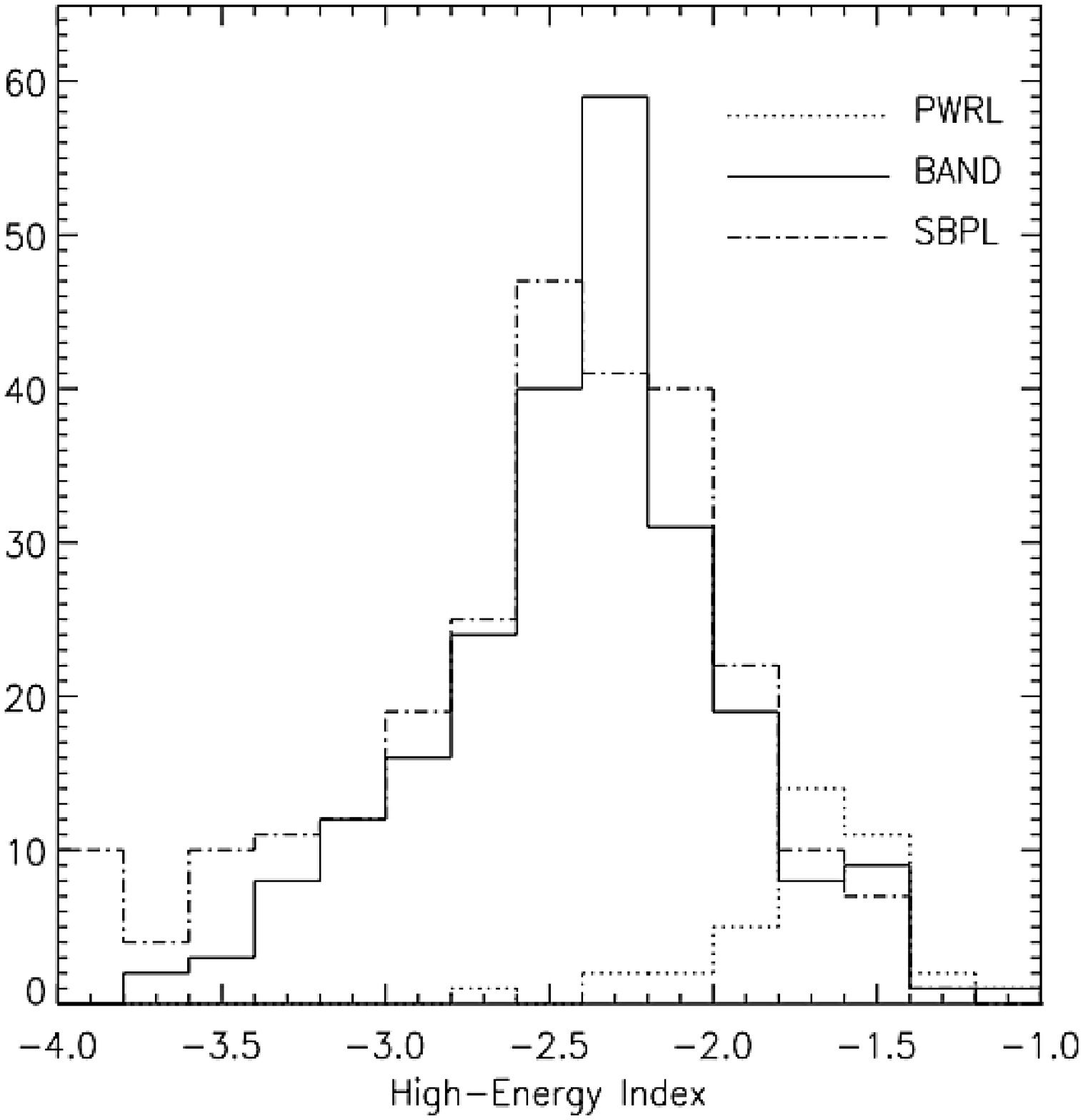}
\caption{Distribution of the high energy index $\beta$ for the best fit model in the Fermi (left) and BATSE (right) burst catalogues. Models are the same as in Fig.~\ref{fig:alpha}. For the BATSE sample the simple power-law index is included here, though it is the same as in Fig.~\ref{fig:alpha}.}
\label{fig:beta}
\end{figure}

Figs.~\ref{fig:alpha} and~\ref{fig:beta} display typical values of high and low spectral indices in the time-integrated spectra of GRBs. The figures show histograms of the values for the BATSE and Fermi-GBM samples, as presented in \cite{2006ApJS..166..298K} and \cite{2014ApJS..211...12G} respectively.

The figures show the indices for the best fit of the four models (PL, SBPL, Band and Comptonised). When a power-law break is observed, the break energy is typically about $100-200$\,keV. In the cases where a single power law is the best fit, the break energy is either outside of the fitted energy interval (see Section~\ref{sec:telescopes}), or the signal-to-noise (S/N) is too low to be able to resolve the break. The latter is indicated by the fact that for both samples, the single power-law index is between typical values for the low- and high index of the other models. Likewise, for faint bursts, sensitivity is not good enough to detect the high-energy power-law component, and the Comp model gives a better fit than the Band model. The Fermi sample has a high ratio of SPL best fits, which could be an indication that the GBM telescope triggers on relatively more faint bursts. For the Fermi sample (which is significantly the larger of the two), only fits with relatively small errors on all parameters are included ($\pm0.4$ for all parameters but the high energy index, for which $\pm1.0$ was allowed).

\section{Radiative Processes}
To determine which emission mechanisms are responsible for the observed $\gamma$-rays, we need to connect these mathematical descriptions with physical interpretations.

\subsection{Synchrotron Radiation}\label{sec:synch}

\begin{figure}[h]
\centering
\includegraphics[width=0.85\textwidth]{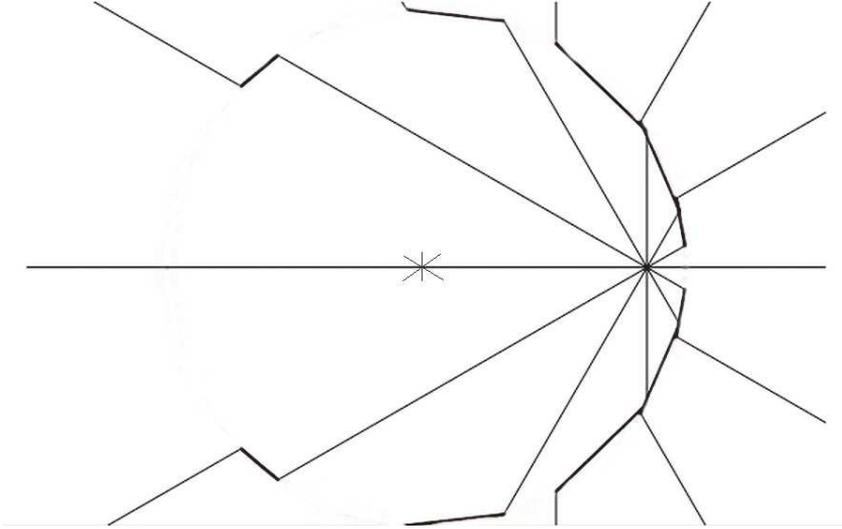}
\caption{A charged particle moving with constant velocity has a uniform electric field. In this figure, the particle experienced an instantaneous acceleration when it was at the position of the cross, causing a change in the field lines, which are now propagating outwards (courtesy of Michael Friis).}
\label{fig:synch}
\end{figure}

In the fireball model described in Section~\ref{sect:model}, the predominant radiation mechanism is synchrotron radiation from electrons in the shocks. Consider a charged particle experiencing acceleration. This acceleration will cause a change to the electric field lines emerging from the particle, and the information of this change is radiated as a dipole perpendicular to the acceleration. This scenario is illustrated in Fig.~\ref{fig:synch}. The total radiated power in a dipole is given by the Larmor formula \citep{larmor}:
\begin{equation}
   P = \frac{dW}{dt\,dA} = \frac{2q^2a^2}{3c^3},
   \label{eq:larmor}
\end{equation}
\noindent where $q$ is the charge of the particle and $a$ is the acceleration.

\begin{figure}[h]
\centering
\includegraphics[width=0.7\textwidth]{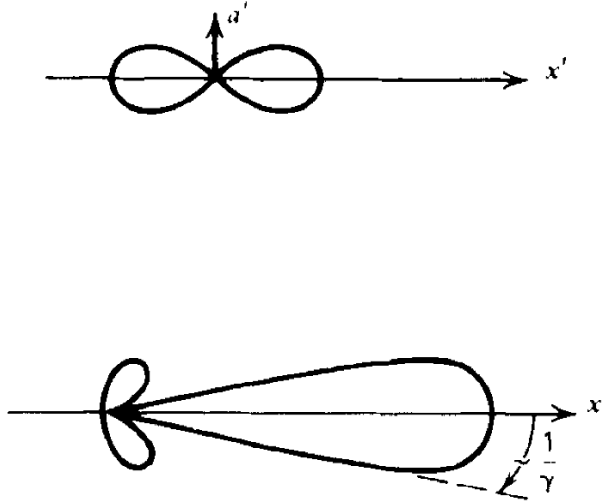}
\caption{Figure from \cite{Rybicki}. The top figure shows dipole radiation from a particle moving at non-relativistic speed. The bottom figure shows the same dipole radiation, but now the particle is moving at relativistic speed, so the radiation is beamed in the direction of motion, with an opening angle of $1/\Gamma$.}
\label{fig:dipole}
\end{figure}

The non-relativistic version of this radiation is called cyclotron radiation. The term synchrotron radiation is used for the specific case of relativistically moving charged particles in a magnetic field. The (relativistic) equation of motion in a magnetic field is:
\begin{equation}
   \frac{d}{dt}(\Gamma m \textbf{v}) = \frac{q}{c}\,\textbf{v $\times$ B},
   \label{eq:motion}
\end{equation}
\noindent where $\textbf{v}$ is the velocity vector, $\textbf{B}$ is the magnetic field vector, $\Gamma$ is the Lorentz factor, and $m$ is the mass (electron mass in this case). The magnetic force is perpendicular to the motion, so the particles will move along a curve, and hence have a radial acceleration component. This acceleration is perpendicular to the velocity, and for a constant field strength will lead to helical motion along the field lines. Since time and energy have the same Lorentz transformation, Equation~\ref{eq:larmor} is invariant, and we can calculate the radiated power in any frame, as long as we know the acceleration. The acceleration can be calculated from Equation~\ref{eq:motion}, and is related to the magnetic gyration frequency (centrifugal acceleration). Using this, we can derive the (beamed) synchrotron spectrum for a single electron \citep[see e.g.][for a more detailed derivation]{Rybicki}:
\begin{equation}
   P(\nu) = \sqrt{3}\,\frac{q\,B\,\text{sin}\alpha}{mc^2}\,F(\frac{\nu}{\nu_{\text{crit}}}),
   \label{eq:single}
\end{equation}
\noindent where $\alpha$ is the angle between the magnetic field and velocity, and $F$ is a dimensionless function peaking just before the frequency $\nu_{\text{crit}}$. As the dipole from the particle is radiated perpendicular to the acceleration, it will in this case be along the line of motion, so synchrotron radiation is heavily boosted due to relativistic motion, see Fig.~\ref{fig:dipole}.

Away from $\nu_{\text{crit}}$ and over a limited frequency range, Equation~\ref{eq:single} can be approximated with a power law. The spectral index $s$ is then defined as:
\begin{equation}
   P(\nu)\propto\nu^{-s}.
   \label{eq:index}
\end{equation}

If we now consider a population of particles with energy distribution:
\begin{equation}
   N(E)dE \propto E^{-p}dE,
   \label{eq:dist}
\end{equation}
\noindent then we can find the total power radiated per volume per frequency by integrating Equation~\ref{eq:dist} times the spectrum for a single particle, Equation~\ref{eq:single}, over all energies. This yields the result:
\begin{equation}
   P_{\text{tot}}(\nu)\propto\nu^{-(p-1)/2},
   \label{eq:total}
\end{equation}
\noindent so that the spectral index for a synchrotron spectrum is given by the particle energy distribution index $p$: 
\begin{equation}
   s = \frac{p-1}{2}.
   \label{eq:p}
\end{equation}

To compare the synchrotron spectrum with the GRB prompt emission, we now need an estimate of $p$, i.e. the energy distribution of the particles. In the GRB jet, synchrotron radiation would be emitted by electrons moving in the electromagnetic field over the shock fronts. The typical electron energy depends on the Lorentz factor of the flow, and the strength of the magnetic field. The characteristic synchrotron frequency is \citep{1998ApJ...497L..17S}:
\begin{equation}
   \nu_m(\Gamma_e) = \Gamma\,\Gamma_e^2\,\frac{e\,B}{2\pi\,m_e\,c},
   \label{eq:char}
\end{equation}
\noindent where $\Gamma$ is the Lorentz factor of the shock front, and $\Gamma_e$ is the Lorentz factor of the electron. 


Electrons moving in a plasma will cool down as they emit radiation. This so-called synchrotron cooling occurs on a time scale given by the energy of the electrons over the rate at which they are radiating away their energy. The energy is: 
\begin{equation}
  E=\Gamma\,m_e\,c^2,
  \label{eq:energy}
\end{equation}
\noindent so taken together with Equations~\ref{eq:single} and~\ref{eq:dist} the cooling time is:
\begin{equation}
  t = \frac{3}{4}\frac{\,m_e\,c^2}{\sigma_T\,\,c\,U_B\,\Gamma\,\beta^2},
  \label{eq:time}
\end{equation}
\noindent where $U_B$ is the magnetic energy density of the field, $\sigma_T$ is the Thomson cross-section, and $\beta=v/c$.

The cooling time sets a lower limit to the time variations we should observe in the GRB light curve. If synchrotron radiation is the dominant emission mechanism, we should not observe individual peaks in the light curve that are shorter than the cooling time; the typical GRB prompt single peak has a fast rise and a slower exponential decline (referred to as FRED). This would mean that the shock heating (and hence acceleration) of the electrons happens rapidly, so the majority of the width of the peak is due to cooling, and hence the width of one peak is roughly equal to the cooling time.

The energy distribution of the electrons depends on whether they are in the \emph{slow} or \emph{fast} cooling regime. This is determined by whether the cooling time scale is longer or shorter than the hydrodynamical time scales of the shock. The large energy requirement of GRBs means that the electrons must be radiating efficiently, and hence are in the fast cooling regime (a short cooling time is also a requirement from the millisecond variations observed). 
Due to the cooling effect, the shock cannot accelerate the electrons to arbitrary high energies. For the maximum electron energy, the acceleration time equals the cooling time. As the electron cools, the synchrotron frequency varies as (see Equation~\ref{eq:char}):
\begin{equation}
   \nu \propto \Gamma_e^2.
   \label{eq:cooling}
\end{equation}

We can use this to find $p$, if we look at the continuity equation for the electron \citep{2014NatPh..10..351U}:
\begin{equation}
  \frac{\partial N(\Gamma_e t)}{\partial t} = Q(\Gamma_e,t) - \frac{\partial}{\partial \Gamma_e}(N(\Gamma_e,t)\frac{d\Gamma_e}{dt}),
  \label{eq:cont}
\end{equation}
\noindent where $Q$ is the injection rate of electrons with a Lorentz factor $\Gamma_e$. This basically gives the relation for the change with time of the number of electrons with $\Gamma_e$, which equals the rate they are injected into the flow, minus the rate at which $\Gamma_e$ changes with time after injection. For the GRB, we assume that the energy is injected almost simultaneously, so after an initial injection, we can set $Q=0$. To maintain a steady distribution in time, then
\begin{equation}
  \frac{\partial}{\partial \Gamma_e}(N(\Gamma_e,t)\frac{d\Gamma_e}{dt}) = 0.
  \label{eq:cont2}
\end{equation}

\noindent Since the change in $\Gamma_e$ is caused purely by synchrotron cooling in the fast cooling regime, then it follows from Equations~\ref{eq:cooling} and~\ref{eq:cont2}, that $\frac{\partial N_e}{\partial \Gamma_e} \propto \Gamma_e^{-2}$. As the Lorentz factor is proportional to the energy of the electron (Equation~\ref{eq:energy}), then $p=2$, where $p$ is defined in Equation~\ref{eq:dist}. 

This means (from Equation~\ref{eq:p}) that the spectral power varies as $\nu^{-1/2}$ until it falls below the cooling frequency, where the cooling time becomes longer than the dynamical time-frame. At lower frequency, the electron does not have time to cool, so only few electrons are at this $\Gamma_e$, and the flux falls quickly off.

For fast cooling the total spectrum is \citep{1999PhR...314..575P}:
\begin{equation}
  F_\nu \propto
  \begin{cases}
  (\frac{\nu}{\nu_c})^{1/3}F_{\nu,max}   &   \nu_c > \nu \\
  (\frac{\nu}{\nu_c})^{-1/2}F_{\nu,max}  &  \nu_m > \nu > \nu_c \\
  (\frac{\nu_m}{\nu_c})^{-1/2}\,(\frac{\nu}{\nu_m})^{-p/2}F_{\nu,max}   &   \nu > \nu_m,
  \end{cases}
  \label{eq:spectrum}
\end{equation}
\noindent where $F_{\nu,max}$ is the observed peak flux. 

To compare this with the Band function, we subtract one from the spectral indices to get the photon indices $\alpha$ and $\beta$ (as the energy spectrum equals the photon spectrum multiplied with the photon energy). Since the high index depends on $p$, the usual method is to compare the observed $\alpha$ with the low frequency part. In general, we expect the spectrum to be somewhere between the $\nu_c > \nu$ and $\nu_m > \nu > \nu_c$ cases, leading to a constraint on the spectral index, $-1/3\leq s \leq 1/2$.

This has led to the so called "line-of-death" problem, as the low energy index is restricted to $\alpha<-2/3$. If we compare to Fig.~\ref{fig:alpha} however, we will see that a significant fraction of bursts have $\alpha$ higher than this, several bursts even have a positive $\alpha$, which cannot be explained purely by synchrotron radiation \citep[see e.g.][]{1998AIPC..428..359C,2004ApJ...613..460B}. For these bursts at least, some other radiation mechanism must contribute.

I have here ignored synchrotron self-absorption, where synchrotron photons, after being emitted, scatters continuously of the synchrotron electrons. For GRB prompt emission it is assumed, at least initially, that the jet is optically thin so we can ignore this effect, as it would also reduce the observed flux significantly, which fits poorly with the large flux observed for GRBs, though it has been argued that it may be used to solve some cases of a positive $\alpha$, see for instance \cite{2000ApJ...543..722L}.

\subsection{Inverse Compton emission}
Even in the optically thin regime however, scattering can alter the spectral shape. A photon scattering off a charged particle and decreasing its energy is called Compton scattering, so inverse Compton (IC) scattering refers to the process where the charged particle loses energy to the photon. 

At the energies of the electrons in the GRB jet, the photons are likely to only scatter off an electron once, as they thereby gain an energy that is so high that the cross-section becomes too small to scatter with another electron (in the rest-frame of the electron). Since the cross-section is energy dependant, IC scattering will predominantly change the spectrum by scattering low energy photons to higher energies. Furthermore, it will influence the cooling time, as the electrons now cool both through synchrotron cooling (Equation~\ref{eq:time}) and through IC scattering. The scattering will boost the photon energy with a factor of $\Gamma_e^2$. A special case of the IC scattering is Synchrotron self-Compton (SSC) radiation. This is the result of IC scattering of synchrotron radiation by the same relativistic electrons that produced the synchrotron radiation in the first place. 

\cite{2004ApJ...613..460B} find that with IC scattering, $\alpha$ could be as high a 0. That could help explain bursts with very hard spectra, such as GRB\,130427A \citep[see][]{2013ApJ...776...95F}. Furthermore, adding IC scattering may help explain the spectral evolution in bursts where $\alpha$ is seen to increase with time (approach zero). Initially IC scattering would be effective, up-scattering photons, so that there are less photons at low frequencies, and $\alpha$ is very negative. At some point, the jet expands, so that the probability for a scattering event decreases, and $\alpha$ hence increases as IC becomes less effective. 

However, there are bursts for which $\alpha$ is observed to \emph{increase} at the beginning of a pulse \citep[e.g.][]{1998AIPC..428..359C}, which would indicate Thomson thickening, which is inconsistent with the simple fireball model. \cite{2000ApJ...544L..17P} suggested that IC scattering of synchrotron self-absorbed radiation could be responsible for the hard spectra, but this would require a scenario where the GRB jet material is in the slow cooling phase, which seems unlikely.

\subsection{Thermal emission}
The inconsistency between non-thermal radiation processes and the spectral observations of GRBs became even more pronounced when the instrumentation became good enough to study the temporal evolution of spectra. Time-resolved spectra have an even higher occurrence of $\alpha$ outside the synchrotron allowed values \citep{2014ApJS..211...12G}. It is not just that the best fit value is outside the synchrotron range, but allowed synchrotron values give a significantly poorer fit to the data.

Still, many bursts do have decent fits to synchrotron models, even for individual time slices of the spectra. The problem though, is that we observe the spectral shape to have a variance in time which is hard to explain from the considerations outlined in Section~\ref{sec:synch}. The conditions in the GRB jet are expected to change over time, so that the electrons move from the fast to the slow cooling regime, but that does not explain the large changes in $\alpha$ observed by for instance \cite{1998AIPC..428..359C}.

One radiation process that is expected to display the soft spectral evolution observed, is thermal emission from a black body. Blackbody radiation was originally expected in the jet model, as the relativistic moving plasma will be thermally coupled by multiple Compton scatterings, if the density is high enough. We would observe this radiation as originating from the photosphere where the jet becomes optically thin. It was hence obvious to look for thermal emission when it became clear that an additional component was needed on top of the synchrotron radiation. 

The models used for the fit are either simple blackbody functions, such as this one (as implemented by the HEASoft program Xspec):
\begin{equation}
    f_{\text{BB}} = \frac{K[E(1+z)]^2}{(1+z)\,k_{\text{B}}\,T^4\,(\text{exp}[E(1+z)/(k_BT)]-1)},
    \label{eq:bb}
\end{equation}
\noindent where $K$ is a normalisation, $k_{\text{B}}$ is the Stefan-Boltzman constant, and $z$ is the redshift, or alternatively, more complex models can be used, such as multi-blackbodies, which consists of a superposition of Planck functions with different temperatures. The latter is probably more physically correct, while the former has the advantage of being conceptually simple. It should be noted that even for the more complex functions, the observed temperature is not the physical blackbody temperature of the flow, but rather the Doppler boosted value:
\begin{equation}
    T_{obs} = D\,T_{co}
    \label{eq:temp}
\end{equation}
\noindent where $T_{co}$ is the temperature in the co-moving frame, and 
\begin{equation}
    D = \frac{1}{\Gamma(1-\beta\mu)}
    \label{eq:doppler}
\end{equation}
is the Doppler factor of the plasma moving with an angle, $\mu$ = cos(angle), to the observer.

\begin{figure}[h]
\centering
\includegraphics[width=0.9\textwidth]{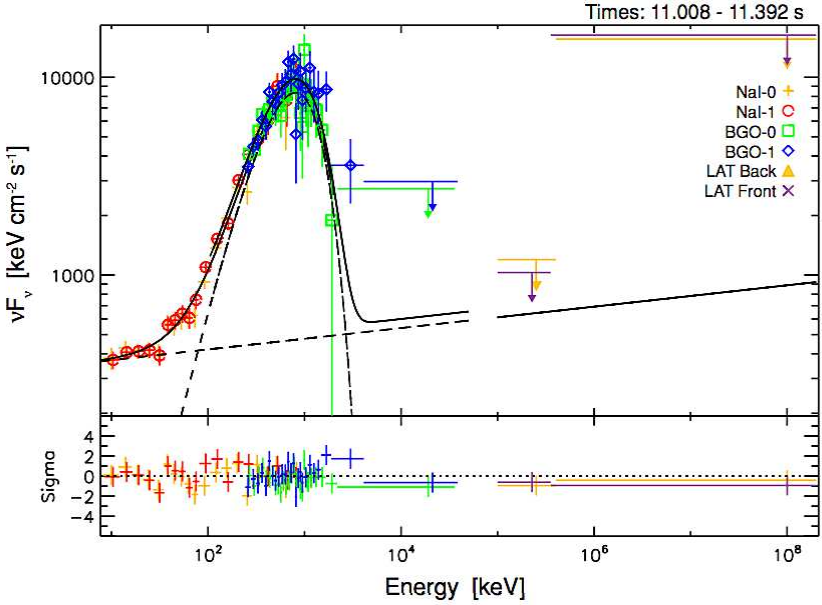}
\caption{Figure from \cite{2010ApJ...709L.172R} showing an example of a blackbody fit to the spectrum of GRB\,090902B. Different colours are data from different \emph{Fermi} detectors. Lower panel show the deviation between data point and fitted model.}
\label{fig:BBspec}
\end{figure}

The difficulty in studying thermal emission in GRBs, and the reason this component was originally ruled out, is that the blackbody temperature and flux will naturally evolve with time, smearing out the signal in a time-integrated spectrum, making it difficult to separate from the non-thermal emission. When individual spectra from shorter time periods are extracted, a model consisting of a blackbody spectrum with an underlying power-law to represent the non-thermal part, often gives a good fit to the data, as seen in Fig.~\ref{fig:BBspec} showing the Fermi spectrum of GRB\,090902B as an example \citep{2010ApJ...709L.172R}.

Since the emitting plasma is moving at relativistic speeds, the blackbody emission is not the classical version, but rather what is often referred to as 'quasi-thermal' or 'blackbody like' radiation. \cite{2008ApJ...682..463P} calculated  the temporal evolution of a thermal component in a relativistically expanding plasma, including the observed radiation from equal arrival time surfaces (including the same high angle emission effects discussed in Chapter~\ref{chapter:grb}), and found that after energy injection stops (the end of inner engine activity), the blackbody flux decays as $F(t)\propto t^{-2}$ and the temperature as $T_{\text{BB}}(t)\propto t^{\alpha}$, with $\alpha\sim 1/2-2/3$. This is found to match observations for a sample of bursts \citep{2004ApJ...614..827R,2005A&A...432..105R,2010arXiv1003.2582P}.

Fitting the thermal component has the advantage over the non-thermal emission, that it allows us to probe the inner engine directly, and determine flow properties such as $\Gamma$, the radius of the photosphere, $R_{\text{ph}}$, and the initial size of the flow, $R_0$ \citep[see][and Chapter~\ref{sec:paper_thermal}]{2007ApJ...664L...1P}.

\section{The Afterglow Spectrum}\label{sec:agspec}
The GRB afterglow (that is, the part of Fig.~\ref{fig:xray_lc} that is marked 'afterglow') is believed to be caused by synchrotron radiation from external shocks when the jet plows into the surrounding medium. The physics is similar to that described in Section~\ref{sec:synch}, except the jet now interacts with the ISM, so the density and type of the ISM is relevant. 

Initially, the jet electrons are still in the fast cooling regime. Later, however, they transit into slow cooling. Furthermore, the density profile of the surrounding medium can be one of several different types, so we have several different physical cases, leading to a multitude of models for the GRB afterglow. Writing the density profile in the form $n(r)\propto r^{-k}$, two typical values of $k$ are often explored.
For long GRBs, the likely progenitor is a massive star, which has likely gone through mass-loss (see Section~\ref{sec:progenitor}), so we expect a stellar wind profile which has $k=2$. Alternatively a constant density medium with $k=0$, as is the case for the typical ISM, is assumed. 

The index $k$ can be determined from the spectral and temporal flux indices, $F_\nu \propto \nu^{-b}\,t^{-a}$, as:
\begin{equation}
    k = \frac{4(3b-2a)}{3b-2a-1},
    \label{eq:density}
\end{equation}
\noindent see \cite{2008ApJ...672..433S}. This relationship is valid for the spectral index in the part of the spectrum where $\nu_{m}<\nu<\nu_{c}$. \cite{2008ApJ...672..433S} do indeed find that $k$ is consistent with a wind for four out of five long GRBs, and a homogeneous circum-burst medium for the last \citep[though other studies find that a constant density medium is overwhelmingly preferred for GRBs, e.g.][]{2006ApJ...647.1269F}. They also determine $p$, the electron energy distribution index, through Equation~\ref{eq:p}. They find a mean value of $2$ in a sample of 10 bursts, consistent with the fireball model \citep[the existence of bursts with $p<2$ is termed the '$p$-problem', as this would cause the integral of Equation~\ref{eq:dist} to diverge, unless we introduce an upper cutoff in the electron energy distribution, see e.g.][]{2001BASI...29..107B}. 

IC scattering may also contribute to the afterglow emission of some GRBs, particularly if the surrounding medium density is high. This could be observed as a break in the light curve \citep{2000A&A...360L..13W}. In general the afterglow fits better with the synchrotron emission of the standard model than the prompt emission, though it is often difficult to isolate the afterglow from late-time contributions from the prompt emission such as flares. \cite{2015ApJS..219....9W} found that at least half of GRB afterglows fit well with the external shock synchrotron radiation, while up to $96\%$ can be well fitted by including additional factors such as a long-lasting reverse shock, see Section~\ref{sec:afterglow}.

In several bursts, thermal emission has also been observed in soft X-rays during the afterglow \citep[e.g.][]{2006Natur.442.1008C,2011MNRAS.416.2078P}. In most models, this emission is not directly associated with the external shock, or fireball, afterglow. To understand the different models for this component, we will first take a look at the GRB-SN connection.

\section{Supernova Connection}
\cite{1993ApJ...405..273W} originally suggested that the progenitor model for GRBs evolved into what he termed a 'failed supernova', as it was thought that not enough $^{56}$Ni would be produced during the core-collapse, so it could not lead to a traditional SN (the half-life of nickel is several days, so radioactive-decay energy can reheat the ejecta and supply energy to the SN). We now know however, that the majority of long GRBs are followed by a Type Ib/Ic SN, which has on several occasions been observed in the optical light curve a few days after the burst, either spectroscopically \citep[e.g. GRB\,030329,][]{2003Natur.423..847H,2003ApJ...591L..17S}, or through a bump in the photometric light curve \citep[e.g. GRB\,980326,][see also Sect~\ref{sec:long}]{1999A&AS..138..449C}.

The highest redshift for a secure GRB-SN connection is $z=0.283$ \citep[GRB\,120422A][]{schulze14}, though GRB\,101219B at $z=0.55$ has a clear bump in the afterglow and some spectroscopic evidence as well \citep{sparre11}. At larger distances, it becomes difficult to observe the SN with sufficiently good S/N. The total energy output in a SN might be comparable to that of a GRB, but the explosion is more isotropic, and not ultra-relativistic, so it cannot be observed out to nearly as high a redshift. Added to that, the $\gamma$-rays from the GRB stands out from the host galaxy emission, and the SN also has to compete with the optical afterglow, which for some GRBs can be very long-lasting. SN bumps in the afterglow light curve have been more or less convincingly detected out to a redshift of about $z\sim1$ \citep{2003A&A...406L..33D}.

Despite this difficulty, there are several long GRBs for which we can rule out an associated SN (or the SN was extreme faint), such as GRBs 060505 and 060614 \citep{2006Natur.444.1047F}. For GRB\,060505, no SN was detected down to a limit several hundred times fainter than that of the 'standard' GRB-SN SN\,1998bw, and even fainter than any Type Ic SN ever observed. This GRB is in all other aspects a typical long GRB, leading to the conclusion that some long GRBs just do not produce a SN, possibly because no radioactive material is formed as the material falls directly in to the black hole before forming an accretion disc, or because there simply is not enough energy left by the time the GRB jet pierces the star \citep[][and references therein]{2012grbu.book..169H}. 

The resulting picture of the GRB-SNe is a diverse population, with a broad dispersion both in peak brightness rise time, and spectral broadness. \cite{2006ARA&A..44..507W} showed that when including the non-detections, GRB-SN span the entire 'normal' Type Ib/Ic population.

Short GRBs are not associated with classical SNe as shown by the clear non-detection for GRBs 050509B \citep{2005ApJ...630L.117H,2006ApJ...638..354B} and 050709 \citep{2005Natur.437..859H}. It was suggested that the merger of compact objects, believed to be the progenitor for short GRBs, should be accompanied by a so-called `kilonova' produced by the decay of radioactive species from the merger, visible in the NIR. This signature was indeed detected in observation of the short GRB\,130603B \citep{2013ApJ...774L..23B,2013Natur.500..547T}, and possibly in the long/short GRB\,060614 \citep[][the burst lasted 102\,s, but if the observation is indeed a kilonova, a compact binary merger is the more likely progenitor model]{2015NatCo...6E7323Y}.

\section{A SN shock breakout}
Several bursts with associated SNe have a detection of the thermal emission observed in the soft X-ray afterglow mentioned in Section~\ref{sec:agspec} \citep[][]{2006Natur.442.1008C,2011MNRAS.411.2792S,2012MNRAS.427.2950S,2011MNRAS.416.2078P}. It was suggested for GRB\,060218 \citep{2006Natur.442.1008C} that this component is due to the SN shock front breaking into the winds surrounding the progenitor, as would be the case if this was a Wolf-Rayet star. This interpretation is motivated by high blackbody temperature, which can be obtained, if the thermal radiation is from shock-heated plasma. Furthermore the radius of the emitting region, which should be where the wind becomes optically thin, is larger than the blackbody radius, indicating asymmetry, which would be the case for a jet, or asymmetric winds. The wind mass-loss rate can be calculated, by considering a shock propagating into the wind. This will compress the wind material into a thin shell, which we know should have an optical depth of about one since we observe thermal radiation. This means we can calculate the mass of the shell, and with the approximate wind velocity and the radius of the shell (the radius of the emitting region), we can determine the mass-loss. \cite{2006Natur.442.1008C} found that for GRB\,060218 this approximately equalled the expected value for a Wolf-Rayet star.

This interpretation of the thermal emission is attractive, as it has large explanatory value. We can even back-extrapolate to find the likely radius of the progenitor. However, there are some problems with the energetics. The blackbody luminosity inferred from the observations of GRB\,060218 seems too large to be explained as the shock break-out \citep[see e.g.][and Section~\ref{sec:SNbreak}]{2007MNRAS.382L..77G,2007MNRAS.375..240L}.

\cite{2007ApJ...656.1001B} interpreted the thermal emission in GRB\,060218 as radiation from L-shell transitions of Fe. He suggested that this line emission originates from a cocoon around the GRB jet as it exits the progenitor stellar envelope. This model, and the one presented in Chapter~\ref{sec:paper_thermal} are inconsistent with the suggestion that weak bursts such as GRB\,060218, which are classified as XRFs, are caused by the jet being 'quenched' in the stellar envelope, so that the $\gamma$-rays cannot escape, and a 'proper' GRB never happens. 

\section{The Cocoon model}
\cite{2001ApJ...556L..37M} postulated the existence of the cocoon around the jet. As the jet moves through the progenitor stellar-envelope, the surrounding material slows it down, and energy is deposited outwards, heating the material that has been pushed out to a hot plasma cocoon. As the jet penetrates the star, the cocoon can escape as well, accelerating up to relativistic velocity (though still slower than the jet, due to the higher density), and the deposited energy is converted into bulk kinetic energy. As the density is high, the optical depth is initially much larger than one, so the radiation from the cocoon is delayed with respect to the prompt emission, as the photons are initially trapped, and need to diffuse before they can escape. \cite{2006ApJ...652..482P} suggested that this is the physical model for the initial steep decay of the X-ray light curve seen in Fig.~\ref{fig:xray_lc}.

The spectrum for this emission has a high energy tail in the form of a power law as the energy is continuously injected, but \cite{2001ApJ...556L..37M} also suggest that we might observe the underlying thermal emission when the cocoon has expanded sufficiently, and that this is what is observed in the X-ray afterglow of for instance GRB\,090618 \citep{2011MNRAS.416.2078P}. This model allows for higher blackbody luminosities and temperatures than SN shock breakout. However there are observations of thermal X-rays for bursts which do not show a steep decay in the light curve, such as GRBs 060218 and 100316D. \cite{2012MNRAS.427.2950S} argue that the underlying mechanism might be different for those low-luminosity bursts, such that the thermal emission for those bursts may be due to the shock breakout, but that the cocoon emission may be responsible for the more energetic bursts. 

In the following chapter, I present a model that could encompass all observed thermal soft X-ray components.

\cleardoublepage
\chapter{Thermal Emission in the Early X-ray Afterglows of GRBs: Following the Prompt Phase to Late Times.}\label{sec:paper_thermal}

\begin{center}
M.~Friis$^1$ \&
D.~Watson$^2$
\end{center}
   
\begin{center}
\emph{Published in ApJ, Volume 771, July 2013, 15F}
\end{center}

\begin{footnotesize}
\begin{itemize}
\addtolength{\itemsep}{-1.25\baselineskip}
  \item[$^1$]	Centre for Astrophysics and Cosmology, Science Institute, University of Iceland, Dunhagi 5, 107 Reykjav\'ik, Iceland \\
  \item[$^2$]	Dark Cosmology Centre, Niels Bohr Institute, University of Copenhagen, Juliane Maries Vej 30, 2100 Copenhagen, Denmark
 \end{itemize}
\end{footnotesize}

\textbf{\begin{large}Abstract\end{large}}
    Thermal radiation, peaking in soft X-rays, has now been
    detected in a handful of GRB afterglows and has to date been interpreted
    as shock break-out of the GRB's progenitor star.  We present a search
    for thermal emission in the early X-ray afterglows of a sample of
    \emph{Swift} bursts selected by their brightness in X-rays at early
    times.  We identify a clear thermal component in eight GRBs and track
    the evolution.  We show that at least some of the emission must come
    from highly relativistic material since two show an apparent
    super-luminal expansion of the thermal component.  Furthermore we
    determine very large luminosities and high temperatures for many of
    the components---too high to originate in a SN shock break-out.  Instead we
    suggest that the component may be modelled as late photospheric emission
    from the jet, linking it to the apparently thermal component observed
    in the prompt emission of some GRBs at gamma-ray and hard X-ray
    energies.  By comparing the parameters from the prompt emission and the
    early afterglow emission we find that the results are compatible with
    the interpretation that we are observing the prompt quasi-thermal
    emission component in soft X-rays at a later point in its evolution.
    
    \section{Introduction\label{introduction}}

Gamma-ray bursts (GRBs) are extremely bright transient sources, completely dominating the gamma-ray sky for milliseconds to half an hour. The nature of these bursts was long unknown, but since discovering an association with type Ic supernovae (SNe) (e.g. \citealt{1998Natur.395..670G,2003Natur.423..847H,2003ApJ...591L..17S}), long GRBs (any burst lasting longer than two seconds) are known to be caused by the core-collapse of massive stars. While the afterglows have provided a lot of information over the past 15 years, there is still a lot of uncertainty associated with the prompt high energy radiation. Recent evidence has been found in some GRBs for an apparently thermal component in the early soft X-ray emission \citep{2006Natur.442.1008C,2011MNRAS.411.2792S,2011MNRAS.416.2078P}. This component is often, though not exclusively \citep[e.g.][]{2007MNRAS.382L..77G}, interpreted as originating from the SN shock-breakout, as associated SNe have been detected spectroscopically for GRB\,060218/ SN2006aj \citep{2006Natur.442.1008C,2006Natur.442.1011P} and GRB\,100316D/SN2010bh \citep{2011MNRAS.411.2792S,2010GCN..10525...1W,2010arXiv1004.2262C,2012ApJ...753...67B}, and photometrically for GRB\,090618 \citep{2011MNRAS.416.2078P,2011MNRAS.413..669C}. 

Interestingly, a weak second component was found to be required statistically in the combined late prompt/early afterglow spectra of a few very soft GRBs \citep{2008A&A...478..409M}, suggesting either the emergence of the afterglow or a thermal component. The clear evidence for individual X-ray thermal components has so far only been detected in low redshift bursts with SNe. \cite{2012MNRAS.427.2965S} searched for this thermal emission in the total sample of \emph{Swift} bursts, but removed all bursts with high redshift, as they argue that the thermal emission detected in these are at high risk of being false positives. We have analysed a sample of bright early X-ray afterglows for the presence of thermal-type emission as well, but without the redshift filter, as we see no reason to rule these out beforehand. We find several new apparent thermal components, including bursts with redshifts $z>1$. In section~\ref{observations} we present the sample and the analysis of the Burst Alert Telescope (BAT) and the X-Ray Telescope (XRT) data. In section~\ref{results} the results of our time-resolved spectroscopy is presented for the best fit time series of each burst. In section~\ref{discussion} we examine the implications of the physical parameters deduced from modeling the emission component.

A flat universe cosmology is assumed with $H_0=70$\,km\,s$^{-1}$\,Mpc$^{-1}$ and $\Omega_{\rm M} = 0.27$. 1\,$\sigma$ errors for each parameter have been used. When only upper limits are given, these are 3\,$\sigma$.
%
%
\section{Observational data and methods}\label{observations} 

The sample presented in this paper consists of the brightest bursts (as of
2011 December 20) in the \emph{Swift} online catalogue
\citep{2007A&A...469..379E}.  We have selected bursts with at least 20\,000
counts in the XRT Window Timing (WT) data as well as reliable spectroscopic
redshifts, to ensure good fit statistics and to be able to study the
evolution in time and the rest-frame properties of the afterglows.  Twenty
nine bursts fit these criteria, but the dataset for
GRB\,100906A had to be discarded as repeated extractions of the spectra
failed to produce reliable results, so the final sample size is twenty eight bursts (see
Table~\ref{tab:best}).

\begin{figure}
\centering
\includegraphics[width=0.43\textwidth]{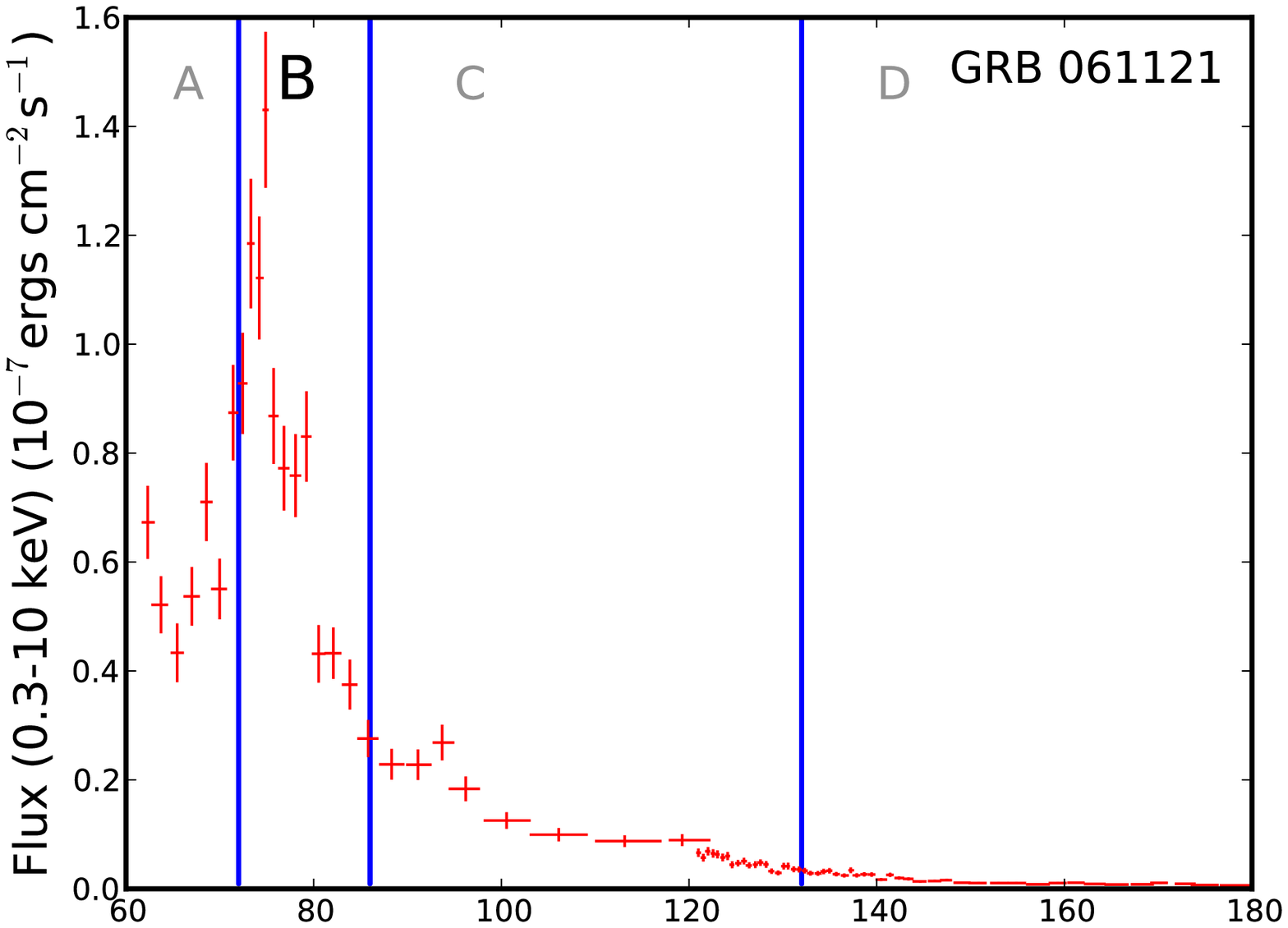}
\includegraphics[width=0.43\textwidth]{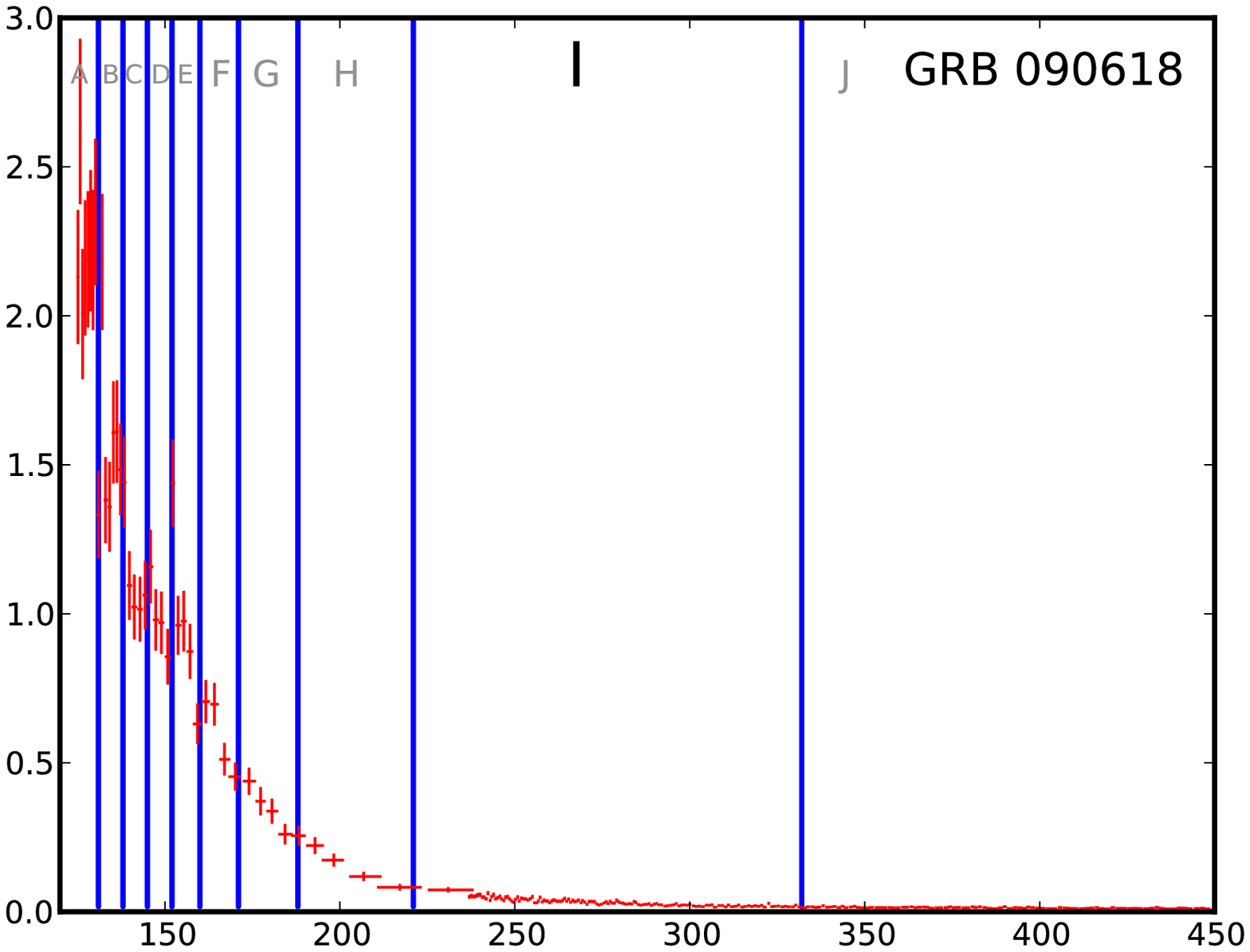}
\includegraphics[width=0.43\textwidth]{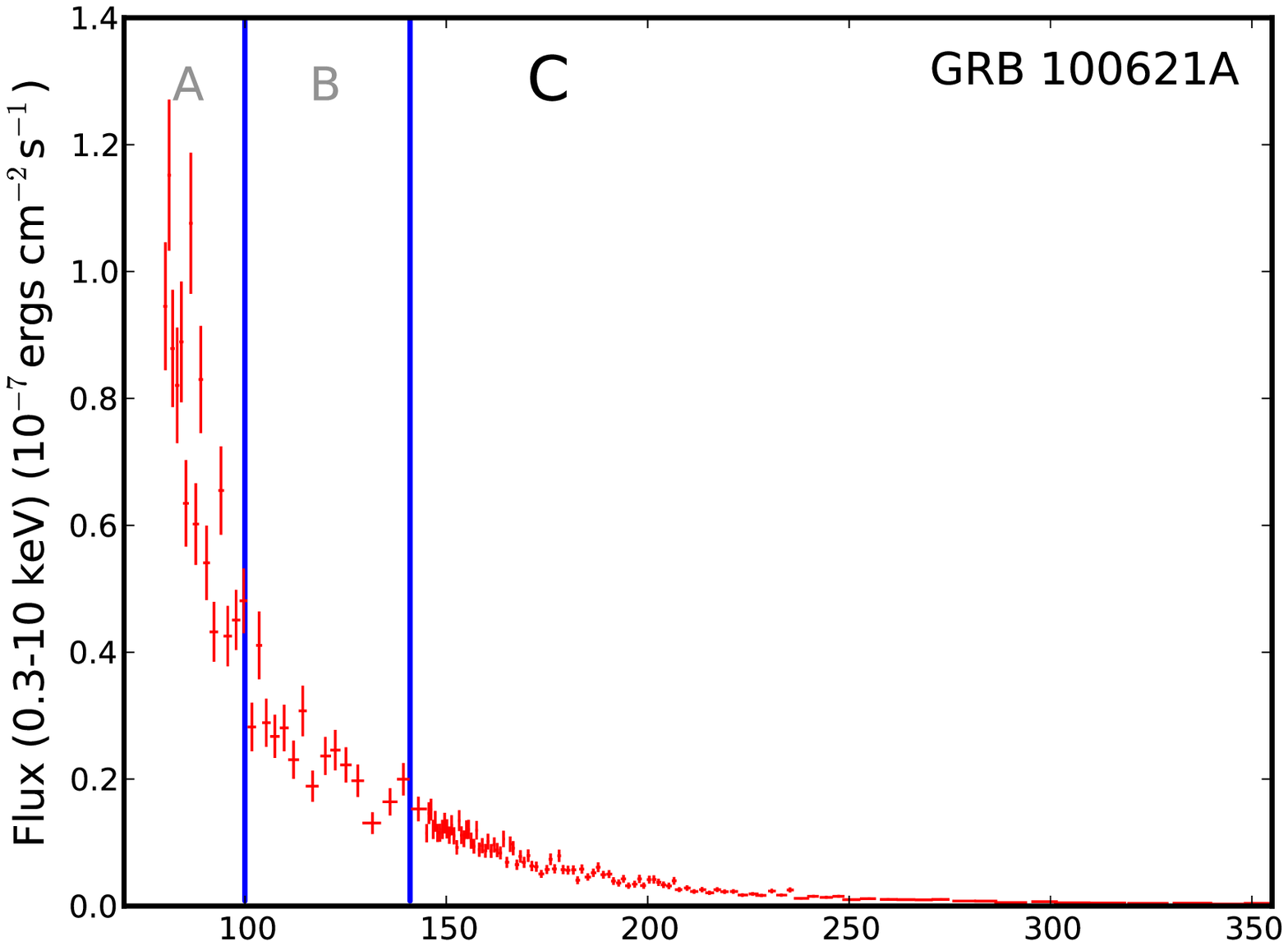}
\includegraphics[width=0.43\textwidth]{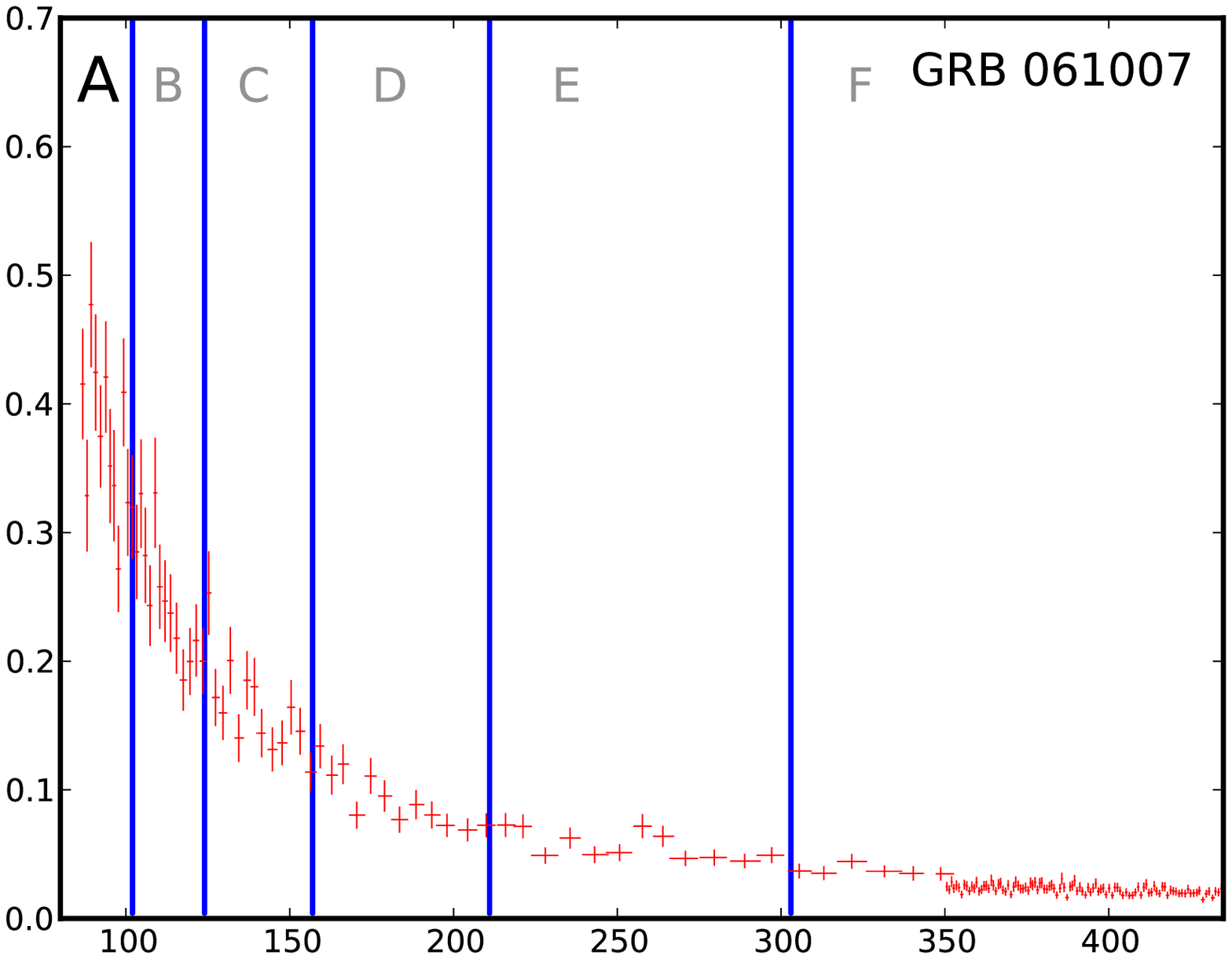}
\includegraphics[width=0.43\textwidth]{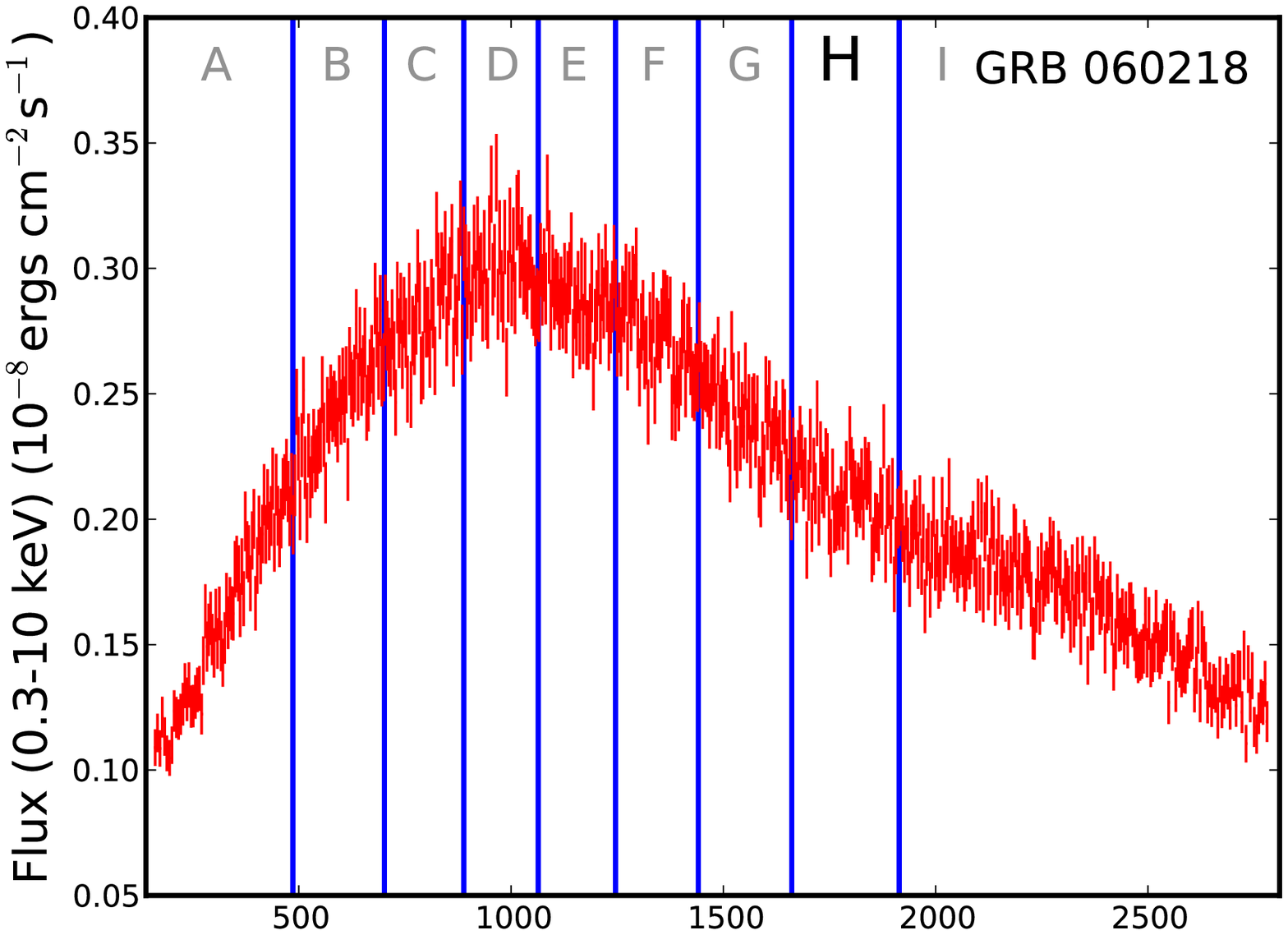}
\includegraphics[width=0.43\textwidth]{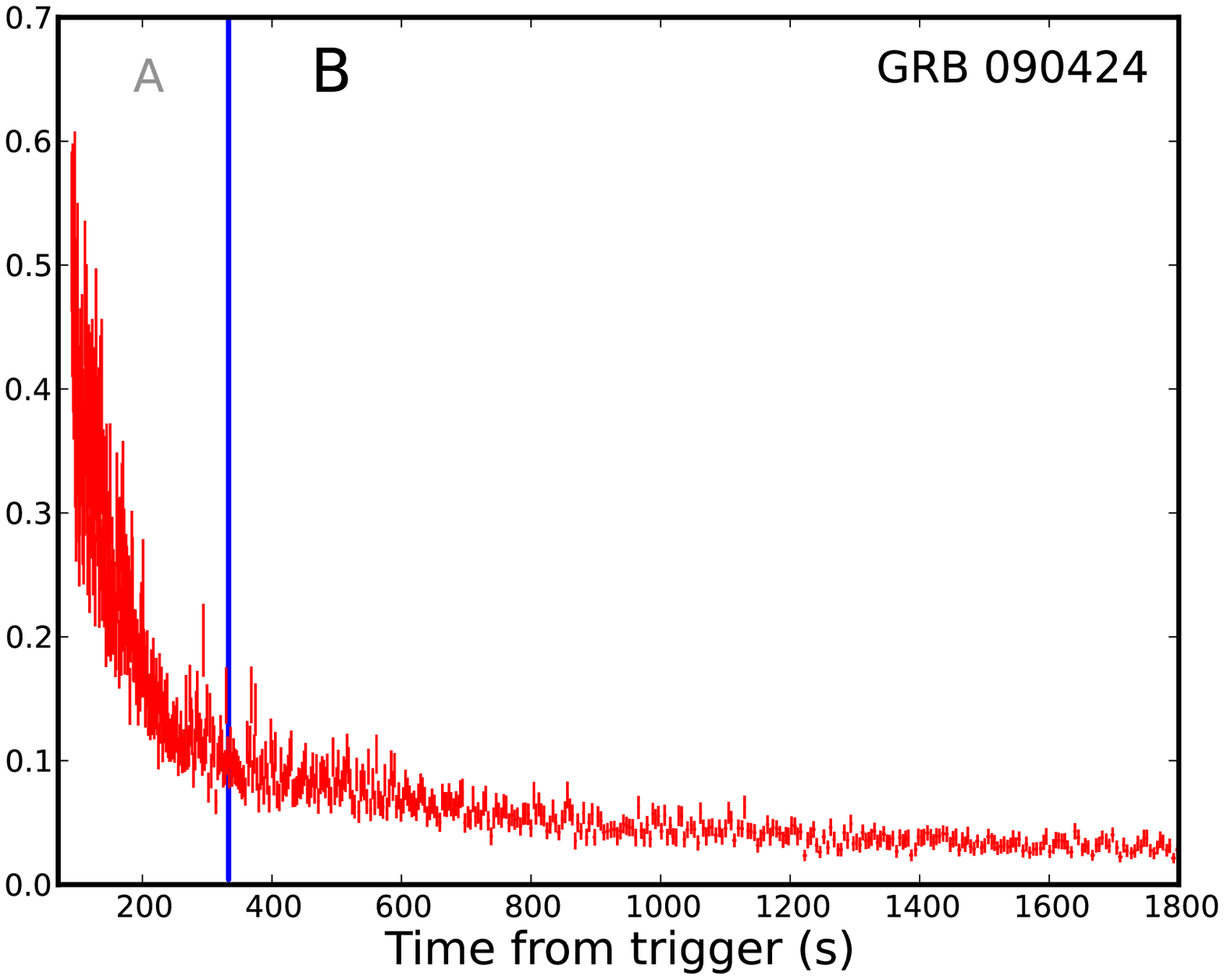}
\includegraphics[width=0.43\textwidth]{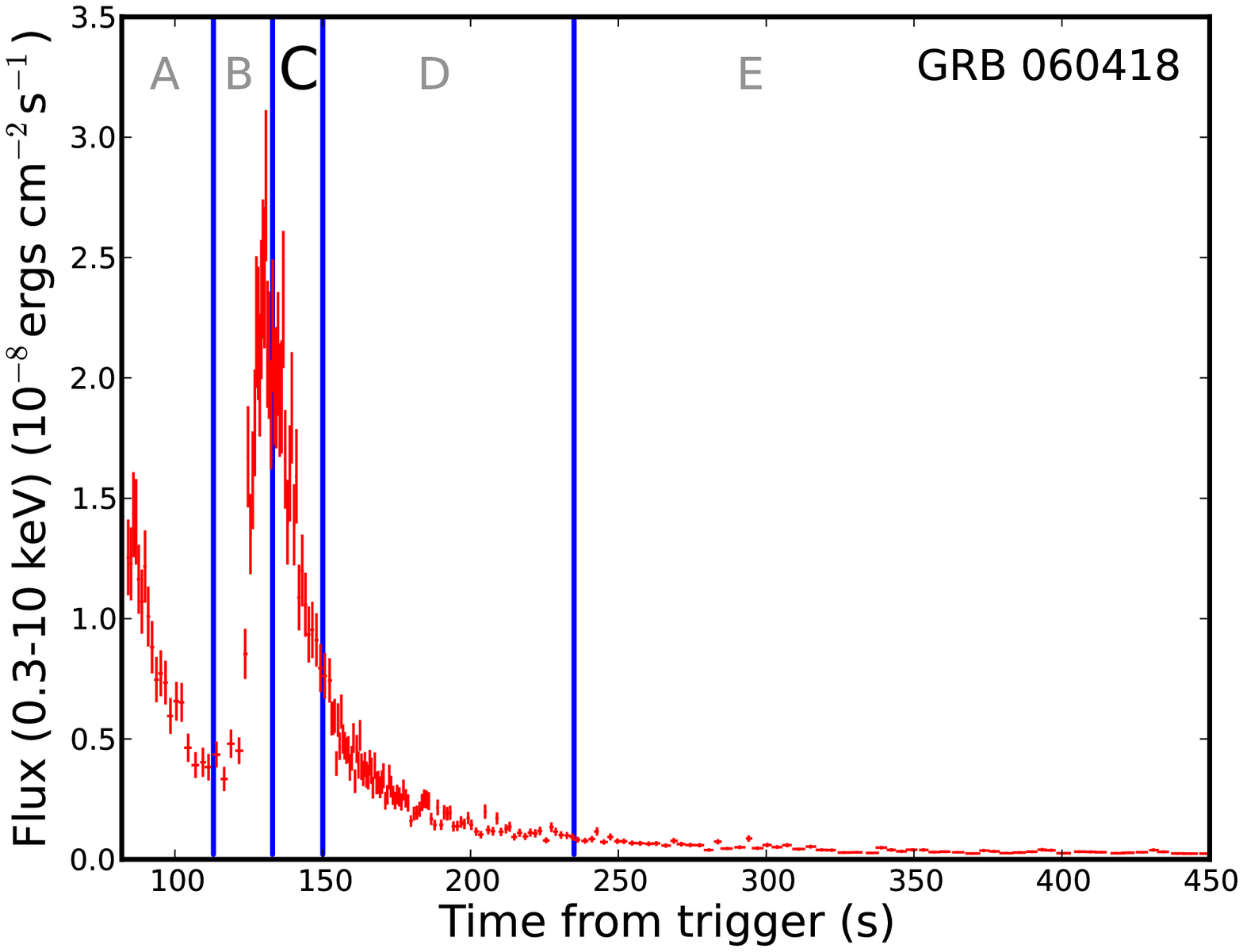}
\includegraphics[width=0.43\textwidth]{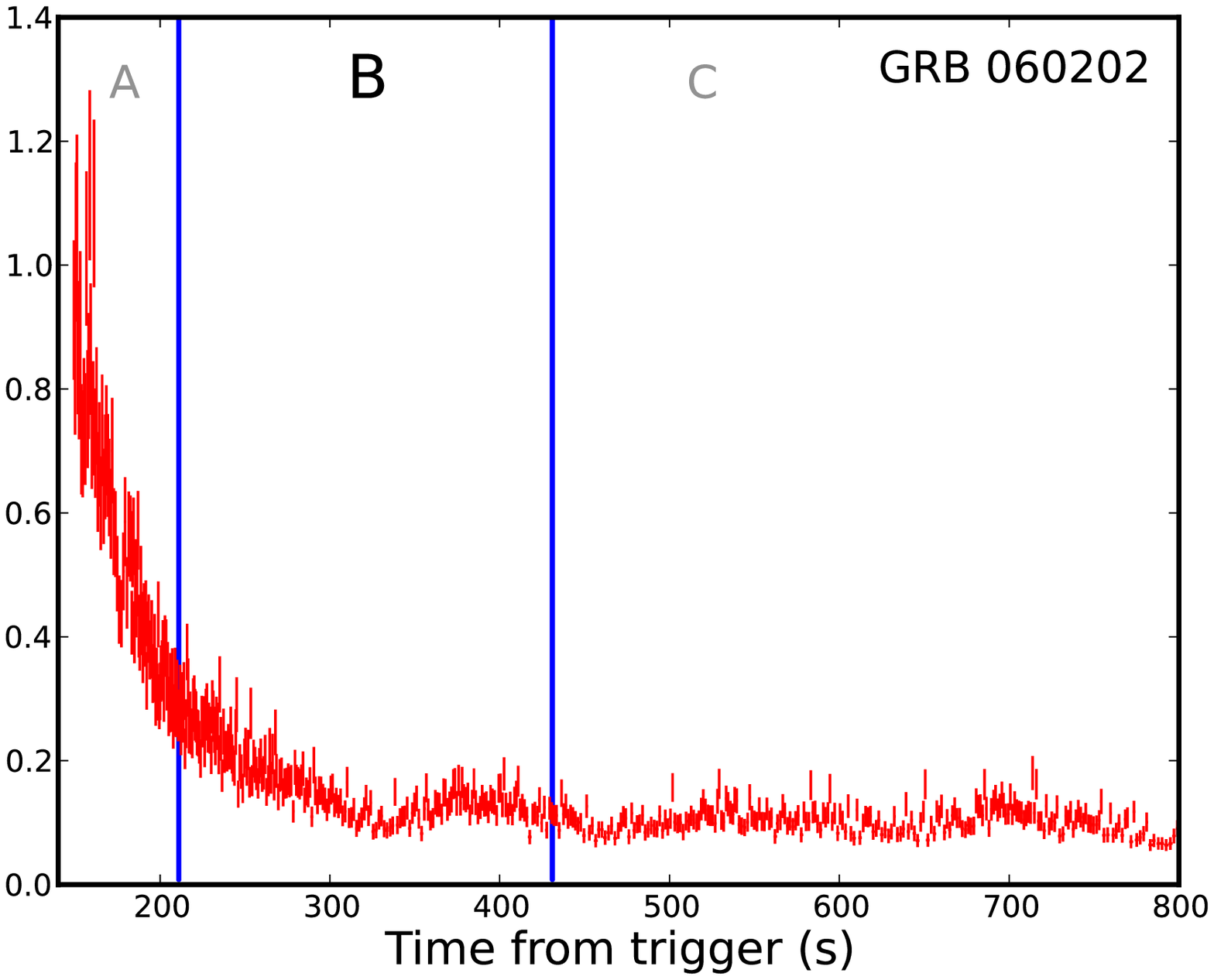}
\caption{Light curves for bursts with apparent thermal emission. The lines indicate where the data has been split into different spectra. Black letters indicate the time series with highest $\Delta\chi^2$.}
 \label{fig:grb061121_lc}
\end{figure}

In this paper, we have processed data from \emph{Swift} BAT and XRT using the analysis tools within HEASOFT. 
BAT data have been included when available and the spectrum extracted using standard procedure, binning the data in the standard energy binning.
The WT mode data have been used from the XRT observations. This has been divided into time periods with
a minimum of 10\,000 counts for each
spectrum, except for GRB\,060418 where 5\,000 counts per spectrum was used.
The XRT light-curves for bursts with a detection of a thermal component are
presented in Fig.~\ref{fig:grb061121_lc} and the time periods delineated.
This way between two and nine spectra have been extracted for each burst. For data reduction the FTOOLs \emph{Swift}-specific sub package "xrtproducts" has been used. As centre-position the output from running "xrtcentroid" on the Photon Counting (PC) mode data has been used. The response files are from the \emph{Swift} repository. The spectra have been pile-up corrected following \cite{2006A&A...456..917R}.

We should note here that this sample of the brightest early-phase
\emph{Swift}-XRT bursts only overlaps with the six candidate bursts with
possible blackbody emission of \citet{2012MNRAS.427.2965S} in GRB\,100621A
as well as GRB\,090618 and GRB\,060218 which were claimed elsewhere.  The
rest of their bursts were not bright enough to enter our sample.  Of the
bursts we claim as possible detections in Table~\ref{tab:best}, the
above-mentioned bursts are detected in common. We also find detections in
GRB\,060202, 060418, 061007, 061121, and 090424. All of these bursts,
except GRB\,061007 are noted as initial possible candidates by
\citet{2012MNRAS.427.2965S}.

%
%
\section{Results}\label{results}

The extracted spectra were fit in \texttt{Xspec} with a Band model
\citep{1993ApJ...413..281B} with photoelectric absorption from both the
Milky Way ($z=0$, fixed) and the GRB host galaxy (variable and at the
redshift of the host).  This model was compared to a similarly absorbed
Band\,+\,blackbody model.  The Galactic foreground column densities were
fixed to values from the Leiden/Argentine/Bonn (LAB) Survey of Galactic
\ion{H}{1} \citep{2005A&A...440..775K} throughout the analysis.  The
equivalent hydrogen column density based on the dust extinction column would
be somewhat different, usually slightly lower \citep{2011A&A...533A..16W}. 
Using this value would typically increase the column density inferred at the
host galaxy redshift ($N_{\rm H}$), but would not affect the other fit
parameters substantially.  All fits have been done in two steps to reduce
the computational cost to a manageable level.  First the spectra for a given
burst were fit simultaneously to determine $N_{H}$ for the host galaxy.  The
absorption was then frozen to this value during individual fitting of the
spectra.  $N_{H}$ for the host was determined separately for the Band and
the Band\,+\,blackbody models.  The optical spectroscopic redshifts of the
GRBs were used to calculate model parameters in the host galaxy rest frame. 
Fit statistics and parameters of the best-fit blackbody for the time series
with the most significant blackbody detection for each burst in the sample
can be seen in Table~\ref{tab:best}.  The temperature and the
luminosity\footnote{The luminosity is given by the blackbody normalisation
in the xspec model: \[norm_{bb} =\frac{L_{39}}{D^2_{10}} =
\frac{L_{39}}{D_{L(10)}^2\,(1+z)^2}\] Here $L_{39}$ is the luminosity in
units of 10$^{39}$\,ergs/s, and $D_{10}/D_{L(10)}$ is the proper
motion/luminosity distance to the source in units of 10\,kpc.} were determined
from the fits.  Using these parameters and the Stefan-Boltzmann eq., the
apparent radiative surface area was determined, assuming a simple,
non-relativistic blackbody.  Under spherical geometry, the radius would then
be: $L = \sigma\,A\,T^4$, where $A = 4 \pi\,R^2$.

\begin{landscape}

\begin{table}
\caption{Blackbody parameters for the best fit time series for each burst in the sample. \\ Parameters have only been included if the improvement for the addition of a blackbody is better than $\Delta \chi^2$\,=\,25.}
\small{
\begin{tabular}{@{} p{1.3cm}  p{1.1cm} p{0.7cm} p{0.7cm} l p{1.0cm} p{2.1cm} p{1.6cm}  p{1.3cm} c c c c c c c c @{}}
\hline\hline
GRB & Redshift & refs. & $\Delta \chi^2$ & bb lum.$^a$  & bb \%$^b$ & kT/keV & R$_{phot}$$^c$ & $\gamma$ & Time$^d$ & Time Series$^e$ \\ 
\hline

060202 & 0.783 & (1) & 28.1 & $2.7^{+0.1}_{-0.2}$ & 13 &  $0.38^{+0.01}_{-0.02}$ & $5.9^{5.1}_{5.2}$ & $60^{+14}_{-18}$ & 211--431 & B \\ 

060218 & 0.0331 & (2) & 73.4 & $0.0116^{+0.0007}_{-0.0006}$ & 0.24 & $0.156\pm{0.004}$ & $0.46^{+0.45}_{-0.46}$ & $40^{+9}_{-12}$ & 1661--1914 & H \\ 

060418 & 1.489 & (3) & 33.9 & $1.6^{+9}_{-0.7}$ & 3.5 & $0.53\pm{0.02}$ & $2.2^{+7.7}_{-1.2}$ & $<$254 & 150--235 & D \\ 

061007 & 1.262 & (4) & 36.1 & $119^{+12}_{-11}$ & 10 & $3.2^{+0.4}_{-0.3}$ & $1.9^{+0.17}_{-0.18}$ & $328^{+100}_{-64}$ & 86--102 & A \\ 

061121 & 1.314 & (4) & 49.4 & $257\pm{26}$ & 0.74 & $2.9\pm{0.2}$ & $6.7\pm{1.6}$ & $669^{+14}_{-18}$ & 72--86 & B \\ 

090424 & 0.544 & (5) & 49.8 & $0.16^{+0.01}_{-0.04}$ & 27 & $0.228\pm{0.006}$ & $19^{+23}_{-17}$ & $26^{+8}_{-9}$ & 333--5554 & B \\ 

090618 & 0.54 & (6) & 46.9 & $1.81^{+0.08}_{-0.08}$ & 17 & $0.74^{+0.08}_{-0.06}$ & $12\pm{2}$ & $<$1058 & 138--145 & C \\ 

100621A & 0.542 & (7) & 36.5 & $0.73^{+0.07}_{-0.09}$ & 23 & $0.38^{+0.39}_{-0.36}$ & $2.8^{+2.6}_{-2.6}$ & $40^{+11}_{-10}$ & 141--40540 & C \\ 

\hline

060124 & 2.300 & (4) & 11.9 & $<$25 & $<$0.25 & --- & --- & --- & 557--635 & B \\ 

060510B & 4.9 & (8) & 9.26 & $<$136 & $<$38 & --- & --- & --- & 127--252 & A \\ 

060526 & 3.221 & (9) & 8.17 & $<$8.8 & $<$4.1 & --- & --- & --- & 81.5--282 & A \\ 

060614 & 0.125 & (10) & 14.2 & $<$0.14 & $<$15 & --- & --- & --- & 97--109 & A \\ 

060729 & 0.543 & (4) & 11.6 & $<$37 & $<$23 & --- & --- & --- & 130--146 & A \\ 

060814 & 1.92 & (11) & 24.0 & $<$0.25 & $<$52 & --- & --- & --- & 168--75055 & C \\ 

060904B & 0.703 & (4) & 15.5 & $<$1.8 & $<$33 & --- & --- & --- & 77--178 & A \\ 

071031 & 2.692 & (4) & 19.3 & $<$70 & $<$22 & --- & --- & --- & 109--178 & A \\ 

080310 & 2.42 & (12) & 14.5 & $<$8.6 & $<$3.1 & --- & --- & --- & 215--287 & B \\ 

080319B & 0.937 & (13) & 20.1 & $<$7.6 & $<$0.05 & --- & --- & --- & 90--120 & B \\ 

080607 & 3.036 & (14) & 7.45 & $<$222 &$<$18 & --- & --- & --- & 123--144 & B\\ 

080928 & 1.692 & (4) & 19.1 & $<$22 & $<$8.0 & --- & --- & --- & 210--248 & B\\ 

081008 & 1.9685 & (15) & 14.7 & $<$35810 & $<$100 & --- & --- & --- & 94--144 & A \\ 

081028 & 3.038 & (16) & 8.60 & $<$3.5 & $<$14 & --- & --- & --- & 341--61920 & B \\ 

090417B & 0.345 & (17) & 2.87 & $<$0.1 & $<$1.8 & --- & --- & --- & 702--1497 & B \\ 

090516A & 4.109 & (18) & 14.6 & $<$325 & $<$46 & --- & --- & --- & 171--136 & A \\ 

090715B & 3.0 & (19) & 9.33 & $<$82 & $<$7.7 & --- & --- & --- & 219--274 & C \\ 

100814A & 1.44 & (20) & 9.17 & $<$41 & $<$3.6 & --- & --- & --- & 94--157 & A \\ 

110205A & 2.22 & (21) & 12.0 & $<$76 & $<$7.1 & --- & --- & --- & 273--325 & D \\ 

110801A & 1.858 & (22) & 18.5 & $<$57 & $<$13 & --- & --- & --- & 380--424 & C \\ 

\hline
\end{tabular}
}
\\
\footnotesize{
$^a$ Blackbody luminosity in $10^{48}$\,ergs\,s$^{-1}$,
$^b$ Percent of total luminosity in the thermal component \\
$^c$ Photospheric radius in $10^{13}$\,cm,
$^d$ Seconds since BAT trigger,
$^e$ Part of light curve that contains best fit improvement \\
References: (1) \cite{2007ApJ...656.1001B}; (2) \cite{2006GCN..4792....1M}; (3) \cite{2006ApJ...648L..93P}; (4) \cite{2009ApJS..185..526F}; (5) \cite{2009GCN..9243....1C}; (6) \cite{2009GCN..9518....1C}; (7) \cite{2010GCN..10876...1M}; (8)\cite{2006GCN..5104....1P}; (9) \cite{Jakobsson2006}; (10) \cite{2006Natur.444.1050D}; (11) \cite{2012ApJ...752...62J}; (12) \cite{2012A&A...545A..64D}; (13) \cite{2008Natur.455..183R}; (14) \cite{2009ApJ...691L..27P}; (15) \cite{2008GCN..8350....1D}; (16) \cite{2008GCN..8434....1B}; (17) \cite{2009GCN..9156....1B}; (18) \cite{2009GCN..9383....1D}; (19) \cite{GCN9673};   (20) \cite{2010GCN..11089...1O}; (21) \cite{2011GCN..11638...1C}; (22) \cite{2011GCN..12234...1C}}
\label{tab:best}
\end{table}

\end{landscape}

In Table~\ref{tab:best}, the $\Delta\chi^2$ values are given for the two models. The actual probabilities as seen in Table~\ref{tab:fit} are inferred from a Monte Carlo analysis, and gives the likelihood that the added blackbody is just a better fit per random chance. As can be seen from Table~\ref{tab:best} the spectral fit seems generally to be improved with the addition of a blackbody. The blackbody model only adds two free parameters, so the fact that $\chi^2$ decrease by more than a factor of two in all but one burst indicates that an extra component is needed in addition to the Band function. 


To test the statistical significance of the fit improvement, we used a Monte
Carlo method, generating 10\,000 artificial spectra from the single Band
function model.  We did this for every burst with a large improvement in
$\chi^2$ ($\Delta\chi^2\geq25$, the 3\,$\sigma$ limit from the first three
time series analysed), since it is extremely unlikely, that a small
$\Delta\chi^2$ value will turn out to be significant, and because the Monte
Carlo analysis is computationally expensive.  The distribution of $\chi^2$
was found to be the same for different time series in the same burst, but
not compatible between different bursts using the Kolmogorov-Smirnoff test.  For
bursts with no match in $\Delta\chi^2$ in the first 10\,000 spectra, a
further 10\,000 were done to improve statistics.  The detection probabilites
for an additional blackbody component resulting from these simulations are
provided in Table~\ref{tab:fit} for each time series. We adopt a thermal component
detection criterion of less than one in 10\,000 over the null hypothesis in
at least one of the time series. This corresponds to a detection probability
of better than $\sim4\sigma$.

Early afterglow thermal emission has already been reported for GRB\,060218
\citep{2006Natur.442.1008C}, and extensively analysed
\citep{2007ApJ...667..351W, 2007MNRAS.375..240L, 2007MNRAS.375L..36G, 2007MNRAS.382L..77G, 2006A&A...454..503S}.  The detection of thermal emission in
GRB\,090618 \citep{2011MNRAS.416.2078P} and GRB\,100621A
\citep{2012MNRAS.427.2965S} has also been reported, though not as
extensively, and so we include our analysis of these bursts here for
comparison.  In addition to these bursts we report the detection of thermal
emission in GRBs 060202, 060418, 061007, 061121, and 090424.

\subsection{GRB\,061007}\label{grb061007}

The spectral fitting for GRB\,061007 is improved for the added blackbody for
time series A to $>4\sigma$ confidence (that is, the Monte Carlo analysis
found no $\Delta\chi^2$ match in 20\,000 simulated spectra).  We also
performed 10\,000 simulations separately for time series B, where 0.16\%
(approximately $3\sigma$ confidence) of the faked data had a $\Delta\chi^2$
as large or larger than the real one.  The component is very bright,
accounting for respectively 10 and 11\% of the total luminosity in series A
and B.  With luminosities of up to $1.9\times10^{50}$\,ergs\,s$^{-1}$ as
well as blackbody temperatures of several keV, this thermal emission is both
more luminous and hotter than those previously reported
\citep{2006Natur.442.1008C,2011MNRAS.416.2078P,2012MNRAS.427.2965S}.  The
spectra are plotted in Fig.~\ref{fig:spectra}, top panel.

\subsection{GRB\,090424}

The thermal emission from GRB\,090424 looks more like the ``classical"
thermal component, with low luminosity and temperatures of $\sim0.2$\,keV,
similar to, for instance, GRB\,060218.  We performed the Monte Carlo
analysis for both series separately.  For time series B no $\Delta\chi^2$
match was found in 20\,000 spectra.  For the time series A about 0.2\%
($>3\sigma$) match the real value.  Compared to the burst's low total
luminosity, the thermal emission is very bright, constituting about one
fourth of the total luminosity.  The spectra can be seen in
Fig.~\ref{fig:spectra}, second panel.

\subsection{GRB\,061121, 060202 and 060418}\label{rest}

\begin{figure*}
\centering
\includegraphics[bb=70 16 555 732, angle=-90, width=0.35\textwidth]{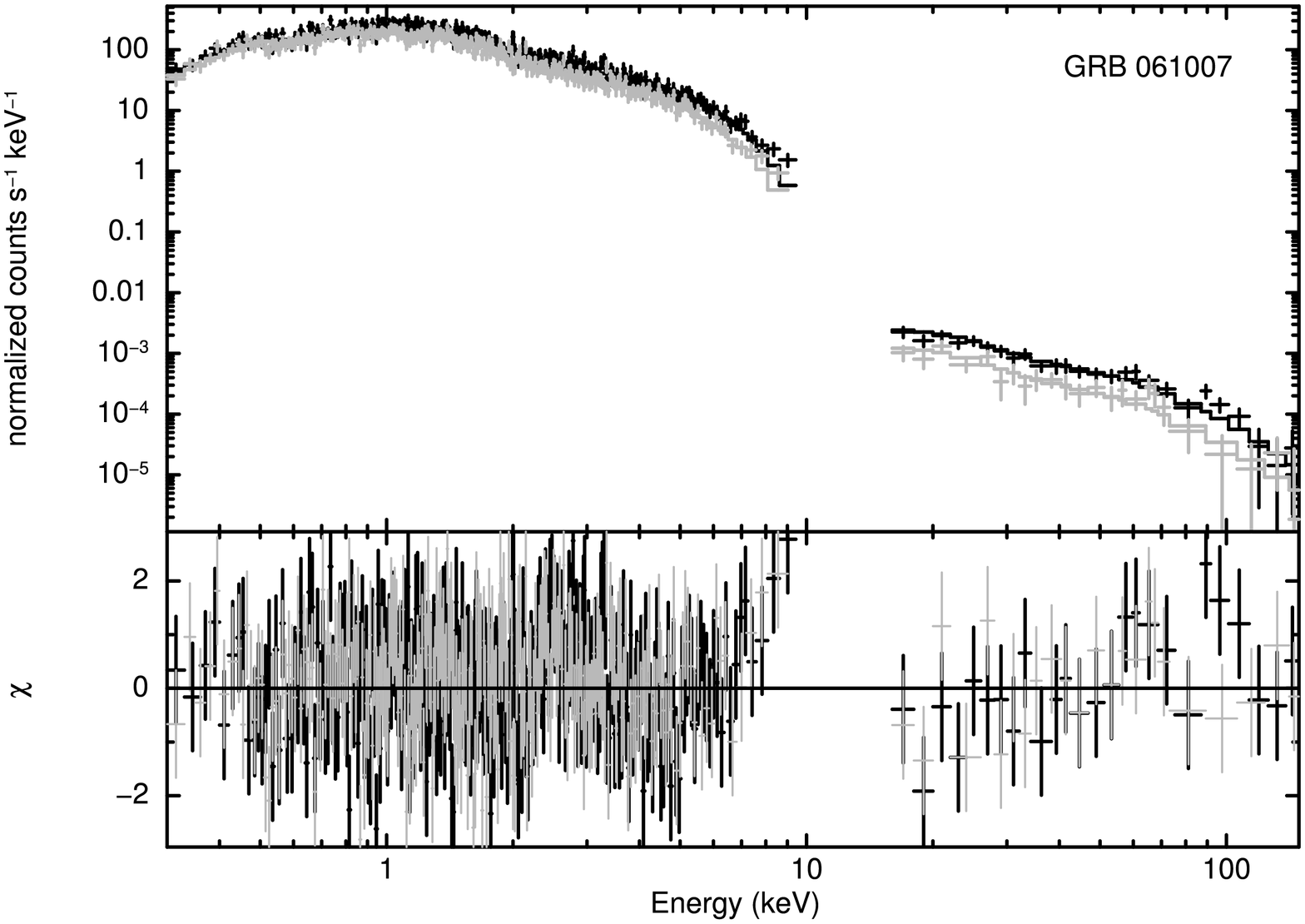}
\includegraphics[bb=70 16 555 732, angle=-90, width=0.35\textwidth]{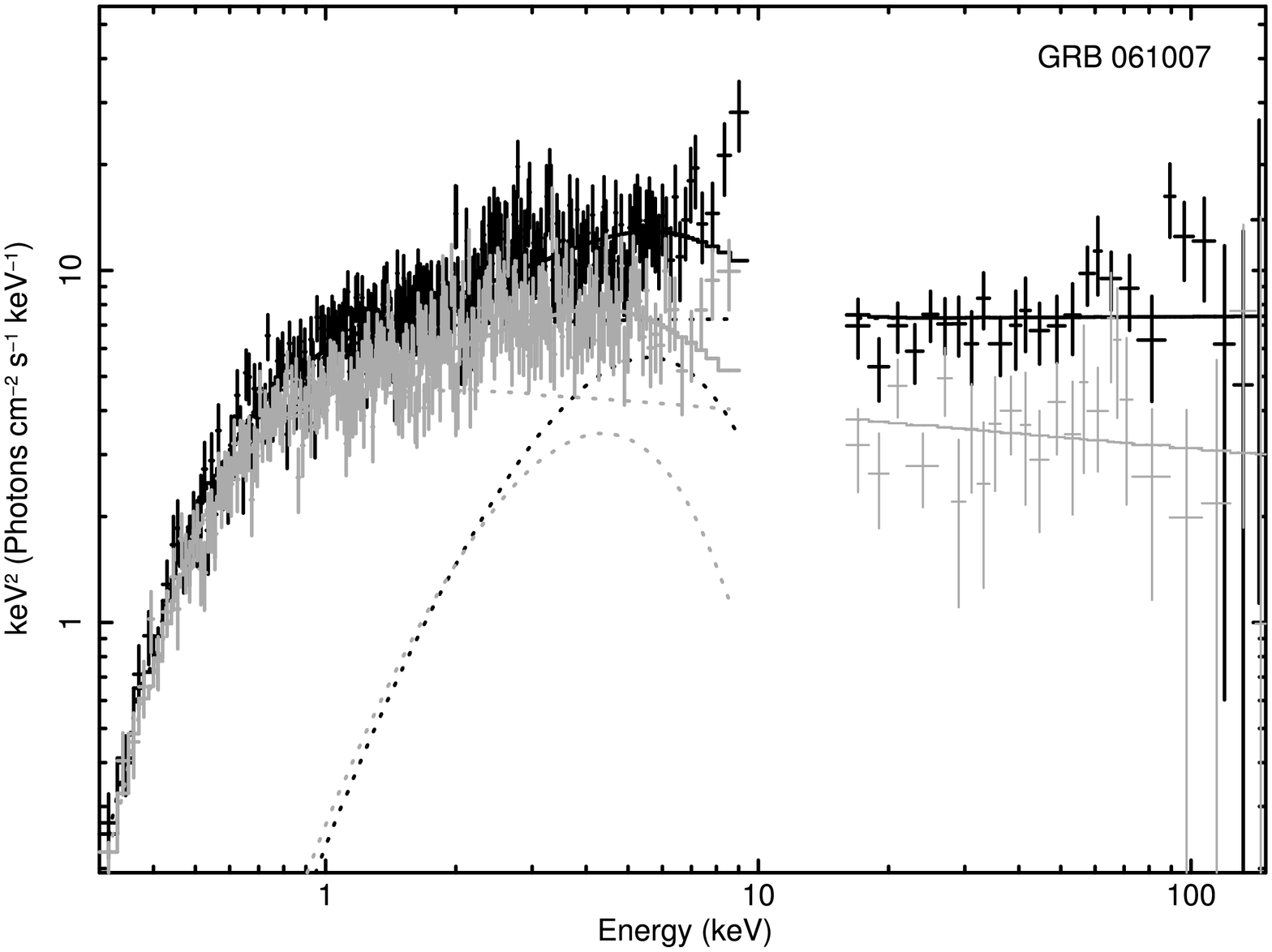}
\includegraphics[bb=70 16 575 732, angle=-90, width=0.35\textwidth]{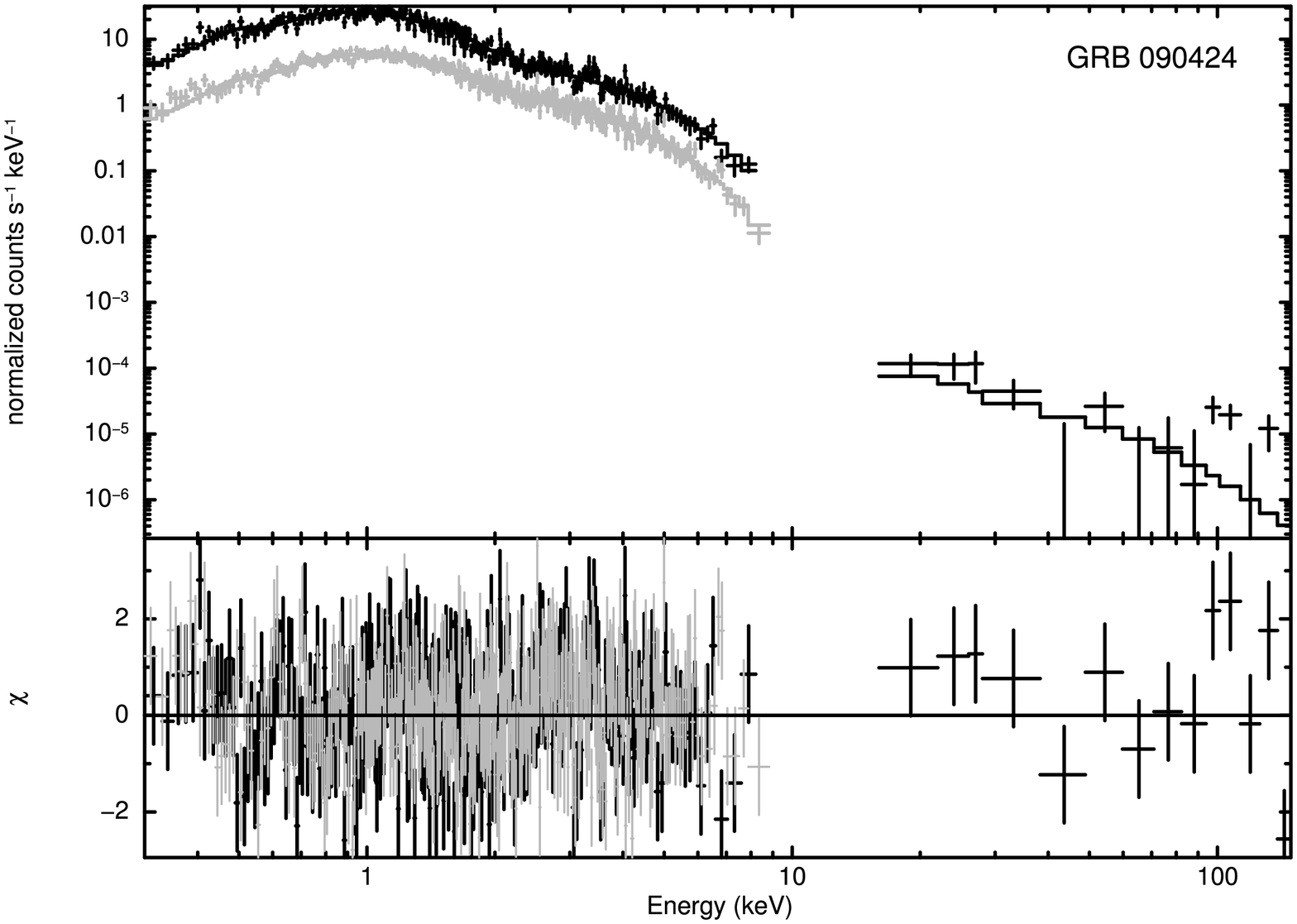}
\includegraphics[bb=70 16 575 732, angle=-90, width=0.35\textwidth]{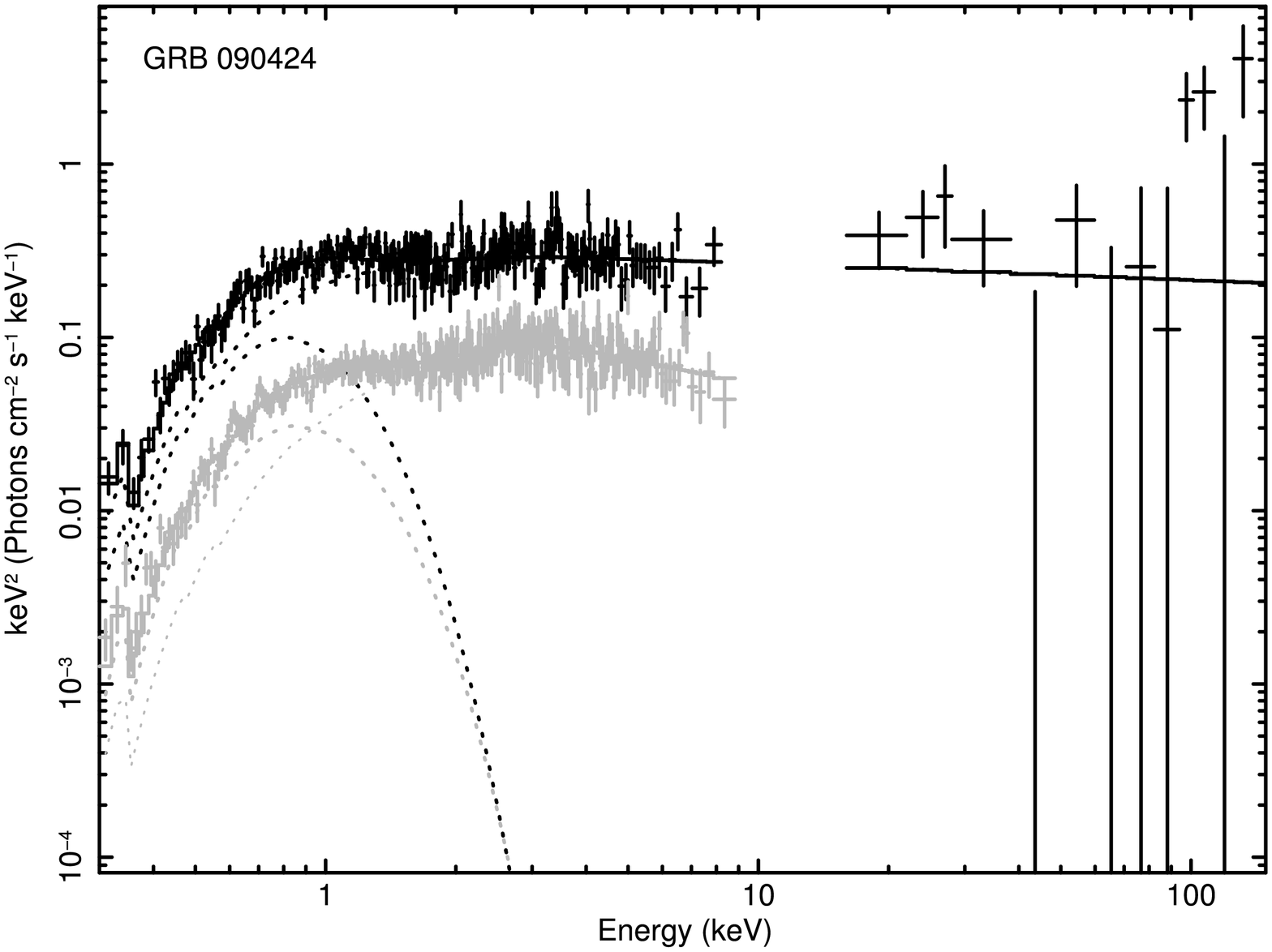}
\includegraphics[bb=70 16 575 732, angle=-90, width=0.35\textwidth]{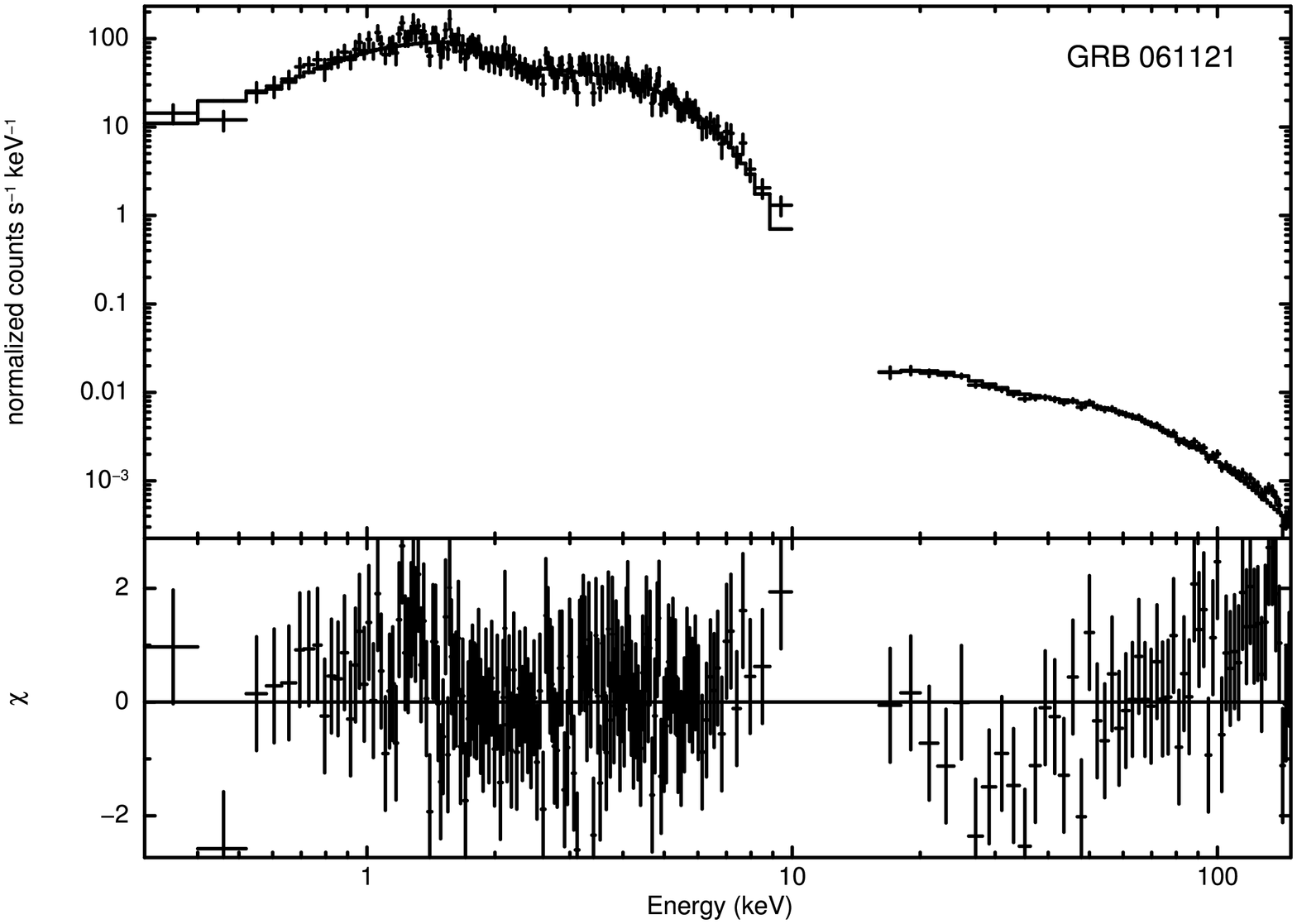}
\includegraphics[bb=70 16 575 732, angle=-90, width=0.35\textwidth]{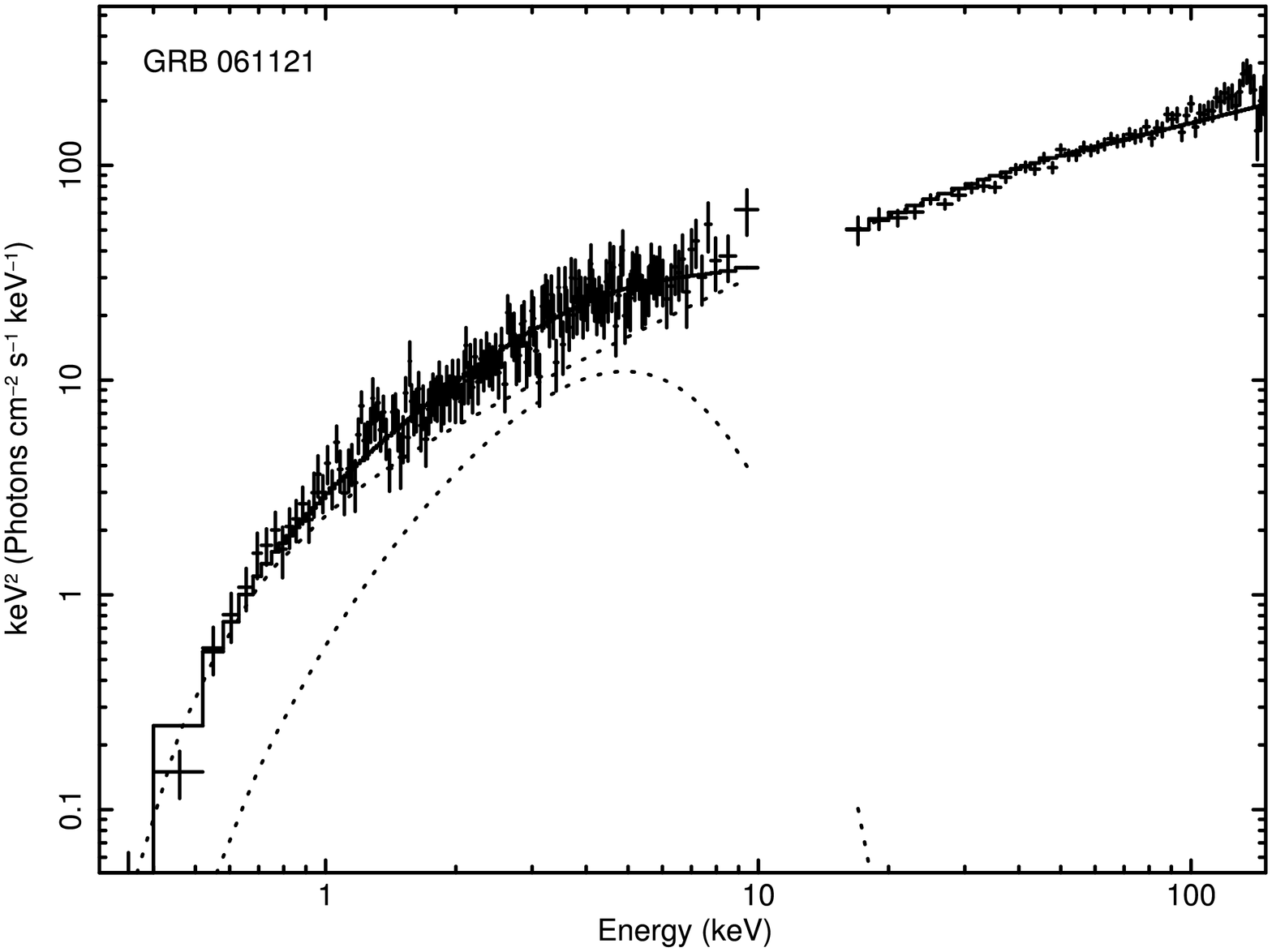}
\includegraphics[bb=70 16 575 732, angle=-90, width=0.35\textwidth]{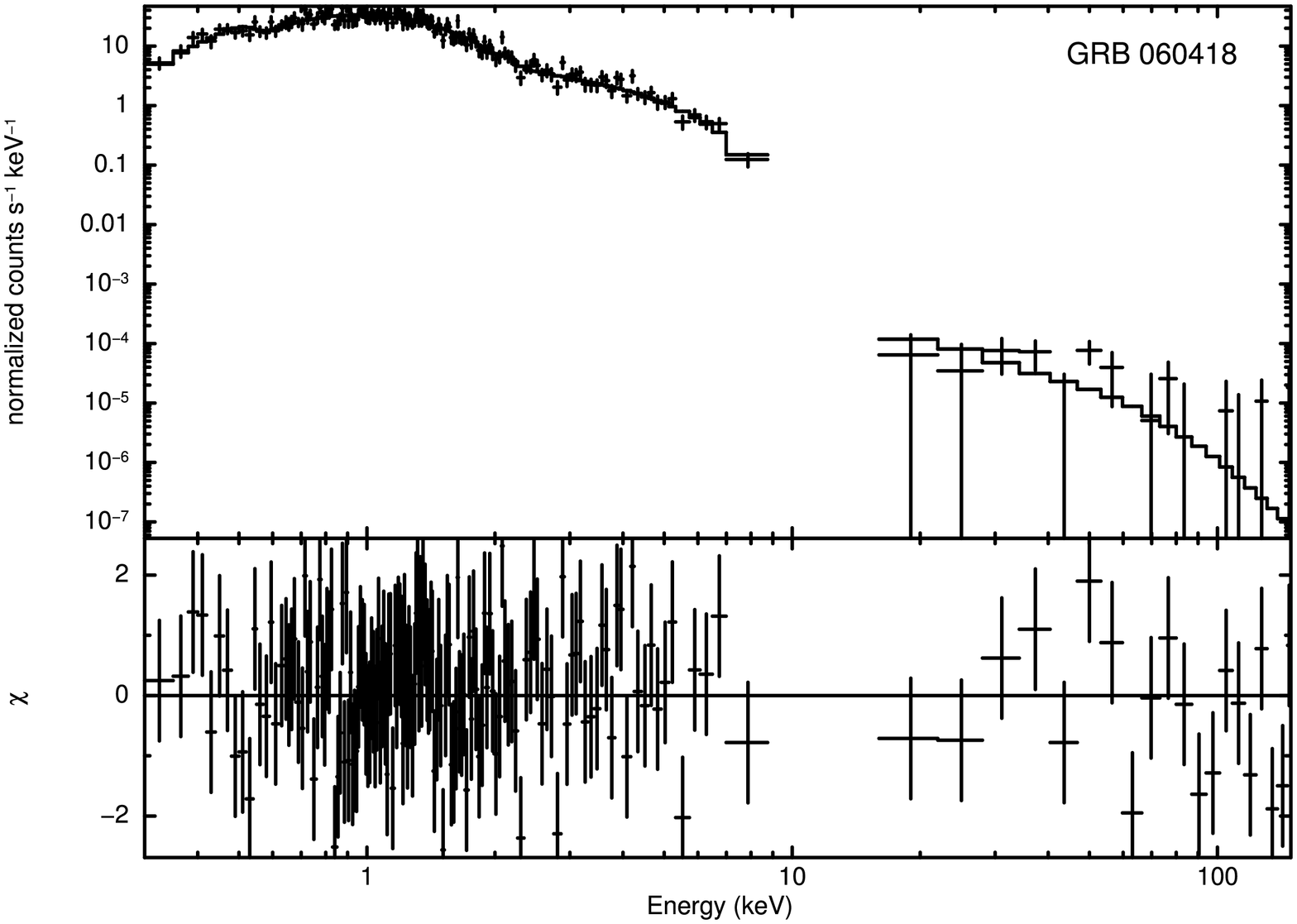}
\includegraphics[bb=70 16 575 732, angle=-90, width=0.35\textwidth]{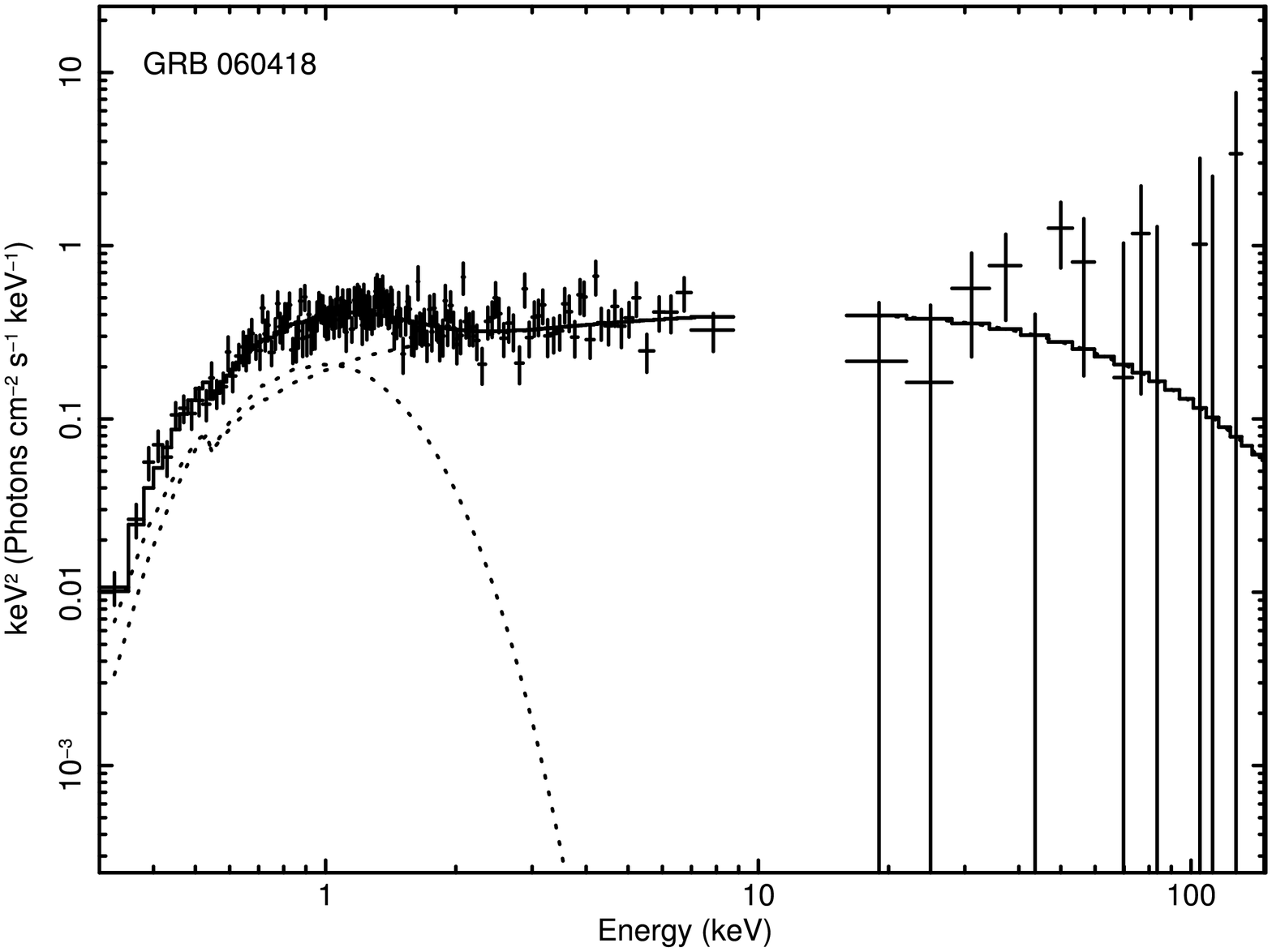}
\includegraphics[bb=70 16 470 732, angle=-90, width=0.35\textwidth]{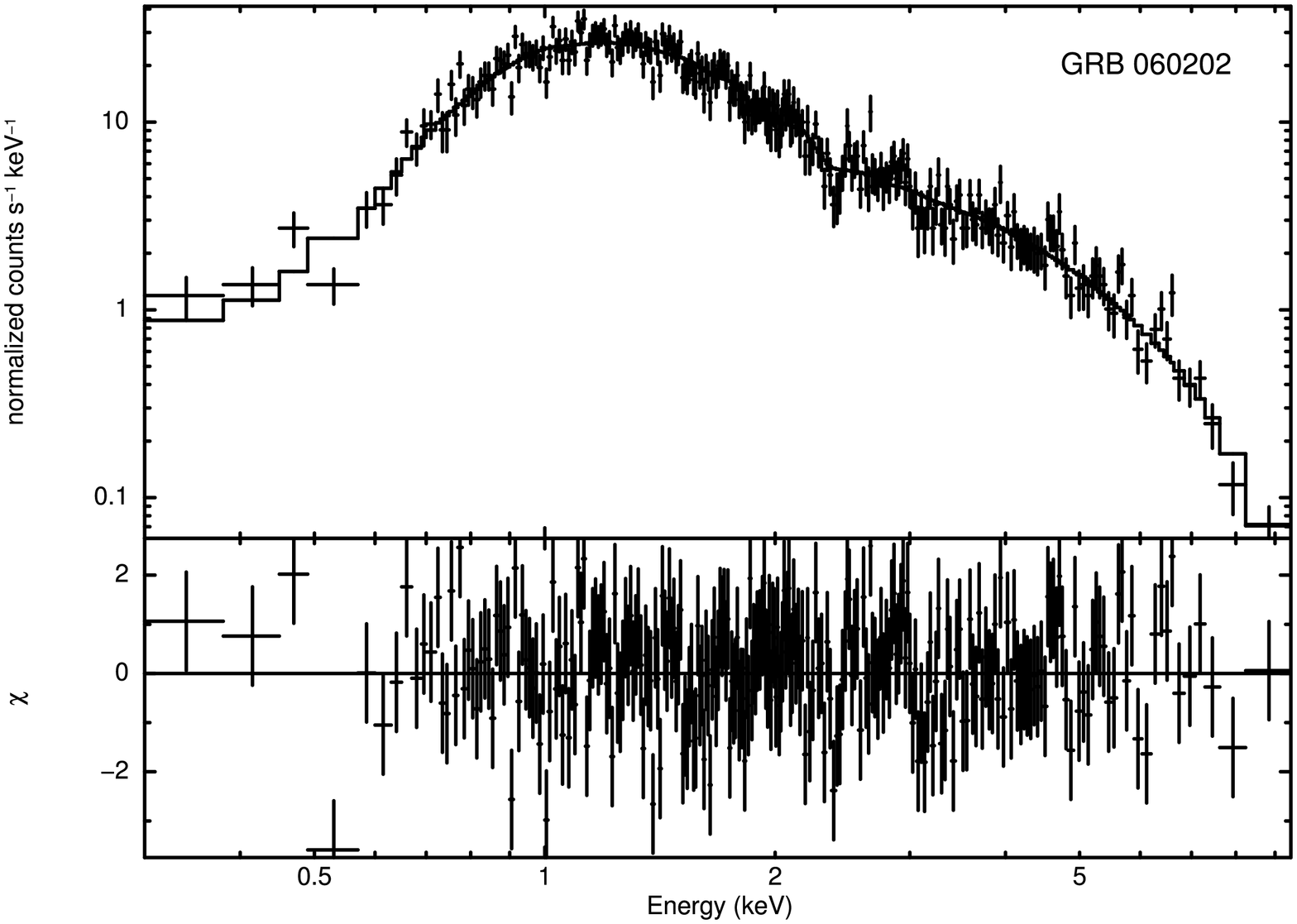}
\includegraphics[bb=70 16 470 732, angle=-90, width=0.35\textwidth]{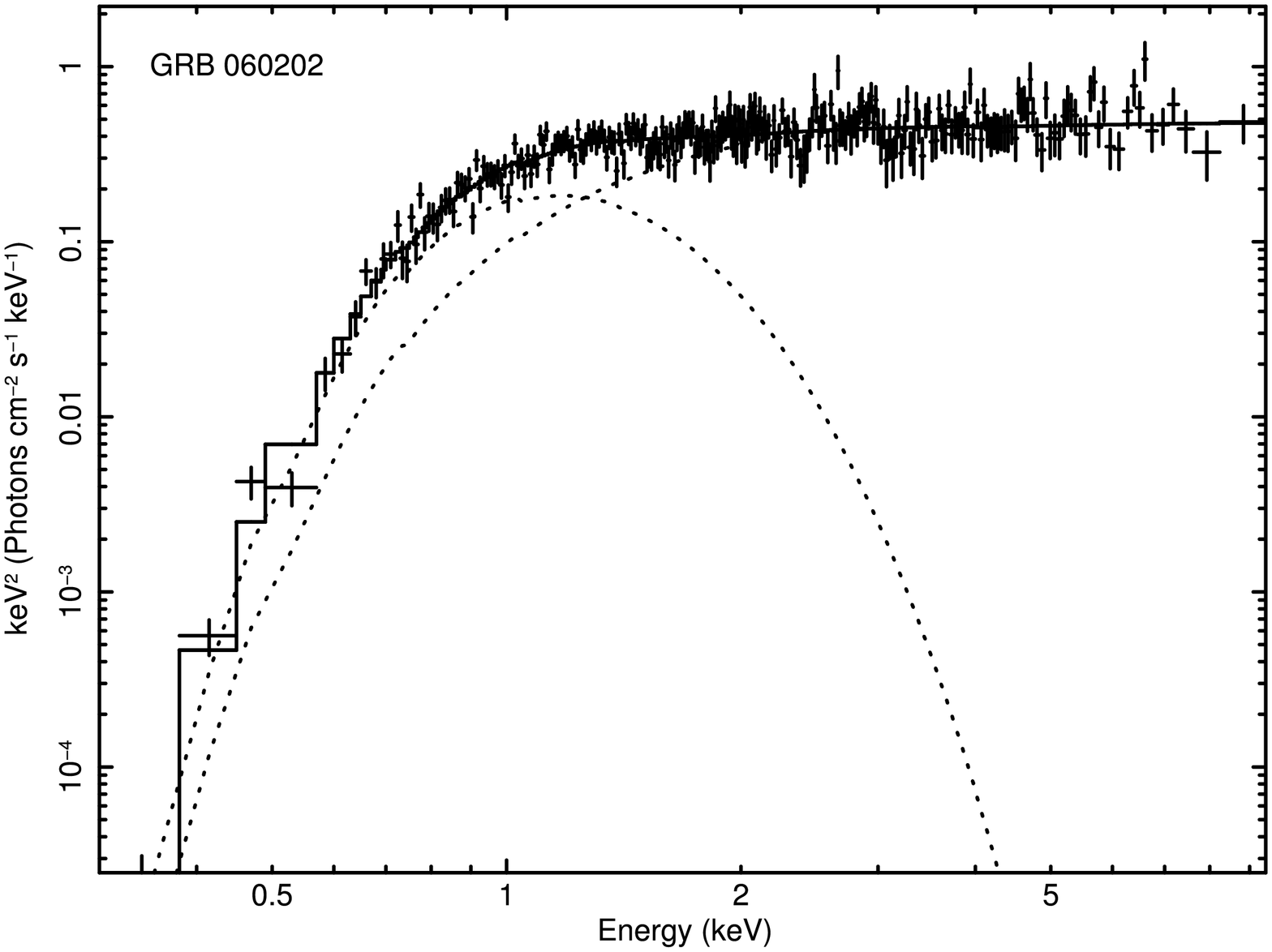}
\caption{Spectra for GRBs 061007 (time series A and B), 090424 (time series A and B), 061121 (time series B), 060418 (time series C) and 060202 (time series B) respectively. Left panel shows the spectra including the residuals, right panel shows the unfolded spectra. For GRBs 061007 and 090424 black lines show time series A, while grey lines show time series B.}
 \label{fig:spectra}
\end{figure*}

GRBs 061121, 060202 and 060418 are only good candidates for thermal emission
in one time series.  Spectra can be seen in Fig.~\ref{fig:spectra}. 
We note that the parameters for GRB\,060418 are not very well constrained.  All
parameters for the detections are similar, with temperatures around 1\,keV,
luminosities corresponding to few percent of the total, and apparent
host-frame blackbody radii around $10^{11}$--$10^{12}$\,cm.
The general tendency seems to be for the thermal component to 
be detected in the declining phase after a flare
(though for GRB\,061121 the time series includes the peak of the flare), see Fig~\ref{fig:grb061121_lc}. 

\subsection{Previous detections}

Our statistical analysis is relatively conservative as we require a
$>4\sigma$ detection.  Furthermore, we use a Band model as our underlying
continuum, which allows for a certain curvature in the spectra without
invoking an additional component.  For example for a burst such as
GRB\,090618 where thermal emission has been claimed previously, we only have
a clear detection by our criteria in the very late phase.  For our
claimed detections, only GRB\,061007 is not noted to possibly have a thermal component (i.e.\ to have a significant deviation from a single power-law) in the analysis of
\citet{2012MNRAS.427.2965S}.  However, this may not be surprising, since we
find a highly significant detection only in the first epoch and the signal
may be washed out over the rest of the burst.  GRBs 060202, 060418, 061121,
and 090424 are all flagged as initial possible candidates in
\citet{2012MNRAS.427.2965S}, but are excluded from their final list for a variety of reasons. We examine these reasons below and discuss why these GRBs have thermal components.  GRBs\,060202 and 061121 were excluded because the
significance of their detections was too sensitively dependent on the column
density used in the fits.  For our analysis, we do not find that varying our
column density, which is based on a simultaneous fit of all the data, has a
notable effect on the statistical significance of the detection for these
bursts. Within a 90\,\% c.l $\Delta\chi^2$ varies with a maximum of 3, while the temperature 
for the blackbody fit remains constant within errors. 
Furthermore it was noted that GRB\,061121 had a high temperature and redshift.  We do not include the
temperature or redshift attributes of the thermal components in our search
criteria.  GRB\,060418 was excluded by \citet{2012MNRAS.427.2965S} because
of strong flaring activity in the light curve.  We do not exclude flaring
light curves in our analysis a priori since we fit the data with a Band model
as the underlying continuum and include the BAT data in our analysis where
it is available.  Finally, GRB\,090424 was found to have a low statistical
significance in their time-sliced spectra.  Our detection comes from a
spectrum over a much wider time range (333--5554\,s compared to
700--1000\,s) and hence has a far higher signal-to-noise ratio.

\begin{figure}[!ht]
\includegraphics[width=\columnwidth]{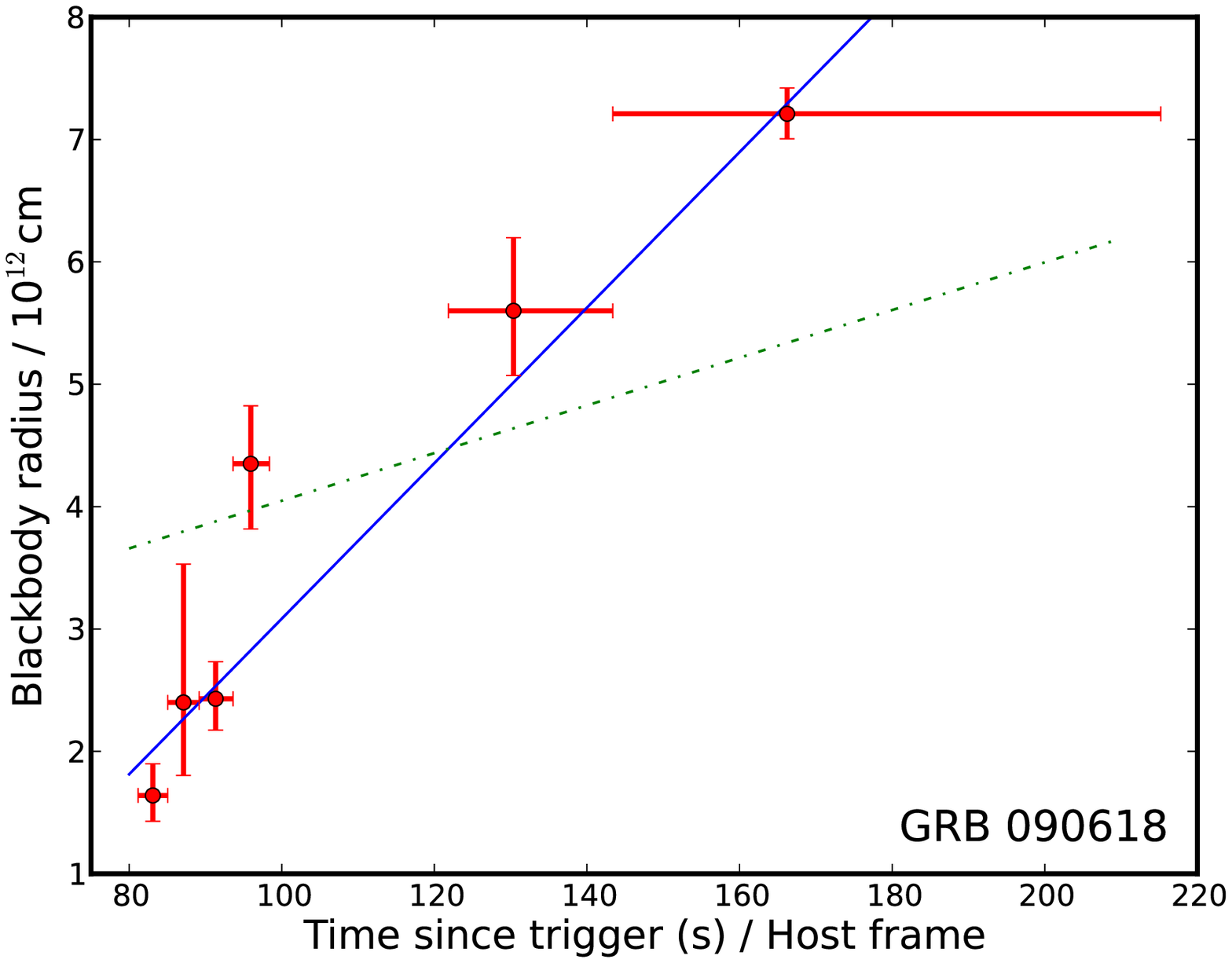}
\caption{Expansion of the blackbody radius for GRB\,090618. The solid line shows the best fit for a constant velocity ($v=3.5\,c$. For comparison a fit with an apparent velocity equal to the speed of light ($v=c$, dot-dashed) is also shown, but does not fit the data well.}
\label{fig:hostframe090618}
\end{figure}

We track the evolution of the temperature and luminosity of GRB\,090618 and
find, in the host galaxy rest-frame an apparent expansion with a best-fit
speed of $10.5^{+9.5}_{-2.1}\times10^{10}$\,cm\,s$^{-1}$ (see Fig.~\ref{fig:hostframe090618}).  This is $3.5^{+3.0}_{-0.7}$ times the speed of light.  We therefore conclude that the thermal component in
GRB\,090618 must be treated relativistically. 

\begin{table}
\caption{$\Delta\chi^2$'s and probabilities for all time series of bursts where at least one series has $\Delta \chi^2$\,$\geq$\,25. (GRB\,060218 has been omitted).}
\renewcommand*{\arraystretch}{1.05}
\begin{tabular}{@{} p{2.1cm} p{1.6cm} p{3.2cm} p{1.7cm} l c c c @{}}
\hline\hline
Time$^{a}$ & Nr.$^{b}$  & Band $\chi^2$/d.o.f & $\Delta\chi^2$ & Probability \\ \hline
\hline \multicolumn{4}{ l }{{GRB\,060202 ($z\,=\,0.783$)}} \\ \hline
171.3 & A & 361.29/329 & 6.7 & 0.12 \\ 
295.3 & B & 342.65/268 & 28 &  0.000068\\ 
684.1 & C & 370.49/342 & 6.4 & 0.13 \\ 
\hline \multicolumn{4}{ l }{{GRB\,060418 ($z\,=\,1.489$)}}  \\ \hline 
 93.00 & A & 174.59/192 & 5.0 & 0.16 \\ 
 128.5 & B & 245.48/202 & 2.1 & 0.50 \\ 
 140.2 & C & 216.15/201 & 13 & 0.0078 \\ 
 171.9 & D & 240.63/211 & 34 & $<5\times10^{-5}$ \\ 
 542.1 & E & 157.76/176 & 6.9 & 0.076 \\ 
\hline \multicolumn{4}{ l }{{GRB\,061007 ($z\,=\,1.261$)}} \\ \hline
94.22 & A & 340.71/316 & 36 &  $<3.3\times10^{-5}$ \\ 
113.1 & B & 336.81/308 & 27 &  0.0016 \\ 
139.5 & C & 379.37/319 & 11 & 0.10 \\ 
181.1 & D & 356.18/317 & 4.0 & 0.56 \\ 
251.8 & E & 330.56/316 & 8.1 & 0.21 \\ 
372.8 & F & 330.56/315 & 12 & 0.070 \\ 
612.8 & G & 387.64/316 & 14 & 0.041 \\ 
1286  & H & 312.30/264 & 24 & 0.0033 \\ 
\hline \multicolumn{4}{ l }{{GRB\,061121 ($z\,=\,1.314$)}} \\ \hline
68.80 & A & 298.45/323 & 2.0 & 0.61 \\ 
78.72 & B & 456.34/339 & 49 &  $<5\times10^{-5}$ \\ 
102.6 & C & 257.15/235 & 4.1 & 0.27 \\ 
371.3 & D & 228.59/194 & 20 & 0.0003 \\ 
\hline \multicolumn{4}{ l }{{GRB\,090424 ($z\,=\,0.544$)}} \\ \hline
165.0 & A & 387.76/305 & 21 &  0.0020 \\ 
618.1 & B & 380.28/328 & 50 & $<3.3\times10^{-5}$ \\ 
\hline \multicolumn{4}{ l }{{GRB\,090618 ($z\,=\,0.54$)}} \\ \hline 
 128.1 & A & 191.41/208 & 11 & 0.034 \\ 
 134.3 & B & 244.24/226 & 19 & 0.0045 \\ 
 140.8 & C & 255.60/241 & 27 &  0.00095 \\ 
 147.8 & D & 254.21/231 & 16 & 0.0074 \\ 
 155.6 & E & 234.05/214 & 5.7 & 0.21 \\ 
 165.1 & F & 214.43/208 & 6.6 & 0.16 \\ 
 178.3 & G & 200.51/185 & 1.3 & 0.83 \\ 
 200.9 & H & 253.39/233 & 29 & 0.00057 \\ 
 256.1 & I & 368.69/300 & 47 &$<3.3\times10^{-5}$ \\ 
\hline \multicolumn{4}{ l }{{GRB\,100621A ($z\,=\,0.542$)}} \\ \hline
88.05 & A & 249.80/311 & 7.6 & 0.041 \\ 
117.3 & B & 322.56/286 & 8.8 & 0.028 \\ 
173.3 & C & 257.35/212 & 37 & $<5\times10^{-5}$  \\ 
 \hline
\end{tabular}
\footnotesize{
\\$^{a}$ Mean time in seconds after BAT trigger time.
\\$^{b}$ Time series}
\label{tab:fit}
\end{table}

%
%

\section{Origin of the thermal component}\label{discussion}
\subsection{SN shock breakout}\label{sec:SNbreak}

It has been suggested that the thermal emission found in previous GRB early
afterglows is due to a SN shock break-out from the stellar winds surrounding
the progenitor.  While this may be a plausible model for low-luminosity
systems, it seems implausible that the very large luminosities discovered
here could possibly be related to a shock breakout from a SN.  Models of SN
shock breakouts confirm this \citep[e.g.][]{2007MNRAS.375..240L}. Typical 
values reported are $10^{47}$\,ergs for the break-out energy and a
blackbody temperature of $1$\,keV. This energy is lower than any we observe in 
the soft X-ray thermal component. Even attempts to explain GRB\,060218's thermal 
component, (the lowest blackbody luminosity in our sample) with an asymmetric 
explosion \citep{2007ApJ...667..351W} have been shown by 
\cite{2007MNRAS.382L..77G} to require a deal of fine tuning. The fact that we obtain very large luminosities for 
both GRB\,061007 and GRB\,061121 (about four orders of magnitude larger than 
GRB060218), but that the properties of the extra component are not qualitatively 
dissimilar to previous thermal components found, suggests strongly that not only is 
a SN shock break-out not the origin in these cases, but it may not be in
most other cases where this has been discovered either. \cite{2008Natur.453..469S} reported the case of possible shock breakout in a SN without an accompanying GRB, SN2008D. It has also been suggested that this was photospheric emission from a mildly relativistic jet \citep{2008Sci...321.1185M}. The case has been throughly studied, \citep{2011ApJ...726...99V,2010A&A...522A..14G,2009ApJ...700.1680T}, but without conclusive results. With a total energy of $E_X\approx2\times10^{46}$\,ergs, this burst is consistent with what could be expected for a break-out, and with the limited photon statistics we cannot distinguish the origin of the emission using the spectra. For the GRBs though, we should look elsewhere for the origin of an apparently thermal component in the late
prompt/early afterglow soft X-ray emission.

\subsection{Alternative models}

While much progress has been made because of the afterglows of GRBs, the
origin of, and mechanism behind the prompt phase and early afterglow are
still uncertain.  Emission from a cocoon surrounding the
jet has been proposed \citep[e.g.][]{2013ApJ...764L..12S}, but the model does not 
seem to explain the energies and expansion velocities reported here. 

The models traditionally used to model the high energy
spectra, a cut-off power-law or smoothly broken power-law
\citep{1993ApJ...413..281B}, are empirical and not strongly motivated from a
physical understanding of the emission process.  The main part of the
radiation is often considered to be non-thermal, with high energy photons
originating from synchrotron and/or inverse Compton processes in the
ultra-relativistic jet \citep[e.g.][]{1996ApJ...466..768T,1997ApJ...488..330C,2011A&A...526A.110D}). 
However, based on the detection of components in the prompt gamma-ray
emission with a blackbody-like spectral shape in addition to a power law
\citep{2005ApJ...625L..95R}, and on difficulties in reproducing low energy
spectral indices with synchrotron models \citep{1998AIPC..428..359C}, a
trend has been growing to attribute much of the prompt-phase emission to the
photosphere at the head of the ultra-relativistic jet
\citep[e.g.][]{2007RSPTA.365.1171P,2010arXiv1003.2582P,2013arXiv1301.3920L}. 
Such emission emerges from an optically thick plasma, generally not in
thermal equilibrium, producing a quasi-thermal spectrum
\citep{2009ApJ...702.1211R}.  Observationally, this high energy photospheric
emission decays in luminosity and temperature as a power-law in time
\citep{2009ApJ...702.1211R}.  If that trend continues, it is not
unreasonable to suppose that it may appear at the end of the prompt phase in
soft X-rays as an apparently thermal component with high apparent luminosity
and temperature.  Under this hypothesis, we can then model the excess
component as late photospheric emission.

Using eqs.~5 and 7 found in \cite{2012MNRAS.420..468P} (modified to our parameters), we can calculate the Lorentz factor and the photospheric radius:
\[r_{ph} = R^{host}_{bb}\times\frac{\gamma}{\,\xi\,(1+z)^{2}}\] 
\[ \gamma = [(1+z)^2\,D^2_L\frac{F^{obs}_{bb}\sigma_T}{2\,m_p\,c^3\,R^{host}_{bb}}]^{1/4}\times(L_{tot}/L^{obs}_{bb})^{1/4}\] 
Taking the model luminosities found for each time series we get values for the Lorentz factors, as seen in Table~\ref{tab:best}. These values should be corrected with a factor expressing the ratio of the blackbody and total luminosity that is in the relevant burst epoch, as the equation uses the values over the entire burst, but as the luminosity ratio is to the power of 1/4, the correcting factor will be close to 1. We then use these values to calculate the photospheric radii (also shown in Table~\ref{tab:best}), where $\xi$ is a geometrical factor close to one (we set $\xi$\,=\,1 exactly).

\begin{figure}
\includegraphics[clip=,width=0.92\columnwidth]{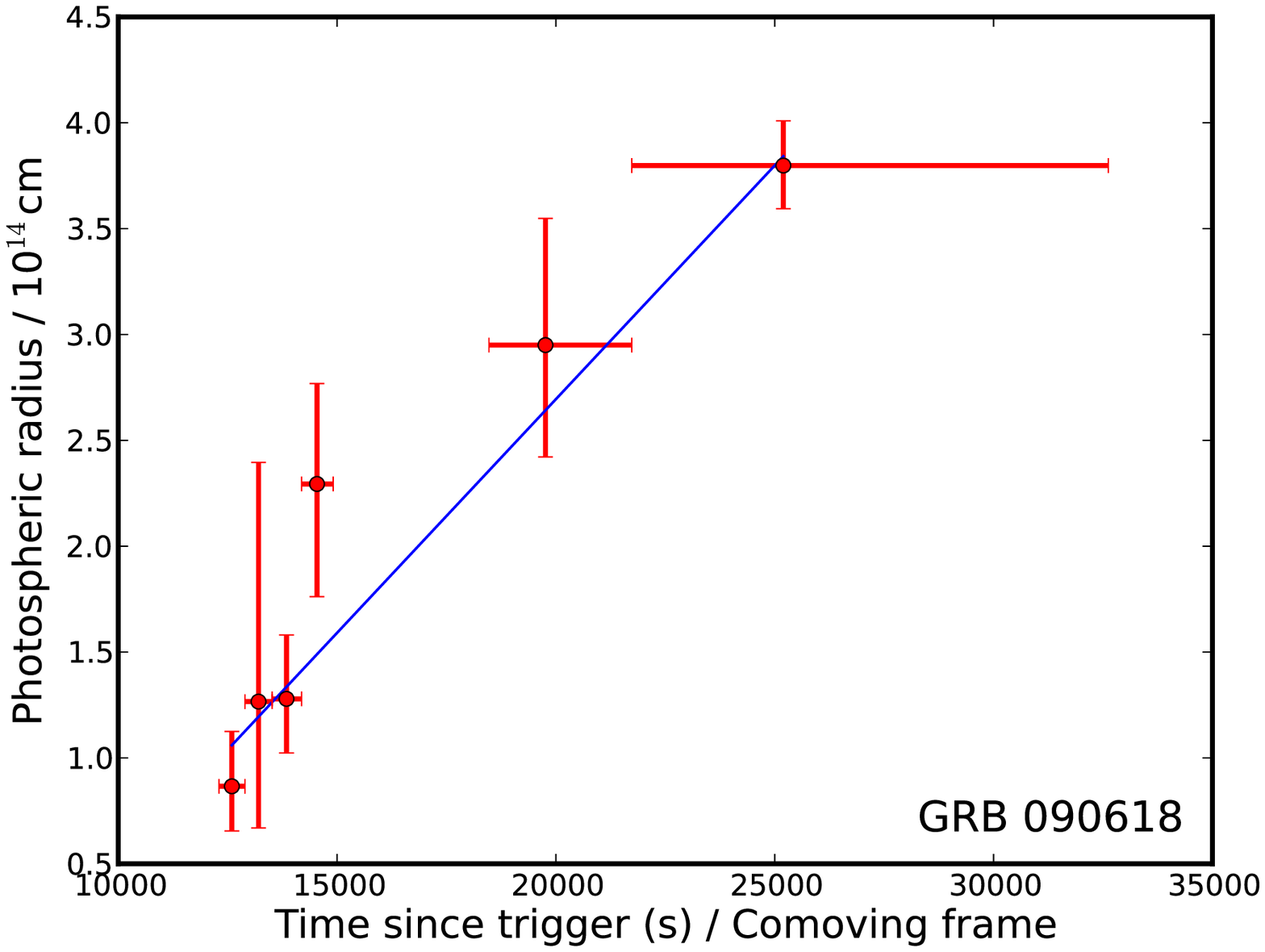}
\includegraphics[clip=,width=0.92\columnwidth]{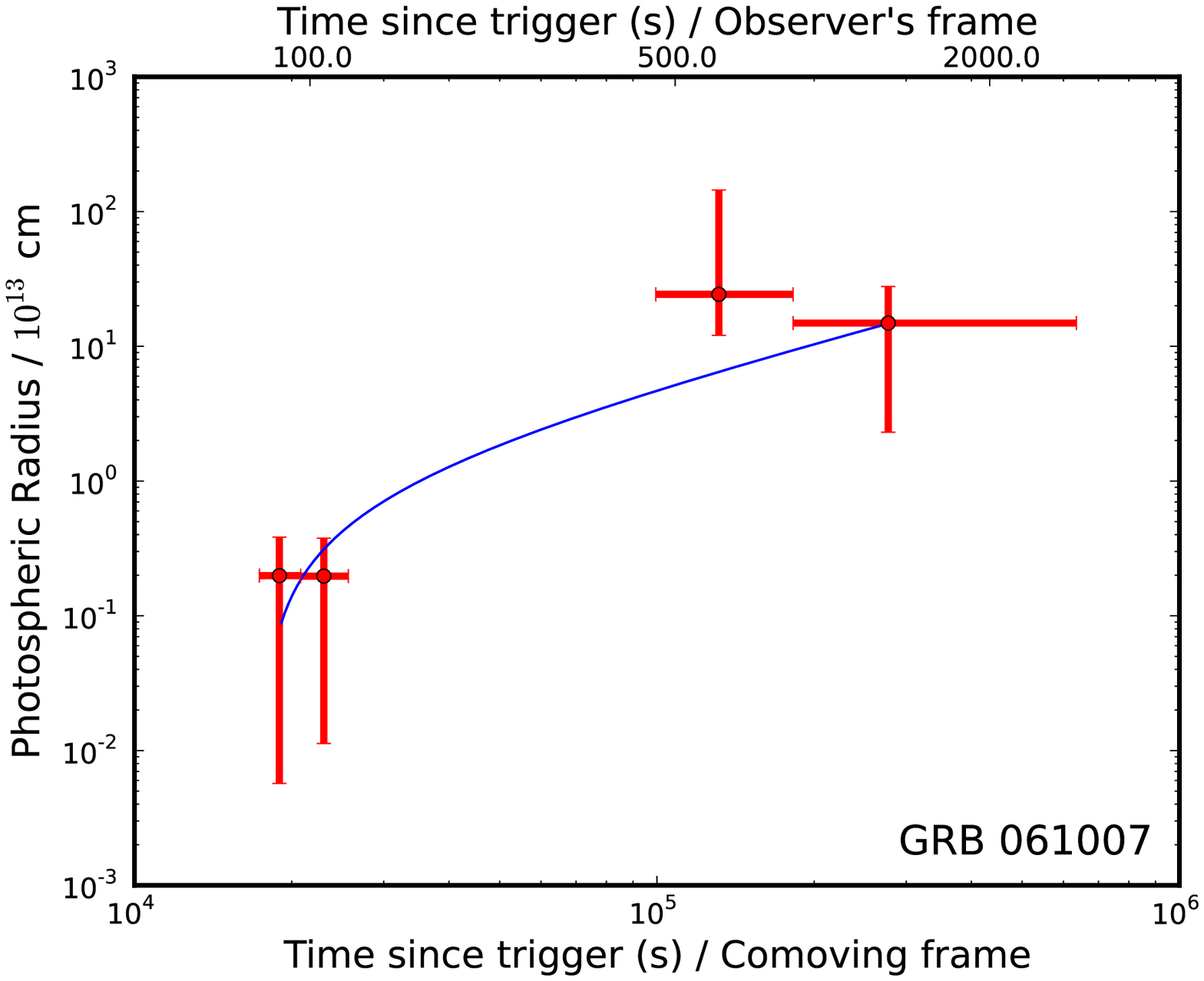}
\caption{Evolution of the photospheric radius for GRBs 090618 and 061007. The photospheric radii are derived from eqs. 5 and 7 in \cite{2012MNRAS.420..468P}. The solid lines show the best fit for constant velocity. Only the velocity of GRB\,090618 is consistent with expansion close to $c$ in the jetframe using the calculated asymptotic Lorentz factor for the burst, indicating a slowdown of the jet in the other bursts by the time of the XRT observations.}
\label{fig:radius}
\end{figure}

The Lorentz factors calculated here are asymptotic values, valid for the
coasting phase of the jet, so they should be similar to the values
calculated from the prompt phase emission.  As seen in Table~\ref{tab:best},
the Lorentz factors we get are consistent with the emission being late
photospheric, with values between ten and several hundreds.  Another
consistency check is apparent superluminal motion of the photospheric radii. 
In order to look for this, we fitted the photospheric expansion assuming this to be constant throughout.  
We checked all bursts with detections better than 2\,$\sigma$ in more than one time series, using these for the fit.  
Fig.~\ref{fig:radius} shows fits for GRBs 090618 and 061007.  

For GRB\,090618 we fit an apparent velocity of $111\,c$, which corresponds
to $0.6^{+0.4}_{-0.2}\,c$ in the jet frame using the calculated asymptotic
Lorentz factor, consistent with a relativistic expansion close to that
Lorentz factor.  For GRBs 061007 and 061121 the fitted non-relativistic
velocities are $93\,c$ and $68\,c$ respectively.  However, using the
asymptotic Lorentz factors in this case yields jetframe velocities only a
small fraction of the speed of light. We conclude that for these bursts the
jet has started to slow down and expand, so the assumption of a constant
Lorentz factor is no longer valid.  Our sample then includes all cases, both
the jet coasting phase, with velocities near the speed of light, as seen in
GRB\,090618, the deceleration phase as seen for GRBs 061007 and 061121, and
the case where the jet has slowed down completely, seen by the constant
radius (photospheric as well as simple blackbody) of GRBs 060418 and 090424.

In principle, the observed evolution of the thermal component in the soft
X-ray should be compatible with those observed at early times in the
gamma-ray regime.  While we have only one GRB in our sample which has a
reported prompt phase gamma-ray thermal component, GRB\,061007 (see below),
as reported by \cite{2009ApJ...702.1211R}, the evolution in temperature and
luminosity behaves consistently across bursts, with an initial increase in
value until a break occurs after a few seconds, and then a decay with a
power-law index between $-4.5$ and $-0.8$ for the luminosity and $-1.3$ and
$-0.3$ for the temperature.  Using the points we have from the bursts with
good fit over several time series to calculate power-law indices, we get
values largely consistent with these for GRB\,061007, 090618, 061121,
090424 and 060218; luminosity: $-0.2^{+0.2}_{-1.1}$, $-0.7^{+0.3}_{-1.9}$,
$-1.9^{+1.0}_{-8.5}$, $-0.8^{+0.6}_{-0.6}$, and $0.7^{+0.2}_{-0.6}$, and temperature: $-1.0^{+0.5}_{-1.0}$,
$-0.4^{+0.2}_{-1.4}$ and $-0.8^{+0.4}_{-4.9}$, $0.04^{+0.05}_{-0.04}$, and $-0.12^{+0.18}_{-0.03}$ respectively. 
As far as we are aware there is no theoretical reason why these power-law
decay trends necessarily have to continue indefinitely, but the fact that
our results for the luminosity and temperature decay rates lie in the range
observed from the gamma-ray prompt phase is encouraging for the model.

GRB\,061007 is of special interest, since \cite{2011MNRAS.414.2642L} find
the prompt emission of this burst to be dominated by apparent thermal
emission.  They find that the evolution in temperature and flux follow the
light curve.  By the time the XRT started observing, the light curve is in
constant decline, so the temperature and flux would be expected to decay, as
we find in our analysis of the XRT data.  We cannot compare decay indices,
as \citet{2011MNRAS.414.2642L} observe before the onset of the declining phase. 
The thermal component started off very strong, accounting for about 75\% of the total
luminosity.  By the time of the XRT observation it has fallen to about 10\%
for the first two time series, and then becomes too weak to be statistically
significant.  

A comparison of Lorentz factors is also instructive. 
\citet{2011MNRAS.414.2642L} find values of $\sim200-600$, which should be
directly comparable to the asymptotic Lorentz factors in our sample, for
which we obtain similar values, between about 30 and 670 (see
table~\ref{tab:best}), from our observations.  For GRB\,061007, at XRT observation time, the
jet seems to be slowing down, but we still find apparent superluminal
expansion of the blackbody radius, which means that the Lorentz factor is
still significant.  Our proposed model is furthermore supported by
observations such as that of \cite{2012ApJ...757L..31A}, who report a
detection of prompt phase thermal emission for GRB\,110721A, following the
blackbody temperature all the way down to 4.9\,keV.

With this interpretation, our discovery of highly luminous quasi-thermal
components in the soft X-ray emission allows the photospheric prompt
emission model to be explored at much later times and with the more
sensitive narrow-field soft X-ray instruments, and thus potentially with
better statistics and spectral resolution than has so far been the case.

%
%
\section{Conclusions\label{conclusions}}

We have examined a sample of the brightest \emph{Swift} GRBs looking for thermal emission in the XRT data. We find clear evidence for this emission in 8 out of 28 bursts, with an indication that such emission exists in the majority of bursts. We track the temperatures and luminosities of these components over time. We find that several of these thermal components are very luminous (three to four orders of magnitude more luminous than the component in GRB\,060218) and the temperature is high. These facts therefore make SN shock breakout an unlikely explanation in the generality of GRB thermal components, since the components uncovered here are physically similar to other thermal components discovered so far. We find that several of the components have apparent superluminal expansion, indicating that they are clearly expanding relativistically and use late photospheric emission from the jet as a physically well-motivated theory, to allow us to determine Lorentz factors for the bursts. We find the decay rates of the luminosity and temperatures as well as the Lorentz factors to be compatible with values obtained elsewhere from prompt gamma-ray thermal emission. This explanation links the emission observed in the prompt phase to that of later times and is supported by the detection of superluminal motion and the observation that the trends observed in the gamma-ray thermal emission reported extend to the soft X-ray regime and may mark a crucial step in understanding the prompt/early phase of GRB emission.

\section{Acknowledgements}
The Dark Cosmology Centre is funded by the DNRF. MF acknowledges support by the University of Iceland Research fund. We thank P\'all Jakobsson and Gunnlaugur Bj\"ornsson for helpful comments. This work made use of data supplied by the UK Swift Science Data Centre at the University of Leicester.

\cleardoublepage
\chapter{Characterising the Environment}\label{chapter:host}
Having looked at the GRB itself, we now turn our attention to the environment in which these bursts are found. Progress in studying the birth sites of GRBs is in large part owed to afterglow follow-up programs providing spectroscopy and photometry of the GRB afterglow and host. In this chapter, I will detail a number of the diagnostics used on afterglow/host observations to characterise the burst environment. I will end the chapter by using the results from one such follow-up program, namely the X-shooter GRB program, to show some statistical properties of the homes of GRBs.

\section{Voigt-Profile Fitting}
One way to study the environment, is by characterising the absorption lines found in GRB afterglow spectra. Since the GRB optical afterglow is often very bright, it serves as a background source for lighting up the line-of-sight between the burst and observer. The absorption lines from the gas will tell us, not only the redshift, see Section~\ref{sec:redshift}, but also the relative abundances of the elements in that gas. Afterglow photons pass through the gas, but are absorbed if their energy matches that of a quantum mechanical transition of a atom/ion/molecule in the gas. The depth and width of a resultant absorption line will depend upon a number of parameters that together form what is known as a Voigt profile. Before I describe this profile, I will go through its individual components.

\subsection{Oscillator strength}
The oscillator strength, $f$, of the line, quantifies the intrinsic strength of the transition, i.e. the strength of the line per excitation. It is given by: 
\begin{equation}
f_{\nu}=B_{12}\frac{h\nu m_ec}{(2\pi e)^2},
\label{eq:f}
\end{equation}
\noindent where $c$ is the speed of light, $h$ is Planck's constant, $\nu$ is the frequency of the transition, and $m_e$ and $e$ are the mass and charge of the electron. Here, $B_{12}$ is the Einstein coefficient giving the probability for a transition from energy state 1 to the higher energy state 2. 

\subsection{Number density}
The depth of the absorption line naturally also depends on how much absorbing material is available. This is related to the optical depth at the given frequency, which is found from the simple solution to the radiative transfer equation: 
\begin{equation}
I_{\nu}(\tau_{\nu})=I_{\nu}(0)e^{-\tau_{\nu}},
\label{eq:transfer}
\end{equation} 
\noindent where $I_{\nu}(0)$ is the specific intensity before absorption and $\tau_{\nu}$ is the optical depth. This solution is derived for the simplified case of a homogenous absorbing material with no scattering. This is not necessarily a realistic assumption, but it provides an analytical solution.

The optical depth is defined as: 
\begin{equation}
\tau_{\nu}(x)=\int^x_{x_0} n\sigma_{\nu}dx=N\sigma_{\nu},
\label{eq:depth}
\end{equation}
\noindent where $n$ is the number density of the absorbing material, which integrated over the path $x$ is given by the column density $N$, and the cross section $\sigma_\nu$ at the transition frequency $\nu$.

\subsection{Doppler broadening}
Having now dealt with the depth of the line; intrinsic strength and strength in numbers, we turn to the main contributors for broadening the line. If the absorbing material, here assumed to be identical atoms, are at rest with respect to each other, then they will all absorb at the same frequency (though see the section below). However, atoms at a temperature $T$ have a velocity distribution determined by the Maxwell--Boltzmann distribution (assuming the gas is ideal) given by: 
\begin{equation}
n(\nu)d\nu=N\sqrt{\frac{m_a}{2\pi k T}}e^{-m_av^2/2kT}d\nu,
\label{eq:Maxwell}
\end{equation} 
\noindent where $m_a$ is the mass of the atom and $k$ is the Boltzmann constant. If $\nu_0$ now is the rest frequency for the transition in question, then photons will be absorbed that have a frequency of: 
\begin{equation}
\nu = \nu_0(1\pm v/c),
\end{equation} 
\noindent where $v$ is the velocity of the atom. Solving for $v$ we then get a distribution of radiation around the centre frequency of: 
\begin{equation}
I(\nu)=I(0)e^{-m_ac^2(\nu-\nu_0)^2/2kT\nu_0^2}.
\label{eq:rad}
\end{equation}
 Equation~\ref{eq:rad} is in the form of a Gaussian with a variance of:
\begin{equation}
\sigma_G=\sqrt{\frac{kT}{m_a}}\frac{\nu_0}{c}.
\label{eq:gauss}
\end{equation}
\noindent The full width at half maximum (\emph{FWHM}) of the line profile can then be found:
\begin{equation}
FWHM = 2\sqrt{2\ln2}\,\sigma_G=\frac{2\nu_0}{c}\sqrt{2\ln2\frac{kT}{m_a}},
\end{equation}
\noindent in frequency space, or substituting back in velocity space:
\begin{equation}
FWHM=2\sqrt{2\ln2\frac{kT}{m_a}}=2\sqrt{2\ln2}\frac{b}{\sqrt{2}}, 
\label{eq:fwhm}
\end{equation}
\noindent where $b$ is the Doppler broadening generally defined as:
\begin{equation}
b=\sqrt{b_{\text{th}}^2+b_{\text{turb}}^2}.
\end{equation}
\noindent Note that $b$ can have a contribution from turbulent motion of the gas. So far we have only dealt with thermal broadening. Turbulence is difficult to describe physically, and usually we have to settle for solving for the total $b$-value.

\subsection{Natural broadening}\label{sec:nat}
\begin{figure}[h]
\includegraphics[width=1.0\textwidth]{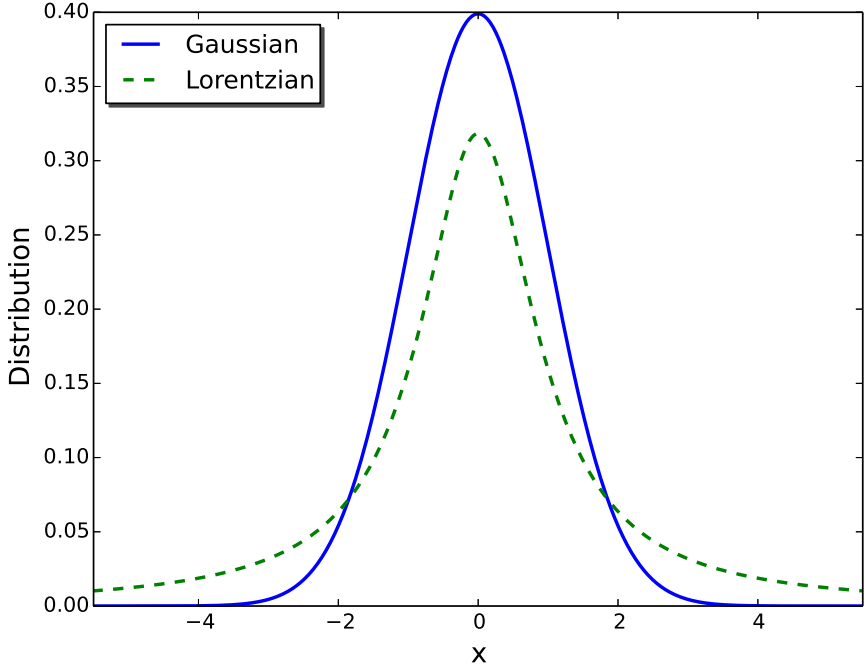}
\caption{Comparison of a Gaussian (blue) and Lorentzian (green) profile. Both distributions have a mean of 0 and a variation of 1.}
\label{fig:dist}
\end{figure}

Besides the classical Doppler broadening, the line also has an intrinsic quantum mechanical width due to the uncertainty principle; $\Delta E\Delta t\geq\hbar/2$. This dictates that the energy difference between lower and upper state of the transition does not correspond to one exact energy, but rather a range of energies $\Delta E$. This results in a natural broadening of the line determined by the Lorentz profile \citep{Rybicki}:
\begin{equation}
\phi_\nu=\frac{\Gamma}{4\pi^2(\nu-\nu_0)^2+(\Gamma/2)^2},
\label{eq:Lorentz}
\end{equation}
\noindent where $\Gamma$ is the quantum mechanical damping constant which is given by the sum of probabilities for natural decay from each energy level in the transition. This profile peaks at $\nu=\nu_0$ which gives a profile maximum of $\phi(\nu_0)=4/\Gamma$. Half of that is $\phi_{HM}=2/\Gamma$ which we put back into Equation~\ref{eq:Lorentz} to solve for the frequency at half max:
\begin{equation}
\nu_{1/2}=\nu_0\pm\Gamma/4\pi => FWHM=\Gamma/2\pi.
\end{equation}
\noindent Compared to a Gaussian profile, a Lorentz profile with the same area under the curve has broader wings, as illustrated in Fig.~\ref{fig:dist}.

\subsection{Instrumental broadening}
The last component we need is not related to the actual absorption line, but to how we observe it. No spectrograph has infinite resolution, and hence the spectral distribution observed is convolved with an instrumental point-spread function (PSF), usually approximated by a Gaussian. 

The spectral resolving power is defined as R\,=\,$\lambda$/$\Delta$$\lambda$, and is usually specified for each instrument. However, under ideal conditions (good seeing), the actual resolution is often higher, so for instruments such as X-shooter (see Section~\ref{sec:xshooter}) we measure the width of atmospheric lines. By using these telluric lines, which should not be resolvable by X-shooter, we can rule out any other contribution to the broadening.

\subsection{The Voigt profile}
We now have all the components needed to describe the absorption line. The convolution of a Doppler and Lorentz profile is known as a Voigt profile:
\begin{equation}
\phi_\nu=\frac{\Gamma}{4\pi^2}\int_{-\infty}^{+\infty}\frac{1}{b\sqrt{\pi}}\frac{e^{-(v/b)^2}}{[\nu-\nu_0(1+v/c)]^2+(\Gamma/4\pi)^2}dv,
\end{equation} 
\noindent \citep{Rybicki}. This is often written in the compact form:
\begin{equation}
\phi_\nu=\frac{1}{(\Delta\nu_D)^{-1}\sqrt{\pi}}H(a,u),
\end{equation}
\noindent where $\Delta\nu_D=b\,\nu_0/c$ is the Doppler width and $H(a,u)$ is the Voigt function, given as:
\begin{equation}
H(a,u)=\frac{a}{\pi}\int_{-\infty}^{+\infty}\frac{e^{-y^2}dy}{a^2+(u-y)^2},
\end{equation}
\noindent with $a=\Gamma/4\pi\Delta\nu_D$, $y=v/b$ and $u=(\nu-\nu_0)/\Delta\nu_D$. We can express the optical depth as:
\begin{equation}
\tau_\lambda=\frac{\sqrt{\pi}e^2}{m_ec}\frac{Nf\lambda}{b}H(a,u).
\label{eq:od}
\end{equation}
\noindent The central optical depth is given by $H(a,u)=1$:
\begin{equation}
\tau_c=\frac{\sqrt{\pi}e^2}{m_e}\frac{Nf}{b\nu}.
\end{equation}

\subsection{Determining column density}
Now that we have a description of the absorption line, we can use this to determine some physical parameters of the gas. One of the most interesting parameters is the column density, $N$, i.e. the number of particles contained in a column with a surface area of 1\,cm$^2$ and length along the full line-of-sight. In principle, all we need to do is fit the absorption line(s), and solve for $N$ and $b$. However, even relatively weak lines (i.e. low $f$) can appear saturated (i.e. unable to appear any deeper despite the column density being high enough) in low resolution spectroscopy. The lines weak enough to avoid this could also be smeared out due to the instrumental broadening.

\begin{figure}
\includegraphics[width=1.0\textwidth]{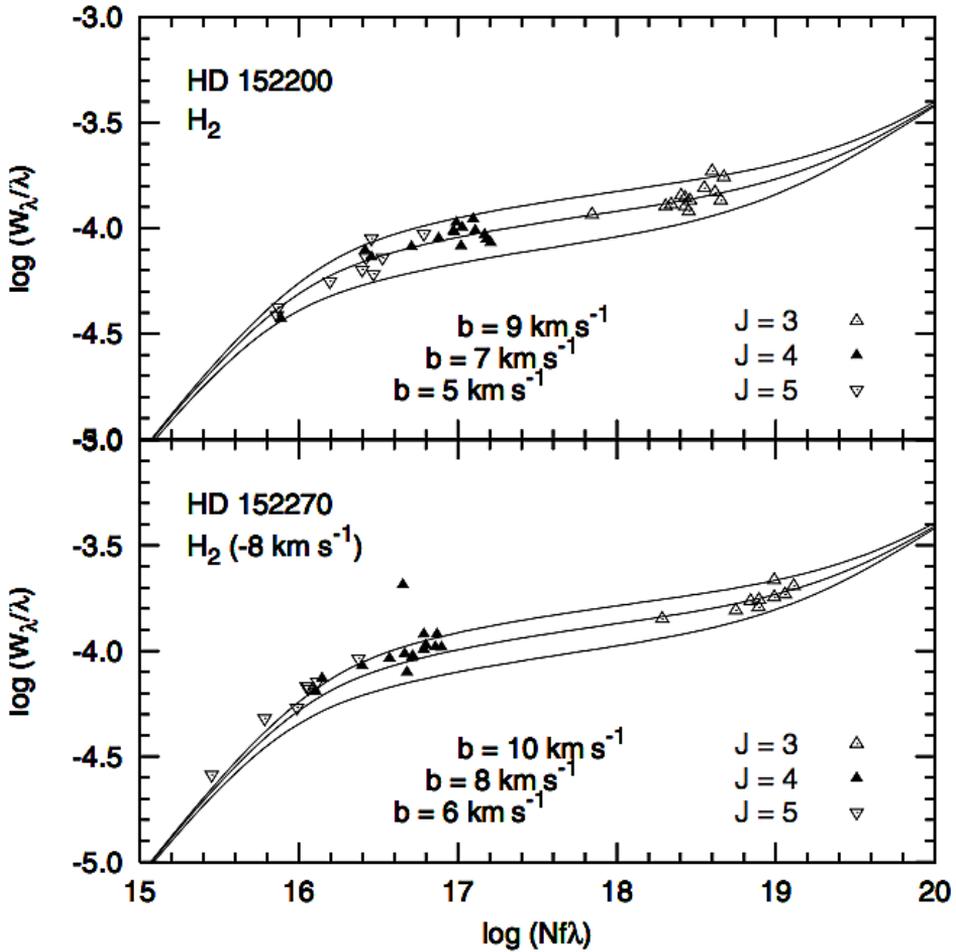}
\caption{Figure from \cite{2004A&A...416..251M} showing the curve of growth for Lyman-Werner transitions towards two sources in the Milky Way ($EW=W_\lambda$). The plot is normalised by wavelength and oscillator strength, to easily show several transitions in the same plot. J gives the quantum number for the excited state of the absorption line, and the three curves illustrates the dependence on $b$ in the saturated regime. The linear regime is below log$(Nf\lambda)\sim$\,16, followed by the flat regime, and then the damping regime starts to the very right of the plot.}
\label{fig:cog}
\end{figure}

For low-resolution spectroscopy, rather than fitting to the Voigt profile, the column density can be determined through the equivalent width, $EW$, of the absorption line. This is defined as the width, in wavelength units, of a rectangular strip of the spectrum having the same area as the absorption line, i.e.:
\begin{equation}
EW_{\text{rest}}=\int\frac{I_{\text{cont}}-I(\lambda)}{I_{\text{cont}}}d\lambda = \int 1 - e^{-\tau(\lambda)}d\lambda.
\label{eq:ew}
\end{equation}
\noindent Solving this, leads to the relationship between column density and $EW$ known as the curve of growth (CoG). A CoG for molecular hydrogen lines is displayed in Fig.~\ref{fig:cog}. It is divided into three segments that each have an approximate analytical solution, so to determine $N$ we need to identify where on the CoG the gas conditions put us. \\
\\

\noindent \textbf{Linear or weak regime} \\
Optical thin region where $\tau_0<1$. The line is not saturated, either because $f$ or $N$ is very low, or a combination of the two. The $EW$ then gives a direct measure of $N$, $EW\propto N$, independent on the value of $b$. The proportionality factor can be found by Taylor expansion of Equation~\ref{eq:ew}, and by using the numerical values we get:
\begin{equation}
N=1.13\times10^{20}\,\frac{EW}{\lambda^2 f}\,\text{cm}^{-2}.
\end{equation}

\noindent \textbf{Flat or saturated regime} \\
As the optical depth nears and exceeds unity the line becomes saturated, and $EW$ is no longer useful to determine the column density. It is however possible to constrain $b$, as seen in Fig.~\ref{fig:cog}. In this regime, the line is predominantly Gaussian, and we can use Equations~\ref{eq:transfer},~\ref{eq:rad}, and~\ref{eq:fwhm} to find $\tau$, and then solve Equation~\ref{eq:ew} in the limit where $\tau_0$, the optical depth at the central wavelength, exceeds 1. This leads to:
\begin{equation}
EW_{\text{rest}}\propto b\sqrt{ln(\tau_0)}.
\end{equation}

\noindent \textbf{Damping regime} \\
At very high optical depth $\tau_0\gg1$, the damping wings can be used to estimate $N$, as more radiation is absorbed, so the wings are clearly visible, but unlike for the centre of the line, this part is not saturated. In this regime, the natural broadening dominates the wings, see also Fig.~\ref{fig:dist}. We can hence treat the Gaussian contribution to the Voigt profile as a $\delta$ function, and set $\tau_\lambda=N\phi_{\text{Lorentz}}(\lambda)$. Using the approximation $(\lambda-\lambda_0)=\Delta\lambda\gg\Gamma\lambda^2/4\pi c^2\equiv\beta$ as the wings makes the line very broad, we can expand $\tau_\lambda$ (Equation~\ref{eq:Lorentz} and~\ref{eq:od}) to:
\begin{equation}
\tau_\lambda\sim\frac{\pi e^2\lambda^2}{mc^2}Nf\beta\frac{1}{\Delta\lambda^2}.
\end{equation}
\noindent Plugging this into Equation~\ref{eq:ew}, the integral solves as a gamma function, and we get:
\begin{equation}
EW\propto\sqrt{N}.
\end{equation}

\noindent Note that while $EW$ is not dependent on $b$ in this regime, if $b$ is large, $N$ needs to be very large for the damping wings to be dominated by natural broadening (i.e. for the line to enter this regime).

Even if the spectral resolution is high, the line can still be saturated, simply because the gas is optically thick to the exciting photons. If the line is in the damping region, we can instead fit the Lorentz profile to the damping wing directly. If there is doubt about whether a line is saturated or not, it is useful to fit more lines from the same species. The line ratio should then correspond to that given by the intrinsic strength $f$, if not one or more lines are saturated. If the line is saturated, a fit will give a lower limit on $N$, as there will in reality have been more of the given species of atoms or ions, but not enough photons to be absorbed. To get a reliable estimate of both $N$ and $b$, lines from more than one species can be simultaneously fit.

It is important to accurately determine the continuum level in the spectrum around the line, otherwise the area of the line will be over- or underestimated. This is especially tricky for lines such as Ly$\alpha$, which is often completely saturated and has extensive damping wings. The continuum then needs to be estimated far from the line to be sure it does not include the wing. 

\section{Fitting Emission Lines}\label{sec:emission}
In addition to absorption lines, afterglow spectra sometimes contain emission lines. These are not stimulated emission, and are not related to the afterglow, but originate from the host galaxy. Hence the gas associated with emission and absorption may be entirely different components of the host. 

Emission lines are less troublesome to fit, as they do not experience saturation (although they could in principle be too bright for the instrument to handle). The emission lines in GRB spectra are often broad due to the fact that they are composed of emission from the entire host galaxy, which comprises several different gas clouds, moving with respect to each other. For that reason all other contributions to the broadening can be ignored, and the line can be fit with a Gaussian, and integrated to find the total flux.

Note though, that while the instrumental PSF may be ignored for the line flux itself, an accurate determination of the continuum is important. This can be done by using the 2D spectrum, and fitting a PSF to regions free of features on both sides of the line, which are then interpolated to find the PSF near the line. This PSF is then subtracted around the line, and the 1D spectrum is extracted and fit with a Gaussian. 

Emission lines are not only from the line-of-sight but from extended regions, and hence we cannot calculate a column density. There are ways to determine abundances from emission line fluxes though, which we will now examine.

\section{Nebular-Line Diagnostics}\label{sec:diag}
It is possible to determine the electron temperature or density from the ratios of certain nebular lines that are insensitive to one of these parameters. To determine the temperature, lines with a large difference in upper energy level (and hence $\Delta$E) are compared, as the ratio will depend only weakly on density. Inversely, to determine the density, lines from energy levels very close to each other are compared, as a change in temperature (and hence a change in number of photons available to excite these levels) will affect the line fluxes roughly the same way. 

The electron temperature is a function of metallicity, as metals serve as the primary cooler of the gas, such that higher electron temperatures correspond to lower metallicities. In astronomy, the most common set of lines observed that are directly temperature dependent are the [\mbox{O\,{\sc iii}}]  $\lambda$4363 line, compared with for instance [\mbox{O\,{\sc iii}}]  $\lambda$5007. This is referred to as the 'direct'  or 'T$_e$' method of determining metallicity. Unfortunately the [\mbox{O\,{\sc iii}}]  $\lambda$4363 line is very weak, and is rarely observed, even in metal-poor environments. Furthermore, temperature gradients within the nebula may cause the metallicity estimate to be significantly off \citep[e.g.][]{2002RMxAC..12...62S,2005A&A...434..507S}. 

Because of the weakness of [\mbox{O\,{\sc iii}}]  $\lambda$4363, empirical calibrations have been established between the direct method and ratios using strong emission lines. Below I briefly describe the three most commonly used. For reference, solar metallicity equals 12 + log(O/H) = 8.69.

\subsection{R$_{23}$}
R$_{23}$ refers to the ratio:
\begin{equation}
R_{23}  \equiv \frac{[\mbox{O\,{\sc ii}}] \lambda3727 + [\mbox{O\,{\sc iii}}]  \lambda\lambda4959, 5007}{\text{H}\beta}.
\label{eq:r23}
\end{equation}

\begin{figure}
\centering
\includegraphics[width=0.8\textwidth]{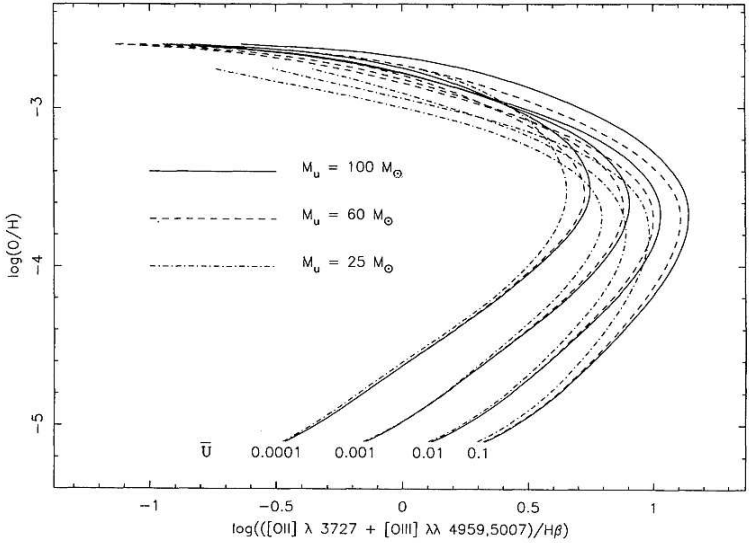}
\caption{Figure from \cite{mcgaugh91} showing the calibration of R$_{23}$ as a metallicity indicator. \={U} is the ionisation parameter.}
\label{fig:r23}
\end{figure}

The metallicity calibration of R$_{23}$ has two branches of solutions. The degeneracy is broken by using one (or both) of the ratios: $[\mbox{N\,{\sc ii}}] / \mathlarger{\text{H}\alpha}$ or $[\mbox{N\,{\sc ii}}] / [\mbox{O\,{\sc ii}]}$. 
Several calibrations are in use, one of the most common being from \cite{mcgaugh91}, which is plotted in Fig.~\ref{fig:r23}. This solution is divided into an upper branch for log(O/H)$ >-3.4$:
\begin{multline}
12+\text{log(O/H)} = 12-2.939-0.2x-0.237x^2-0.305x^3-0.0283x^4 \\ -y(0.0047-0.0221x-0.102x^2-0.0817x^3-0.00717x^4),
\end{multline}
\noindent and a lower branch log(O/H)$ <-3.9$:
\begin{equation}
12+\text{log(O/H)} = 12-4.944+0.767x+0.602x^2-y(0.29+0.332x-0.331x^2),
\end{equation}
\noindent where $x$ = logR$_{23}$ and $y$ = log($[\mbox{O\,{\sc iii}}]  \lambda\lambda4959, 5007/[\mbox{O\,{\sc ii}}] \lambda3727$), and then a turnover region in between. This calibration is based on theoretical models of H\,{\sc ii} regions. Other calibrations, for instance \cite{2001A&A...374..412P} for dwarf irregular galaxies, are done empirically through a comparison to the direct method.

\subsection{O3N2}
The O3N2 indicator is defined as:
\begin{equation}
\text{O3N2} \equiv \text{log} \frac{[\mbox{O\,{\sc ii}}] \lambda5007 /\text{H}\beta}{[\mbox{N\,{\sc ii}}] \lambda6583 /\text{H}\alpha}.
\label{eq:o3n2}
\end{equation}

A commonly used calibration is \cite{PP} who find:
\begin{equation}
12 + \text{log(O/H)} = 8.73-0.32\times \text{O}3\text{N}2,
\end{equation}

\noindent from a sample of 137 extragalactic H\,{\sc ii} regions.

O3N2 has the advantage over R$_{23}$ of not being degenerate with metallicity. Furthermore it relies on ratios of emission lines which are close in wavelength, making it less depending on a correct extinction correction. R$_{23}$ is still more often used though, as the [\mbox{N\,{\sc ii}}]$\lambda$6583 line is weaker than the others, and has the longest wavelength, so it is the first to be shifted out of the optical spectrum with increased redshift.

\subsection{N2}
N2 is the ratio:
\begin{equation}
N2 \equiv \text{log} \frac{[\mbox{N\,{\sc ii}}]}{\text{H}\alpha}.
\label{eq:o3n2}
\end{equation}

\cite{PP} find a best linear fit of:
\begin{equation}
12 + \text{log(O/H)} = 8.90 + 0.57\times \text{N}2,
\end{equation}
\noindent with only a very minor improvement when including higher order terms. N2 has the same advantages and drawbacks as O3N2. Both are less sensitive to metallicity at low [\mbox{N\,{\sc ii}}]/H$\alpha$ values, and should preferentially be used at metallicities approaching solar.

Besides using one or more of these methods, results are reported from a simultaneous fit to all three, see e.g. \cite{2011MNRAS.414.1263M}. This however, can complicate the calibration issues, as discussed below.

\subsection{Calibration issues}
The calibrations given here for the three methods are by far not the only ones in use. And as illustrated in Fig.~\ref{fig:r23} where the $R_{23}$ solution is plotted for different values of ionisation parameter, model predictions strongly depend on the assumed physical parameters of the H\,{\sc ii} regions. Furthermore models are for the most part still limited to plane or spherical geometries and neglect to account for the possibility of a clumpy gas distribution. Empirical calibrations fare little better however, as they share the same bias mentioned above for the direct method, namely that temperature gradients or fluctuations within the gas could mean that the determined electron-temperature metallicity is significantly underestimated. On top of this, recombination (free electrons recombining with ions which then quickly cascade to the ground state) could contribute to the excitation of the lines, possibly even dominating the line ratio at low enough temperatures.

Since all methods are affected by errors, we do not have a 'correct' version to calibrate against. Nebular-line diagnostics are therefore only useful for comparing metallicities found by using the same method and calibration, as \cite{KE} find that the errors are mainly systematic (for the theoretical calibrations at least) and hence affect all measurements the same. They also supply conversions between some of the most used calibrations, concluding that the discrepancy can be as large as $0.7$\,dex. The metallicity we calculate is hence only relative to others calculated by strong-line methods.

This large discrepancy makes nebular-line diagnostics an uncertain method for measuring metallicity. Unfortunately, in many cases no alternative exists. In Chapter~\ref{sec:paper_121024A} the simultaneous metallicity from absorption and emission lines is reported. But as argued, these two methods probe different physical regions, and may not be expected to yield the same result.


\section{Fitting the Stellar Population}\label{sec:spsm}
GRB host galaxies can also be studied through photometry, by modelling the stellar population. Stellar population synthesis modelling compares the spectrophotometric observations with a grid of models over populations of stars having a range of different physical conditions such as stellar ages and mass, specific star-formation rate (sSFR), metallicity, and extinction, as well as the redshift.

\begin{figure}
\centering
\includegraphics[width=1.0\textwidth]{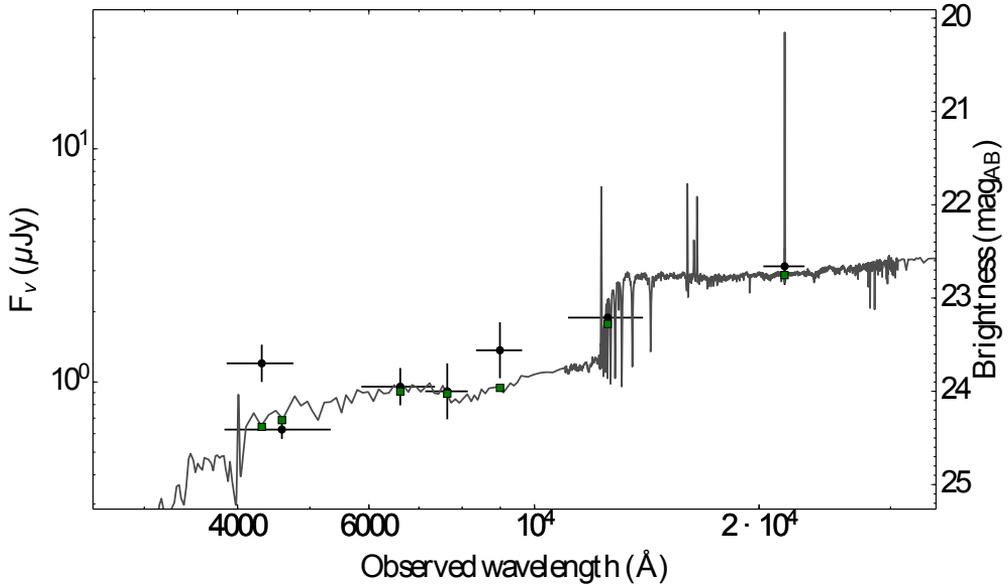}
\caption{The result of modelling the stellar population of the host galaxy of GRB\,121024A (see Chapter~\ref{sec:paper_121024A}). Black points show the actual observed photometric magnitudes, while the green points are equivalent magnitudes calculated from the model spectra to fit the band-widths used for the observations.}
\label{fig:spsm}
\end{figure}

A stellar population is modelled by integrating the contribution from the individual stars in the population. If we want to model the spectral energy distribution (SED) of a population of a certain age, mass and metallicity, we assume an initial stellar mass function (IMF; see Section~\ref{sec:imf}), and then use stellar evolutionary tracks to follow the stars to a given age. A star's evolutionary track is the movement with time through the Hertzsprung-Russel diagram, i.e. the evolution of luminosity and effective temperature (colour) of a star with a given mass and chemical composition. Models of the stellar atmosphere are then used to calculate the bolometric luminosity, i.e. the star's SED. From the IMF and individual stellar models, we can then sum up the contribution and get the total SED. Stellar population synthesis modelling then fit the observed SED to a large range of these models with different inputs, to find the best possible combination of physical parameters. An example of a best-fit model to data is shown in Fig.~\ref{fig:spsm}.

One of the most widely used set of models today are those by \cite{BC03}. Like most other models, these are tested on stellar clusters within the Milky Way and Magellanic Clouds, for which we know the physical parameters. Stellar clusters provide a good test of stellar models, since all stars have the same age \citep[as recently cemented by][]{2014Natur.516..367L} and are born from the same gas (i.e. have same initial chemical composition), and hence the only variable governing the stellar evolution is mass.

Below we take a deeper look at some of the more complicated in- and output parameters of stellar population synthesis modelling.

\subsection{The Initial Mass Function}\label{sec:imf}
The most important characteristic describing a star is its mass. The initial mass a star is born with (i.e. the mass of the star as it enters the main-sequence and starts burning hydrogen in the core), governs the further life of the star. The IMF is an empirical function that describes the distribution of initial masses of a population of stars. This is an important piece of information if we want to study collections of stars such as stellar clusters or whole galaxies. It is also important for the theory of star formation, as most stars form in clusters or groups. The exact form of the IMF, and whether there exists a universal form is still very much a matter of debate. Theory predicts a dependence on the environment, but up until recently observations seem to indicate only a very small difference \citep[though see for instance][]{2003MNRAS.339L..12C,2010ApJ...709.1195T}. These observations are not of the IMF directly, but rather the distribution of luminosity, i.e. the luminosity function as, unlike the stellar mass, this quantity can actually be measured, provided that we know the distance to the star. We then assume a relationship between the luminosity and mass, and use stellar models to extrapolate back to the initial mass distribution. The IMF is usually expressed as a simple power law or series thereof, with a cut-off in mass both at the low and high end. 

\begin{figure}
\centering
\includegraphics[width=0.8\textwidth]{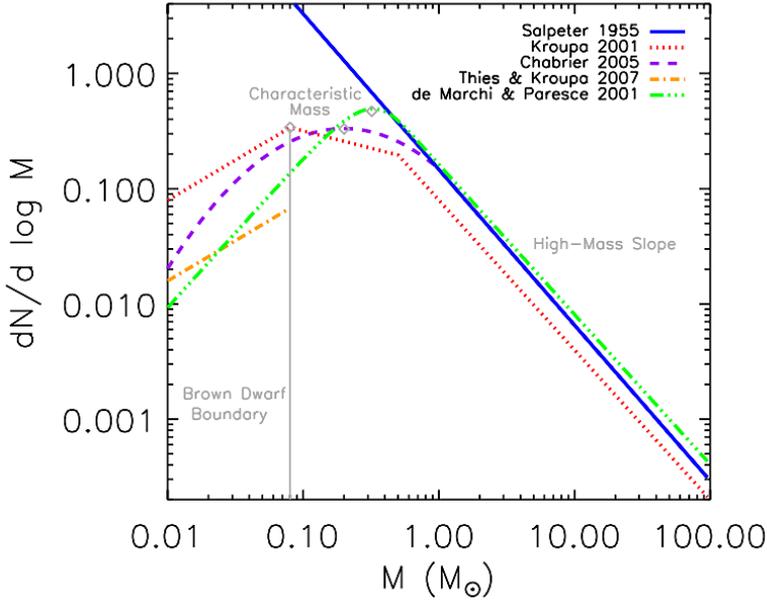}
\caption{Figure from \cite{2014prpl.conf...53O} showing different initial mass functions.}
\label{fig:imf}
\end{figure}

To illustrate the importance of IMFs on physical parameters, we look at SFRs. Section~\ref{sec:emsfr} explains how emission line fluxes can be used to estimate an SFR, following \cite{1998ARA&A..36..189K}. In that review, he assumes a Salpeter IMF for the SFR relations. This was determined by \cite{1955ApJ...121..161S} for stars in the Solar neighbourhood, and was for many years widely used. The Salpeter IMF is a power law with a steep increase towards low masses on the form $\xi(M)\propto M^{-2.35}$, where $\xi(M)\Delta M$ is the number of stars with masses in the range $M$ to $M+\Delta M$. It predicts the formation of numerous solar-mass (and below) stars, and only few stars of very high masses (10--100\,M$_{\odot}$). More recent work has shown that while the Salpeter function works reasonably well at high masses, the IMF appears to be flatter at lower masses, and actually starts to decrease as the masses decrease towards sub-stellar (brown dwarfs), though the variations in IMFs over the Galaxy increase at low masses. Salpeter himself never observed stars with $M<0.1$M$_{\odot}$. Current functions include (among many others) the Kroupa \citep{2001MNRAS.322..231K} or Chabrier \citep{chabrier03} IMFs. These are piecewise functions with different exponents for different mass intervals (the low mass piece of Chabrier is a log-normal function). Some observations even seem to suggest what is called a top-heavy IMF, meaning that there are actually more stars at high mass \citep[see e.g.][]{2010ApJ...717..342H}. A comparison of a handful of IMFs can be seen in Fig.~\ref{fig:imf}. 

To convert the SFRs of \cite{1998ARA&A..36..189K} to newer IMFs, \cite{2007ApJS..173..267S} used the \cite{BC03} models (see Section~\ref{sec:spsm}) to find conversion factors between the different functions. This relies on fitting the expected H$\alpha$ luminosity for populations with given SFRs, and comparing the results for the different IMFs. They find a conversion factor of $1.58$ between the Salpeter and Chabrier IMFs and $1.06$ from Chabrier to Kroupa. 

\subsection{Age}
The stars in a stellar population such as a galaxy do not all have the same age. Rather, stars are formed continuously or in bursts as long as there is gas available. For the modelling, the usual strategy to handle this, is to sum up single-age stellar populations (often with a single metallicity to go with that age, although in theory there could be metallicity gradients in the galaxy). When giving the age of the stellar population, there is no one obvious choice of how to define an age parameter. One choice is to give the luminosity-weighted mean stellar age, that is, the average age of the stars in the population weighted by luminosity. This is not the most physically meaningful parameter, as massive-young stars are more luminous and hence will dominate the age, but it is directly connected to the SED. 

One problem with determining the age, is the age/metallicity degeneracy. A higher metallicity will make a star colder, but so will increasing the age. The associated colour change cannot be disentangled by standard photometric observations. One way to distinguish between the two is to compare the strength of Balmer lines, which are sensitive to age, to the strength of emission lines from metals. This require either spectral data or specific narrow-band photometry. Photometric observations can detect a strong Balmer break however, see Section~\ref{sec:balmer}, which can rule out a very young population, as the Balmer absorption line widths increase with stellar age until about a few hundred Myr, where the stellar population is dominated by A type stars.

Regarding metallicity, a further problem is the possibility of non-Solar relative abundances, see Section~\ref{sec:abundances}. Different abundance ratios have yet to be implemented in most models.

\subsection{Mass-loss}
Stars lose mass through stellar winds and pulsation throughout their lives. Stellar mass loss is a very critical part of the input models, as it impacts the colour by a large amount. Despite this, it is one of the parameters  that models handle extremely badly, and hence it is often put in 'by hand' from calibration of clusters. Specifically, the contribution from thermally pulsing asymptotic giant branch (AGB) stars are implemented very differently in different models, causing a large uncertainty \citep[e.g.][]{2007ASPC..374..303B,2009ApJ...699..486C}.

\subsection{Gas and dust}
A galaxy of course consists of more than stars. Non-baryonic matter is not going to affect the SED, so we can ignore that, but gas and dust also interacts with light. Gas is mainly ignored as it affects the spectra through lines that for the most part are smeared out in the SED. Strong emission lines such as H$\alpha$ will affect the magnitude, and are part of the models. Dust is even more important, as it attenuates the light from the stars, contributing to a change in the SED. To account for this an extinction or attenuation curve is assumed. This contribution is explained in detail in Chapter~\ref{sec:dust}. \\
\\
\indent It is worth noting that the subject of stellar evolution theory, despite being a very fundamental part of astronomy, still has a lot of unsolved problems, and that the different models available can give significantly different results. The uncertainties on the resultant fit parameters are therefore large, especially when little other information is available, so that the fit has many unknown parameters for a relatively small data set.

\section{Methods for Determining the Star Formation Rate}\label{sec:starfr}
One of the parameters fitted in stellar population synthesis modelling is the SFR. This can be determined through a range of indicators, some of the most used of which I will go through here.

\subsection{UV luminosity}
The ultraviolet (UV) part of the SED is dominated by light from young, massive stars. With the assumption of an IMF, the UV luminosity can hence, in principle, provide a direct measure of the star-formation rate, provided that this has been constant on time-scales comparable to the life of these young stars. \cite{1998ARA&A..36..189K} gives the relation
\begin{equation}
\text{SFR}(M_{\odot}\,\text{year}^{-1})=1.4\times10^{-28}L_\nu(\text{erg\,s}^{-1}\text{Hz}^{-1}),
\end{equation}
\noindent assuming a Salpeter IMF and with $L_\nu$ in the range $1500$--$2800$\,\AA. 

The UV luminosity is a widely used method of determining the SFR, especially at higher redshifts ($z>1$) where the UV part of the spectrum is shifted into the atmospheric window, and can hence be observed by ground-based telescopes. It is also an appealing method for its simplicity, but suffers several drawbacks. First, as we are probing the massive population of young stars, which in terms of numbers is a minority of the population, the method is very sensitive to the extrapolation to less massive stars, and hence the choice of IMF. Secondly, the UV part of the SED is highly affected by dust, and the determined SFR is therefore very dependent on a correct attenuation curve, as demonstrated in Chapter~\ref{chap:ext}.

\subsection{Emission line fluxes}\label{sec:emsfr}
Nebular recombination lines provide an indirect measure of the same young massive stellar population through re-emission of the ionising photons emitted by these stars. Only the most massive stars, $M>10M_{\odot}$, contribute significantly to the ionisation, so the SFR measured is almost instantaneous. This however makes the method even more dependent on the assumed IMF. One of the most observed lines is H$\alpha$, for which \cite{1998ARA&A..36..189K} gives the conversion (again for a Salpeter IMF):
\begin{equation}
\text{SFR}(M_{\odot}\,\text{year}^{-1})=7.9\times10^{-42}L(\text{H}\alpha)(\text{erg\,s}^{-1}),
\end{equation}
\noindent assuming a Case B recombination at $T_e=10,000$\,K. Case B recombination is for an optically thick nebula, where the emitted photons from recombination to the ground state of hydrogen are re-absorbed again by other hydrogen atoms. Similar relations exists for other lines.

SFRs from recombination lines also suffers from uncertainties in dust extinction, and furthermore assumes a low escape fraction for ionising photons. For a galaxy as a whole, this is usually taken as a fairly good assumption, but for an individual H\,{\sc ii} region, the escape fraction can be high \citep[see e.g.][]{1993AJ....106.1797H}.

H$\alpha$ is the most suitable recombination line to determine SFRs, as it is the least affected by underlying stellar absorption. Weaker lines such as H$\beta$ suffers from absorption, so that the line flux is not a representation of the ionising flux. Instead, forbidden lines have been used which are at shorter wavelengths than H$\alpha$, and are hence redshifted out of the optical spectrum later, and can be used out to larger distances. 

The [\mbox{O\,{\sc ii}}]$\lambda$3727 doublet is often used, as it has a short wavelength, while being relatively bright. The luminosity of forbidden lines is not directly linked with the ionising photons, and are sensitive to metallicity and ionisation conditions. [\mbox{O\,{\sc ii}}]$\lambda$3727 has been shown to calibrate well with H$\alpha$, but the calculated SFR is still associated with a large error. The \cite{1998ARA&A..36..189K} relation is:
\begin{equation}
\text{SFR}(M_{\odot}\,\text{year}^{-1})=(1.4\pm0.4)\times10^{-41}L[\mbox{O\,{\sc ii}}](\text{erg\,s}^{-1}).
\end{equation}

SFRs from emission lines rely on a correct aperture correction and correct continuum extraction, see Section~\ref{sec:emission}. 

\subsection{FIR luminosity}
SFRs are also being determined from far-infrared (FIR) luminosities. At these wavelengths the spectrum is dominated by dust emission (see Section~\ref{sect:em}). Star-forming regions are usually dusty, and since the dust absorbs the most in UV, the FIR luminosity can be used to estimate the amount of light from young massive stars that has been re-processed by dust. However, older stars also contribute to heating the dust. Furthermore, not all the UV light is absorbed, and how much depend heavily on the distribution and amount of dust in the observed system. 

FIR SFRs are hence most useful when provided together with UV SFRs, in which case both the escaping UV light and dust absorbed light is included. This has the added advantage of not depending on an accurate extinction correction. \cite{1998ARA&A..36..189K} gives:
\begin{equation}
\text{SFR}(M_{\odot}\,\text{year}^{-1})=4.5\times10^{-44}L_{\text{FIR}}(\text{erg\,s}^{-1}),
\end{equation}
\noindent valid for starburst galaxies (and a Salpeter IMF). $L_{\text{FIR}}$ is the integrated luminosity over the wavelength range $8$--$1000$\,$\mu$m.
The FIR SFR shares the dependance on IMF. It has the disadvantage compared to the UV and optical lines, that less telescope time is available to perform the observations, as it requires space-based telescopes, at least out to a redshift where FIR is shifted into the sub-millimeter range. Furthermore, the whole FIR range is rarely observed, so certain monochromatic luminosities are used to represent the integrated range, see e.g. \cite{2013ApJ...774...62L}.

\subsection{Radio Emission}
Radio emission is correlated with star formation through thermal electrons in the H\,{\sc ii} regions emitting through bremsstrahlung, as well as from accelerated electrons within SN shocks. Since SNe are the results of the death of massive (and hence not very old) stars, they trace star formation. SFRs are determined from calibrations to the other star-formation tracers, see for instance \cite{2014AJ....147..103H}. Radio emission has the advantage of not being absorbed by dust. However, it is difficult to perform radio observations deep enough to detect ordinary star-forming galaxies outside of the local Universe.

\subsection{X-ray emission}
In a few cases, X-ray emission has been used as an SFR indicator as well \citep[e.g.][]{2004A&A...425L..33W}. This depends on the galaxy having no active galactic nucleus (AGN), as this otherwise dominates the emission. In the absence of an AGN, a large amount of a galaxies' X-rays originate from young stars, specifically X-ray binaries. However, this method is little tested, and likely has a very large uncertainty, so I will not go into further detail here.

\section{The X-Shooter Sample}\label{sec:xshooter}
We will now look into the practical results of these diagnostics, by taking a look at the sample from the dedicated X-shooter GRB program. The X-shooter is a multi-wavelength, medium resolution spectrograph mounted on the VLT. It has 3 spectroscopic arms which provide spectroscopy from near UV to NIR simultaneously. 

The ultimate goal of the GRB program is to produce a well-defined, homogeneous sample of GRB afterglows. The current sample (as of December 31st, 2013), consists of 58 afterglow observations. The sample is defined so that the GRBs targeted are easy to observe, with Galactic absorption in the line-of-sight at $A_V\leq0.5$\,mag, a visibility above airmass (path length through the atmosphere) 2 for more than 60\,min and an X-ray afterglow observation within 10\,min of the burst trigger. These criteria are not related to the host conditions and should hence minimise any bias, but maximise optical afterglow detections and redshift completeness. 

\begin{figure}[h]
\centering
\includegraphics[width=0.9\textwidth]{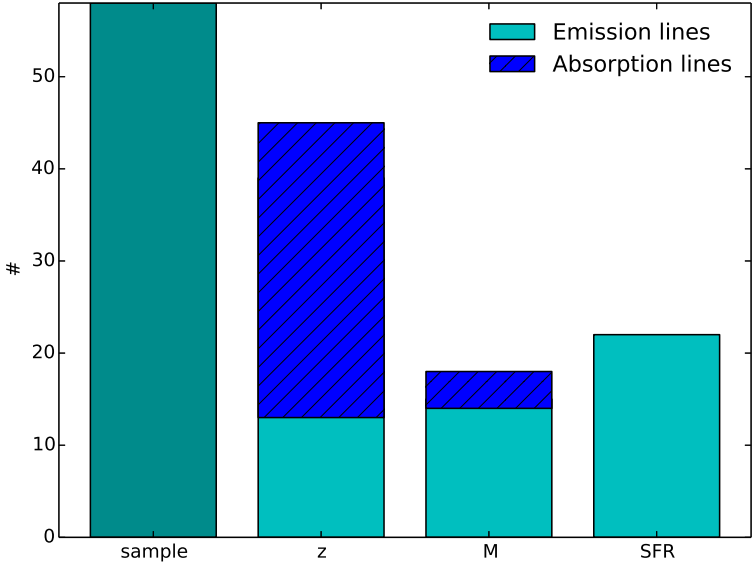}
\caption{Bar chart of the X-shooter GRB sample showing total burst size, number of redshift determinations, number of metallicity determinations, and number of SFR determinations. For redshift and metallicity it is marked how many determinations are from absorption and emission lines respectively. Especially for the redshift determinations there is an overlap, where $z$ can be determined through both emission and absorption.}
\label{fig:sample}
\end{figure}

Fig.~\ref{fig:sample} shows the number of bursts in the sample, those for which it is possible to determine redshift and the number of metallicity and star-formation rates determined from the X-shooter spectrum. The success rate for determining $z$ is quite high, while roughly one third of the bursts allow for a determination of metallicity either through absorption or emission lines (depending on redshift), and similarly about one third of the bursts have a well-determined star-formation rate determined from nebular line fluxes.

One of the science goals of the X-shooter GRB program, is to study star-forming galaxies across the Universe. Lines-of-sight towards GRBs are often observed to pass through Damped Lyman-$\alpha$ absorbers (DLAs). These systems are characterised by a column density of neutral hydrogen $N(\text{H\,{\sc i}})>2\times10^{20}$, and for GRBs sight lines we regularly observe a column density of about log\,$N(\text{H\,{\sc i}})=22$, allowing us to trace the reservoirs of the Universe' neutral hydrogen, i.e. the material available to form stars, out to large redshifts (five bursts in the sample have a redshift $z>4$, with the highest at $z=5.9$). In order to use GRBs to study the evolution in SFR density, or the metallicity bias, see Section~\ref{sec:tracesfr} (though the absorption systems are, in some cases, located at kpc distances from the GRB), we must have a high redshift completeness at these large redshifts.

\begin{figure}[h]
\centering
\includegraphics[width=0.9\textwidth]{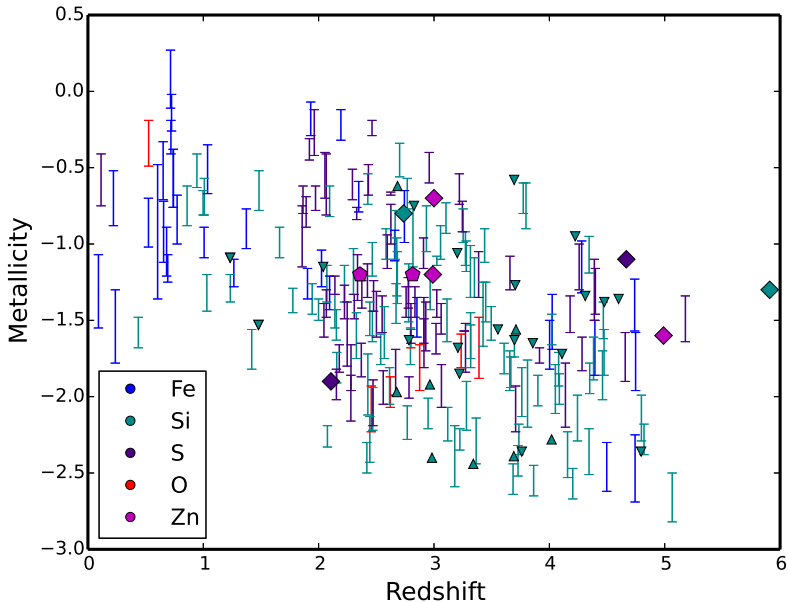}
\caption{Absorption metallicity along the line of sight to quasars and GRBs. GRB metallicities are marked with diamonds (except the two bursts not in the sample, which are marked with pentagons), while QSO-DLAs are marked with bars giving the uncertainty, and QSO-DLA upper and lower limits are marked with triangles. The QSO-DLA sample is taken from \cite{2012ApJ...755...89R} and references therein, while the GRB data are from Friis et al., in prep. Metallicities determined from Fe lines have been increased with 0.3, to account for dust depletion, see Section~\ref{sec:ap}.}
\label{fig:metallicity}
\end{figure}

By examining the line-of-sight metallicity (i.e. in absorption), we may be able to constrain any metallicity preference for the GRB environment. Fig.~\ref{fig:metallicity} shows a comparison to QSO-DLA metallicity evolution with redshift. Though the statistics are still poor, it appears that the GRBs trace environments with similar metallicities, and have a potential to push metallicity detections out to a higher redshift than that observed using quasars as background sources. In the figure, I have included X-shooter data that are newer than 2013 (or older than the official start date of the sample), to increase the sample size, as well as included two bursts not part of the official sample, but observed because they were seen to be unusually bright or interesting. This may introduce a small bias. At lower redshift metallicities are secured for the host galaxies (i.e. not along the line-of-sight), from emission line diagnostics, as described in Section~\ref{sec:diag}.

The X-shooter GRB program has also more than doubled the number of  molecular hydrogen detections towards GRBs. For a discussion of this see Section~\ref{GRBmol}, where the detection towards GRB\,121024A is discussed. These detections expand on those from QSO sight-lines, and hence might be used to test, for instance, the relation between Cl\,{\sc i} and H$_2$ column densities (theoretically the two should be linked through chemical reactions) reported by \cite{balashev15}. For GRB\,121024A (see Chapter~\ref{sec:paper_121024A}) the non-detection of Cl\,{\sc i} gives an upper limit of log\,$N(\text{Cl\,{\sc i}})=14.3$, which together with log\,$N(\text{H}_2)\sim19.8$ fits with the relation found for QSOs. The observations of GRB afterglows also offer a unique opportunity to study the dust content of the burst environment, both through depletion line analysis, and through the change in SED, see Chapters~\ref{sec:dust} and~\ref{chap:ext}.

\cleardoublepage
\chapter{The Warm, the Excited, and the Molecular Gas: GRB\,121024A Shining Through its Star-Forming Galaxy.}\label{sec:paper_121024A}

\begin{center}
 M.~Friis$^1$,
 A.~De Cia$^2$, 
 T.~Kr\"{u}hler$^{3,4}$, 
 J. P. U.~Fynbo$^4$, 
 C.~Ledoux$^3$, 
 P. M.~Vreeswijk$^2$,  
D. J.~Watson$^4$, 
D.~Malesani$^4$, 
J.~Gorosabel$^{5,6,7}$, 
R. L. C.~Starling$^8$, 
P.~Jakobsson$^1$,
K.~Varela$^9$, 
K.~Wiersema$^8$, 
A. P.~Drachmann$^{4}$, 
A. Trotter$^{10,11}$, 
C. C.~Th\"{o}ne$^5$,
A.~de Ugarte Postigo$^{5,4}$, 
V.~D'Elia$^{12,13}$, 
J.~Elliott$^9$, 
M.~Maturi$^{14}$, 
P.~Goldoni$^{15}$,
J.~Greiner$^9$, 
J.~Haislip$^{10}$, 
L.~Kaper$^{16}$, 
F.~Knust$^9$, 
A.~LaCluyze$^{10}$, 
B.~Milvang-Jensen$^4$,
D.~Reichart$^{10}$, 
S.~Schulze$^{17,18}$, 
V. Sudilovsky$^9$, 
N. Tanvir$^{8}$, 
S. D.~Vergani$^{19}$
\end{center}

\begin{center}
\emph{published in MNRAS, Volume 451, July 2015, 167}
\end{center}

\begin{footnotesize}
\begin{itemize}
\addtolength{\itemsep}{-1.25\baselineskip}
  \item[$^1$]	Centre for Astrophysics and Cosmology, Science Institute, University of Iceland, Dunhagi 5, 107 Reykjav\'ik, Iceland \\
  \item[$^2$]	Department of Particle Physics and Astrophysics, Faculty of Physics, Weizmann Institute of Science, 76100, Rehovot, Israel \\
  \item[$^3$]	European Southern Observatory, Alonso de C\'ordova 3107, Casilla 19001, Santiago 19, Chile \\
  \item[$^4$]	Dark Cosmology Centre, Niels Bohr Institute, University of Copenhagen, Juliane Maries Vej 30, 2100 Copenhagen, Denmark \\
  \item[$^5$]	Instituto de Astrof\'isica de Andaluc\'ia (IAA-CSIC), Glorieta de la Astronom\'ia s/n, 18008 Granada, Spain \\
  \item[$^6$]	Unidad Asociada Grupo Ciencia Planetarias UPV/EHU-IAA/CSIC, Departamento de F\'isica Aplicada I, E.T.S. Ingenieria,  \\ Universidad del Pais Vasco UPV/EHU, Alameda de Urquijo s/n, 48013 Bilbao, Spain \\
  \item[$^7$]	Ikerbasque, Basque Foundation for Science, Alameda de Urquijo 36-5, 48008 Bilbao, Spain \\
  \item[$^8$]	Department of Physics and Astronomy, University of Leicester, University Road, Leicester LE1 7RH, UK \\
  \item[$^9$]	Max-Planck-Institut f\"{u}r extraterrestrische Physik, Giessenbachstra\ss e 1, 85748 Garching, Germany \\
  \item[$^{10}$]	Department of Physics and Astronomy, University of North Carolina at Chapel Hill, Campus Box 3255, Chapel Hill, NC 27599, USA \\
  \item[$^{11}$]	Department of Physics, NC A\&T State University, 1601 E. Market St, Greensboro, NC 27411, USA \\
  \item[$^{12}$]	INAF/Rome Astronomical Observatory, via Frascati 33, I-00040 Monteporzio Catone (Roma), Italy \\
  \item[$^{13}$]	ASI-Science Data Center, Via del Politecnico snc, I-00133 Rome, Italy \\
  \item[$^{14}$]	Zentrum f\"{u}r Astronomie der Universit\"{a}t Heidelberg, Institut f\"{u}r Theoretische Astrophysik, Philosophenweg 12, 69120 Heidelberg, Germany \\
  \item[$^{15}$]	APC, Astroparticule et Cosmologie, Universite Paris Diderot, CNRS/IN2P3, CEA/Irfu, Observatoire de Paris, Sorbonne Paris Cit\'{e}, \\ 10 Rue Alice Domon et Leonie Duquet, F-75205 Paris, Cedex 13, France \\
  \item[$^{16}$]	Anton Pannekoek Institute for Astronomy, University of Amsterdam, Science Park 904, 1098 XH, Amsterdam, The Netherlands \\
  \item[$^{17}$]	Millennium Institute of Astrophysics, Casilla 306, Santiago 22, Chile \\
  \item[$^{18}$]	Instituto de Astrof\'isica, Facultad de F\'isica, Pontif\'icia Universidad Catolica de Chile, Casilla 306, Santiago 22, Chile \\
  \item[$^{19}$]	Laboratoire GEPI, Observatoire de Paris, CNRS-UMR8111, Universit\'{e} Paris Diderot, 5 place Jules Janssen, F-92195 Meudon, France
\end{itemize}
\end{footnotesize}

\textbf{\begin{large}Abstract\end{large}}
We present the first reported case of the simultaneous metallicity determination
of a gamma-ray burst (GRB) host galaxy, from both afterglow absorption lines as well as
strong emission-line diagnostics. Using spectroscopic and imaging observations of the afterglow and
host of the long-duration \emph{Swift} GRB\,121024A at $z\,=\,2.30$, we give
one of the most complete views of a GRB host/environment to date. We observe a
strong damped Ly$\alpha$ absorber (DLA) with a hydrogen column density of log\,$N(\text{H\,{\sc
i}})\,=\,21.88\pm0.10$, H$_2$ absorption in the Lyman-Werner bands
(molecular fraction of log($f$)\,$\approx-1.4$; fourth solid detection of molecular hydrogen in a GRB-DLA), 
the nebular emission lines H$\alpha$, H$\beta$, [\mbox{O\,{\sc ii}}], [\mbox{O\,{\sc iii}}] and [\mbox{N\,{\sc ii}}], as well as metal absorption lines. 
We find a GRB host galaxy that is highly star-forming (SFR\,$\sim$\,40\,M$_\odot$\,yr$^{-1}$), with a dust-corrected metallicity along the line of sight of
[Zn/H]$_{\rm corr} =-0.6\pm0.2$ ($\text{[O/H]}\sim-0.3$ from emission lines), and a depletion factor \lbrack Zn/Fe\rbrack\,=\,$0.85\pm0.04$. 
The molecular gas is separated by 400\,km\,s$^{-1}$ (and 1--3\,kpc) from the gas that is photo-excited by the GRB. This implies a fairly massive host, in agreement with the derived stellar mass of log(M$_*$/M$_\odot$) = $9.9^{+0.2}_{-0.3}$. 
We dissect the host galaxy by characterising its molecular component, the excited gas, and the line-emitting star-forming regions. 
The extinction curve for the line of sight is found to be unusually flat ($R_V\sim15$). We discuss the possibility of an anomalous grain size distributions.
We furthermore discuss the different metallicity determinations from both absorption and emission lines, which gives consistent results for the line of sight to GRB\,121024A.

\section{Introduction}

The study of gamma-ray burst (GRB) afterglows has proven to be a powerful tool
for detailed studies of the interstellar medium (ISM) of star-forming galaxies,
out to high redshifts
\citep[e.g.][]{vreeswijk04,prochaska07,2009A&A...506..661L,sparre13}. With quickly fading
emission spanning the entire electromagnetic spectrum, GRB afterglows offer
a unique opportunity to probe the surrounding environment. The intrinsic
spectrum of the afterglow is well fitted with simple power-law segments, so the
imprints of the intergalactic medium (IGM) as well as the ISM surrounding the burst
are relatively easy to distinguish from the afterglow in the observed spectrum. Moreover, with absorption and emission-line analysis it 
is possible to determine parameters such as H\,{\sc i} column density, metallicity, dust
depletion, star-formation rate (SFR) and kinematics of the GRB host galaxy.

Metallicity is a fundamental parameter for characterising a galaxy and it
holds important information about its history. Metallicity might also play a crucial role in 
the GRB production mechanism. For GRB hosts, the metallicity is
measured either from hydrogen and metal absorption lines, or by using
diagnostics based on the fluxes of strong nebular emission lines, calibrated in
the local Universe. Different calibrations are in use leading to some
discrepancy \citep[e.g.][]{Kudritzki}, and the different diagnostics have
their strengths and weaknesses (e.g. less sensitive to reddening, multiple solutions, 
or more sensitive at high metallicities). 
The absorption lines probe the ISM along the line of sight,
while the nebular line diagnostics determine the integrated metallicity of the H\,{\sc ii} regions of the
host. For GRB damped Ly$\alpha$ absorbers \citep[GRB-DLAs, $N$(H\,{\sc i})$>2\times10^{20}$\,cm$^{-2}$][]{2005ARA&A..43..861W}, a direct comparison of
metallicity from the two methods is interesting because it can either provide a 
test of the strong-line methods or alternatively allow a measurement of a possible
offset in abundances in H\,{\sc ii} regions and in the ISM. So far, this 
comparison has only been carried out for a few galaxy counterparts of DLAs found in the line of sight of background QSOs 
\citep[QSO-DLAs, e.g.][]{bowen05,peroux,noterdaeme12,fynbo13,JW14}. To our knowledge, a comparison for GRB-DLAs has
 not been reported before. For both emission and absorption measurements to be feasible with current instrumentation, the observed
host needs to be highly star-forming, to have strong nebular lines, and at the
same time be at a redshift high enough for the Ly$\alpha$ transition to be
observed (at redshifts higher than $z\approx1.5$ the
Ly$\alpha$ absorption line is redshifted into the atmospheric transmission
window). GRB\,121024A is a $z=2.30$ burst hosted by a highly star-forming galaxy.
We measure abundances of the GRB host galaxy in absorption and compare them with the metallicity determined by
strong-line diagnostics using observed nebular lines from [\mbox{O\,{\sc ii}}],
[\mbox{O\,{\sc iii}}], [\mbox{N\,{\sc ii}}] and the Balmer emission lines.


Apart from the absorption features from metal lines, we also detect the
Lyman-Werner bands of molecular hydrogen. Molecular hydrogen is hard to detect in absorption, because it requires high S/N and mid-high resolution. 
As long duration GRBs (t$_{\text{obs}}>\,$2\,s) are thought to be associated with the death of
massive stars \citep[e.g.][]{2003Natur.423..847H,2003ApJ...591L..17S,sparre11,cano13,schulze14}, they are
expected to be found near regions of active star formation, and hence molecular
clouds. In spite of this, there are very few detections of molecular absorption
towards GRBs  \citep[see e.\,g.][]{tumlinson07}. \cite{2009A&A...506..661L} found that this is likely due to the low
metallicities found in the systems observed with high resolution spectrographs
($R=\lambda/\Delta\lambda\gtrsim40000$). Typically, mid/high-resolution spectroscopy at a sufficient S/N is only possible for the brighter sources.
As is the case for QSO-DLAs, lines of sight with
H$_2$ detections will preferentially be metal-rich and dusty. The observed spectra are therefore
UV-faint and difficult to observe \citep[GRB\,080607 is a striking exception, where observations were possible 
thanks to its extraordinarily intrinsic luminosity and 
rapid spectroscopy, see][]{2009ApJ...691L..27P}. Now with X-shooter \citep{xshooter}
on the Very Large Telescope (VLT) we are starting to secure spectra with 
sufficient resolution to detect H$_2$ for fainter systems resulting in additional detections \citep{thomas1,delia14}.

Throughout this paper we adopt a flat $\Lambda$CDM cosmology with $H_0\,=\,71$\,km\,s$^{-1}$ and $\Omega_\text{M}\,=\,0.27$, and report $1\,\sigma$ errors ($3\,\sigma$ limits), unless otherwise indicated. Reference solar abundances are taken from \cite{asplund09}, where either photospheric or meteoritic values (or their average) are chosen according to the recommendations of \cite{lodders09}. Column densities are in cm$^{-2}$. In Sect.~\ref{obs} we describe the data and data reduction used in this paper, in Sect.~\ref{results} we present the data analysis and results, which are then discussed in Sect.~\ref{discussion}.

\section{Observations and Data Reduction}\label{obs}
On 2012 October 24 at 02:56:12 UT the Burst Alert Telescope \citep[BAT,][]{Barthelmy} onboard the \emph{Swift} satellite \citep{swift} triggered on GRB\,121024A. The X-Ray Telescope (XRT) started observing the field at 02:57:45 UT, 93 seconds after the BAT trigger. About one minute after the trigger, Skynet observed the field with the PROMPT telescopes located at CTIO in Chile and the $16"$ Dolomites Astronomical Observatory telescope (DAO) in Italy \citep{prompt} in filters $g'$,\,$r'$,\,$i'$,\,$z'$ and $BRi$. Approximately 1.8 hours later, spectroscopic afterglow measurements in the wavelength range of 3000\,\AA \ to 25\,000\,\AA \ were acquired (at 04:45 UT), using the cross-dispersed, echelle spectrograph X-shooter \citep{xshooter} mounted at ESO's VLT. Then at 05:53 UT, 3 hours after the burst, the Gamma-Ray burst Optical/NIR Detector \citep[GROND,][]{grond1,grond2} mounted on the 2.2 m MPG/ESO telescope at La Silla Observatory (Chile), performed follow-up optical/NIR photometry simultaneously in $g', r', i', z'$ and $JHK$. About one year later (2013 November 07), VLT/HAWK-I imaging of the host was acquired in the $J$ (07:02:13 UT) and $K$ (06:06:47 UT) band. To supplement these, $B$, $R$ and $i$ band imaging was obtained at the Nordic Optical Telescope (NOT) at 2014 January 06 ($i$) and February 10 ($R$) and 19 ($B$). Gran Telescopio Canarias (GTC) observations in the $g$ and $z$ band were optioned on 2014 February 28. For an overview see Tables~\ref{tab:xshooter},~\ref{tab:res} and~\ref{tab:phot}. Linear and circular polarisation measurements for the optical afterglow of GRB\,121024A have been reported in \cite{wiersema}.

\begin{table}
\centering
\caption{X-shooter observations}
\renewcommand*{\arraystretch}{1.3}
\begin{tabular}{@{} l c c c c c @{}}
\hline\hline
$t_{\rm obs}$ (UT)$^a$	&	$t_{\rm GRB}$ (min)$^b$	&	$t_{\rm exp.}$ (s)&	Mean Airmass	&	Seeing				\\ \hline
04:47:01				&	116					&	600			&	1.23			&	0\parcsec6-0\parcsec7	\\
04:58:35				&	127					&	600			&	1.19			&	0\parcsec6-0\parcsec7	\\
05:10:12				&	139					&	600			&	1.16			&	0\parcsec6-0\parcsec7	\\
05:21:46				&	151					&	600			&	1.13			&	0\parcsec6-0\parcsec7	\\ \hline
\end{tabular}
\\
 $^a$ Start time of observation on October 24, 2012. \\
 $^b$ Mid-exposure time in seconds since GRB trigger.
\label{tab:xshooter}
\end{table}

\subsection{X-shooter NIR/Optical/UV Spectroscopy}
The X-shooter observation consists of four nodded exposures with exposure times of 600\,s each, taken simultaneously by the ultraviolet/blue (UVB), visible (VIS) and near-infrared (NIR) arms. The average airmass was 1.18 with a median seeing of $\sim$0\parcsec7. The spectroscopy was performed with slit-widths of 1\parcsec0, 0\parcsec9 and 0\parcsec9 in the UVB, VIS and NIR arms, respectively. The resolving power $R=\lambda$/$\Delta$$\lambda$ is determined from telluric lines to be $R\,=\,13000$ for the VIS arm. This is better than the nominal value due to the very good seeing. Following \cite{fynbo11} we then infer $R\,=\,7100$ and $R\,=\,6800$ for the UVB and NIR arms, see Table~\ref{tab:res} for an overview.

\begin{table}
\centering
\caption{X-shooter resolution}
\renewcommand*{\arraystretch}{1.3}
\begin{tabular}{@{} l c c @{}}
\hline\hline
Arm		&	Slit			&	R\,=\,$\lambda$/$\Delta$$\lambda$	\\ \hline
NIR		&	0\parcsec9	&	6800							\\
VIS		&	0\parcsec9	&	13000						\\
UVB		&	1\parcsec0	&	7100							\\
\end{tabular}
\label{tab:res}
\end{table}

X-shooter data were reduced with the ESO/X-shooter pipeline version 2.2.0 \citep{pipeline}, rectifying the data on an output grid with a dispersion of 0.15\,\AA/pixel in the UVB, 0.13\,\AA/pixel in the VIS and 0.5\,\AA/pixel in the NIR arm. The wavelength solution was obtained against arc-lamp frames in each arm. Flux-calibration was performed against the spectrophotometric standard GD71 observed during the same night. 
We further correct the flux-calibrated spectra for slit-losses by integrating over filter curves from GROND photometry shifted to X-shooter observation times (assuming a slope of $\alpha=0.8$). For the UVB arm, only the $g'$ band photometry is available, which covers the DLA (see Sect.~\ref{abs}), making this calibration less secure. Wavelengths are plotted in vacuum and corrected for heliocentric motion.

\begin{table}[!ht]
\centering
\caption{Photometric Observations}
\renewcommand*{\arraystretch}{1.3}
\begin{tabular}{@{} l c c c c c @{}}
\hline\hline
Instrument 				&	Time$^{a}$	&	Filter					&	Exp. time (s)		&		Seeing			&	Mag. (Vega)		\\ \hline
MPG/GROND 				&	3.0 h		 	&	$g'$					&	284				&	$1\parcsec55$			&	$20.79\pm0.07$	\\ 
MPG/GROND 				&	3.0 h		 	&	$r'$					&	284				&	$1\parcsec40$			&	$19.53\pm0.05$	\\ 
MPG/GROND 				&	3.0 h		 	&	$i'$					&	284				&	$1\parcsec26$			&	$19.05\pm0.07$	\\ 
MPG/GROND				&	3.0 h		 	&	$z'$					&	284				&	$1\parcsec39$			&	$18.66\pm0.08$	\\
MPG/GROND 				&	3.0 h		 	&	$J$					&	480				&	$1\parcsec36$			&	$17.84\pm0.09$	\\
MPG/GROND 				&	3.0 h		 	&	$H$					&	480				&	$1\parcsec29$			&	$16.98\pm0.10$	\\
MPG/GROND 				&	3.0 h		 	&	$K_s$				&	480				&	$1\parcsec21$			&	$16.07\pm0.11$	\\ \hline
VLT/HAWK-I				&	355.2 d		&	$J$					&	$240\times10$		&	$0\parcsec6$			&	$22.4\pm0.1$		\\
VLT/HAWK-I				&	355.1 d		&	$K$					&	$240\times10$		&	$0\parcsec5$			&	$20.8\pm0.2$		\\
NOT/ALFOSC				&	483.9 d		&	$B$					&	$5\times480$		&	$1\parcsec3$			&	$24.2\pm0.2$		\\
NOT/ALFOSC				&	475.0 d		&	$R$					&	$9\times265$		&	$1\parcsec1$			&	$23.8\pm0.3$		\\
NOT/ALFOSC				&	440.0 d		&	$i$					&	$9\times330$		&	$0\parcsec9$			&	$23.8\pm0.3$		\\
GTC/OSIRIS				&	491.3 d		&	$g'$					&	$3\times250$		&	$1\parcsec6$			&	$24.9\pm0.1$		\\
GTC/OSIRIS				&	491.3 d		&	$z'$					&	$10\times75$		&	$1\parcsec4$			&	$23.2\pm0.3$		\\
\hline
\end{tabular}
\flushleft{$^a$ Time since the GRB trigger (observer's time frame). \\ For the afterglow measurements time is given in hours, while for the host galaxy, it is shown in days.}
\label{tab:phot}
\end{table}


\subsection{NOT, GTC and VLT/HAWK-I imaging}
To derive physical parameters of the host of GRB\,121024A via stellar population synthesis modelling, we obtained late-time photometry from VLT/HAWK-I, NOT and GTC. Exposure times and seeing can be found in Table~\ref{tab:phot}.

$J$ and $K$ band images were observed with HAWK-I on the Yepun (VLT-UT4) telescope at the
ESO Paranal Observatory in Chile. HAWK-I is a near-infrared imager with a pixel scale of 0\parcsec106/pix and a total field of view of $7\parcmin5 \times 7\parcmin5$.
$B$, $R$ and $i$ images were obtained with the ALFOSC optical camera on the NOT. The photometric
calibration was carried out by observing the standard star GD71 at a similar airmass to the GRB field.
$g^\prime$ and  $z^\prime$-band host galaxy images were taken with the
10.4m GTC. The images were acquired with the OSIRIS instrument which provides an
unvignetted field of view of $7\parcmin8 \times 7\parcmin8$ and a pixel
scale of 0\parcsec25/pix \citep{cepa00}. Images were taken
following a dithering pattern. The $z^\prime$-band images
were defringed by subtracting an interference pattern which was
constructed based on the dithered individual frames. The photometric
calibration was carried out by observing the standard star SA95-193
\citep{smith02}. NOT and GTC are located at the observatory of
Roque de los Muchachos, La Palma, Spain.

All images were dark-subtracted and flat-fielded using IRAF standard routines.

\subsection{GROND and Skynet Photometry}
GROND data was reduced using standard IRAF tasks \citep{tody97,thomas08}. The afterglow image was fitted using a general point spread function (PSF) model obtained from bright stars in the field. The optical images in $g', r', i', z'$ were calibrated against standard stars in the SDSS catalogue, with an accuracy of $\pm0.03$\,mag. The NIR magnitudes were calibrated using stars of the 2MASS catalogue, with an accuracy of $\pm0.05$\,mag. Skynet obtained images of the field of GRB\,121024A on 2012 October 24-25 with four $16''$ telescopes of the PROMPT array at CTIO, Chile, and the $16''$ DAO in Italy. Exposures ranging from 5 to 160\,s were obtained in the $BVRI$ (PROMPT) and $g', r', i'$ (DAO) bands, starting at 02:57:07\,UT ($t = 55$\,s since the GRB trigger) and continuing until $t = 7.3$\,h on the first night, and continuing from $t = 20.7 - 25.5$\,h on the second night. Bias subtraction and flat-fielding were performed via Skynet's automated pipeline. Post-processing occurred in Skynet's guided analysis pipeline, using both custom and IRAF-derived algorithms. Differential aperture photometry was performed on single and stacked images, with effective exposure times of 5\,s to 20\,min on the first night, and up to $\sim$4\,h on the second night. Photometry was calibrated to the catalogued $B, V, g', r', i'$ magnitudes of five APASS DR7 stars in the field, with $g', r', i'$ magnitudes transformed to RI using transformations obtained from prior observations of Landoldt stars (Henden, A. et al., in preparation). The Skynet magnitudes can be seen in Appendix~\ref{appendix}.

\section{Analysis and Results}\label{results}

\subsection{Absorption Lines}\label{abs}

\begin{figure}
\centering
\resizebox{\hsize}{!}{\includegraphics[bb=10 0 548 358,clip]{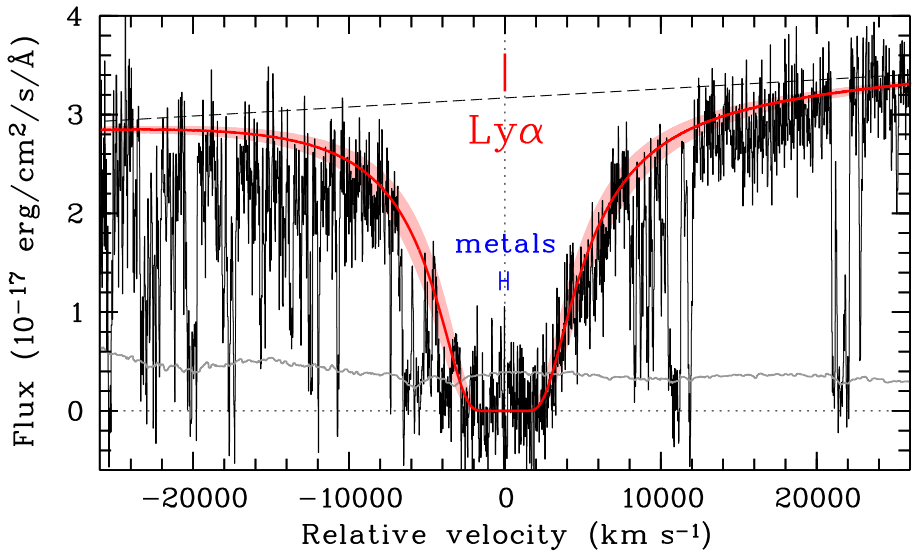}}
\caption{The UVB spectrum centred on the damped Ly$\alpha$ absorption line at the GRB host galaxy redshift. For clarity purposes, the spectrum has been smoothed with a median filter with a sliding window width of 3 pixels. A neutral hydrogen column density fit (log\,$N(\text{H\,I})\,=\,21.88\pm0.10$) to the damped Ly$\alpha$ line is shown with a solid line (red), while the 1$\sigma$ errors are shown with the shaded area (also red). In blue is shown the velocity range of the metal absorption lines. The dashed line shows the continuum placement, while the grey line near the bottom shows the error spectrum.}
\label{fig:dla}
\end{figure}

The most prominent absorption feature is the Ly$\alpha$ line. We plot the spectral region in Fig.~\ref{fig:dla}. Over-plotted is a Voigt-profile fit to the strong Ly$\alpha$ absorption line yielding log\,$N(\text{H\,{\sc i}})=21.88\pm0.10$. The error takes into account the noise in the spectrum, the error on the continuum placement and background subtraction at the core of the saturated lines. Tables~\ref{tab:components_res} and~\ref{tab:components_fine} shows the metal absorption lines identified in the spectrum.

\begin{landscape}

\begin{table}
\renewcommand{\thetable}{\arabic{table}a}
\caption{Ionic column densities of the individual components of the line profile. The transitions used to derive column densities are reported in the second column. Transitions marked in bold are those unblended and unsaturated lines that we use to determine the line-profile decomposition. Velocities given are with respect to the [\mbox{O\,{\sc iii}}] $\lambda$5007 line ($z=2.3015$). (b) / (s) indicate that the line is blended / saturated. The error on the redshifts of each component is $0.0001$}
\renewcommand*{\arraystretch}{1.2}
\begin{tabular}{@{} l c c c c c c @{}}
\hline\hline
Component	&			Transition								&		a 			&		b			&		c			&		d			&		e 			\\ \hline
$z$			&				---								&		2.2981		&		2.2989		&		2.3017		&		2.3023		&		2.3026		\\
$b$ (km\,s$^{-1}$)&				---								&		26			&		21			&		20			&		22			&		35			\\
$v$ (km\,s$^{-1}$)&				---								&		$-264$		&		$-191$ 		&		$64$    		&		$118$		&		$145$		\\ \hline
log($N$)		&												&					&					&					&					&					\\ \hline
Mg\,{\sc i}		&	$\lambda$1827, $\lambda$2026(b)					&	$13.97\pm0.05$	&	$13.57\pm0.05$	&	$<13.4$			&	$13.57\pm0.07$	&	$<13.4$			\\
Al\,{\sc iii}		&	$\lambda$1854(s), $\lambda$1862(s)				&		---			&		---			&		---			&		---			&		---			\\
Si\,{\sc ii}		&	$\lambda$1808(s)								&		---			&		---			&		---			&		---			&		---			\\
S\,{\sc ii}		&	$\lambda$1253(s)								&		---			&		---			&		---			&		---			&		---			\\
Ca\,{\sc ii}		&	$\lambda$3934, $\lambda$3969					&  $13.25\pm{0.16}^{a}$  &  $12.50\pm{0.16}^{b}$   &	$12.20\pm{0.16}$	&	$11.90\pm{0.16}$	&	$11.20\pm{0.16}$	\\
Cr\,{\sc ii}		&	\textbf{$\boldsymbol{\lambda}$2056}, $\lambda$2062(b), \textbf{$\boldsymbol{\lambda}$2066}	&$13.47\pm0.05$	&	$13.48\pm0.05$	&	$13.39\pm0.05$	&	$13.67\pm0.05$	&	$13.34\pm0.09$	\\
Mn\,{\sc ii}		&	\textbf{$\boldsymbol{\lambda}$2576}, \textbf{$\boldsymbol{\lambda}$2594}, \textbf{$\boldsymbol{\lambda}$2606}	&	$13.15\pm0.05$&	$13.07\pm0.05$&	$12.71\pm0.05$&	$13.21\pm0.05$&	$12.93\pm0.05$	\\
Fe\,{\sc ii} 		&	$\lambda$1611, \textbf{$\boldsymbol{\lambda}$2260}, \textbf{$\boldsymbol{\lambda}$2249}	&$15.15\pm0.05$	&	$15.09\pm0.05$	&	$14.81\pm0.05$	&	$15.27\pm0.05$	&	$15.12\pm0.05$	\\
Ni\,{\sc ii}		&	$\lambda$1345, $\lambda$1454, $\lambda$1467.3, $\lambda$1467.8, $\lambda$1709		&	$13.91\pm0.10$	&	$13.88\pm0.10$	&	$13.95\pm0.09$	&	$14.17\pm0.10$	&	$13.73\pm0.29$	\\
Zn\,{\sc ii}		&	$\lambda$2026(b), $\lambda$2062(b)				&	$13.14\pm0.05$	&	$13.05\pm0.05$		&	$12.50\pm0.08$		&	$13.40\pm0.05$	&	$12.19\pm0.40$ \\ \hline
\end{tabular}
\\
\flushleft{
$^{a}$ redshift: 2.2979, $b$-value: 30\,km\,s$^{-1}$, see main text
\\$^{b}$ redshift: 2.2989, $b$-value: 23\,km\,s$^{-1}$, \ \ --
\\$^{c}$ redshift: 2.2979, $b$-value: 20\,km\,s$^{-1}$, \ \ --
\\$^{d}$ redshift: 2.2987, $b$-value: 30\,km\,s$^{-1}$, \ \ --}
\label{tab:components_res}
\end{table}

\end{landscape}

\begin{table}
\centering
\addtocounter{table}{-1}
\renewcommand{\thetable}{\arabic{table}b}
\caption{Same as for Table~\ref{tab:components_res}, but for excited levels.}
\renewcommand*{\arraystretch}{1.2}
\begin{tabular}{@{} l c c c c c c @{}}	
\hline\hline
Component	&												&	$\alpha$ 			&	$\beta$			\\ \hline
$z$			&				---								&	2.2981			&	2.2989			\\
$b$ (km\,s$^{-1}$)&				---								&	28				&	30				\\
$v$ (km\,s$^{-1}$)&				---								&	$-264$   			&	$-191$			\\ \hline
log($N$)		&												&					&					\\ \hline
Fe\,{\sc ii}*	&	$\lambda$2389, $\lambda$2396(b)					&	$13.25\pm0.05$	&	$13.16\pm0.05$	\\
Fe\,{\sc ii}**	&	$\lambda$2396(b), $\lambda$2405(b), $\lambda$2607	&	$12.92\pm0.05$	&	$12.80\pm0.05$	\\
Fe\,{\sc ii}***	&	$\lambda$2405(b), $\lambda$2407, $\lambda$2411(b)	&	$12.63\pm0.05$	&	$12.58\pm0.07$	\\
Fe\,{\sc ii}****	&	$\lambda$2411(b), $\lambda$2414, $\lambda$2622	&	$12.53\pm0.06$	&	$12.61\pm0.05$	\\
Fe\,{\sc ii}*****	&	$\lambda$1559, $\lambda$2360					&	$13.95\pm0.08$	&	$13.68\pm0.13$	\\
Ni\,{\sc ii}**	&	$\lambda$2166, $\lambda$2217, $\lambda$2223		&	$13.43\pm0.05$	&	$13.47\pm0.05$	\\
Si\,{\sc ii}*		&	$\lambda$1309, $\lambda$1533, $\lambda$1816$^{a}$	&$14.98\pm0.11$$^{c}$	&	$14.39\pm0.05^{d}$ \\
\hline
\end{tabular}
\\
\flushleft{
$^{a}$ The column density of the $\alpha$ component of Si\,{\sc ii}* has been determined solely from the $\lambda$ 1816 line.}
\label{tab:components_fine}
\end{table}

\begin{figure}
\includegraphics[width=\textwidth]{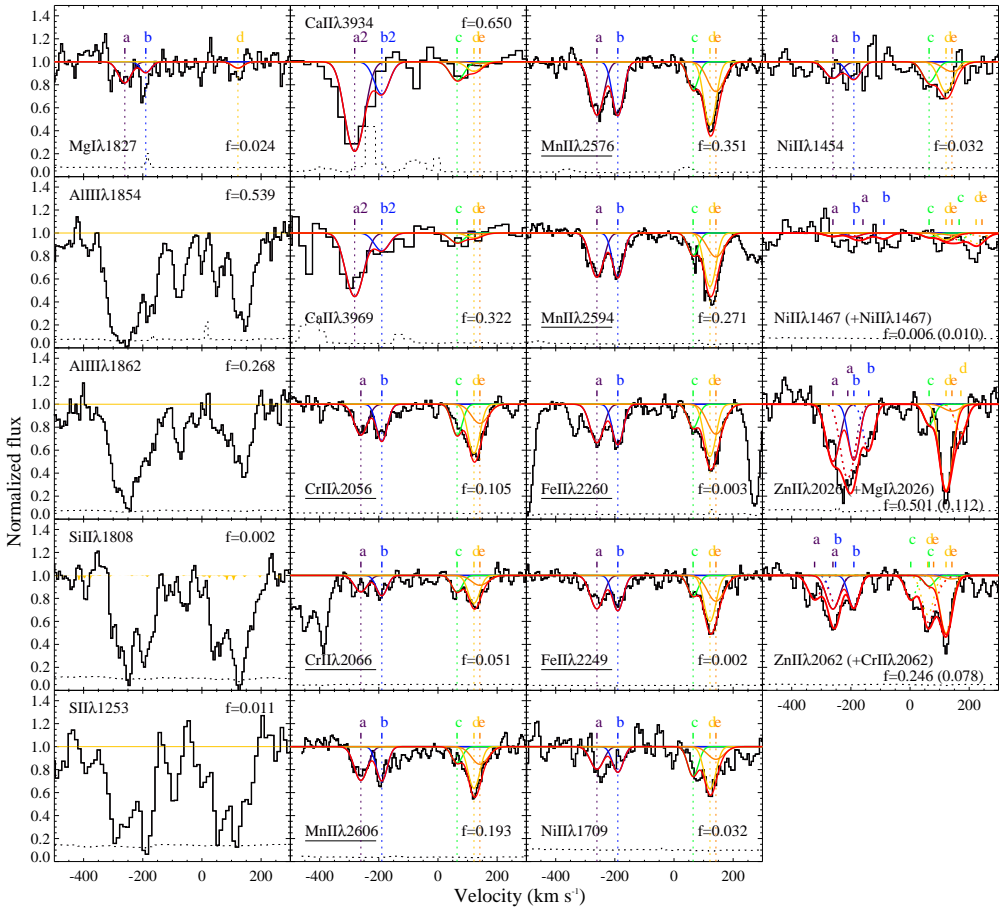}
\caption{Velocity profiles of the metal resonance lines. Black lines show the normalised spectrum, with the associated error indicated by the dashed line at the bottom. The Voigt-profile fit to the lines is marked by the red line, while the single components of the fit are displayed in several colours (vertical dotted lines mark the centre of each component). The decomposition of the line-profile was derived by modelling only the underlined transitions. The oscillator strength 'f', is labelled in each panel. Saturated lines have not been fitted with a Voigt-profile, so for these we show only the spectrum. See online version for colours.}
\label{fig:absorption}
\end{figure}

\begin{figure*}
\includegraphics[width=\textwidth]{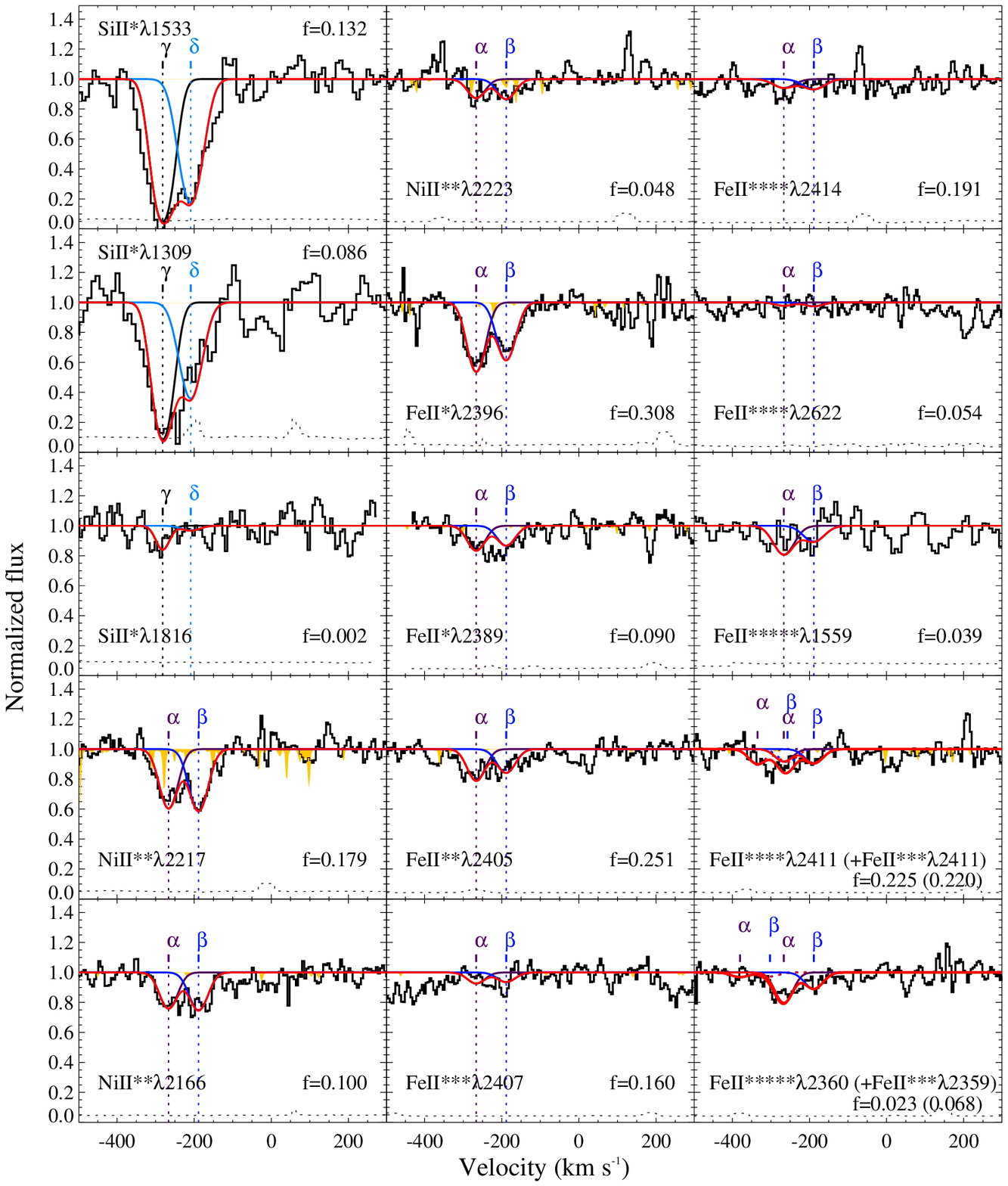}
\caption{The same as Fig.~\ref{fig:absorption}, but for fine-structure lines. Telluric features are highlighted in yellow.}
\label{fig:fine}
\end{figure*}

To determine the ionic column densities of the metals, we model the identified absorption lines with a number of Voigt-profile components, as follows. We use the Voigt-profile fitting software VPFIT\footnote{\url{http://www.ast.cam.ac.uk/~rfc/vpfit.html}} version 9.5 to model the absorption lines. We first normalise the spectrum around each line, fitting featureless regions with zero- or first-order polynomials. To remove the contribution of atmospheric absorption lines from our Voigt-profile fit, we compare the observed spectra to a synthetic telluric spectrum. This telluric spectrum was created following \cite{smette} as described by \cite{2012A&A...545A..64D} and assuming a precipitable water-vapour column of $2.5$\,mm. We systematically reject from the fit the spectral regions affected by telluric features at a level of $>1$ per cent\footnote{This procedure does not aim at reproducing the observed telluric spectrum, but simply reject suspect telluric lines from the Voigt-profile fit.}. None of the absorption lines that we include are severely affected by telluric lines. The resulting column densities are listed in Tables~\ref{tab:components_res},~\ref{tab:components_fine} and~\ref{tab:metal} for lines arising from ground-state and excited levels, respectively. We report formal 1-$\sigma$ errors from the Voigt-profile fitting. We note that these do not include the uncertainty on the continuum normalisation, which can be dominant for weak lines \citep[see e.g.][]{2012A&A...545A..64D}. We hence adopt a minimum error of 0.05\,dex to account for this uncertainty. The error on the redshifts of each component is $0.0001$. The Voigt-profile fits to the metal lines are shown in Figs.~\ref{fig:absorption} and~\ref{fig:fine}.

\begin{landscape}

\begin{table*}
\caption{Total column densities (summed among individual velocity components and including excited levels) and abundances with respect to H and Fe.}
\renewcommand*{\arraystretch}{1.3}
\begin{tabular}{@{} l c c c c c c c c @{}}
\hline\hline
Ion		& log($N$/cm$^{-2}$)$_\text{{tot}}$ 	& log($N$/cm$^{-2}$)$_{\text{a+b}}$& log($N$/cm$^{-2}$)$_{\text{c+d+e}}$ & \lbrack X/H\rbrack$_{\text{tot}}$	& \lbrack X/Fe\rbrack&  \lbrack X/Fe\rbrack$_{\text{a+b}}$	& \lbrack X/Fe\rbrack$_{\text{c+d+e}}$\\ \hline
H\,{\sc i}	&		$21.88\pm0.10$		&		---				&		---				&		---				&		---		&		---				&		---				\\
Mg\,{\sc i}	&		$<14.31$				&	$14.11\pm0.03$		&	$<13.86$				&		---				&		---		&		---				&		---				\\
Al\,{\sc iii}	&		$>14.11$				&		---				&		---				&		---				&		---		&		---				&		---				\\
Si\,{\sc ii}	&		$>16.35$				&	 	---				&		---				&		$>-1.0$			&	$>0.53$		&		---				&		---				\\
S\,{\sc ii}	&		$>15.90$				&		---				&		---				&		$>-1.1$			&	$>0.46$		&		---				&		---				\\
Ca\,{\sc ii}	&		$13.37\pm0.12$		&	$13.32\pm0.13$$^{a}$	&	$12.40\pm0.12$		&		$-2.9\pm0.2$		&	$-1.29\pm0.13$&	$-0.97\pm0.14$$^{a}$	&	$-2.02\pm0.11$		\\
Cr\,{\sc ii}	&		$14.18\pm0.03$		&	$13.78\pm0.04$		&	$13.97\pm0.03$		&		$-1.3\pm0.1$		&	$0.22\pm0.05$	&	$0.18\pm0.05$			&	$0.24\pm0.04$			\\
Mn\,{\sc ii}	&		$13.74\pm0.03$		&	$13.41\pm0.04$		&	$13.47\pm0.03$		&		$-1.6\pm0.1$ 		&     $-0.01\pm0.05$ &	$0.03\pm0.05$			&	$-0.04\pm0.04$		\\
Fe\,{\sc ii}	&		$15.82\pm0.05$		&	$15.45\pm0.05$		&	$15.58\pm0.03$		&		$-1.6\pm0.1$		&		---		&		---				&		---				\\
Ni\,{\sc ii}	&		$14.70\pm0.06$		&	$14.33\pm0.05$		&	$14.47\pm0.06$		&		$-1.4\pm0.1$		&	$0.17\pm0.08$	&	$0.02\pm0.08$			&	$0.16\pm0.06$			\\
Zn\,{\sc ii}	&		$13.74\pm0.03$		&	$13.40\pm0.03$		&	$13.47\pm0.04$		&		$-0.7\pm0.1$		&	$0.85\pm0.06$	&	$0.88\pm0.05$			&	$0.83\pm0.05$			\\
\hline
\end{tabular}
\\
\flushleft{
$^a$\,Different a and b broadening parameter and redshift for Ca\,{\sc ii}, see Sect.~\ref{abs}}
\label{tab:metal}
\end{table*}
\end{landscape}

The fit to the absorption lines from ground-state levels is composed of five components (a-e). We consider the redshift of the [\mbox{O\,{\sc iii}}] $\lambda$5007 emission-line centroid 	$z=2.3015$, as the reference zero-velocity. Components 'a' through 'e' are shifted $-264$, $-191$, $64$, $118$ and $145$\,km\,s$^{-1}$, respectively. Given the resolution of the instrument of 23\,km\,s$^{-1}$ (VIS arm), the individual components are blended, and therefore the profile decomposition is not unequivocal. However, regardless of the properties (and numbers) of the individual components, they are clearly divided into two well separated groups: a+b and c+d+e. When forcing more components to the fit of each group, the resultant total column density are consistent with the previous estimate for each of the two groups. We stress that the resultant $b$-values are not physical, but likely a combination of smaller unresolved components. First we determine redshift $z$ and broadening parameter $b$ (purely turbulent broadening) of the individual components of the line profile, by considering only a master-sample of unblended and unsaturated lines (shown in bold in Table~\ref{tab:components_res}), with $b$ and $z$ tied among transitions of different ions. Values for $z$ and $b$ were then frozen for the rest of the absorption lines, and the column densities were fitted. We report 3-$\sigma$ lower and upper limits for the saturated and undetected components, respectively. For the saturated lines  Al\,{\sc iii}, Si\,{\sc ii} and S\,{\sc ii} we do not report column densities from the Voigt-profile fit, but instead from the measured equivalent widths (EWs), converted to column densities assuming a linear regime. For these, we only report the total column density for all the components together.

At the H\,{\sc i} column density that we observe, we expect most elements to be predominantly in their singly ionised state \citep{2005ARA&A..43..861W}. We hence expect much of the Mg to be in Mg\,{\sc ii} (for this reason we do not report the abundance of Mg\,{\sc i} in Table~\ref{tab:metal}). Ca\,{\sc ii} seems to have a different velocity composition than the rest of the lines. One possibility is that Ca\,{\sc ii} may extend to a slightly different gas phase, as its ionisation potential is the lowest among the observed lines (less than 1\,Ryd = 13.6\,eV). Alternatively, since the Ca\,{\sc ii} lines are located in the NIR arm, a small shift in the wavelength solution with respect to the VIS arm could cause the observed difference. However, a positive comparison between the observed and synthetic telluric lines rules out any shift in the wavelength calibration. We have allowed $z$ and $b$ to have different values for the two Ca\,{\sc ii} lines. This resulted in a slightly different a+b component, but the same c+d+e component as for the rest of the sample.

The fine-structure lines show a different velocity profile composed only of two components, $\alpha$ and $\beta$, see Table~\ref{tab:components_fine}. The redshift of $\alpha$ and $\beta$ are the same as for component a and b found for the resonance lines (but different broadening parameters). Remarkably, no fine-structure lines are detected at the position of components c+d+e. The Si\,{\sc ii}* lines are poorly fitted when tied together with the rest of the fine-structure lines, so we allow their $z$ and $b$ values to vary freely. These components are then referred to as $\gamma$ and $\delta$, which are quite similar to components $\alpha$ and $\beta$, respectively, see Fig.~\ref{fig:absorption}. The column density for component $\gamma$ of the stronger Si\,{\sc ii}* line appears strongly saturated, so only the $\lambda$1816 line has been used to determine the column density in this component. 

The total ionic column densities (summed over individual components and including excited levels when necessary) are given in Table~\ref{tab:metal}. We also report the column densities of the groups of component a+b and c+d+e, which are well resolved from each other, unlike the individual components. Our first metallicity estimate is from Zn, as this element is usually not heavily depleted into dust \citep[see e.g.][]{pettini94}. We derive \lbrack Zn/H\rbrack=$-0.7\pm0.1$ (the other non-refractory elements Si and S are saturated, but the limits we find are consistent). This is in agreement with the value reported in \cite{cucchiara14}.

\begin{figure}
\centering
\includegraphics[width=0.5\columnwidth]{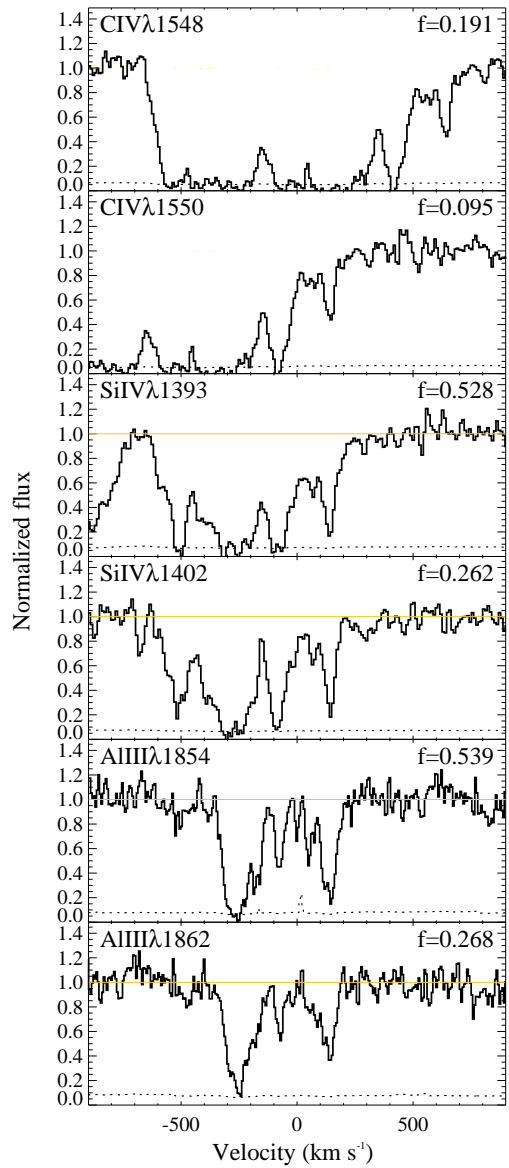}
\caption{High ionisation lines. These lines are highly saturated. See Fig.~\ref{fig:absorption} for details.}
\label{fig:high}
\end{figure}

We note that high ionisation lines from Si\,{\sc iv} as well as C\,{\sc iv} are detected, but are highly saturated, see Fig.~\ref{fig:high}.

\subsection{Dust Depletion}\label{depl}
\label{extinction}
Refractory elements, such as Fe, Ni, and Cr, can be heavily depleted into dust grains \citep[e.g.][De Cia et al. in prep.]{SS96,2002A&A...385..802L}, and thus can be missing from the gas-phase abundances. A first indicator of the level of depletion in the ISM is the relative abundance [Zn/Fe] (referred to as the depletion factor), because Zn is marginally if not at all depleted into dust grains, and its nucleosynthesis traces Fe. We measure [Zn/Fe] $= 0.85 \pm 0.06$. This value is among the highest for QSO-DLAs, but typical at the observed metallicity of [Zn/H] $=-0.7\pm0.1$ \citep[e.g.][De Cia et al. in prep.]{noterdaeme08}. Following \cite{2013A&A...560A..88D} we calculate a column density of Fe in dust-phase of $\log N(\mbox{Fe})_{\rm dust} = 16.74\pm0.17$ and a dust-corrected metallicity of [Zn/H]$_{\rm corr} = -0.6\pm0.2$, indicating that even Zn is mildly depleted in this absorber, by $\sim0.1$\,dex. This is not surprising given the level of depletion, as also discussed by \cite{jenkins09}.

\begin{figure}
\includegraphics[width=0.95\columnwidth]{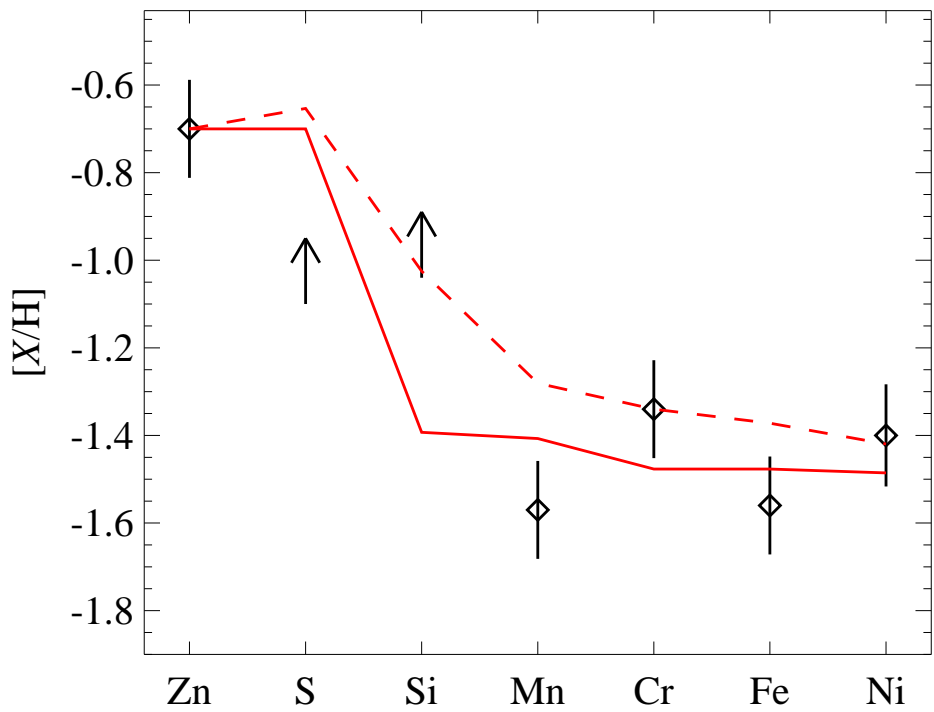}
\caption{The dust-depletion pattern fit for a cold disk (red solid curve) and a warm disk (red dashed curve) to the observed abundances measured from absorption-line spectroscopy (diamonds and arrows, for the constrained and $3\,\sigma$ limits, respectively).}
\label{fig:depletion}
\end{figure}

We also compare the observed abundances of a variety of metals (namely Zn, S, Si, Mn, Cr, Fe, and Ni) to the depletion patterns of a warm halo (H), warm disk+halo (DH), warm disk (WD) and cool disk (CD) types of environments, as defined in \cite{SS96}. These are fixed depletion patterns observed in the Galaxy and calculated assuming that Zn is not depleted into dust grains. We fit the observed abundances to the depletion patterns using the method described in \cite{savaglio01}. We find that none of the environments are completely suitable to describe the observed abundances. The fits to cool- and warm-disk patterns are displayed in Fig.~\ref{fig:depletion} ($\chi^{2}_\nu$=1.18 and 1.58, respectively, with 4 degrees of freedom). For the cool disk the lower limit on the Si column density is not very well reproduced, while the fit for the warm disk overestimates the Mn abundance. The real scenario could be somewhere in between these two environments. Alternatively, the actual depletion pattern is different than what has been observed by \cite{SS96}, or there are some nucleosynthesis effects which we cannot constrain for our case.



Another quantity that is very useful to derive from the observed dust depletion is the dust-to-metals ratio (\dtm{}, normalised by the Galactic value). Constraining the \dtm{} distribution on a variety of environments can indeed shed light on the origin of dust \citep[e.g.][]{mattsson}. Based on the observed [Zn/Fe] and following \cite{2013A&A...560A..88D}, we calculate \dtm{} $=1.01\pm0.03$, i.e. consistent with the Galaxy. From the depletion-pattern fit described above we derive similar, although somewhat smaller, \dtm{} $=0.84\pm0.02$ (CD) and \dtm{} $=0.89\pm0.02$ (WD). These values are in line with the distribution of the \dtm{} with metallicity and metal column densities reported by \cite{2013A&A...560A..88D}, and are also consistent with those of \cite{zafar13}. Following \cite{zafar13}, we calculate \dtm{} $=0.1$ now based on the dust extinction $A_V$ that we model from the SED fit (Sect.~\ref{sed}). Due to the small amount of reddening in the SED, this \dtm{}(A$_V$) value is a factor of 10 lower than expected at the metal column densities observed. This will be discussed further in Sect.~\ref{ext}. 


At the metallicity of GRB\,121024A ($\sim1/3$ solar), it is not possible to draw further conclusions on the dust origin based on the \dtm{}. Both models of pure stellar dust production and those including dust destruction and grain growth in the ISM converge to high (Galactic-like) \dtm{} values at metallicities approaching solar \citep{mattsson}.

\subsection{Distance between GRB and Absorbing Gas}
The most likely origin of the fine-structure transitions observed in the a+b ($\alpha$+$\beta$) component, is photo-excitation by UV photons from the GRB afterglow itself \citep[see e.g][]{prochaska06,2007A&A...468...83V}. Assuming the afterglow to be the only source of excitation, we model the population of the different levels of Fe and Ni, closely following \cite{2013A&A...549A..22V}. Using an optical light curve to estimate the luminosity of the afterglow, we can then determine how far the excited gas must be located from the GRB site, for the afterglow to be able to excite these levels. We model the total column density from component a+b ($\alpha$+$\beta$) of all observed levels (ground state and excited states) of  Ni\,{\sc ii} and Fe\,{\sc ii}. We input the optical light curve from Skynet, see Fig.~\ref{fig:skynet} and Tables in Appendix~\ref{appendix}, which is extrapolated to earlier times using the power-law decay observed. We use the broadening parameter $b$ from the Voigt-profile fits, and the same atomic parameters \citep[see][]{2013A&A...549A..22V}.

\begin{figure}[!ht]
\centering
\includegraphics[width=1.10\columnwidth]{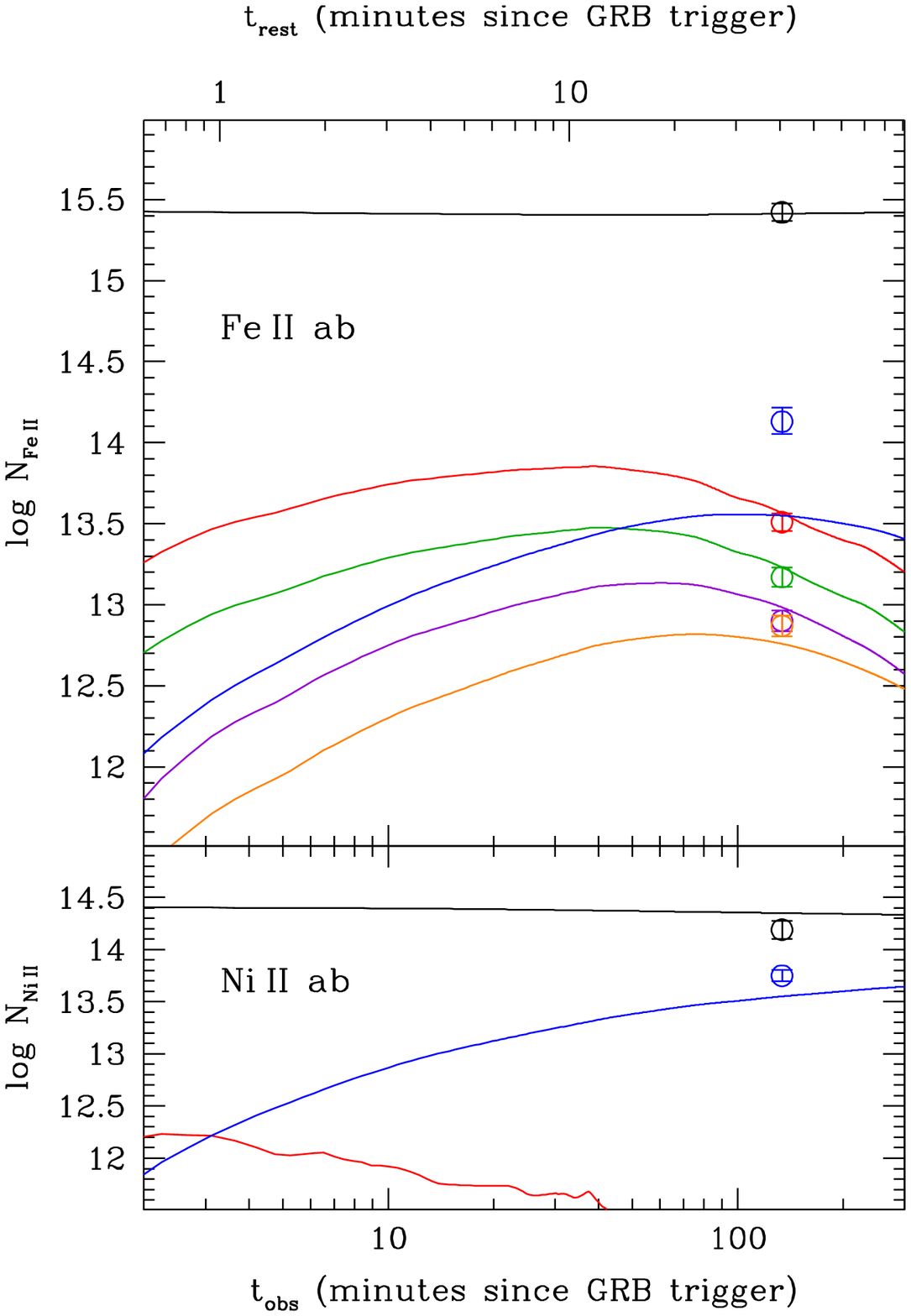}
\caption{Best-fit model for the excited-level populations of the a+b ($\alpha$+$\beta$)
column densities of Fe\,{\sc ii} (top panel) and Ni\,{\sc ii} (bottom). Black lines show the fit to the resonance level. For Fe\,{\sc ii}, from the lower levels and up, excited-level population are shown with red, green, purple, orange and blue. For Ni\,{\sc ii} the red line shows the first excited level, while the blue line shows the second. Open circles show the actual values from Voigt-profile fits. See online version for colours.}
\label{fig:distance}
\end{figure}

The best fit (see Fig.~\ref{fig:distance}) is obtained with a distance of $590\pm100$\,pc between the cloud and burst, and a cloud size of $<333$\,pc (1$\sigma$). The resultant fit is rather poor ($\chi^2$/d.o.f$=40.6/4$). As can be seen in Fig.~\ref{fig:absorption} and~\ref{fig:fine}, the column densities of the ground level of Ni\,{\sc ii} (as probed by Ni\,{\sc ii} $\lambda\lambda\lambda$ 1709, 1454, 1467) and 5th excited level of Fe\,{\sc ii} (as probed by Fe\,{\sc ii} 5s $\lambda\lambda$ 1559, 2360) are not very well constrained due to the observed spectrum having a low S/N ratio near those features. The formal errors from the Voigt profile fit are likely an underestimate of the true error for these column densities. This, in turn, results in the $\chi^2$ of the excitation model fit being overestimated. Furthermore the lack of spectral time series means the resultant parameters are not well constrained. For the c+d+e component we are able to set a lower limit of 1.9\,kpc on the distance to the burst using Fe\,{\sc ii}, and 3.5\,kpc using Si\,{\sc ii} (3$\sigma$). Since Si\,{\sc ii} is saturated, we use the EW to determine the column density, but that only gives the total value of all components together. Hence, for the c+d+e component we fitted using VPFIT and compared the total column density with what we get from the EWs. After establishing that both methods yield the same result, we feel confident in using the column density of log$N$(Si\,{\sc ii})$_{\text{c+d+e}} > 15.99$ together with a detection limit log$N$(Si\,{\sc ii}*)$_{\text{c+d+e}} < 12.80$ on the 1265\,\AA \ line, as this is the strongest of the Si\,{\sc ii}* lines. The lack of vibrationally-excited H$_2$ in the spectra, see below, is in agreement with a distance $\gg100$\,pc, see \cite{draine00}.

\begin{figure}[!ht]
\centering
\includegraphics[width=0.95\columnwidth]{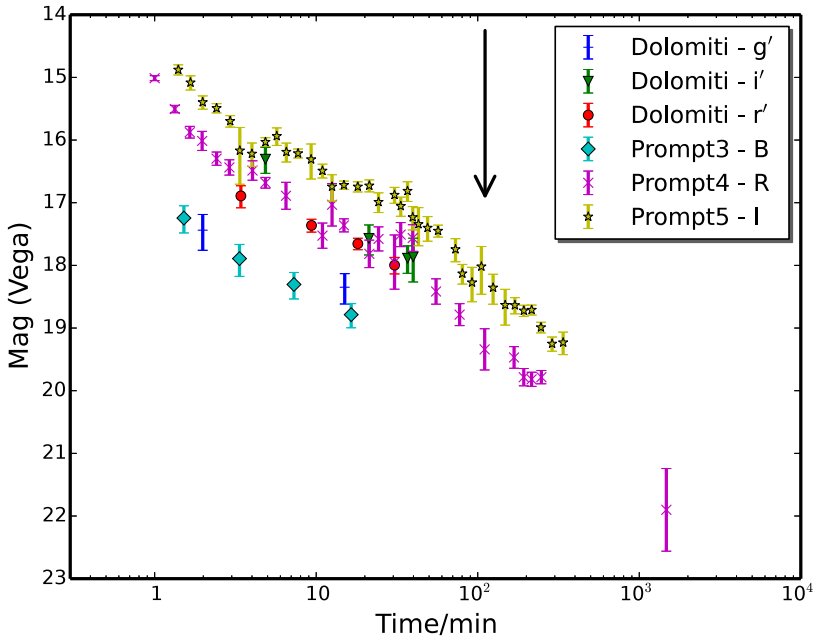}
\caption{GRB-afterglow light curve from the Skynet instruments, used as input for the population modelling. The legend gives the instrument and observational band. The black arrow indicates the starting point of the X-shooter observations. Observations started 55\,s after the GRB trigger. See online version for colours.}
\label{fig:skynet}
\end{figure}

\subsection{Molecular Hydrogen}\label{molecules}

\begin{figure}[!ht]
\includegraphics[width=\textwidth]{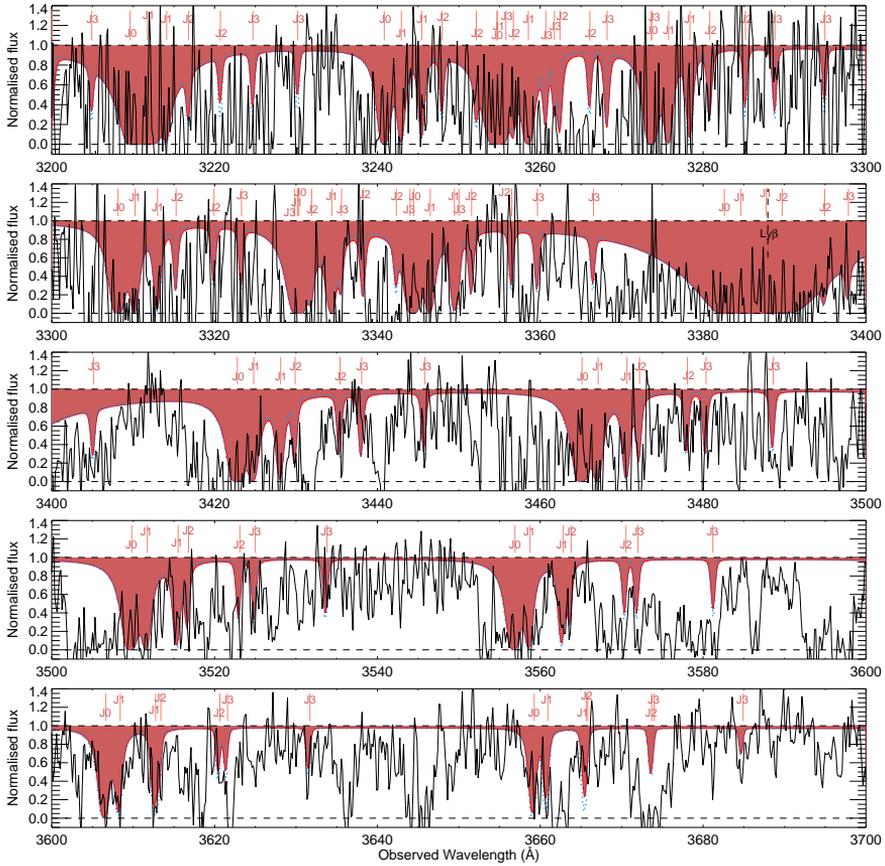}
\caption{X-shooter spectrum showing Lyman- and Werner-band absorption. The shaded area shows the synthetic spectrum from a fit with Doppler parameter $b=1$\,km\,s$^{-1}$, while the blue-dotted line shows the fit for $b=10$\,km\,s$^{-1}$. J0 marks the transitions from the J = 0 rotational level, and likewise for higher J. See online version for colours.}
\label{fig:H2}
\end{figure}

We detect Lyman- and Werner-band absorption lines of molecular hydrogen at redshift $z\,=\,2.3021$ (corresponding to metal-line component "c+d+e") in rotational levels J=0, 1, 2 and 3, see Fig.~\ref{fig:H2}. The fitting and analysis of the molecular hydrogen transition lines follow \cite{2002A&A...385..802L,ledoux03} and \cite{thomas1}. We performed a Voigt-profile fit of lines mainly from the Lyman bands L0-0 up to L3-0, as these are found in the less noisy part of the spectrum (a few J\,=\,2 and 3 lines from the Lyman bands L4-0 and L5-0 were also fitted). $J=0$ and 1 lines are strong and fairly well constrained by the presence of residual flux around them, hinting at damping wings in $L\ge 1$. Given the low spectral resolution of the data and the possibility of hidden saturation, we tested a range of Doppler parameters. The estimated H$_2$ column densities, log $N$(H$_2$) are given in Table~\ref{tab:molecules} for Doppler parameter values of $b=1$ and $10$\,km\,s$^{-1}$ resulting in log\,$N(\text{H}_2)$\,=\,19.8--19.9. Using the column density of neutral hydrogen for component 'c+d+e' of log\,$N(\text{H\,I})=21.6$, calculated assuming the same Zn metallicity for the two main velocity components ('a+b' and 'c+d+e'), this results in a molecular fraction in the order of log $f\sim-1.4$, where $f$\,$\equiv$\,2$N$(H$_2$)/($N$(H\,I)+2$N$(H$_2$)). For the component 'a+b' at redshift $z\sim2.2987$, we report log\,$N(\text{H}_2)<18.9$ as a conservative upper limit on detection. A more detailed analysis is not possible because of the high noise-level. The implications of this detection are discussed in Sect.~\ref{GRBmol}. 


We searched for vibrationally-excited H$_2$ by cross-correlating the observed spectrum with a theoretical model from \cite{draine00} and \cite{2002ApJ...569..780D} similar to the procedure outlined in \cite{thomas1}. There is no evidence for H$_{2}^{*}$ in our data, neither through the cross-correlation nor for individual strong transitions, and we set an upper limit of 0.07 times the optical depth of the input model. This approximately corresponds to $\log N$(H$_{2}^{*})<15.7$. A column density of H$_2^*$ as high as seen in e.g. GRBs\,120815A or 080607 \citep{sheffer09} would have been clearly detected in our data.

We furthermore note that CO is not detected. We set a conservative limit of log\,$N(\text{CO})<14.4$, derived by using four out of the six strongest CO AX bandheads with the lowest 6 rotational levels of CO. The wavelength range of the other two bandheads are strongly affected by metal lines, and thus do not provide constraining information.

\begin{table}
\centering
\caption{Estimated column densities for H$_2$ for broadening parameter values $b=10$ and $b=1$ (km\,s$^{-1}$).}
\renewcommand*{\arraystretch}{1.3}
\begin{tabular}{@{} l >{\centering\arraybackslash}p{2.5cm} >{\centering\arraybackslash}p{2.5cm} @{}}
\hline\hline
Rotational level	 	& \multicolumn{2}{c}{log($N$(H$_2$)/cm$^{-2}$)}		\\ 
					&	$b=10$\,km\,s$^{-1}$	&	$b=1$\,km\,s$^{-1}$	\\ \hline
$J=0$				& 		19.7				&   	19.7				\\
$J=1$ 				& 		19.2 				&  	19.3				\\
$J=2$				& 		16.1				&   	18.3				\\
$J=3$	 			&		16.0 				&	18.2				\\ \hline
Total					&		19.8				&	19.9				\\
\hline
\end{tabular}
\label{tab:molecules}
\end{table}

\subsection{Emission Lines}
\begin{figure}
\centering
\includegraphics[bb=90 162 522 630,width=0.35\columnwidth]{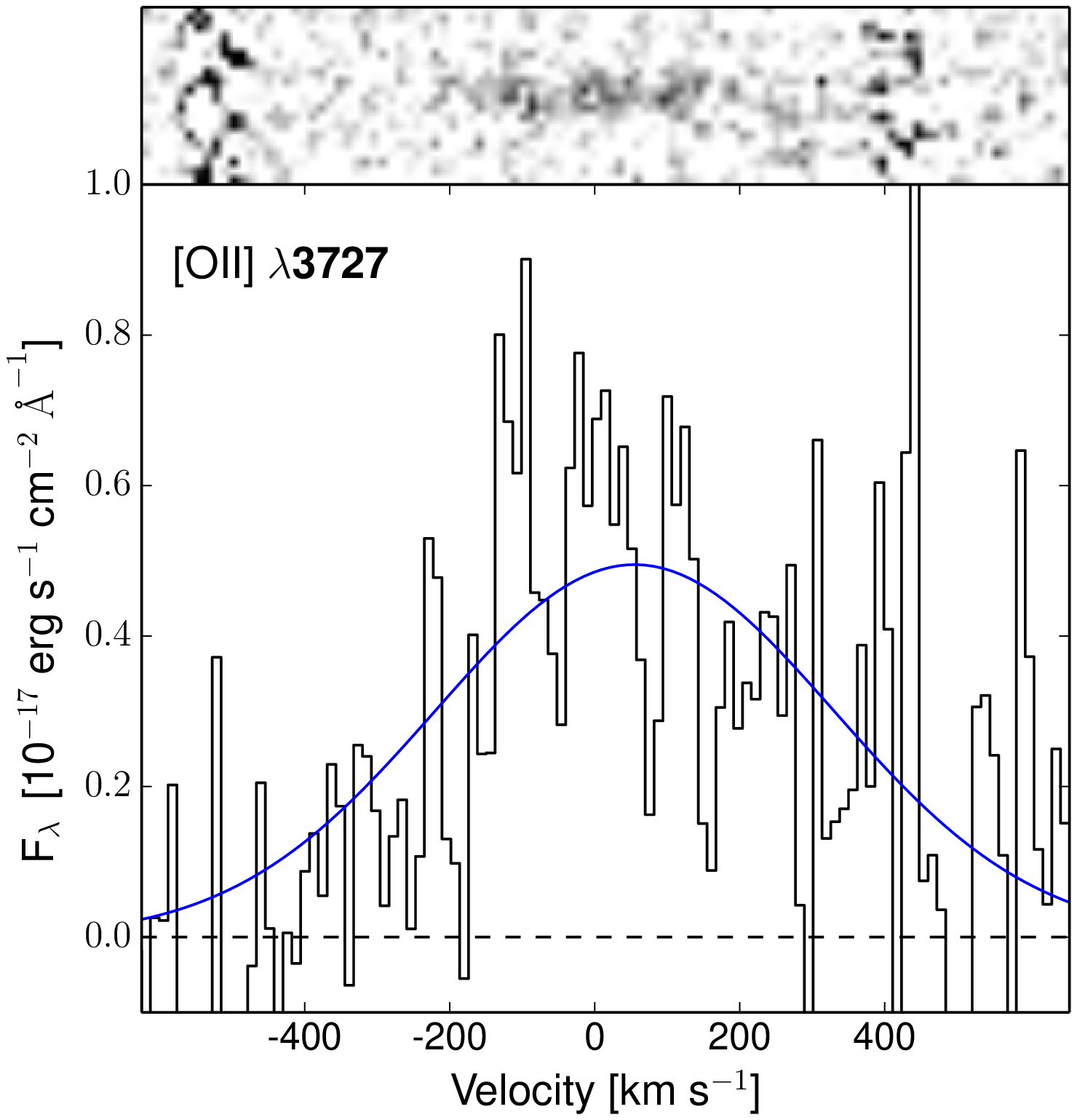}
\includegraphics[bb=90 162 522 630,width=0.35\columnwidth]{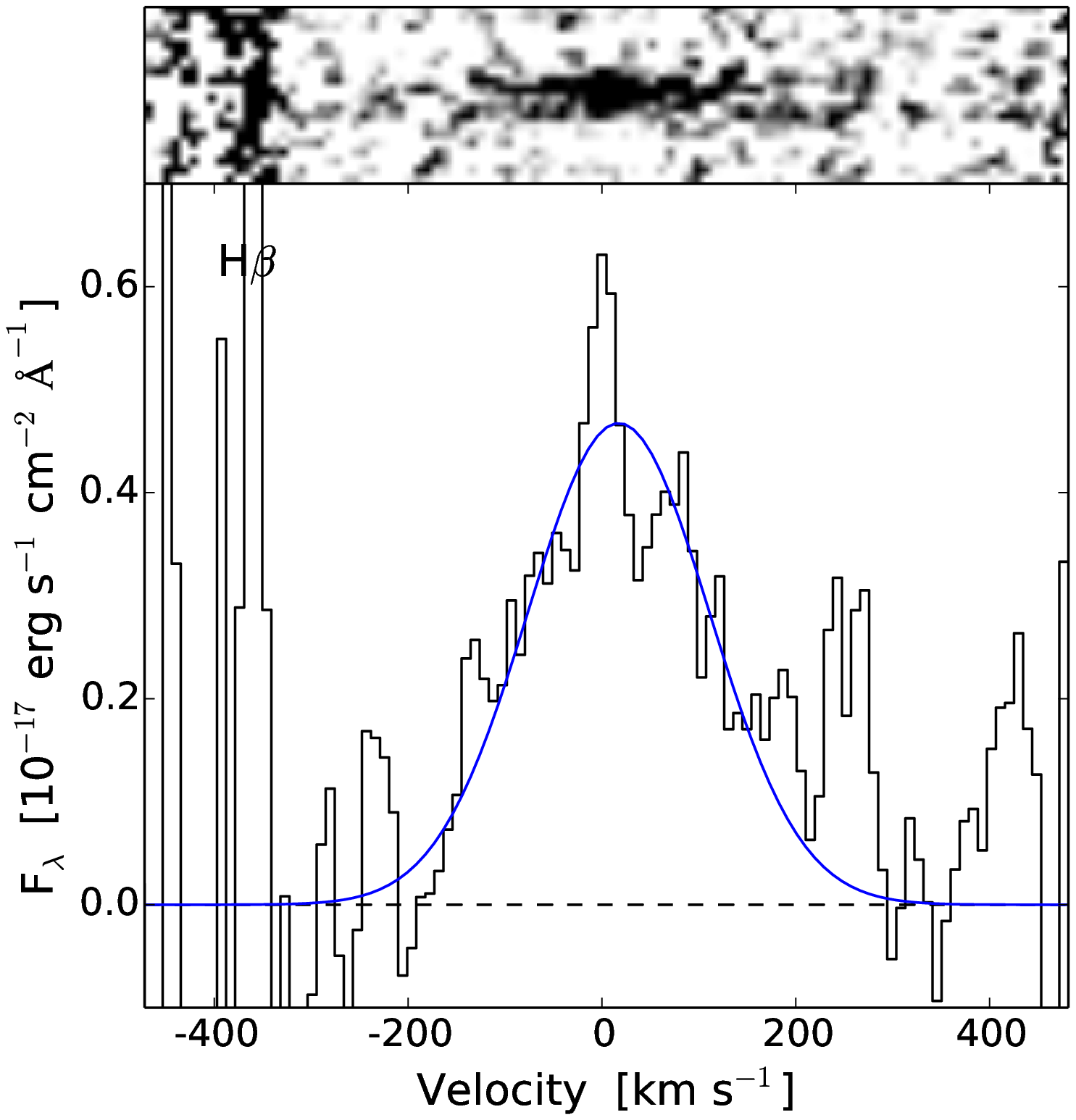}
\includegraphics[bb=90 162 522 630,width=0.35\columnwidth]{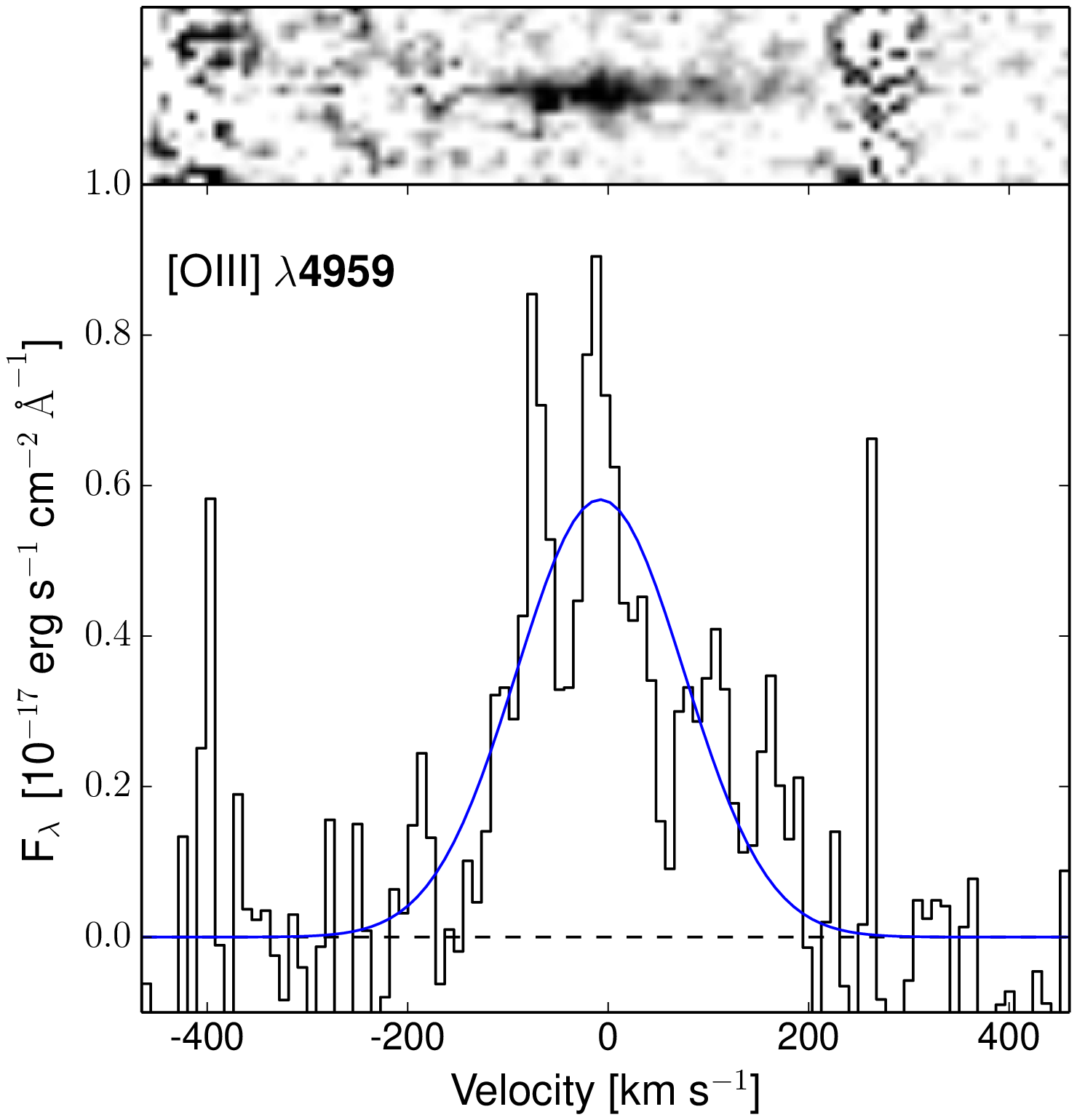}
\includegraphics[bb=90 162 522 630,width=0.35\columnwidth]{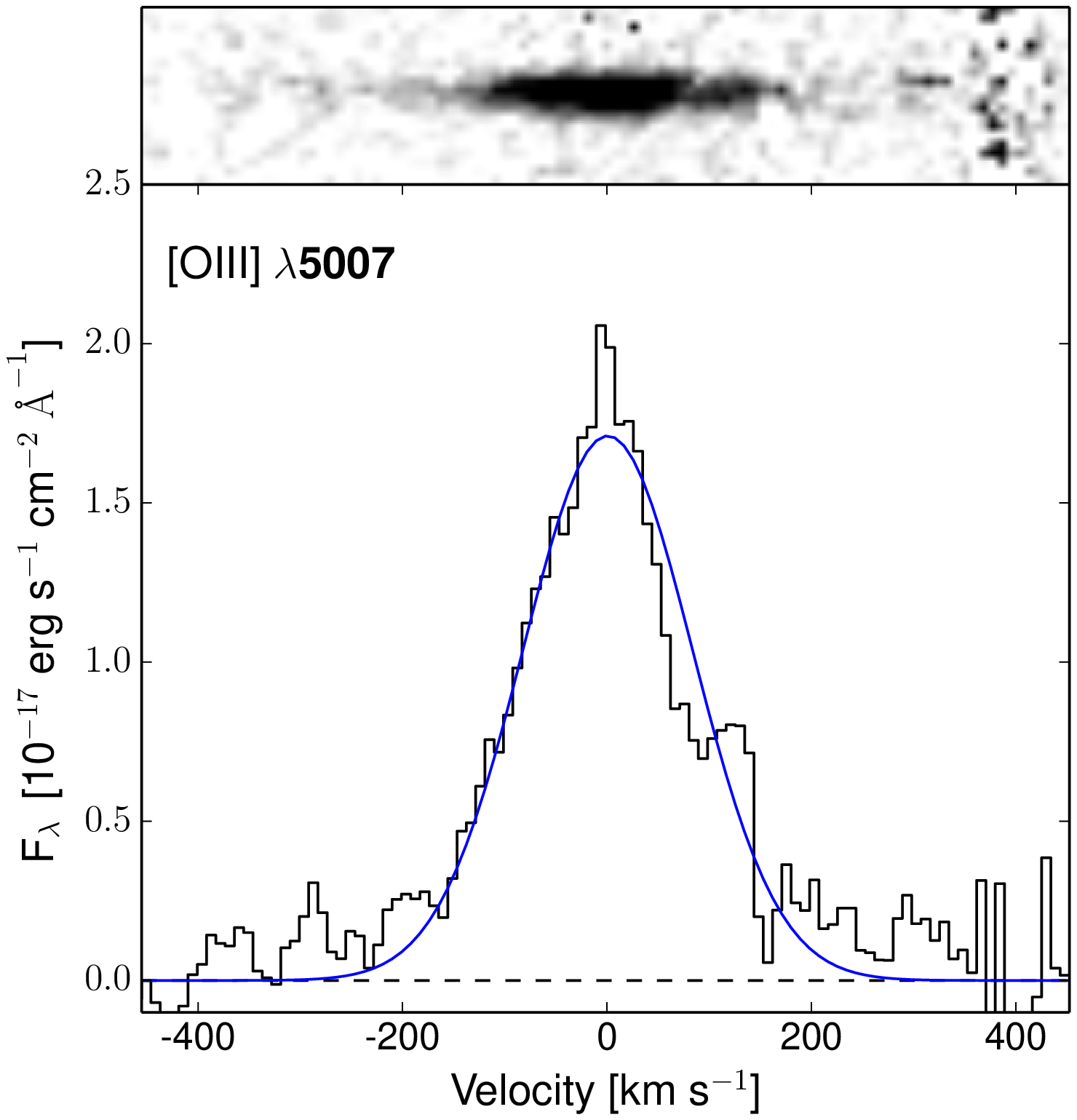}
\includegraphics[bb=90 162 522 630,width=0.35\columnwidth]{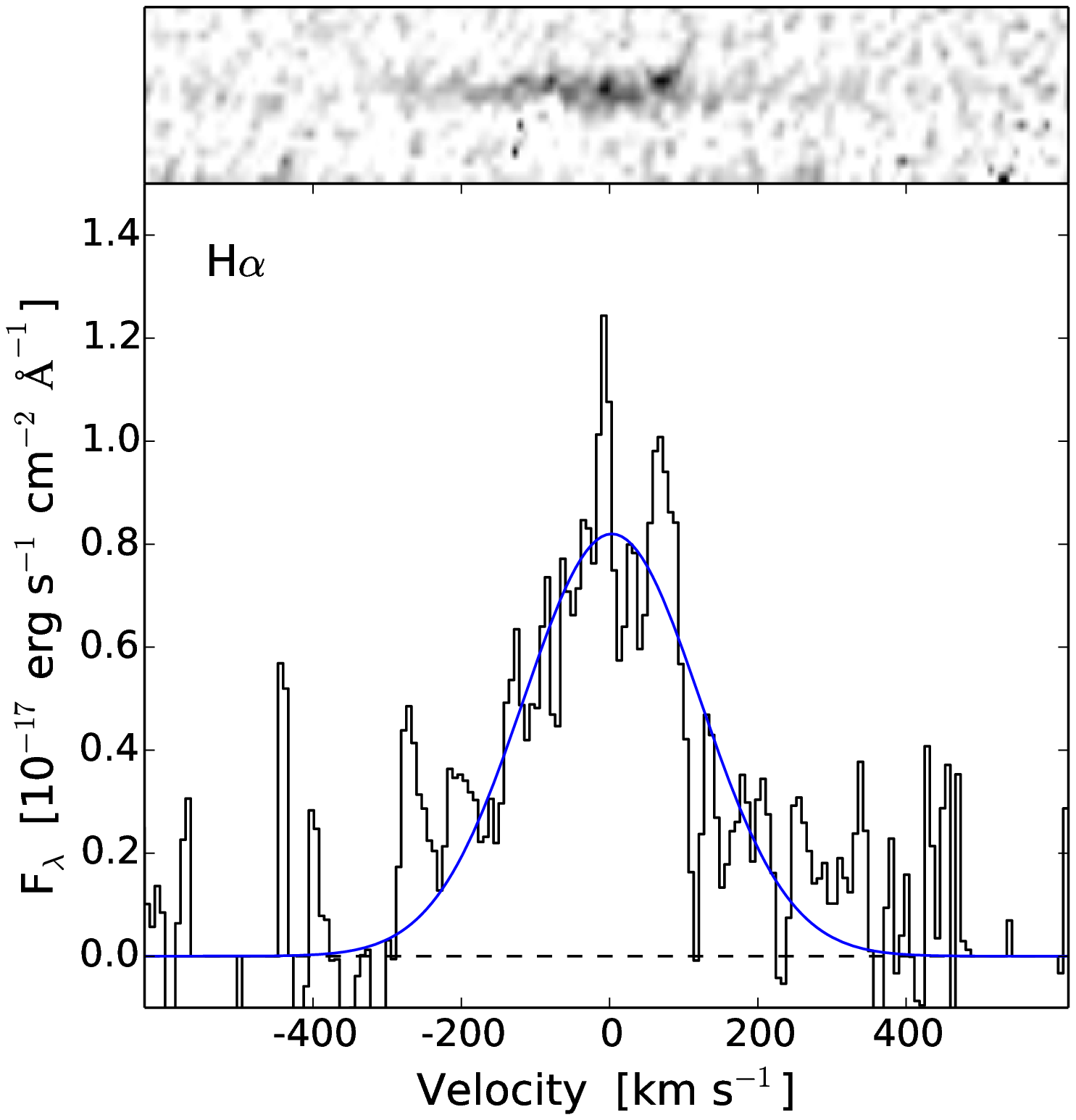}
\includegraphics[bb=0 0 425 425,width=0.36\columnwidth]{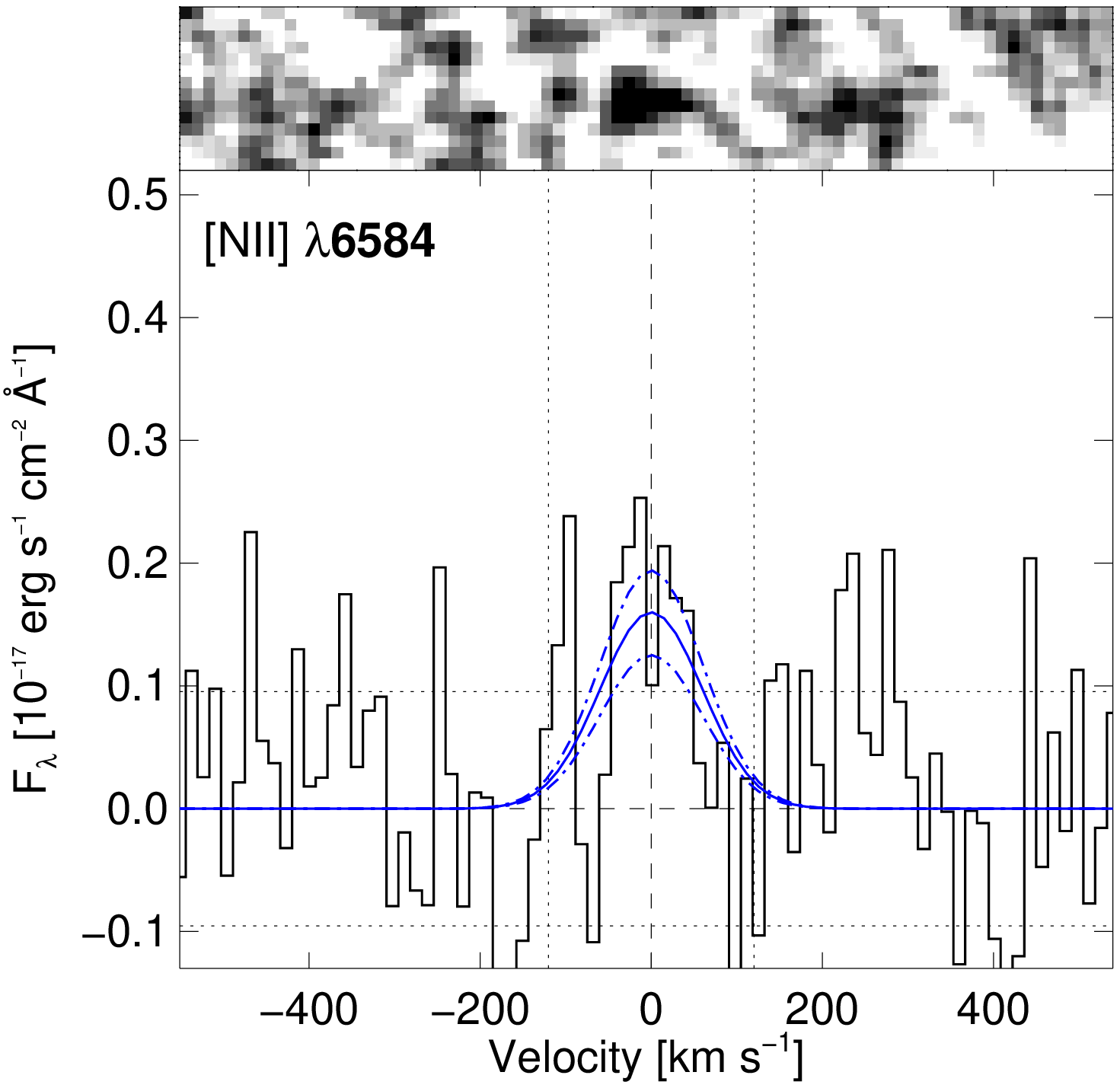}
\caption{Emission lines detected from the GRB\,121024A host. Each panel shows the 2D spectrum after continuum PSF subtraction on top. The bottom part shows the extracted 1D spectrum. The blue line shows the Gaussian fit to the line profile. The abscissa shows the velocity dispersion with respect to the  [\mbox{O\,{\sc iii}}] $\lambda$5007 reference frame. The  [\mbox{N\,{\sc ii}}]  spectrum has been smoothed and binned differently than the other lines, and the fit has been performed with the Gaussian profile centre frozen at 0\,km\,s$^{-1}$ with respect to the reference frame, as indicated with the dashed line in the figure. [\mbox{O\,{\sc ii}}] has been fit as a doublet for the flux estimate.}
\label{fig:emission}
\end{figure}

In the NIR spectrum, we detect H$\alpha$, H$\beta$, the [\mbox{O\,{\sc ii}}] $\lambda$$\lambda$3727, 3729 doublet, [\mbox{N\,{\sc ii}}] $\lambda$6583 (highest-redshift [\mbox{N\,{\sc ii}}] detection published
for a GRB host) and the two [\mbox{O\,{\sc iii}}]  $\lambda$$\lambda$4959, 5007. Table \ref{tab:flux} shows the fluxes (extinction-corrected, see Sect.~\ref{bd}). The reported fluxes are derived from Gaussian fits, with the background tied between the [\mbox{O\,{\sc iii}}] doublet and H$\beta$, and between H$\alpha$ and [\mbox{N\,{\sc ii}}], assuming a slope of the afterglow spectrum of 0.8. [\mbox{O\,{\sc ii}}] is intrinsically a doublet, so we fit a double Gaussian with a fixed wavelength spacing based on the wavelength of the rest-frame lines. Using the GROND photometry, we estimate a slit-loss correction factor of $1.25\pm0.10$.
Fig. \ref{fig:emission} shows the emission-line profiles, the 2D as well as the extracted 1D spectrum. The figure shows a Gaussian fit to the lines, after subtracting the PSF for the continuum \cite[done by fitting the spectral trace and PSF as a function of wavelength locally around each line, see][for details]{moller}. For the weaker [\mbox{N\,{\sc ii}}], a formal $\chi^2$ minimisation is done by varying the scale of a Gaussian with fixed position and width. The noise is estimated above and below the position of the trace (marked by a horizontal dotted line in Fig. \ref{fig:emission}). We assign the zero-velocity reference at the redshift of the [\mbox{O\,{\sc iii}}] $\lambda$5007 line. For the weaker [\mbox{N\,{\sc ii}}] line, we fix the Gaussian-profile fit to be centred at this zero-velocity. 

\begin{table}
\caption{Measured emission-line fluxes}
\renewcommand*{\arraystretch}{1.3}
\begin{tabular}{@{} p{1.9cm} >{\centering\arraybackslash}p{2.7cm} >{\centering\arraybackslash}p{1.9cm} >{\centering\arraybackslash}p{1.9cm} >{\centering\arraybackslash}p{1.9cm} @{}}
\hline\hline
Transition			&	Wavelength$^{a}$	&	Flux$^b$			&	Width$^{c}$	&	Redshift		\\ \hline
[\mbox{O\,{\sc ii}}]  	&	3726.03, 3728.82	&	14.5$\pm$1.2		&	---$^{d}$		&	2.3015$^{e}$	\\
H$\beta$			&	4861.33			&	7.4$\pm$0.4		&	218$\pm$12	&	2.3012		\\
$[\mbox{O\,{\sc iii}}]$&	4958.92			&	9.0$\pm$0.4	  	&	194$\pm$28	&	2.3017		\\
$[\mbox{O\,{\sc iii}}]$&	5006.84			&	27.2$\pm$0.7	 	&	192$\pm$7	&	2.3010		\\
H$\alpha$		&	6562.80			&	21.0$\pm$1.5		&	279$\pm$17	&	2.3010		\\
$[\mbox{N\,{\sc ii}}]$ 	&	6583.41 			&	1.9$\pm$0.7		&	$\sim140$	&	2.3015$^{f}$	\\
\hline
\end{tabular}
\\ $^{a}$ Wavelengths in air in units of \AA.
\\ $^{b}$ Extinction corrected flux in units of 10$^{-17}$\,erg\,s$^{-1}$\,cm$^{-2}$.
\\ $^{c}$ FWHM of line (after removing instrumental broadening) in units of km\,s$^{-1}$. Errors do not include uncertainty in continuum.
\\ $^{d}$ [\mbox{O\,{\sc ii}}] is intrinsically a doublet, which is not fully resolved here, so we do not give the width.
\\ $^{e}$ Calculated using a weighted wavelength average of 3727.7\,\AA.
\\ $^{f}$ The Gaussian fit shown of [\mbox{N\,{\sc ii}}] has a redshift frozen to that of the [\mbox{O\,{\sc iii}}] $\lambda$5007 line.
\label{tab:flux}
\end{table}

\subsection{Star-Formation Rate}\label{sfr}
The SFR can be derived from the emission line fluxes of H$\alpha$ and [\mbox{O\,{\sc ii}}]. Using conversion factors from \cite{1998ARA&A..36..189K}, but converted from a Salpeter initial mass function (IMF) to Chabrier \citep{treyer07}, we report extinction corrected (see Sect.~\ref{bd}) values of $\text{SFR}_{H\alpha}=42\pm11$\,M$_\odot$\,yr$^{-1}$ from the H$\alpha$ flux and a $\text{SFR}_{[\mbox{O\,{\sc ii}}]}=53\pm15$\,M$_\odot$\,yr$^{-1}$ derived from [\mbox{O\,{\sc ii}}]. For a comparison with results from the stellar population synthesis modelling see Sect.~\ref{pop}.

\subsection{Metallicity from Emission Lines}
We determine the gas-phase metallicity of the GRB host galaxy using the strong-line diagnostics R$_{23}$ (using the ratio ([\mbox{O\,{\sc ii}}] $\lambda$$\lambda$3727 + [\mbox{O\,{\sc iii}}]  $\lambda$$\lambda$4959, 5007)/H$\beta$), O3N2 (using ([\mbox{O\,{\sc ii}}]/H$\beta$)/(\nitrogen/H$\alpha$)) and N2 \citep[using \nitrogen/H$\alpha$; for a discussion of the different diagnostics see e.\,g.][]{KE}. Note that different metallicity calibrators give different values of metallicity. R$_{23}$ appears to be consistently higher than O3N2 and N2. The R$_{23}$ diagnostic has two branches of solutions, but the degeneracy can be broken using the ratios [\mbox{N\,{\sc ii}}]/H$\alpha$ or [\mbox{N\,{\sc ii}}]/[\mbox{O\,{\sc ii}}]. In our case [\mbox{N\,{\sc ii}}]/H$\alpha$\,=\,0.09$\pm0.02$ and [\mbox{N\,{\sc ii}}]/[\mbox{O\,{\sc ii}}]\,=\,$0.13\pm0.03$, which places the R$_{23}$ solution on the upper branch (though not far from the separation). Because of the large difference in wavelength of the emission lines used for R$_{23}$, this method is sensitive to the uncertainty on the reddening. Both O3N2 and N2 use lines that are close in wavelength, so for these we expect the reddening to have a negligible effect. Instead, they then both depend on the weaker [\mbox{N\,{\sc ii}}] line, which has not got as secure a detection. We derive 12\,+\,log(O/H)\,=\,$8.6\pm0.2$ for R$_{23}$ \citep{mcgaugh91}, 12\,+\,log(O/H)\,=\,$8.2\pm0.2$ for O3N2 and 12\,+\,log(O/H)\,=\,$8.3\pm0.2$ for N2 \citep[both from][]{PP}. The errors include the scatter in the relations \citep[these values are from][and references therein]{KE}, though the scatter in N2 is likely underestimated. See Sect.~\ref{abund} for a comparison with absorption-line metallicity.

\subsection{Balmer Decrement}\label{bd}
The ratio of the Balmer lines H$\alpha$ and H$\beta$ can be used to estimate the dust extinction. We use the intrinsic ratio found I(H$\alpha$)/I(H$\beta)=2.86$ \citep{balmer}, for star-forming regions (and case B recombination, meaning photons above 13.6\,eV are not re-absorbed), where we expect GRBs to occur. The ratio we measure is 2.98 which, assuming the extinction law of \cite{2000ApJ...533..682C}\footnote{We use the \cite{2000ApJ...533..682C} law, which is an attenuation law for star burst galaxies, where the \cite{pei} laws are relevant for lines of sight towards point-sources inside galaxies where light is lost due to both absorption and scattering out of the line of sight.}, results in $E(B-V)=0.04\pm0.09$\,mag. We note that adopting a different extinction law \citep[from e.g.][]{pei} results in the same reddening correction within errors, because there is little difference within the wavelength range of the Balmer lines.

\subsection{Broad-Band Spectral Energy Distribution}\label{sed}

\begin{figure}
\centering
\includegraphics[width=0.7\columnwidth,angle=-90]{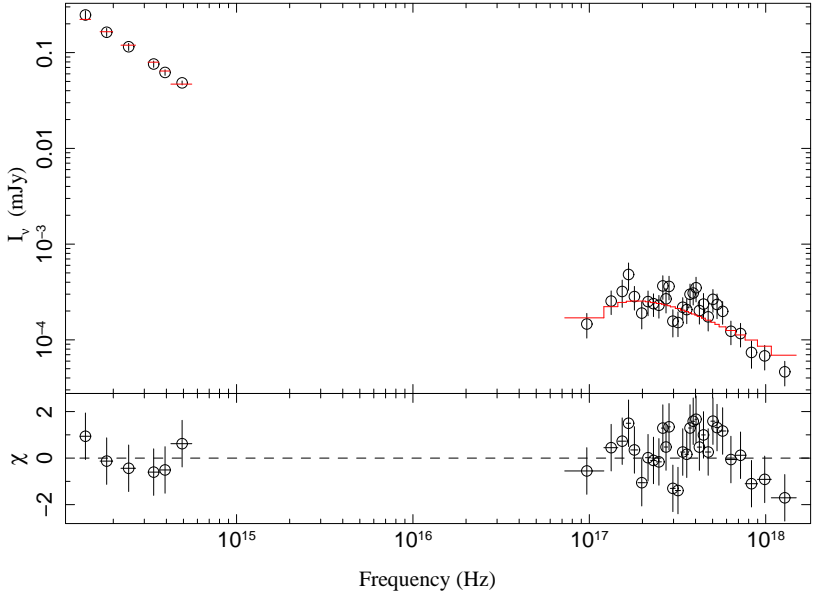}
\caption{NIR-to-X-ray spectral energy distribution and model for the afterglow at 11\,ks after the trigger. The solid red line shows the model. $g'$-band magnitude is not included in the fit, due to possible contribution from the Ly$\alpha$ transition.}
\label{fig:SED}
\end{figure}

We fitted the broad-band afterglow data from XRT and GROND (without the $g'$-band, due to possible DLA contamination), where simultaneous data exist (11\,ks after the trigger). The fit was perform within the ISIS software \citep{isis} following the method of \cite{2007ApJ...661..787S}. The XRT data were extracted using \emph{Swift} tools. We use single and broken power-law models. For the broken power-law, we tie the two spectral slopes to a fixed difference of 0.5. Such spectral feature is known as the "cooling break" of GRB afterglows \citep[e.g.][]{sari98}, and is observed to be the best-fit model for most burst \cite{zafar11}, with the exception of GRB\,080210 \citep{zafar11,annalisa11}. We fit with two absorbers, one Galactic fixed at $N(\text{H})_{\text{X}}^{\text{Gal}}=7.77\times10^{20}$\,cm$^{-2}$ \citep{willingale13}, and one intrinsic to the host galaxy\footnote{We assume solar metallicity, not to provide a physical description of the absorbers, but purely to let $N$(H)$_\text{X}$ conform to the standard solar reference. The reference solar abundances used are from \cite{wilms}.}. An SMC dust-extinction model (the average extinction curve observed in the Small Magellanic Cloud) was used for the host, while the reddening from the Milky Way was fixed to $E(B-V)=0.123$ \citep{schlegel}. A single power-law is preferred statistically ($\chi^2$\,/\,d.o.f = 1.07), see Fig.~\ref{fig:SED}, but the two models give similar results.

The best fit parameters for the single power-law SMC absorption model are $N(\text{H})_\text{X}=(1.2^{+0.8}_{-0.6})\times10^{22}$\,cm$^{-2}$ and $E(B-V)=0.03\pm0.02$\,mag at a redshift of $z=2.298$, and a power-law index of $\beta=0.90\pm0.02$ (90 per cent confidence limits), see Table~\ref{tab:sed}. LMC and MW (the average extinction curves observed in the Large Magellanic Cloud and the Milky Way) model fits result in the same values within errors. For a discussion on the extinction see Sect.~\ref{ext}.

\begin{table}
\centering
\caption{Best fit parameters from the broad-band spectral energy distribution, for a single power-law SMC absorption model.}
\renewcommand*{\arraystretch}{1.5}
\begin{tabular}{@{} p{4.2cm} >{\centering\arraybackslash}p{3.6cm} @{}}
\hline\hline
$N(\text{H})_\text{X}$					&	$(1.2^{+0.8}_{-0.6})\times10^{22}$\,cm$^{-2}$		\\
$E(B-V)$								&	$0.03\pm0.02$\,mag		\\
Power-law index $\beta$					&	$0.90\pm0.02$		\\
\hline
\end{tabular}
\label{tab:sed}
\end{table}

\subsection{Stellar Population Synthesis Modelling}\label{pop}
Using our photometry of the host, see Table~\ref{tab:phot}, we perform stellar population synthesis modelling of the host galaxy. We use a grid of stellar evolution models with different star formation timescales, age of stellar population and extinction, to compute theoretical magnitudes and compare them to the observed photometry. For the model input, we assume stellar models from \cite{BC03}, based on an IMF from \cite{chabrier03} and a Calzetti dust attenuation law \citep{2000ApJ...533..682C}. Table~\ref{tab:pop} lists the galaxy parameters resulting from the best-fit to the HAWK-I, NOT and GTC data. The best fit is obtained with a $\chi^2 = 8$ for the 7 data points used in the modelling. Most of the contribution to the $\chi^2$ comes from the $B$-band observations. This data point lies $\approx$ 3$\sigma$ above the best-fit and the $g$-band measurement, which probes a very similar wavelength range. The reported value of the SFR takes into account the uncertainty in the dust attenuation, and thus has large error bars. We observe a significant Balmer break, which is well fit with star-burst ages between 50 and 500 Myr. The SFR of $\sim40$\,M$_\odot$\,yr$^{-1}$ is consistent with the results from Sect.~\ref{sfr}.


\begin{table}
\centering
\caption{Host galaxy parameters from stellar population synthesis modelling}
\renewcommand*{\arraystretch}{1.5}
\begin{tabular}{@{} p{4.2cm} >{\centering\arraybackslash}p{3.6cm} @{}}
\hline\hline
Starburst age (Myr)						&	$\sim250$		\\
Extinction	(mag)						&	$0.15\pm0.15$		\\
M$_\text{B}$							&	$-22.1\pm0.2$		\\
$\rm log(M_*/M_{\odot})$					&	$9.9^{+0.2}_{-0.3}$	\\
SFR (M$_\odot$\,yr$^{-1}$)				&	$40^{+80}_{-25}$	\\
\hline
\end{tabular}
\label{tab:pop}
\end{table}

\subsection{Kinematics}\label{kinematic}
The X-shooter spectrum contains information both on the kinematics of the absorbing gas along the line-of-sight to the location of the burst inside the host galaxy, as well as kinematics of the emitting gas in H\,{\sc ii} regions probed by the emission lines. The emission lines have a full-width-at-half-maximum (FWHM) of around 210\,km\,s$^{-1}$ from a Gaussian fit, see Table~\ref{tab:flux}. We do not observe signs of rotation in the 2 dimensional spectrum. One possibility is that the galaxy could be dominated by velocity dispersion, as observed for galaxies of similar mass and properties, \cite{sins}.
The velocity width that encloses 90\% of the optical depth \citep[as defined by][]{ledoux06} is 460 km s$^{-1}$ based on the Si\,{\sc ii} $\lambda1808$ line. This is consistent with the correlation between absorption line width and metallicity for GRB host galaxies of \cite{arabsalmani}. The velocity for each absorption component, with respect to the emission lines, is given in Table~\ref{tab:components_res}. The characteristics of different gas components are discussed in Sect.~\ref{gas}.

\subsection{Intervening Systems}\label{intervening}
We identify three intervening systems along the line of sight, at redshifts $z\,=\,2.0798$, $z\,=\,1.959$, and $z\,=\,1.664$. Table~\ref{tab:ew} lists the observed lines along with the measured EWs. Furthermore only at $z\,=\,1.959$ do we observe the Ly$\alpha$ line, but blended with a Si\,{\sc iv} line.
The intervening systems will not be discussed further in this work.


\begin{table}
\centering
\caption{Equivalent widths of intervening systems.}
\renewcommand*{\arraystretch}{1.3}
\begin{tabular}{@{} l >{\centering\arraybackslash}p{1.8cm} >{\centering\arraybackslash}p{1.8cm} >{\centering\arraybackslash}p{1.8cm} @{}}
\hline\hline
					 	& \multicolumn{3}{c}{EWs / \AA }									\\ 
Transition					&	$z=2.0798$		&	$z=1.959$		&	$z=1.664$		\\ \hline
C\,{\sc iv}, $\lambda$1548	& 	$0.43\pm0.10$		&   	$3.64\pm0.15$		&	---				\\
C\,{\sc iv}, $\lambda$1550 	& 	$0.38\pm0.10$		&  	$3.11\pm0.14$		&	---				\\
Fe\,{\sc ii}, $\lambda$2382	& 	---				&	$0.72\pm0.10$		&	$0.36\pm0.05$		\\
Mg\,{\sc ii}, $\lambda$2796	&	---	 			&	$3.73\pm0.08$		&	$1.29\pm0.05$		\\
Mg\,{\sc ii}, $\lambda$2803	&	---				&	$2.53\pm0.11$		&	$1.02\pm0.05$		\\
\hline
\end{tabular}
\label{tab:ew}
\end{table}

\section{Discussion and Implications}\label{discussion}
 
\subsection{Abundance Measurements from Absorption and Emission Lines}\label{abund}

The metallicity of GRB hosts is usually determined either directly through absorption line measurements, or via the strong-line diagnostics using nebular-line fluxes. The two methods probe different physical regions; the ISM of the host galaxy along the line of sight as opposed to the ionised star-forming H\,{\sc ii} regions emission - weighted over the whole galaxy. Hence, the two methods are not necessarily expected to yield the same metallicity, see for instance \cite{JW14}. The line of sight towards the GRB is expected to cross star-forming regions in the GRB host. Thus, the absorption and emission lines may probe similar regions. Local measurements from the solar neighbourhood show a concurrence of the two metallicities in the same region, see e.\,g. \cite{esteban04}. Only a few cases where measurements were possible using both methods have been reported for QSO-DLAs \citep[see e.g.][]{bowen05,JW14} but never for GRB-DLAs. The challenge is that the redshift has to be high enough ($z \gtrsim 1.5$) to make Ly$\alpha$ observable from the ground, while at the same time the host has to be massively star-forming to produce sufficiently bright emission lines. Furthermore, the strong-line diagnostics are calibrated at low redshifts, with only few high redshift cases available \citep[see for instance][]{christensen12}. 

The spectrum of GRB\,121024A has an observable Ly$\alpha$ line as well as bright emission lines. We find that the three nebular line diagnostics R$_{23}$, O3N2 and N2 all find a similar oxygen abundance of $12+\text{log(O/H)}=8.4\pm0.4$. Expressing this in solar units we get a metallicity of \lbrack O/H\rbrack\,$\sim-0.3$ (or slightly lower if we disregard the value found from the R$_{23}$ diagnostic, given that we cannot convincingly distinguish between the upper and lower branch). This is indeed consistent with the absorption line measurement from the low-depletion elements (dust-corrected value) [Zn/H]$_{\rm corr}=-0.6\pm0.2$, though the large uncertainty in the strong-line diagnostics hinders a more conclusive comparison.

\cite{krogager13} find a slightly lower metallicity from absorption lines in the spectrum of quasar Q2222-0946, compared to the emission-line metallicity. However, this is easily explained by the very different regions probed by the nebular lines (6\,kpc above the galactic plane for this quasar) and the line of sight, see also \cite{peroux}. QSO lines of sight intersect foreground galaxies at high impact parameters, while the metallicity probed with GRB-DLAs are associated with the GRB host galaxy. Interestingly, \cite{noterdaeme12} find different values for the metallicities, even with a small impact parameter between QSO and absorber, for a QSO-DLAs. A comparison of the two metallicities is also possible for Lyman break galaxies (LBGs), see for instance \cite{pettini02} for a discussion on the metallicity of the galaxy MS 1512--cB58. They find that the two methods agree for a galaxy with an even larger velocity dispersion in the absorbed gas than observed here ($\sim1000$\,km\,s$^{-1}$). The line of sight toward GRB\,121024A crosses different clouds of gas in the host galaxy, as shown by the multiple and diverse components of the absorption-line profiles. The gas associated with component a+b is photo-excited, indicating that it is the closest to the GRB. Given the proximity, the metallicity of this gas could be representative of the GRB birth site. Assuming the GRB exploded in an H\,{\sc ii} region, the emission- and absorption-metallicities are expected to be similar, though if other H\,{\sc ii} region are dominating the brightness, the GRB birth sight might contribute only weakly to the emission line-flux, see Sect~\ref{gas}. Building a sample of dual metallicity measurements will increase our understanding of the metallicity distribution and evolution in galaxies.


\subsection{The Mass-Metallicity Relation at $z\sim2$}
Having determined stellar mass, metallicity and SFR of the GRB host, we can investigate whether the galaxy properties are consistent with the mass-metallicity relation at the observed redshift. Appropriate for a redshift of $z\approx2$, we use equation 5 from \cite{mannucci10}: \\
\\$12 + \text{log(O/H)} = 8.90 + 0.47 \times (\mu_{0.32} - 10)$ \\
\\
where $\mu_{0.32} = \text{log(M}_*[\,\text{M}_\odot]) - 0.32\times\text{log(SFR}_{\text{H}\alpha}[\text{M}_\odot]\,\text{yr}^{-1})$. Using the stellar mass from Sect.~\ref{pop} and the SFR from H$\alpha$ we find an equivalent metallicity of \\ 12\,+\,log(O/H)\,=\,$8.6\pm0.2$ ([O/H]\,=\,$-0.1\pm0.2$). The error does not include a contribution from the scatter in the relation, and is hence likely underestimated. This value is consistent with the metallicity derived from the emission lines, but given the large uncertainty this is perhaps not that illustrative. Instead, we use the mass-metallicity relation determined in \cite{christensen14} for QSO-DLAs for absorption-line metallicities remodelled to GRBs by \cite{arabsalmani}. This results in a metallicity [$M$/H]\,=\,$-0.3\pm0.2$, not including scatter from the relation, and using the mean impact parameter of 2.3\,kpc calculated in \cite{arabsalmani}. This is consistent with the dust-corrected metallicity of [Zn/H]$_{\rm corr}=-0.6\pm0.2$.

\subsection{Grey Dust Extinction?}\label{ext}
We determine the dust extinction/attenuation of the host galaxy of GRB\,121024A both from the Balmer decrement (Sect.~\ref{bd}) and a fit to the X-ray and optical spectral energy distribution (SED, see Sect.~\ref{sed}), as well as from the stellar population synthesis modelling (Sect.~\ref{pop}). The first method determines the attenuation of the host H\,{\sc ii} regions (from the X-shooter spectrum alone), while the SED fitting probes the extinction along the line of sight (using XRT+GROND data). The stellar population synthesis modelling models the host attenuation as a whole (using host photometry). All methods determine the amount of extinction/attenuation by comparing different parts of the spectrum with known/inferred intrinsic ratios, and attribute the observed change in spectral form to dust absorption and scattering. We find values that agree on a colour index $E(B-V)\sim0.04$\,mag. This value is small, but falls within the range observed for GRB-DLA systems. However, low $A_V$'s are typically observed for the lowest metallicities. For our case we would expect a much higher amount of reddening at our determined H\,{\sc i} column density and metallicity. Using the metallicity of [Zn/H]$_{\rm corr}=-0.6\pm0.2$, column density log\,$N(\text{H\,{\sc i}})=21.88\pm0.10$, dust-to-metal ratio \dtm{}$=1.01\pm0.03$ (see Sect. 3.2), and a reference Galactic dust-to-metal ratio $A_{V, \rm{Gal}}/N_{(H, \rm{Gal})}=0.45\times 10^{-21}$ mag cm$^2$ \citep{2011A&A...533A..16W}, we expect an extinction of $A_V=0.9\pm0.3$\,mag \citep[De Cia et al. in prep. and][]{2003ApJ...585..638S}. This is incompatible with the determined reddening, as it would require $R_V>15$ ($R_V$ for the Galaxy is broadly in the range 2--5). For the Balmer decrement and SED fitting we have examined different extinction curves (MW and LMC besides the SMC) and we have tried fitting the SED with a cooling break, neither option changing the extinction significantly. In an attempt to test how high a fitted reddening we can achieve, we tried fitting the SED with a lower Galactic $N(\text{H\,{\sc i}})$ and reddening. While keeping reasonable values (it is unphysical to expect no Galactic extinction at all), and fitting with the break, the resulting highest colour index is $E(B-V)\sim0.06$\,mag. This is still not compatible with the value derived from the metallicity, so this difference needs to be explained physically.

One possibility to consider is that the host could have a lower dust-to-metals ratio, and hence we overestimate the extinction we expect from the metallicity. However, we see no sign of this from the relative abundances, see Sect.~\ref{depl}. The metallicity is robustly determined from Voigt-profile fits and EW measurement of several lines from different elements (including a lower limit from the none Fe-peak element Si). The lines are clearly observed in the spectra, see Fig.~\ref{fig:absorption}, and the metallicity that we find is consistent with the mass-metallicity relation.

To examine the extinction curve, we perform a fit to the XRT (energy range: 0.3--10\,keV) X-ray data alone and extrapolate the resultant best-fit power-law to optical wavelengths. We try both a single power law, as well as a broken power law with a cooling break in the extrapolation. The latter is generally found to be the best model for GRB extinction in optical fit \citep[e.g.][]{zafar11,2011A&A...526A..30G,schady12}. We calculate the range of
allowed $A_\lambda$ by comparing the X-shooter spectrum to the extrapolation, within the 90\,\% confidence limit of the best-fit photon index, and a break in between
the two data sets (X-ray and optical). The resulting extinction does not redden the afterglow strongly, so we cannot constrain the total extinction very well directly from the SED. However, the optical spectroscopy indicates a high metal column density. The strong depletion of metals from the gas phase supports the presence of dust at $A_V=0.9\pm0.3$\,mag (see Sect~\ref{depl}). Fixing $A_V$ at this level, allows us to produce a normalised extinction curve (Fig.~\ref{fig:curves}). This extinction curve is very flat, much flatter than any in the local group \citep{FM}, with an $R_V>9$.

\begin{figure}[!ht]
\centering
\includegraphics[width=1.05\columnwidth]{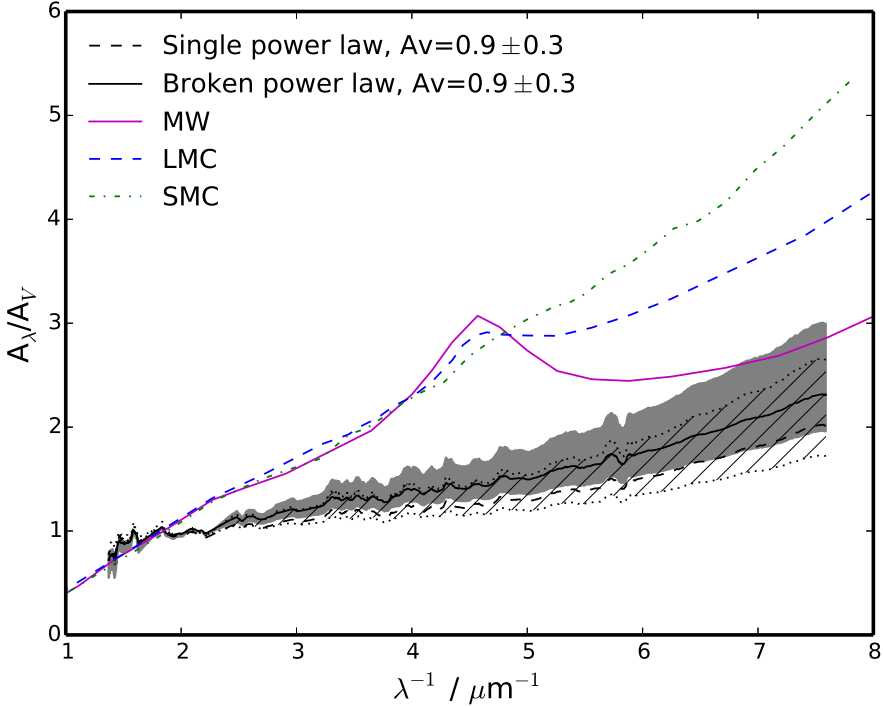}
\caption{Extinction curves for the line of sight to GRB\,121024A. We plot the extinction law assuming $A_V=0.9\pm0.3$\,mag as expected from the measured metallicity, H\,{\sc i} column density and dust-to-metal ratio. The solid black curve shows the extinction curve for a broken power law with $A_V=0.9$, while the grey-shaded area corresponds to the $A_V$ error-space. Likewise, the extinction curve for a single power law is plotted with the dashed black curve, and the hatched area displays the error-space. Over-plotted, in colours, are extinction curves from Pei (1992).}
\label{fig:curves}
\end{figure}

The most likely physical scenario that can explain this shape of the curve is grey dust. If the dust extinction is 'grey', i.e., has a much weaker dependence on wavelength than in the local extinction laws, then a given visual extinction will be much less apparent in the SED ('flat' extinction curve) and thus underestimated in our analysis. Such grey extinction corresponds to larger $R_V$ and is physically interpreted with large grain sizes. A weak wavelength dependance in the extinction for GRBs has been suggested before. \cite{2004ApJ...614..293S} reported a MW-like depletion pattern, but a very low reddening in the SED. \cite{li08} likewise claim a grey extinction law for GRB lines of sight determined by comparing observed spectra to intrinsic ones (by extrapolating from X-rays), arguing for grain growth through coagulation in the dense molecular clouds surrounding GRBs. The larger grains have an extinction that is less dependent on wavelength, because of the contribution of their physical cross-section to the opacity. Preferable destruction of the smaller grains by the GRB would be another possibility, but is unlikely in our case, because the absorbing gas is far from the GRB. We note that the other GRB-DLAs with molecular-hydrogen detection show the expected amount of reddening (using a standard extinction curve), though anomalies do exist in GRB observations. The most notable example to date is reported by \cite{2008ApJ...672..449P} for GRB\,061126. As for the GRB\,121024A afterglow, they observe a very flat optical--to--X-ray spectral index, arguing for large quantities of grey dust, or a separate origin of the optical and X-ray afterglow. To fit the extinction curve for GRB\,061126, an $R_V\sim10$ is needed. We find an $R_V$ even higher than this, making this case even more extreme than previously observed. We refer to future work on this problem (Friis et al., in prep), as a deeper analysis is beyond the scope of this paper.



\subsection{Molecular Hydrogen in GRB-DLAs}\label{GRBmol}
The lack of detection of molecular hydrogen towards GRBs has puzzled astronomers \citep[see e.\,g][]{tumlinson07}, given that long GRBs are associated with active star formation, and hence are expected to show signatures of molecular clouds. Compared to QSO-DLA line of sights then, we would expect the presence of H$_2$ to be more common for GRB-DLAs, because the QSO-DLA line of sights have a higher probability to intersect the outskirts/halo of the intervening galaxy, where we would anticipate a low molecular content. Recently, a number of H$_2$ detections in GRB afterglows have been reported \citep{2009ApJ...691L..27P,thomas1,delia14}, making GRB121024A the fourth definite case. This detection supports the emerging picture that dust has played a major role in biasing past observations against molecular detection \citep[e.g.][]{2009A&A...506..661L}. Molecules are thought to form on the surface of dust grains, and once formed, shielded from Lyman-Werner photons by the grains. \cite{thomas1} suggest that it is likely this connection that is responsible for the low number of H$_2$ detections towards GRB-DLAs. The high dust column density makes the GRB afterglow UV-faint, preventing high-resolution and high S/N spectroscopy, which is needed to identify the presence of molecular gas. Thus, the lack of H2 detections in most GRB-DLAs can be explained with an observational bias. They illustrate this argument by investigating the metallicity, N(H\,{\sc i}) and dust depletion parameter space, showing that the GRB-DLAs with unsuccessful molecular searches fall outside the region where we would expect detections (with the only exception being GRB\,050829A). This argument is supported by the observed log\,$N(\text{H\,{\sc i}})$, metallicity and depletion factor of GRB\,121024A, which lies inside the parameter space where molecular detections are expect.

The high level of dust depletion observed in this GRB-DLA (see Sect.~\ref{depl}), is consistent with molecular detections in QSO-DLAs \citep{noterdaeme08,thomas1}, where there is a strong preference for H$_2$-bearing DLAs to have significant depletion factors. The dependence on the total neutral hydrogen column density is weak (although intrinsically-weak molecular lines are better constrained in strong DLAs), whereas the parameter that seems to determine whether H$_2$ is detected, is the column density of iron locked into dust. The $\log N({\rm Fe})_{\rm dust}$ that we measure is 2\,dex higher than the column density above which a significant presence of molecules has been observed in QSO-DLAs \citep{noterdaeme08}. Indeed \cite{2013A&A...560A..88D} studied $\log N({\rm Fe})_{\rm dust}$ and concluded that GRB hosts are promising sites for molecular detections.

\cite{delia14} find a molecular fraction for GRB\,120327A of log($f$) between $-7$ and $-4$ with a depletion factor of  \lbrack Zn/Fe\rbrack\,=\,$0.56\,\pm\,0.14$, while for the dustier line of sight towards GRB\,120815A \cite{thomas1} reports a value of log($f$)\,=\,$-1.14\pm0.15$ (\lbrack Zn/Fe\rbrack\,=\,$1.01\,\pm\,0.10$). For GRB\,121024A we find intermediate values, although with the high noise-level the numbers are consistent with those reported for GRB\,120815A. For GRB\,080607 \cite{2009ApJ...691L..27P} only report limits on both the molecular fraction and the Zn+Fe column densities. Although the sample is too small to infer anything statistically, it appears that the H$_2$ detection criteria in GRB afterglows follow the trend observed for QSO-DLAs. For a fair estimate of the molecular fraction, the column densities of both H$_2$ and H\,{\sc i} should be constrained for individual velocity components, while this is hardly the case for H\,{\sc i}. Recent work \citep[e.g.][]{balashev15} indicates that the molecular fraction in QSO-DLAs can possibly be much higher than the line-of-sight average values usually measured.

\subsection{Gas kinematics: dissecting the host components}\label{gas}
One of the striking features of the metal absorption-line profiles observed 
towards GRB\,121024A is that they consist of two widely separated groups of 
velocity components (a+b and c+d+e, see Sect.~\ref{abs}). The separation is 
of about 340\,km\,s$^{-1}$, which lies at the high end of the velocity 
distribution of \cite{moller13} and \cite{ledoux06}. The latter is for QSO-DLAs, 
however \cite{arabsalmani} showed that GRB-DLAs follow the velocity-metallicity 
distribution of QSO-DLAs. This suggests that either the two components belong to separate 
galaxies \citep[see for instance][on the GRB\,090323 systems]{savaglio12}, or that this galaxy is fairly 
massive compared to the average GRB host of $\sim$$10^{9}\text{M}_\odot$ 
(\citealt{2009ApJ...691..182S,ceron10}, but see also \citealt{perley13} and 
\citealt{hunt14}). The scenario with separate galaxies is disfavoured, because 
the two absorption components show very similar relative abundances (see 
Table~\ref{tab:components_res}) and also because the emission lines are centred 
in between the two absorption components \citep[unlike for GRBs\,050820A and 
060418; see][]{hw}. Thus, a likely possibility is that the two absorption 
components are probing different regions within the host. This is in agreement 
with the mass found in Sect.~\ref{pop}, of almost $10^{10}\text{\,M}_\odot$.

Furthermore, the blue (a+b) and the red (c+d+e) absorption components are 
associated with gas at different physical conditions. On one hand, Fe\,{\sc ii}, 
Ni\,{\sc ii} and Si\,{\sc ii} fine-structure lines are detected only in the blue 
component. These lines are photo-excited by the GRB radiation at a distance of 
$\sim$$600$\,pc. On the other hand, H$_2$ molecules are detected in the red 
component only, indicating a gas that is not disturbed by the GRB (at a distance 
of minimum 3.5\,kpc). Through absorption-line spectroscopy at the X-shooter resolution, 
these two different gas components could be located inside the host 
(with respect to the GRB) and characterised.

The observed emission component (arbitrarily set at $v = 0$) traces the brightest 
star-forming regions. Since GRBs tend to reside around the brightest star-forming 
regions in their host \citep{2006Natur.441..463F}, one might expect to observe absorption 
components at velocities close to that of the emission as the line of sight passes 
through this gas. However, the gas around the GRB can be photo-ionised out to 
hundreds of parsecs \citep[e.g.][]{2009A&A...506..661L,2013A&A...549A..22V}; it is thus highly unlikely that 
the optical/UV absorption lines are probing the actual GRB environment. For 
GRB\,121024A, this is further supported by the fact that the a,b component is 
located $\sim$$600$\,pc away from the GRB. Given that giant molecular clouds have a 
maximum radius of $\sim$$200$\,pc \citep{murray11}, the a,b component is undoubtedly 
unrelated to the GRB surroundings.

Although GRBs are most often associated with the brightest star-forming regions, 
this is not always the case. GRB\,980425 \citep[e.g.][]{michal} is an example
where the star-forming region in which the GRB occurred is quite faint compared to 
the larger and brighter star-forming regions in the host. A potentially similar 
scenario could hold for GRB\,121024A as well, in which case the burst should not be 
identified with $v = 0$. In this situation, the possible interpretations of the 
kinematics would be different and lead to other geometric setups compared to those 
conceivable were the GRB localised close to $v = 0$. It should also be noted that we 
have not discussed transverse motion which could complicate the interpretation even 
further. Finally, since the host is most likely an irregular galaxy, indicating a 
3D perturbed environment without a rotating disk, we find it appropriate not to draw 
further conclusions.

While the available data do not allow us to discriminate between possible scenarios, 
this work demonstrates how powerful GRB afterglow observations can be to start 
dissecting individual building-block components of star-forming galaxies at $z\sim2$ 
and above. This is especially true once we have gathered enough data to compile a 
statistical sample; see \citep[e.g.][]{2008A&A...491..189F} for previous work on VLT/UVES data and
Fynbo et al. (in prep) for upcoming VLT/X-shooter results on a large afterglow 
sample.

\section*{Acknowledgments}
MF acknowledges support from the University of Iceland Research fund. ADC acknowledges support by the Weizmann Institute of Science Dean of Physics Fellowship and the Koshland Center for Basic Research. The Dark Cosmology Centre is funded by the DNRF. RLCS is supported by a Royal Society Dorothy Hodgkin Fellowship. JPUF acknowledge support from the ERC-StG grant EGGS-278202. CCT is supported by a Ram\'{o}n y Caj\'{a}l fellowship. The research activity of J. Gorosabel, CCT, and AdUP is supported by Spanish research project AYA2012-39362-C02-02. Ad.U.P. acknowledges support by the European Commission under the Marie Curie Career Integration Grant programme (FP7-PEOPLE-2012-CIG 322307). SS acknowledges support from CONICYT-Chile FONDECYT 3140534, Basal-CATA PFB-06/2007, and Project IC120009 "Millennium Institute of Astrophysics (MAS)" of Iniciativa Cient\'{\i}fica Milenio del Ministerio de Econom\'{\i}a, Fomento y Turismo. Part of the funding for GROND (both hardware as well as personnel) was generously granted from the Leibniz-Prize to Prof. G. Hasinger (DFG grant HA 1850/28-1). We thank Alain Smette for providing the telluric spectrum, and the referee for very constructive feedback.

\section{Appendix}
\subsection{Skynet magnitude tables}\label{appendix}

Tables~\ref{tab:skynetR},~\ref{tab:skynetI},~\ref{tab:skynetB},~\ref{tab:skynetgp},~\ref{tab:skynetrp} and~\ref{tab:skynetip} give the magnitudes (not corrected for reddening) used for the optical light-curve input to model the distance between the excited gas (component a+b) and the burst itself.
 
 \newpage
 
\begin{table}
\centering
\caption{Skynet - Filter $R$}
\renewcommand*{\arraystretch}{1.3}
\begin{tabular}{@{} l c c c c @{}}
\hline\hline
Filter				&	Time (h)	&	Exposure time		&	S/N		&	Mag. (Vega) 	\\ \hline
$R$				&	0.01656	&	$1\times10$\,s		&	27.7		&	$15.01\pm0.04$	\\ 
$R$ 				&	0.02232 	&	$1\times10$\,s		&	18.9		&	$15.51\pm0.06$	\\ 
$R$ 				&	0.02760	&	$1\times10$\,s		&	12.1		&	$15.88\pm0.09$	\\ 
$R$				&	0.03288	&	$1\times10$\,s		&	7.58		&	$16.0^{+0.2}_{-0.1}$	\\
$R$				&	0.04032	&	$1\times20$\,s		&	10.7		&	$16.3\pm0.1$		\\
$R$				&	0.04824	&	$1\times20$\,s		&	9.35		&	$16.4\pm0.1$		\\
$R$				&	0.06720	&	$1\times40$\,s		&	7.45		&	$16.5^{+0.2}_{-0.1}$	\\
$R$				&	0.08088	&	$1\times40$\,s		&	13.1		&	$16.68^{+0.09}_{-0.08}$	\\
$R$				&	0.10824	&	$1\times40$\,s		&	5.47		&	$16.9\pm0.2$		\\
$R$				&	0.18216	&	$1\times80$\,s		&	5.82		&	$17.5\pm0.2$		\\
$R$				&	0.20880	&	$1\times80$\,s		&	3.58		&	$17.0\pm0.3$		\\
$R$				&	0.24744	&	$1\times160$\,s	&	10.9		&	$17.4\pm0.1$		\\
$R$				&	0.35520	&	$1\times160$\,s	&	5.37		&	$17.8\pm0.2$		\\
$R$				&	0.40536	&	$1\times160$\,s	&	6.08		&	$17.6\pm0.2$		\\
$R$				&	0.50928	&	$1\times160$\,s	&	2.86		&	$17.9^{+0.4}_{-0.3}$	\\	
$R$				&	0.55584	&	$1\times160$\,s	&	6.08		&	$17.5\pm0.2$		\\
$R$				&	0.66000	&	$1\times160$\,s	&	5.53		&	$17.6\pm0.2$		\\
$R$				&	0.91992	&	$3\times160$\,s	&	5.79		&	$18.4\pm0.2$		\\
$R$				&	1.28712	&	$4\times160$\,s	&	6.64		&	$18.79\pm0.2$		\\
$R$				&	1.84008	&	$9\times160$\,s	&	3.68		&	$19.3\pm0.3$		\\
$R$				&	2.79768	&	$7\times160$\,s	&	6.72		&	$19.5^{+0.2}_{-0.1}$		\\
$R$				&	3.21240	&	$7\times160$\,s	&	8.31		&	$19.8\pm0.1$		\\
$R$				&	3.59184	&	$7\times160$\,s	&	10.1		&	$19.8\pm0.1$		\\
$R$				&	4.12824	&	$12\times160$\,s	&	10.7		&	$19.8\pm0.1$		\\
$R$				&	4.9908	&	$18\times160$\,s	&	7.18		&	$20.0\pm0.1$		\\
$R$				&	24.5124	&	$40\times160$\,s	&	1.94		&	$21.9^{+0.7}_{-0.4}$	\\
\hline
\end{tabular}
\label{tab:skynetR}
\end{table}

\begin{table}
\centering
\caption{Skynet - Filter $I$}
\begin{tabular}{@{} l c c c c @{}}
\hline\hline
Filter				&	Time (h)	&	Exposure time		&	S/N		&	Mag. (Vega) 	\\ \hline
$I$				&	0.02328	&	$1\times5$\,s		&	13.4		&	$14.88\pm0.08$	\\ 
$I$				&	0.02784	&	$1\times5$\,s		&	9.57		&	$15.1\pm0.1$	\\ 
$I$				&	0.03312	&	$1\times10$\,s		&	10.2		&	$15.4\pm0.1$	\\ 
$I$				&	0.04032	&	$1\times20$\,s		&	14.6		&	$15.49^{+0.08}_{-0.07}$	\\ 
$I$				&	0.04896	&	$1\times20$\,s		&	11.8		&	$15.70^{+0.1}_{-0.09}$	\\ 
$I$				&	0.05616	&	$1\times10$\,s		&	2.34		&	$16.2^{+0.5}_{-0.4}$	\\ 
$I$				&	0.06672	&	$1\times40$\,s		&	5.75		&	$16.2\pm0.2$	\\ 
$I$				&	0.08088	&	$1\times40$\,s		&	15.4		&	$16.03\pm0.07$	\\ 
$I$				&	0.09504	&	$1\times40$\,s		&	7.74		&	$15.9\pm0.1$	\\
$I$				&	0.10896	&	$1\times40$\,s		&	7.16		&	$16.2^{+0.2}_{-0.1}$	\\
$I$				&	0.12864	&	$1\times80$\,s		&	15.1		&	$16.21\pm0.07$	\\
$I$				&	0.15504	&	$1\times80$\,s		&	3.84		&	$16.3^{+0.3}_{-0.2}$	\\
$I$				&	0.18216	&	$1\times80$\,s		&	9.82		&	$16.5\pm0.1$	\\
$I$				&	0.20880	&	$1\times80$\,s		&	5.14		&	$16.7\pm0.2$	\\
$I$				&	0.24744	&	$1\times160$\,s	&	16.3		&	$16.72\pm0.07$	\\
$I$				&	0.30144	&	$1\times160$\,s	&	13.3		&	$16.74^{+0.09}_{-0.08}$	\\
$I$				&	0.35520	&	$1\times160$\,s	&	11.3		&	$16.73^{+0.1}_{-0.09}$	\\
$I$				&	0.40536	&	$1\times160$\,s	&	6.89		&	$17.0^{+0.2}_{-0.1}$	\\
$I$				&	0.50928	&	$1\times160$\,s	&	8.51		&	$16.9\pm0.1$	\\
$I$				&	0.55632	&	$1\times160$\,s	&	7.10		&	$17.1^{+0.2}_{-0.1}$	\\
$I$				&	0.61248	&	$1\times160$\,s	&	6.87		&	$16.8^{+0.2}_{-0.1}$	\\
$I$				&	0.66000	&	$1\times160$\,s	&	5.78		&	$17.2\pm0.2$	\\
$I$				&	0.71592	&	$1\times160$\,s	&	3.53		&	$17.3\pm0.3$	\\
$I$				&	0.81528	&	$2\times160$\,s	&	5.50		&	$17.4\pm0.2$	\\
$I$				&	0.94704	&	$2\times160$\,s	&	11.4		&	$17.45^{+0.1}_{-0.09}$	\\
$I$				&	1.21368	&	$1\times160$\,s	&	5.87		&	$17.7\pm0.2$	\\
$I$				&	1.33872	&	$2\times160$\,s	&	6.90		&	$18.1^{+0.2}_{-0.1}$	\\
$I$				&	1.53672	&	$2\times160$\,s	&	3.92		&	$18.3^{+0.3}_{-0.2}$	\\
$I$				&	1.75560	&	$1\times160$\,s	&	2.80		&	$18.0^{+0.4}_{-0.3}$	\\
$I$				&	2.07528	&	$5\times160$\,s	&	4.51		&	$18.4^{+0.3}_{-0.2}$	\\
$I$				&	2.46480	&	$5\times160$\,s	&	3.76		&	$18.6\pm0.3$	\\
$I$				&	2.83080	&	$6\times160$\,s	&	8.36		&	$18.64\pm0.1$	\\
$I$				&	3.21504	&	$7\times160$\,s	&	12.1		&	$18.72\pm0.09$	\\
$I$				&	3.59496	&	$7\times160$\,s	&	13.7		&	$18.71\pm0.08$	\\
$I$				&	4.10328	&	$11\times160$\,s	&	12.8		&	$18.99^{+0.09}_{-0.08}$	\\
$I$				&	4.81200	&	$12\times160$\,s	&	9.54		&	$193\pm0.1$	\\
$I$				&	5.63112	&	$16\times160$\,s	&	6.02		&	$19.2\pm0.2$	\\
$I$				&	24.3192	&	$47\times160$\,s	&	1.89		&	$21^{+0.7}_{-0.4}$	\\
\hline
\end{tabular}
\label{tab:skynetI}
\end{table}

\begin{table}
\centering
\caption{Skynet - Filter $B$}
\renewcommand*{\arraystretch}{1.3}
\begin{tabular}{@{} l c c c c @{}}
\hline\hline
Filter				&	Time (h)	&	Exposure time		&	S/N		&	Mag. (Vega) 	\\ \hline
$B$				&	0.02520	&	$2\times10$\,s		&	4.98		&	$17.2\pm0.2$	\\ 
$B$				&	0.05592	&	$2\times20$\,s		&	4.22		&	$17.9^{+0.3}_{-0.2}$	\\ 
$B$				&	0.12144	&	$3\times40$\,s		&	5.10		&	$18.3\pm0.2$	\\ 
$B$				&	0.27456	&	$2\times160$\,s	&	5.61		&	$18.8\pm0.2$	\\ 
$B$				&	3.87000	&	$37\times160$\,s	&	5.96		&	$21.2\pm0.2$	\\ 
\hline
\end{tabular}
\label{tab:skynetB}
\end{table}

\begin{table}
\centering
\caption{Skynet - Filter $g^\prime$}
\renewcommand*{\arraystretch}{1.3}
\begin{tabular}{@{} l c c c c @{}}
\hline\hline
Filter				&	Time (h)	&	Exposure time		&	S/N		&	Mag. (Vega) 	\\ \hline
$g^\prime$			&	0.03312	&	$1\times20$\,s		&	3.76		&	$17.4\pm0.3$	\\ 
$g^\prime$			&	0.25032	&	$1\times80$\,s		&	4.45		&	$18.3^{+0.3}_{-0.2}$	\\ 
$g^\prime$			&	0.69912	&	$1\times80$\,s		&	2.16		&	$19.5^{0.6}_{-0.4}$	\\ 
\hline
\end{tabular}
\label{tab:skynetgp}
\end{table}

\begin{table}
\centering
\caption{Skynet - Filter $r^\prime$}
\renewcommand*{\arraystretch}{1.3}
\begin{tabular}{@{} l c c c c @{}}
\hline\hline
Filter				&	Time (h)	&	Exposure time		&	S/N		&	Mag. (Vega) 	\\ \hline
$r^\prime$			&	0.05688	&	$1\times20$\,s		&	6.15		&	$16.9\pm0.2$	\\ 
$r^\prime$			&	0.15576	&	$1\times80$\,s		&	10.6		&	$17.4\pm0.1$	\\ 
$r^\prime$			&	0.30168	&	$1\times160$\,s	&	11.8		&	$17.66^{+0.1}_{-0.09}$	\\ 
$r^\prime$			&	0.50952	&	$1\times160$\,s	&	8.21		&	$18.0\pm0.1$	\\ 
$r^\prime$			&	0.56088	&	$1\times80$\,s		&	5.14		&	$18.0\pm0.2$	\\ 
\hline
\end{tabular}
\label{tab:skynetrp}
\end{table}

\begin{table}
\centering
\caption{Skynet - Filter $i^\prime$}
\renewcommand*{\arraystretch}{1.3}
\begin{tabular}{@{} l c c c c @{}}
\hline\hline
Filter				&	Time (h)	&	Exposure time		&	S/N		&	Mag. (Vega) 	\\ \hline
$i^\prime$			&	0.08088	&	$1\times20$\,s		&	5.22		&	$16.3\pm0.2$	\\ 
$i^\prime$			&	0.35328	&	$1\times80$\,s		&	4.50		&	$17.6^{+0.3}_{-0.2}$	\\ 
$i^\prime$			&	0.61272	&	$1\times160$\,s	&	4.91		&	$17.9\pm0.2$	\\ 
$i^\prime$			&	0.66408	&	$1\times80$\,s		&	3.06		&	$17.9^{+0.4}_{-0.3}$	\\ 
$i^\prime$			&	21.99288	&	$38\times160$\,s	&	1.13		&	$20.9^{+1.2}_{-0.6}$	\\ 
\hline
\end{tabular}
\label{tab:skynetip}
\end{table}

\cleardoublepage
\chapter{Interstellar Dust}\label{sec:dust}

Interstellar dust refers to the tiny, but macroscopic, particles that exist in between stars in our and other galaxies. The word 'dust' is slightly misleading, as the dust-grain size is more like that of soot or sand particles. For many years, dust was mainly a hindrance to astronomers, as the grains absorb light at optical wavelengths, making accurate observations of objects behind the dust more difficult. But as our knowledge grew, particularly with the dawn of infrared astronomy, we have learned that dust plays an integral part in many astrophysical processes, such as star and planet formation \citep[e.g.][]{1997ASPC..122...25C,2012MNRAS.423L..60S} and the formation of molecules \citep[e.g.][]{2004ASPC..309..529P,2011ApJ...741L...9J}.

Today, the study of dust is driven as much from a need to understand the physical details of the formation and composition of the dust itself, as from the desire to determine the contribution from dust to absorption and emission, so that this can be removed from observations (dust absorbs up to 50\% of all starlight in the Galaxy). Despite these needs, many aspects of interstellar dust are still highly uncertain, or simply not known at all. This difficulty stems from the complexity of the particles compared to interstellar gas and the common molecules, so rather than dealing with simple quantum mechanical states, a full solid state treatment is needed.

In this chapter, I highlight what we do know about dust, along with the most prevalent theories of dust formation and composition, and the observations that these are based upon. 

\section{Properties of Dust}
The basic observational properties of dust is an overall absorption and scattering in the UV/optical, usually leading to a substantial reddening of the background spectrum, along with specific features at certain wavelengths, and thermal emission in the infrared (as well as some IR band-emission). Below I describe specific spectral features from dust. The general absorption/scattering and thermal emission mechanisms are described in Sections~\ref{sect:em} and~\ref{sect:ext} respectively.

\subsection{Absorption Features}

\subsubsection{The 2175\,\AA \ bump}
The strongest spectroscopic feature in the dust spectrum is a wide 'bump' in absorption at 2175\,\AA. Despite this feature being ubiquitous throughout the Milky Way, the origin is not well understood. The central wavelength is observed to be very constant while the width of the bump varies a lot. The only physical quantity observed to correlate with the width is the mean gas density along the line-of-sight \citep{1986ApJ...307..286F}.

Given the strength of the bump it must be caused by a very abundant material. Carbon is the most abundant material in the Universe, next after hydrogen and helium, and is hence the prime candidate for being responsible for the bump. \cite{1989IAUS..135..313D} calculated that the profile corresponds to an oscillator strength per hydrogen nucleon $n_\text{X} f_\text{X}/n_\text{H}\approx9.3\times10^6$. Assuming $f_\text{X}\leq0.3$, and that the bump is caused by either of the abundant elements Fe, Si, or Mg, it would require that more metals are locked up in dust, than are available in the gas phase in order to produce the bump. Furthermore, graphite, an ordered and stable form of carbon, has a resonance very close to 2175\,\AA. For this reason many models for the bump predict graphite, although we would then expect a correlation between the central wavelength and width, which is not observed \citep{1993ApJ...414..632D}. Besides, observations suggest that amorphous carbon, not graphite, is injected into the ISM (see Section~\ref{sect:form}). 

It is perhaps possible that small graphite particles can be produced from the amorphous carbon later on, but \cite{1986ApJ...305L..23W} reported measurements of scattering in a few cases, indicating that the particles responsible for the bump must, at least in some cases, be relatively large (compared with the photon wavelength). 

An alternative to graphite is polycyclic aromatic hydrocarbon (PAH) molecules (i.e. a large molecule, rather than dust grains). PAHs have a very similar atomic structure to graphite sheets, as the carbon atoms which are not on the surface have very similar electronic transitions to those in graphite. We hence also expect a transition in this wavelength interval. The observed variation of width (without a matching variation in the central wavelength), could possibly be explained by differences in the PAH mix from one sightline to another. However, this has not been tested, and so the exact origin of the bump is still considered an open question.

The 2175\,\AA \ bump is a strong function of the metallicity of the gas, with the UV bump appearing slightly weaker in the LMC extinction curve (metallicity $\sim50$\,\% solar), and essentially absent in the SMC extinction curve (metallicity $\sim10$\,\% solar). 

\subsubsection{Mid-IR silicate features}
We observe bands of absorption lines in the IR, most notably a set of broad bands centred roughly at $9.7\,\mu$m and $18\,\mu$m. The $9.7\,\mu$m features are generally identified as resonance lines from silicate minerals, as observed in the lab from Si--O bending and stretching modes. A bend bond in chemistry is the electron sharing between atoms within small ring molecules, so that the bond is literally 'bend'. Bending is then a change in angle, while stretching is a lengthening of the chemical bond. 

This is supported by the fact that we observe these features in outflows from oxygen-rich stars but not from carbon-rich stars, as the first is expected to form silicate dust, while in the latter all the oxygen (which is needed to form the silicates) is bound up in CO molecules. The bands are observed to be free of subfeatures, unlike in laboratory crystal measurements, probably indicating that the silicate is mainly amorphous rather than crystalline in nature.

The $18\,\mu$m band is likely due to O--Si--O bending modes in silicates. The polarisation of both features can be measured, putting constrains on the ratio of crystal to amorphous silicates.

If a sufficiently high amount of dust is present it is also possible to observe an absorption feature around $3.4\,\mu$m. This is likely from C--H stretching from some kind of hydrocarbon grain, but the precise composition has not been determined.

\subsubsection{DIBs}
Besides the well defined absorption features, a long range of diffuse interstellar bands (DIBs) has also been observed in UV, visible and IR. The origin of these are still debated, and, despite the large number, not one has as yet been securely identified. The fact that the strength of the lines are not observed to be correlated, means that the bands are not all from the same absorbing material.

Observation of fine-structure \citep{1998ApJ...495..941K} in the bands, points towards a large molecule rather than dust grains being the material responsible. Given the large amount of PAHs needed for the 2175\,\AA \ bump, these molecules are theorised to be responsible for at least a subset of the DIBs.

\subsection{Emission Features}\label{sec:emFeat}
\begin{figure}
\centering
\includegraphics[width=0.8\textwidth]{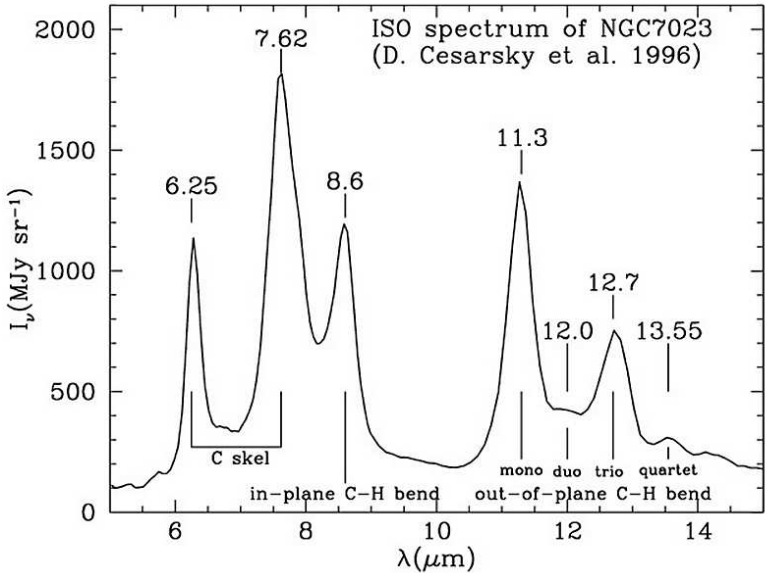}
\caption{PAH emission features in the spectrum of the reflection nebula NGC\,7023. "Mono", "duo", "trio", and "quartet" refers to the number of adjacent H-atoms (in the molecular structure) to the C--H bond that is bending.}
\label{fig:pah}
\end{figure}

Interstellar dust shines in the infrared. The spectrum has 5 clear emission peaks, at wavelengths $\sim12.7$, $11.3$, $8.6$, $7.6$, and $6.25\,\mu$m, all shown in Fig.~\ref{fig:pah} displaying the spectrum of nebula NGC\,7023. Several other (weaker) peaks are also frequently observed. These peaks coincide in wavelength with vibrational transitions (see specific labels in Fig.~\ref{fig:pah}) observed in the laboratory for PAH molecules. The observed strength of these lines supports a large quantity of PAH in the dust. \\

\indent Taken together, all these observational features indicate that dust grains are predominantly composed of silicates, carbon, ice, and/or iron compounds. The exact composition is to a large extent still unknown, and I will not go further into it here, except to note that for absorption models such as Xspec's TBabs used in this thesis, the dust assumed is what is called the MRN \citep{1977ApJ...217..425M} model; spherical and homogeneous grains, which consists of graphite, SiC, (Fe,Mg)SiO$_3$, (Fe,MG)$_2$SiO$_4$+Fe+Fe$_3$O$_4$. Newer models include PAH molecules.

\section{Dust Formation}\label{sect:form}
Dust is believed to originate predominantly in the ejected material from dying stars, either in SNe outflows or in the stellar winds of AGB stars, and in stellar outflows from massive stars. The composition and size distribution however, seem to be predominantly determined by physical processes in the ISM which alter the dust grains. The basis of the theory of dust formation in stellar outflows, is that we need high densities to have a chance of atoms and molecules colliding, so dust cannot be formed in the ISM. The cool expanded envelopes of stars in their late-life phases are much better suited, but we do see the dust in the ISM though, so outflows must be needed to move the dust.

Dust formation is divided into two parts; the nucleation of critical clusters, from atomic gas to solid particles, and the subsequent growth of these clusters to macroscopic dust grains. There are two groups of theories on how dust nuclei form; what is called the standard nucleation theory \citep[historically most used,][]{becker35}, and what is referred to as chemical kinetic nucleation \citep{kinetic}. Classical nucleation theory was first developed to model the condensation of water in the earth's atmosphere. The model assumes that thermodynamical equilibrium is reached. This assumption means that the properties of the dust clusters are given by extrapolation from bulk properties of the gas. This has the advantage that the thermodynamics can be described analytically, but it has been questioned whether this assumption is reasonable for the environment in space, where the density is much lower than in the atmosphere. Furthermore, thermodynamical equilibrium is likely a bad assumption in dynamic systems such as stellar outflows \citep[e.g.][]{1985ApJ...288..187D}. Chemical kinetic nucleation describes nucleation by mapping all chemical reactions in the gas that could lead to formation of dust, in order to limit the number of elements available from which dust and molecules can be formed. This method attempts to take into account the dynamics of stellar winds. The two methods give different dust masses, and observations have yet to fully reach a stage where one can be ruled out.

If we follow the evolution of a population of stars, the first stars to reach a stage where dust is produced are red supergiants (RSGs) and Wolf-Rayet (WR) stars. RSG are evolved O or B stars that have expanded as they burn helium in their core, and are hence relatively cool ($<5000$\,K) on the surface, so that dust can condensate in the winds flowing from the star. \cite{2005ApJ...634.1286M} estimate that RSGs contribute dust grains at a surface density rate of $3\times10^{-8}$\,M$_\odot$\,yr$^{-1}$kpc$^{-2}$.
WR stars are a subset of successors to massive RSGs which undergo heavy mass loss. This drives away the dust created prior to the WR phase, so WR stars appear dust-less, unless in a binary system where the ejecta interacts with the wind from a companion O-star \citep[e.g.][]{2002ApJ...567L.137M}. The overall contribution from evolved O or B stars is estimated to be less than 1\% of that of AGBs and SNe \citep{GHA}.

Later in the stellar evolution, core-collapse SNe are responsible for dust production. SN ejecta are thought to be ideal places for dust production as we know SNe form metals, which are needed. However, some mechanisms in a SN explosion are still not understood, and this complicates the understanding of dust formation as well. We do know however, that dust can be formed quite rapidly after the explosion, as observed by \cite{2014Natur.511..326G}.
SN ejecta contain layers of relatively pure oxygen and carbon, so both silicates and carbon type dust can be formed. Late-time observations of dust in SNe, show a smaller amount of dust than models predict. A solution to this could be that dust is destroyed by reverse shocks, i.e. the shocks caused by the interaction of the ejecta with the ambient medium \citep[e.g.][]{2008ApJ...682.1055N,2014A&A...564A..25B}.

Stars with masses $<8$\,M$_\odot$ go through an AGB phase during the late evolutionary stages. AGB stars burn shells of hydrogen and helium around a degenerate carbon/oxygen core. They have increasingly high mass-loss, losing up to $\sim80\%$ of their total mass during the AGB phase. During these stages, a large amount of dust is formed (and the newly formed dust then help drive mass-loss), and the dust is injected into the ISM, making AGB stars prime candidates for dust formation, as additionally, they are rich in molecules \citep[e.g.][]{2004A&A...414.1049Y}, which are needed to form dust. Whether carbon or silicon material is formed is governed by the C/O ratio.

After the grains are injected into the ISM, physical processes will alter the size and composition. The dust grains formed in AGB stars and SNe are rather large, but at a certain metallicity, $Z_{\text{crit}}$, shattering (fragmentation of the dust grains) dominates and a large amount of small grains are produced. These small grains might grow again, by accreting gas-phase metals from the ISM, or by coagulation (sticking grains together) in dense molecular clouds. \cite{2003ARA&A..41..241D} calculate that in fact most Si atoms in the dust grains are accreted to the dust in the ISM. Grains can also be destroyed in the ISM by UV light from stars.

All this results in complex dust grains that consist of carbon or silicate cores created in the stellar component, with outer layers consisting of a diverse number of elements added subsequently in the ISM. The effects of shattering and accretion/coagulation under different conditions in the ISM lead to a large range in dust size, from complex molecules to grains the size of $\sim0.5\,\mu$m. Models show that dust produced in SNe can even be as large as $1\,\mu$m \citep{2015A&A...575A..95S}.

Having described the formation and composition of dust, I now turn to the methods used to quantify the contribution from dust in astrophysical observations. 

\section{Dust Emission}\label{sect:em}
\begin{figure}[h]
\centering
\includegraphics[width=0.9\textwidth]{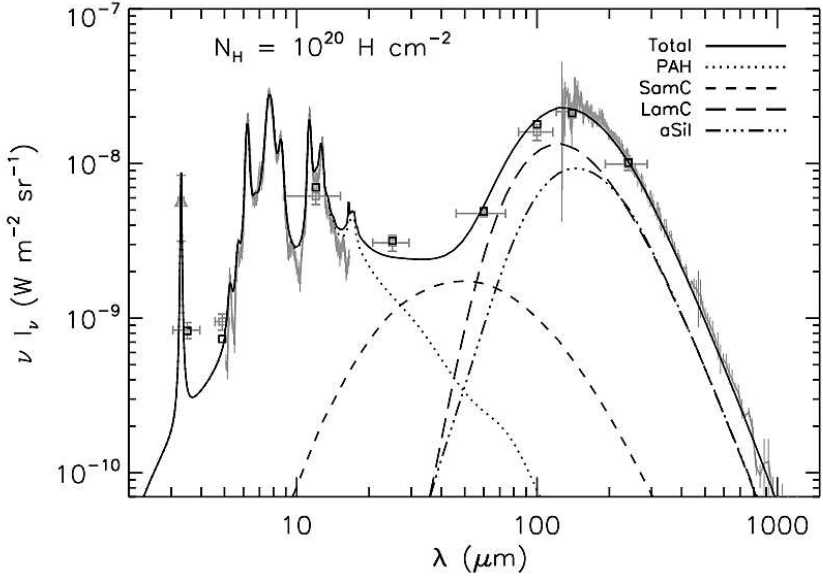}
\caption{Figure from \cite{2011A&A...525A.103C} showing emission from the diffuse ISM at high-Galactic latitude. The data are fitted with a dust model containing silicates, carbonaceous dust and PAH molecules (SamC and LamC is short for small and large carbon-based grains respectively, while aSil is short for amorphous silicates).}
\label{fig:thermal}
\end{figure}

As mentioned above, interstellar dust absorbs up to 50\% of all starlight in a galaxy. This large amount of energy is primarily used to heat up the dust grains, which then radiates with a thermal spectrum. Fig.~\ref{fig:thermal} shows an example of the dust emission spectrum, including the features described in Section~\ref{sec:emFeat}. 

The total luminosity from one dust grain can be divided into a contribution from a pure blackbody spectrum depending only on temperature, and the emissivity Q($\lambda$) which depends on the material; $L_{\text{dust}}\propto Q_{\lambda}\times$\,BB$_{\lambda}(T)$. The emissivity is given as the effective over the geometric cross section of the dust grain; Q$_{\lambda}=\sigma_{\lambda}/\sigma_{\text{g}}$.
It can be approximated as Q$_{\lambda}\propto\lambda^{-\beta}$, where $\beta=1$ for amorphous material, and $\beta=2$ for crystals.
In the special case where $\lambda\sim\,a$, $a$ being the size of the grain, Q$_{\lambda}\sim\,a/\lambda$. When $\lambda\gg a$, the photon is not affected by the dust and Q$_{\lambda}=0$. For $\lambda\ll a$, Q$_{\lambda}\sim$\,constant \citep[see e.g.][]{2011piim.book.....D}. 

Dust radiates predominantly in the mid- and far-infrared (as the grain size is in $\mu$m). The near-infrared wavelength area is relatively free of dust contribution as the Planck spectrum has fallen off here, but grains do not significantly absorb at such large wavelengths. Apart from the thermal emission, if the dust grains have an electric dipole, we would expect emission from rotational transitions in the microwave area \citep[e.g.][]{1957ApJ...126..480E,1994ApJ...427..155F}. Maps of the Milky Way microwave emission does indeed seem to confirm this, see \cite{2004ApJ...614..186F}.

\section{Dust Absorption and Scattering}\label{sect:ext}
At optical and UV wavelengths dust largely affects the background light. When quantifying the amount of light missing at different wavelengths (see e.g. Section~\ref{sect:ext}), it is important to distinguish between the terms extinction and attenuation.  

Extinction refers to the combined absorption and scattering (out of the line-of-sight) of photons by the dust. The light is from a background point-source so the distribution of dust is irrelevant. Attenuation refers to the net effect of dust when the light source is within the dust (an extended source, for instance a galaxy). The scattering contribution now includes scattering both into and out of the line-of-sight. The net effect of dust hence depends on the relative geometry of the source and dust.

\begin{figure}
\centering
\includegraphics[width=0.8\textwidth]{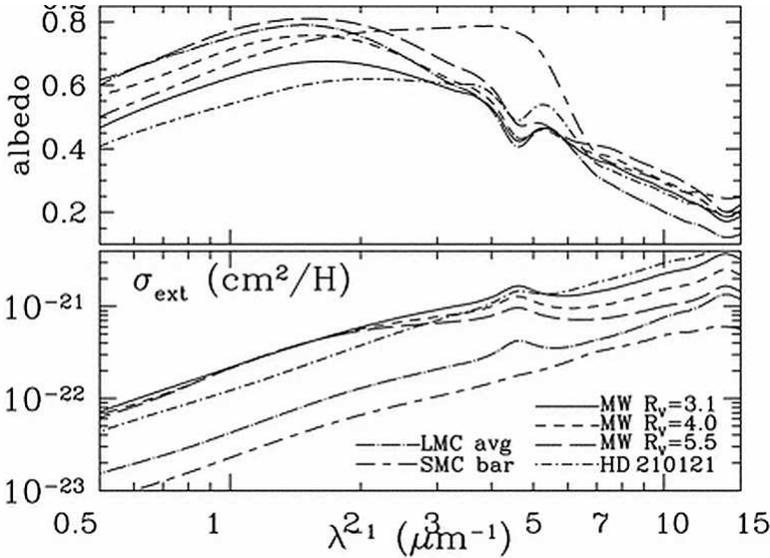}
\caption{Figure from \cite{2003ApJ...598.1017D} showing extinction cross-section (lower panel; normalised per H) and albedo (upper panel) as a function of wavelength.}
\label{fig:albedo}
\end{figure}

Dust scattering gives rise to reflection nebulae, as the grains surrounding a star reflect the starlight. Dust in the ISM of the Milky Way also reflects starlight in general causing what is called the diffuse Galactic light. Fig.~\ref{fig:albedo} show extinction cross-section and albedo (normalised per H) as a function of wavelength. The figure illustrates how adding scattering-effects to the extinction will give a less wavelength dependent results (in the optical region), as the two effects peak at different wavelengths. Scattering also polarises light, but I will not go further into this here, and instead refer to for instance \cite{2003ARA&A..41..241D}.

For the simple case of a point-source behind the dust, the radiative transfer equation gives (Equation~\ref{eq:transfer}):
\begin{equation}
I_\lambda=I_{\lambda,0}\,e^{-\tau_\lambda},
\end{equation}
\noindent where $\tau_\lambda$ is the optical depth of the dust along the line-of-sight, and $I_\lambda$ is the intensity, where 0 indicates the extinction free value. The extinction at wavelength $\lambda$ is often given in magnitudes, defined as the change caused by the dust in the apparent magnitude of the background source:
\begin{equation}
A_\lambda=m_\lambda-m_{\lambda,0}=-2.5\,\text{log}\frac{I}{I_0}=2.5\,\text{log}(e)\,\tau_\lambda.
\end{equation}
\noindent The total contribution of extinction (or attenuation) to the spectrum of a background source, is referred to as an extinction (attenuation) curve.

\section{Extinction and Attenuation Curves}\label{sect:ext}
\begin{figure}
\centering
\includegraphics[width=0.95\textwidth]{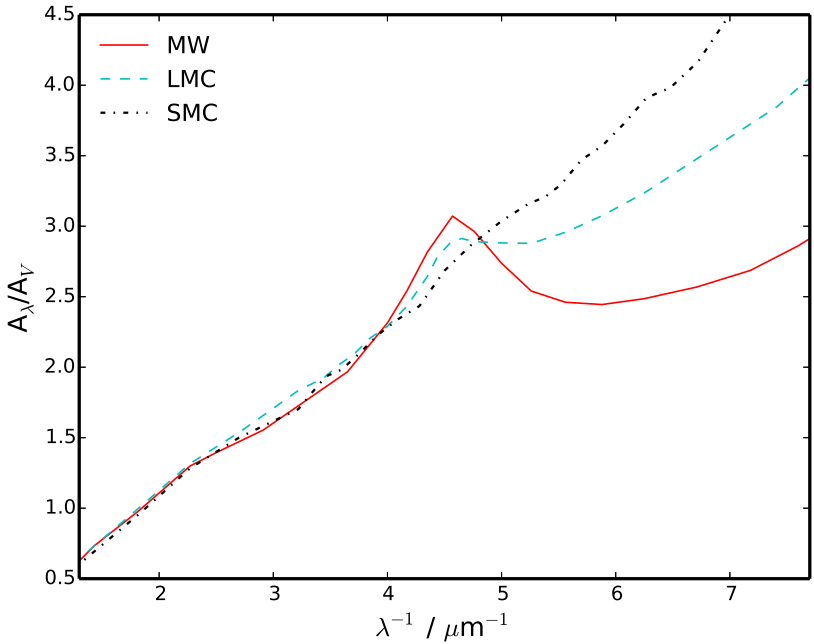}
\caption{\cite{pei} extinction curves normalised at the V band.}
\label{fig:pei}
\end{figure}

Examples of extinction curves are given in Fig.~\ref{fig:pei}. The figure shows the widely used extinction curves for the average line-of-sight through the Milky Way as well as the Large and Small Magellanic Clouds determined by \cite{pei}. Plots often show the curve  normalised to the extinction in the V band (peaks at 540\,nm), $A_V$.

Deriving the extinction curve requires knowledge of the spectrum of the dust-free source. This knowledge is rarely present, and exact extinction curves are therefore only known for the local universe. Here, they are usually determined by comparing pairs of stars with the same spectral type, one extinct by dust, and one with a clear line-of-sight, see e.g. \cite{1990ApJS...72..163F} and \cite{pei}. This can be done because the spectral classification of a single star has no significant dependance on reddening effects. Outside the local galaxy group, single stars cannot be resolved. When correcting for extinction in general, we hence have to make assumptions about the dust along the line-of-sight, and apply the known extinction curve that most likely corresponds to the conditions along the line-of-sight. Common extinction curves include the \cite{pei} SMC curve, the \cite{1990ApJS...72..163F,2007ApJ...663..320F} MW and LMC curves, and the \cite{2000ApJ...533..682C} attenuation curve for starburst galaxies. As can be seen in Fig.~\ref{fig:pei}, even locally, there are large differences between the curves. In most sight-lines through the MW, we observe a large 2175\,\AA \ bump. This is also seen in the LMC (though weaker), but only in few lines-of-sight in the SMC. Furthermore there is a large difference in steepness for the three average curves. 

If the intrinsic colour of the background object is known, the colour excess is measured, rather than the extinction. This relates to the extinction as:
\begin{equation}
E_{\lambda_1-\lambda_2}=A_{\lambda_1}-A_{\lambda_2}.
\end{equation}
\noindent Often the magnitudes used are the visual and blue, in which case the extinction curve is given as:
\begin{equation}
A_\lambda=k(\lambda)E_{B-V}=\frac{k(\lambda)\,A_V}{R_V},
\end{equation}
\noindent where $R_V$ is the ratio of total to selective extinction, and $k(\lambda)$ is the reddening curve. $1/R_V$ is then related to the slope of the extinction curve, and $R_V$ is often given as a measure of the wavelength dependance of the dust absorption. A typical value for the ISM in the MW is $R_V=3.1$ \citep{pei}.

Attenuation curves depend on the geometry of the dust. Generally they are flatter (i.e. have a higher $R_V$) than extinction curves, because the effect of including scattering in and out of the line-of-sight, is to blur the wavelength dependance. 

\section{Methods of Determining Dust Extinction}
There are different methods of estimating the amount of dust extinction in the observed line-of-sight, when the shape of the extinction curve, and/or the intrinsic spectrum of the source is not known. Here I will explore some of the most common methods.

\subsection{Balmer Decrement}
One such method is called the Balmer decrement, which relies on observations including more than one emission line from the Balmer series. Assuming a Case B recombination (usually a good assumption for star-forming regions, see Section~\ref{sec:emsfr}), we know the relative intrinsic strength of the Balmer lines. Any deviation from this ratio is then assumed to be due to reddening by dust. The intrinsic ratios of the first three lines in the series is given by \cite{balmer} as: \[\frac{\text{H}\alpha}{\text{H}\beta}=2.86 \ \ \text{and} \ \ \frac{\text{H}\gamma}{\text{H}\beta}=0.47.\] To translate the observed change in ratios into an amount of dust, the extinction curve must be known, but at the wavelengths of the first Balmer lines, most extinction curves are fairly similar. Further constrains can also be put on the curve, if more than two lines are observed. The colour excess can then be calculated by dividing:
\begin{equation}
F_{\text{int}}(\lambda)=F_{0}(\lambda)\times10^{0.4E_{B-V}k(\lambda)},
\end{equation}
\noindent for e.g. H$\alpha$ with the same equation for H$\beta$, and solve for $E_{B-V}$. \cite{2011MNRAS.414.2793W} has expanded on this method, using multiple H and He recombination lines simultaneously to determine the reddening. 

The Balmer decrement is a very useful tool, under the condition that the actual extinction curve of the dust behaves similar to what is commonly found in the local Universe, but see Chapter~\ref{chap:ext} for an exception. 

\subsection{IRX-$\beta$ Relation}\label{sec:irx}
At higher redshifts, the Balmer line fluxes are usually too weak to be observed. Furthermore, the lines are shifted out of the atmospheric window, where we can observe from the ground. The easiest accessible wavelength area then becomes the rest-frame UV. To determine the dust extinction from the UV spectrum/SED, many observations rely on an empirical relation between the infrared excess (IRX\,=\,$L_{IR}/L_{UV}$) due to dust emission and absorption and the slope $\beta$ of the UV spectrum, see e.g. \cite{1999ApJ...521...64M} and \cite{2011ApJ...726L...7O}. The idea is that the infrared excess traces the total dust attenuation, while $\beta$ traces the reddening. A higher amount of dust should lead to an increase in both parameters.

\begin{figure}[h]
\centering
\includegraphics[width=0.9\textwidth]{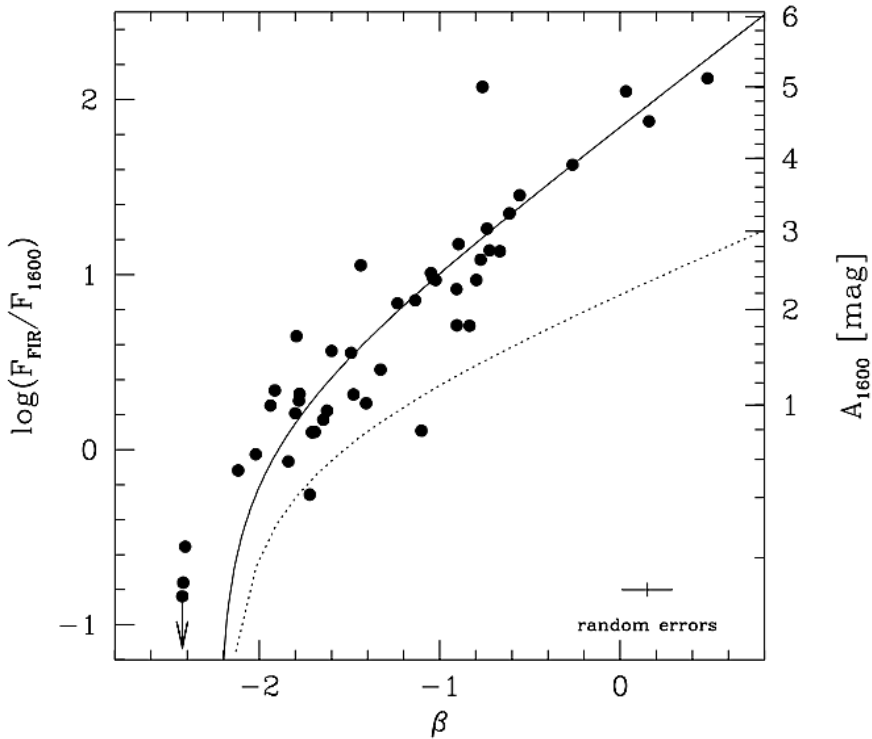}
\caption{IRX-$\beta$ relation for local UV bright starburst galaxies. Figure taken from \cite{1999ApJ...521...64M}. The dotted line shows the dust-absorption/population model proposed by \cite{1998ApJ...508..539P}.}
\label{fig:IRX}
\end{figure}

Fig.~\ref{fig:IRX} shows an example of the IRX-$\beta$ relation for UV-selected starburst galaxies. Galaxies of different classifications do not follow the same relation, as their intrinsic colours are different, and hence relate differently to the total amount of dust. Less active star-forming galaxies have a larger spread in the relation \citep[e.g.][]{2007ApJ...655..863D}. To use a given IRX-$\beta$ relationship for a galaxy, we have to assume that this galaxy has a similar attenuation law to the galaxies used to derive the relation (so a given amount of dust always results in the same amount of reddening). Hence, we need to know the properties of the galaxy we want to study quite well, which makes it uncertain whether we can use this relation at higher redshifts and compare to the local Universe, though it has been suggested that this is indeed possible for LBGs \citep[see e.g.][]{2011ApJ...726L...7O}. \\
\\
\noindent \textbf{Note on age/dust degeneracy} \\
One problem for many dust estimators, including the IRX-$\beta$ relation, is the age--dust degeneracy. Since dust reddens a population of stars, it is difficult to distinguish between a young (intrinsic blue) dusty population from an older, more evolved (intrinsic more red), but dust-free population. To break this degeneracy, we need additional information on either parameters which does not depend on the colour excess, such as for instance the Balmer decrement, or equivalent hydrogen column density, see below.

\subsection{Dust to Gas/Metal Ratio}
Observations of the Milky Way show that the ratio between gas and dust mass is relatively constant throughout the Galaxy. This gives a relationship between $A_V$ and the column density of hydrogen $N_{\text{H}}$, for the observed dust-to-gas ratio. One relation that is often used is that determined by \cite{1978ApJ...224..132B}:
\begin{equation}
A_V/N_{\text{H}} = 5.35\times10^{-21}\,\text{mag}\,\text{cm}^2.
\end{equation}
\noindent For a given sightline through the MW, the extinction can then be determined from the measured hydrogen column density alone (often using a correction factor for molecular hydrogen). The relation is not universal however, and even in the Magellanic clouds, the dust-to-gas ratio is significantly different, see e.g. \cite{2014ApJ...797...86R}. 

Another parameter that is perhaps more tightly linked with dust extinction is the metallicity, as the available metals directly constrain how much dust can be formed. The total metal column density can be determined from X-ray absorption, and is often given by the equivalent hydrogen column density. This is the hydrogen column density the metal absorption would imply, if assuming solar metallicity. Using GRB X-ray afterglows as background sources, \cite{2011A&A...533A..16W} finds the relation:
\begin{equation}
A_V/N_{\text{H}} = 2.2\times10^{21}\,\text{mag}\,\text{cm}^2.
\end{equation}

Observations seem to indicate that this relation, based on a constant dust-to-metal ratio, is valid outside the local Universe, and over a large range of metallicities, with only a small scatter \citep[see][]{zafar13}. If this is the case, then we can use measurements of the metal column density to determine the dust extinction.

\subsection{Abundance Patterns}\label{sec:ap}
If the column densities of several different elements along the line-of-sight is known, then an analysis of the relative abundances can be used to determine the amount of dust, see for example Section~\ref{depl}. If we can determine the intrinsic relation between the abundances of sets of elements, then we can calculate the amount of elements bound up into dust, and from there get an estimate of how much dust this equates to. There are three processes that can cause the relative abundance pattern to be different from that of the solar environment; non-solar nucleosynthesis, which would mean that the intrinsic abundance ratios are different, photoionisation effects, which would mean that the column density we measure from a specific ionisation state might not account for the total amount of the given element, and dust depletion, which is the result of some elements being more easily depleted into dust than others.

Different elements are sensitive to the different processes. To determine whether our observed system has gone through non-solar nucleosynthesis, we can compare the ratio of elements believed to be formed through different processes, for instance an $\alpha$-element such as S or Si, to an iron-peak elements, such as Fe or Cr. 

We can rule out ionisation effects, if we can observe lines of all relevant states, and add up the column densities. This is often not feasible however, and instead we compare the ratio of two elements with the same nucleosynthetic origin, but different sensitivity to ionisation such as O and Si.

To determine the amount of dust depletion, elements are compared which are non-refractory, i.e. rarely depleted into dust, such as S and Zn, with heavily depleted elements, such as Fe or Cr.

The more elements we observe, the better we can disentangle the different effects. Often we have to rely on only a few available elements, and then make assumptions about e.g. the dominant ionisation state in the observed environment. For particular environments, such as DLAs, statistical studies have been carried out, which can be used to help identify which effects are likely dominant for the different ratio-values \citep[see e.g.][]{2002ApJ...566...68P,2006A&A...445...93D}. Fig.~\ref{fig:depletion} shows an example of fitting the entire abundance pattern to a known environment, such as the Milky Way disc. 

\section{Gamma-Ray Bursts Extinction Curves}
Having now looked at some methods to estimate the dust extinction without knowing the extinction curve, we look at one unique method to actually determine the extinction curve for lines-of-sight outside the local group, namely using GRB afterglows. 

GRB afterglow spectra have an intrinsically simple form consisting of a simple power law across the entire observed spectrum, with the possibility of a cooling break, see Section~\ref{sec:agspec}. This means that it is relatively easy to distinguish the contribution from dust to the SED, especially since we can include X-rays, which are hardly affected by dust extinction.

\begin{figure}[h]
\centering
\includegraphics[width=0.9\textwidth]{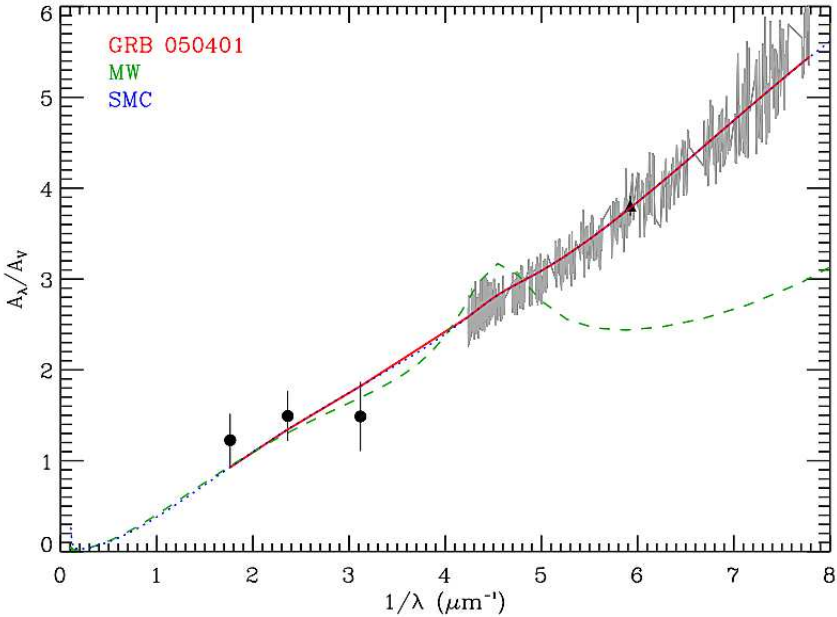}
\caption{Example of extinction curve fits to the afterglow SED of GRB\,050401 from \cite{zafar11}. The best fit for both Milky Way and SMC type extinction is plotted.}
\label{fig:GRBcurve}
\end{figure}

The usual approach to determining the extinction curve for a given GRB line-of-sight, is to assume that one of the local curves can be used to describe the environment, and then fit the SED with an absorbed single or broken power law (including Galactic absorption), and find the curve that results in the best fit. This is the approach used in works such as \cite{zafar11} and \cite{schady12}, which use curves from \cite{pei} for the LMC and SMC and curves from either \cite{pei} or \cite{FM} for the MW, see Figs.~\ref{fig:pei} and~\ref{fig:GRBcurve}. 

The general result from GRB extinction curve fits is that the average SMC extinction curve provides a good fit in the majority of cases \citep[see e.g.][]{2001A&A...370..909J,2001A&A...373..796F,2006ApJ...641..993K,2011A&A...526A..30G,zafar11,schady12}. This confirms other studies \cite[e.g.][]{1991ApJ...378....6P,2006ApJS..166..443E}, finding that the 2175\,\AA \ bump is only rarely seen outside the Galaxy, although it should be noted that it often falls outside the observed wavelength range.

The SMC curve generally goes together with a relatively low/moderate extinction, which is found in the line-of-sight of many bursts. More dusty lines-of-sight have been observed though \citep[e.g.][]{2011A&A...534A.108K}, including several detections of the bump, see for example \cite{2009ApJ...697.1725E} and \cite{2011AJ....141...36P}. It has been speculated that the large preference for an SMC like extinction curve is at least in part due to a bias towards dusty lines-of-sight. A significant fraction of detected GRBs are optically dark \citep[][see Section~\ref{sec:long}]{2004ApJ...617L..21J,2005ApJ...624..868R}, which prohibits a study of their extinction curves. The lack of an optical afterglow is likely due to dust obscuration \citep[e.g.][]{2008ApJ...681..453J,2013MNRAS.432.1231C}, meaning that we are missing the dustiest part of the extinction curve sample.

There is a growing amount of evidence (see for instance Chapter~\ref{chap:ext}) that a large subset of GRBs are in fact poorly fit with local extinction curves. The discrepancy is rarely apparent unless an alternative estimate of the extinction exists, such as a measure of the depletion or X-ray equivalent hydrogen column density. While it is possible that the extinction derived from e.g. the dust depletion could be wrong, if the environment of the GRB is significantly different from that of other DLAs (see Section~\ref{sec:abundances}), it is conspicuous how, for almost every GRB sight line with a dust depletion measure, these are in disagreement with the dust extinction derived from the extinction curve fit, see e.g. \cite{2004ApJ...614..293S} and Chapter~\ref{chap:ext}. 

This has led to suggestions that local extinction curves are inadequate to describe the dust properties in GRB hosts \citep[e.g.][]{2008ApJ...685.1046L}. Instead, the allowed extinction curve should be parameterised, using for instance a \cite{FM} parametrisation, see Section~\ref{sec:curve}. This of course means that the extinction curve is much less well constrained, unless other measures of the dust are available.

In Chapter~\ref{chap:ext}, it is speculated that a part of the GRB sightlines that are well fitted with an SMC curve with very little dust extinction, could in fact have a significant amount of dust, in the form of what is called grey dust.

\section{Grey Dust}
The term 'grey' is used in astronomy to nominate something that is (relatively) wavelength independent. An object that emits like a blackbody, will have a colour, or spectral peak wavelength. An object that does not have a specific colour, is then referred to as grey.

For interstellar dust, grain size is what dominates the wavelength dependance, at least for the silicates, see Fig.~\ref{fig:syntext}. Grey absorption corresponds to large dust grains. So the observation of grey dust, must mean that some mechanism is responsible for unusually large grains. When dust is first formed in stellar ejecta, particularly in SNe, theory predicts that the grains are indeed quite large, but are broken down and partially destroyed by reverse SN shocks, and shattering in the ISM. Subsequently the grains grow again by accreting metals out of the gas, or by coagulation.

To end up with a population of large grains then, either the dust shattering must be ineffective compared to the grain growth in the ISM, or the destruction in shocks has worked primarily on smaller grains, so the original large grain population has survived (or a combination of the two).

The latter has been postulated for dust in the surroundings of a GRB. \cite{2000ApJ...537..796W} showed how the prompt optical-UV flash, sometimes observed to accompany the GRB, can be responsible for destroying dust by sublimation out to a distance $\sim10$\,pc from the burst. \cite{2003ApJ...585..775P} showed that the radiation will preferentially destroy silicates and small graphite grains, leading to a greyer extinction (they suggest this process to be partially responsible for the lack of 2175\,\AA \ bump detections). 

\cite{1991ApJ...381..137F} and \cite{1995AZh....72..650S} investigated how dust swept out of the Galaxy, is affected by sputtering (i.e. the stripping of atoms from the dust grains by bombardment of energetic particles) from hot halo gas. The sputtering is largely ineffective on grains of the size $a\sim0.1\,\mu$m, but has a large effect on grains with $a\sim0.01\,\mu$m, leading to a skewed size distribution for extragalactic dust. If the IGM is indeed rich in large dust grains, that would mean an added term of dust extinction on observations, which does not show as a reddening effect \citep{1999ApJ...525..583A}.

Alternatively, the grains may survive the reverse SN shocks. Observations of SN\,1987A show a population of large grains being ejected. \cite{2015A&A...575A..95S} suggest that the larger grain size distribution is the result of a more clumpy SN ejecta. By clumping together, a bigger part of the original large-sized grains may survive, and enter the ISM.

Detection of thermal emission from dust in high-$z$ QSO observations imply a far higher dust mass in the early Universe than previously thought \citep{2001A&A...374..371O,2003A&A...398..857O,2003A&A...406L..55B}. This has posed a problem for current dust production models, when compared to the observed redshift for re-ionisation and the first galaxies \citep[e.g.][]{2001AJ....122.2850B,2009Natur.461.1254T,2011ApJS..192...18K}, as this means it will have taken only about 500\,Myr, or possibly much less, to build up a very high amount of dust \citep{GHA}. A fast and very efficient production mechanism is needed. In SN models, large grain sizes will decrease the amount of dust needed to fit the data, and hence, may help to solve the problem of where the high amount of dust comes from \citep[e.g.][]{2015MNRAS.446.2089W}. Furthermore large grains will help keep down the temperature needed to heat the grains, which would make the dust spectrum from SNe fit better with SN theory (the heating source here is the radioactive decay of short-lived isotopes).

Grey dust has also been observed locally; coreshine in molecular clouds for instance, has been investigated by e.g. \cite{2014A&A...572A..20L}. Coreshine is the process of scattering of large grains, resulting in emission at $3.6$ and $4.5\,\mu$m. For the grains to be able to scatter radiation at these wavelength, the size needs to be considerable. Another example of local grey dust is reported by \cite{2014MNRAS.445...93D}. They fit the extinction law for the 30 Doradus nebula in the LMC. Due to the fact that the dust is distributed unevenly, they have observations of red giants both unobscured and obscured by dust to compare. They find a good fit using an extinction curve for the diffuse Galactic ISM plus an extra grey extinction component.

Grey dust has also been observed in dust clouds passing in front of the T Tauri star RW Aur A. In this case, we know the total extinction, as we have measurements before and during a passage. \cite{2015A&A...577A..73P} observe this dust to be grey, suggesting that maybe large dust grains have been stirred up from the inclined disk of the star through interaction with the stellar wind.

In conclusion, there is ample evidence for the existence of grey dust both locally, and at higher redshifts. In the next chapter, grey dust in a GRB host is investigated, and the consequences for cosmology and astronomy, if this component is more ubiquitous than assumed, is explored.

\cleardoublepage
\chapter{Is Grey Dust Common in the Universe?}\label{chap:ext}
\begin{center}
\large{I. Evidence for grey extinction from GRB afterglows}
\end{center}
\vspace{0.5cm}

\begin{center}
 M.~Friis$^1$,
 D. J.~Watson$^2$, 
 A. C.~Andersen$^2$
 R. L. C.~Starling$^3$, 
 P.~Jakobsson$^1$,
 A.~De Cia$^4$, 
  J. P. U.~Fynbo$^2$, 
 T.~Kr\"{u}hler$^{5}$, 
 J.~Greiner$^{6}$
\end{center}
\vspace{0.7cm}

\begin{footnotesize}
\begin{itemize}
\addtolength{\itemsep}{-1.25\baselineskip}
  \item[$^1$]	Centre for Astrophysics and Cosmology, Science Institute, University of Iceland, Dunhagi 5, 107 Reykjav\'ik, Iceland \\
  \item[$^2$]	Dark Cosmology Centre, Niels Bohr Institute, University of Copenhagen, Juliane Maries Vej 30, 2100 Copenhagen, Denmark \\
  \item[$^3$]	Department of Physics and Astronomy, University of Leicester, University Road, Leicester LE1 7RH, UK \\
  \item[$^4$]	European Southern Observatory, Karl-Schwarzschild-Str. 2, D-85748 Garching, Germany \\
  \item[$^5$]	European Southern Observatory, Alonso de C\'ordova 3107, Casilla 19001, Santiago 19, Chile \\
  \item[$^6$]	Max-Planck-Institut f\"{u}r extraterrestrische Physik, Giessenbachstra\ss e 1, 85748 Garching, Germany \\
\end{itemize}
\end{footnotesize}
\vspace{1.5cm}

\noindent \textbf{\begin{large}Abstract\end{large}}
  {}
   {\\ \textbf{Aims}: To investigate the occurrence and consequences of a dust grain-size distribution skewed to large grains in GRB host galaxies. We then extrapolate to young, star-forming galaxies at high redshift. The implications of our results for the star-formation history of the Universe is discussed.}
   {\\ \textbf{Methods}: We quantify the metallicity and amount of dust extinction expected along the line-of-sight to GRB\,121024A using depletion measurements from absorption line analysis. We then determine constraints on the extinction curve from fitting the spectral energy distribution. These results are then compared with the literature for GRB extinction curves to determine how common the dust component responsible for such an extinction curve is, and we tentatively suggest an extrapolation of this result to all young galaxies with properties similar to GRB hosts.}
   {\\ \textbf{Results}: We find constraints on the extinction curve towards GRB\,121024A leading to the most extreme $R_V\sim16$ determined to date. Going through the GRB line-of-sight extinction curves in the literature, we find an upper limit of $\sim30\%$ to the amount of systems that could potentially have such a flat extinction curve. We show that the dust responsible for this extinction could be a standard Milky Way dust composition, but skewed towards large significantly grain sizes. We furthermore demonstrate that incorporating the assumption that a large grain component may have been more common in the earlier Universe leads to changes in the standard picture of the Universal star-formation history.}
   {\\ \textbf{Conclusions}: Our results indicate that great care is needed when correcting observations for attenuation outside of the local Universe.}

\section{Introduction}

The study of extinction towards astrophysical sources is of importance both on account of what we might learn about the composition of interstellar dust, as well as enabling us to remove this dust contribution from the underlying source we might wish to study. Well quantified extinction curves exist solely for lines-of-sight within the local group of galaxies, where individual pairs of stars can be observed and compared, and the dust content mapped extensively. When correcting for extinction at higher redshifts it is customary to use the average extinction curve of the Milky Way (MW), or one of the Large and Small Magellanic Clouds (LMC and SMC). These three curves are quite different in shape, with the average MW curve being quite flat and with a large 2175\,\AA \ bump \citep[see e.g.][]{1989IAUS..135..313D}, while the average SMC curve is steep and without any indication of the bump. The LMC curve is between the two in shape. 

The large difference between these curves means that in most cases we will find a reasonable fit when applied to lines-of-sight towards higher redshift sources. On the other hand, the fact that, in these three galaxies alone, we observe such a large difference, should make us consider carefully before they are applied universally. Even the individual lines-of-sight within each of the galaxies vary, with some lines in the SMC actually displaying the 2175\,\AA \ bump, while some lines within the MW do not. For a short review see \cite{2009ApJ...697.1725E} and references therein. To study the extinction along discreet lines-of-sight outside the local group, quasars and gamma-ray bursts (GRBs) have been used as background sources due to their brightness and the existence of canonical GRB spectra \citep[e.g.][]{zafar11,schady12,2012MNRAS.419.1028K}. Spectral energy distributions (SEDs) for GRB afterglows are in the majority of cases well fit with an SMC extinction curve, though exceptions are not uncommon. The 2175\,\AA \ bump has been observed in several cases \citep[][]{2009ApJ...697.1725E,2011AJ....141...36P,2012ApJ...753...82Z}, and an increasing number of unusually flat extinction curves have been reported. \cite{2003ApJ...585..638S} first suggested grey extinction for the GRB environment, to reconcile the fact that they found a large visual extinction, evident from dust depletion measurements in three GRB afterglows, while GRB afterglow SEDs are in general only very slightly reddened. This suggestion was supported by \cite{2004ApJ...614..293S} who, this time, reported high dust depletion in the same burst, GRB\,020813, for which virtually no curvature is apparent in the SED. Since then a possible grey dust component has been reported for several GRBs. Examples include GRB\,020405 towards which \cite{2005A&A...441...83S} reported an extinction law only weakly dependent on wavelength to be the best solution if the simultaneous NIR--to--X-ray spectrum should fit the external shock model for GRB afterglows. Similarly \cite{2008ApJ...672..449P} suggest grey dust towards GRB\,061126 ($z=1.16$) to explain that while the optical--to--X-ray slope, $\beta_{\text{ox}}$ would classify this as a dark burst \citep{2004ApJ...617L..21J}, implying large quantities of dust, an SED fit to local extinction laws indicate no host-frame extinction. We note that the opposite, i.e. an unusually steep extinction curve, has also been reported along lines-of-sight to GRBs, e.g. GRR\,080605 \citep{2012ApJ...753...82Z} and GRB\,140506A \citep{2014A&A...572A..12F}.

Grey dust is known from direct measurements in the local group as well. While  \cite{2012ApJ...752...57R} report grey dust in the disc of a nearby star HD 15115, \cite{2014MNRAS.445...93D} studied the extinction towards the 30 Doradus nebula in the LMC, reporting the absolute extinction of individual stars. They find that in addition to the standard Galactic diffuse interstellar media, an extra grey component is needed to explain the measurements.

In this paper, we present, in detail, the evidence for grey dust in the line-of-sight to GRB\,121024A as reported in \cite{2015MNRAS.451.4686F} (F15, hereafter). While flat extinction curves have been reported in singular cases before, sample studies of extinction curves towards GRBs generally find no need for such a solution. But in the majority of cases, an alternative determination of the extinction does not exist. When this is the case, the lack of reddening in the SED will be interpreted as equal to a lack of dust. In this paper, we go through a sample of GRB extinction curves to determine how often the dust extinction could potentially be underestimated as a consequence of ignoring grey dust. We then go on to explore the ramifications, if a significant ratio of GRB environments does indeed host grey dust, generalising to studies of all star-forming, high-$z$ galaxies. In Sect.~\ref{121024A} we report the results for, and discussion of, the extinction along the line-of-sight to GRB\,121024A. In Sect.~\ref{sec:sample} we discuss the GRB extinction curve samples with regards to grey dust, and in Sect.~\ref{grey} we discuss the origin of grey dust along with the potential consequences for missing this dust in extinction/attenuation corrections.


\section{GRB\,121024A}\label{121024A}
GRB\,121024A is a $z=2.30$ burst first observed by the Burst Alert Telescope \citep[BAT,][]{Barthelmy} onboard the \emph{Swift} satellite \citep{swift} on 2012 October 24 at 02:56:12 UT. In this paper we make use of afterglow data acquired for and described in F15, where it was found that a very flat (R$_V\sim16$) extinction curve was needed to fit the dust extinction. Below we highlight and expand on the analysis leading to that result.

\subsection{Metallicity and dust depletion of the birth site of GRB\,121024A}\label{sec:abundances}
The afterglow spectrum of GRB\,121024A contains a large number of absorption lines from several elements, including hydrogen (the column densities and abundances are reported in Tables 4 and 5 of F15). Most importantly here, we measured [Zn/Fe] $= 0.85 \pm 0.06$. This abundance ratio is referred to as the depletion factor. It is used as an indicator of the amount of dust in a system, as Fe is usually found to be heavily depleted into dust, while Zn is only a mild refractory element. Based on this [Zn/Fe] value and following \cite{2013A&A...560A..88D}, we calculated \dtm{} $=1.01\pm0.03$, i.e. consistent with the Galaxy. Taken together with the determined metallicity of [Zn/H]$_{\rm corr}=-0.6\pm0.2$ and column density log\,$N(\text{H\,{\sc i}})/\text{cm}^{-2}=21.88\pm0.10$, we used the Galactic dust-to-metal ratio $A_{V, \rm{Gal}}/N_{(H, \rm{Gal})}=0.45\times 10^{-21}$\,mag\,cm$^2$ of \cite{2011A&A...533A..16W}, to determine an expected extinction of $A_V=0.9\pm0.3$\,mag.

As described in Sect.~\ref{sec:curve}, the spectral data do not agree with the predicted visual extinction using a standard extinction component. Here we first examine whether the abundance pattern could be explained by other effects, i.e non-solar nucleosynthesis or ionisation. 

[\mbox{Zn/Fe}] is often used to probe the level of dust depletion, as the two elements are observed to have a similar nucleosynthetic origin, but while Zn is observed to be only mildly depleted, Fe is normally heavily depleted into dust. Zn is not an Fe-peak element though, so there could potentially be a difference in nucleosynthesis. Damped Ly$\alpha$ absorber (DLA) samples studies, e.g. \cite{2002A&A...385..802L} and \cite{2006A&A...445...93D}, have shown that while at low dust depletion levels, Zn can be slightly overabundant intrinsically with [\mbox{Zn/Fe}]$\leq0.2$, a larger difference can with high certainty be attributed to Fe being locked in dust. In fact \cite{2015MNRAS.452.4326B} find that the majority of DLAs have sub-solar [\mbox{Zn/Fe}]. For GRB\,121024A we observe [\mbox{Zn/Fe}]$=0.85\pm0.06$, which if indicative of an intrinsic different abundance ratio, would be more extreme than ever observed before. We hence conclude that this large ratio is almost certainly due to significant dust depletion (in fact, if [\mbox{Zn/Fe}] is sub-solar, we underestimate the amount of dust in this line-of-sight). 

We furthermore observe the two $\alpha$ elements S and Si. Unfortunately those lines are all saturated so we only have lower limits on the abundances, but these are in agreement with the Zn metallicity, supporting a negligible contribution from differential nucleosynthesis to [\mbox{Zn/Fe}]. 
At the observed column density of hydrogen most elements are expected to be primarily in the single ionised state, as their neutral states have ionisation potentials lower than hydrogen (1\,Ryd\,$=13.6$\,keV), i.e. the DLA system is optically thin to ionising photons \citep{2005ARA&A..43..861W}. This is supported by the fact that we observed the same velocity profile for all species (with the exception of Ca\,{\sc ii}, see details in F15). 

In addition to these considerations, we also point out that the detection of a significant amount of molecular hydrogen also indicates a relatively high amount of dust, as it is believed that H$_2$ forms on the surface of dust and, once formed, is shielded from Lyman-Werner photons by the dust grains. Detections of molecular hydrogen and a high amount of dust are also seen to coincide for QSO-DLAs, see e.g. \cite{2011MNRAS.413.2481F} and \cite{2015arXiv151004695K}. In conclusion, it is very likely that the observed [\mbox{Zn/Fe}] is indeed due to dust depletion, but in what follows, we allow for a small ([\mbox{Zn/Fe}]$\leq0.2$) intrinsic difference in the uncertainty of the derived $A_{\text{V}}$ values.

\subsection{Extinction Curve}\label{sec:curve}

\begin{figure}[ht]
\includegraphics[width=1.05\columnwidth]{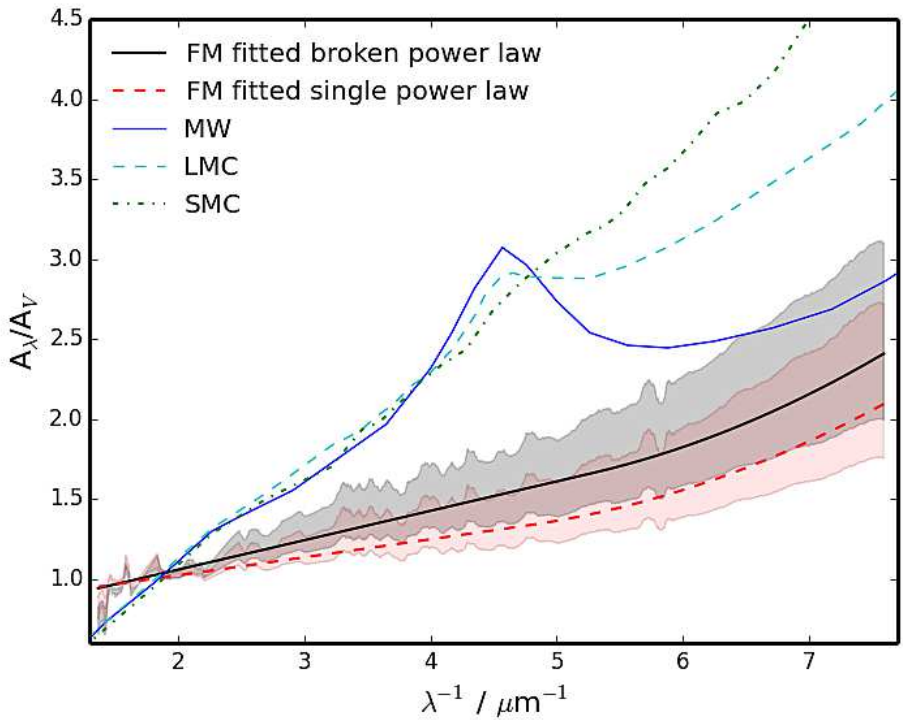}
\caption{Extinction curves for the line of sight to GRB\,121024A. We plot the best fit to a \cite{2007ApJ...663..320F} parameterisation assuming $A_V=0.9$\,mag as expected from the measured metallicity, H\,{\sc i} column density and dust-to-metal ratio. The solid black curve shows the best fit extinction curve for a broken power law, while the best fit for a single power law is plotted with the dashed red curve. The underlying shaded areas display the allowed extinction values within the $A_V$ error-space of $\pm0.3$.}
\label{fig:curves}
\end{figure}

We fitted the simultaneous optical/NIR and X-ray spectral data in F15, including the extinction models of \cite{pei} in the fits. We found that none of these curves can account for the amount of dust we know is present as determined from depletion patterns through a fit to absorption lines in the spectrum. As described in F15, we performed a fit to the XRT X-ray data alone and extrapolated the resultant best-fit power-law to optical wavelengths, which we then compared with the observed optical spectrum to create an extinction law. We applied both a single power law, as well as a broken power law with a cooling break ($\beta=0.5$) placed between the optical and X-ray wavelengths, see Fig.~\ref{fig:curves}. Here, we use the \cite{2007ApJ...663..320F} parameterisation to describe the extinction curve. This model provides a high degree of freedom in the fit, which is suited to our needs, as we simply want to describe the extinction curve in a way that would allow others to use it as a template. To improve fit statistics, we fix the model parameter c$_3=0$ in the final fit, after having determined that there is no indication of the 2175\,\AA \ bump, the strength of which is determined by c$_3$ in the model.

Table~\ref{tab:FM} gives the resulting parameters both for the single power law fit and for the broken power law. The errors are dominated by the error on $A_\text{V}=0.9\pm0.3$ as determined from the depletion pattern.

\begin{table}
\centering
\caption{Parameters of the FM fits}
\renewcommand*{\arraystretch}{1.3}
\begin{tabular}{@{} l c c c c @{}}
\hline\hline
Model			&	c$_1$			&	c$_2$			&	c$_4$				&	c$_5$				\\ \hline
Single power law	&	$-3^{+1}_{-2}$		&	$1.7^{+1.2}_{-0.6}$	&	$0.8^{+0.5}_{-0.2}$		&	$4.8\pm0.1$			\\
Broken power law	&	$-5^{+1}_{-2}$		&	$2.8^{+1.5}_{-0.8}$	&	$0.9^{+0.5}_{-0.2}$		&	$5.31^{+0.07}_{-0.04}$	\\ \hline
\end{tabular}
\label{tab:FM}
\end{table}

\subsection{A different origin for the X-rays?}

\begin{figure}[ht]
\includegraphics[width=1.05\columnwidth]{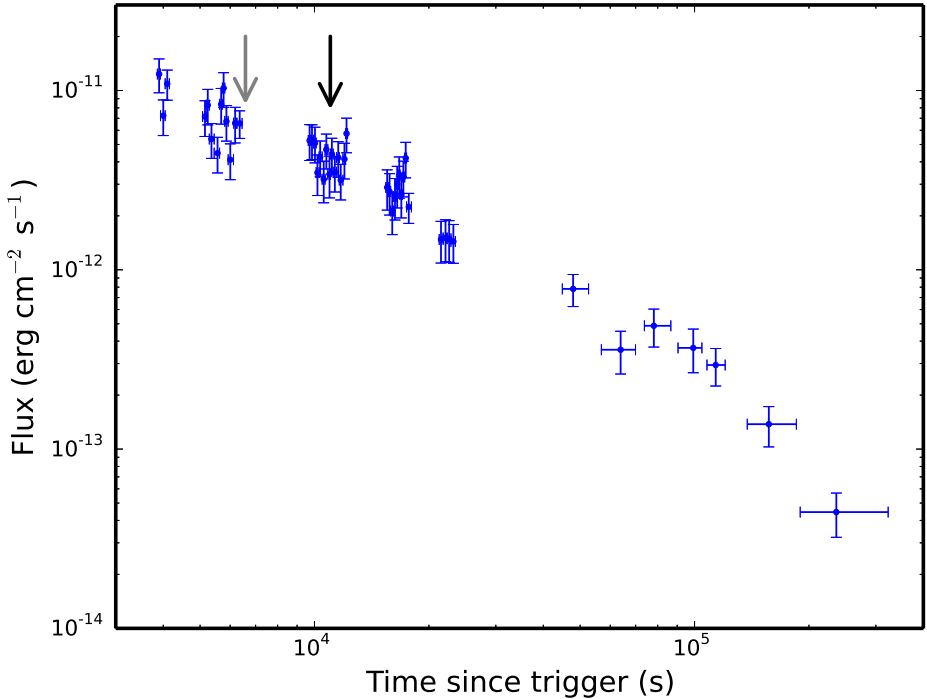}
\caption{Late X-ray light curve for GRB\,121024A. The grey arrow indicates the mid-time for the X-shooter data. Likewise the black arrow indicates mid-time for the GROND data used to flux-calibrate the X-shooter spectra. We assumed a slope of $\alpha=0.8$ for the optical light curve to extrapolate the X-shooter data to GROND mid-time, and extracted simultaneous X-ray data.}
\label{fig:lc}
\end{figure}

As an alternative to a flat extinction curve, \cite{2008ApJ...672..449P} discuss the possibility of a separate origin for the X-rays to explain the discrepancy between X-ray and optical fluxes for the afterglow of GRB\,061126. This possibility has been invoked previously \citep[e.g.][]{2007MNRAS.380..270O,2007MNRAS.380..374P} to explain the \emph{Swift} X-ray breaks. For GRB\,121024A the high amount of dust present is evident in the X-shooter data alone, from the depletion. However, the aforementioned extinction curve (Section~\ref{sec:curve}) is derived assuming that the X-ray and optical data are all dominated by the GRB afterglow. As seen in Fig.~\ref{fig:lc}, the XRT light curve does show evidence of flaring at the time of the X-shooter spectrum. Hence, we cannot rule out that the XRT data is dominated by another component than the afterglow.

If the XRT data are flare-dominated, we would expect them to be well fitted by an absorbed synchrotron model. Furthermore, the low energy tail should fall off fast, as we see no sign of this tail in the optical spectrum. To test the plausibility of this theory for GRB\,121024A, we have fitted the X-ray data to a Band model \citep{1993ApJ...413..281B} using \texttt{Xspec} and including photoelectric absorption from both the Milky Way \citep[$z=0$, fixed at $N(\text{H})_{\text{X}}^{\text{Gal}}=5.36\times10^{20}$\,cm$^{-2}$, from the Leiden/Argentine/Bonn (LAB) Survey;][]{2005A&A...440..775K} and the GRB host galaxy (left free to vary). Using the results found by \cite{2004ApJ...613..460B}, we constrained the low energy spectral index to $\alpha<-2/3$ as expected for a synchrotron or SSC (synchrotron self-compton) spectrum.

\begin{figure}[ht]
\includegraphics[width=1.05\columnwidth]{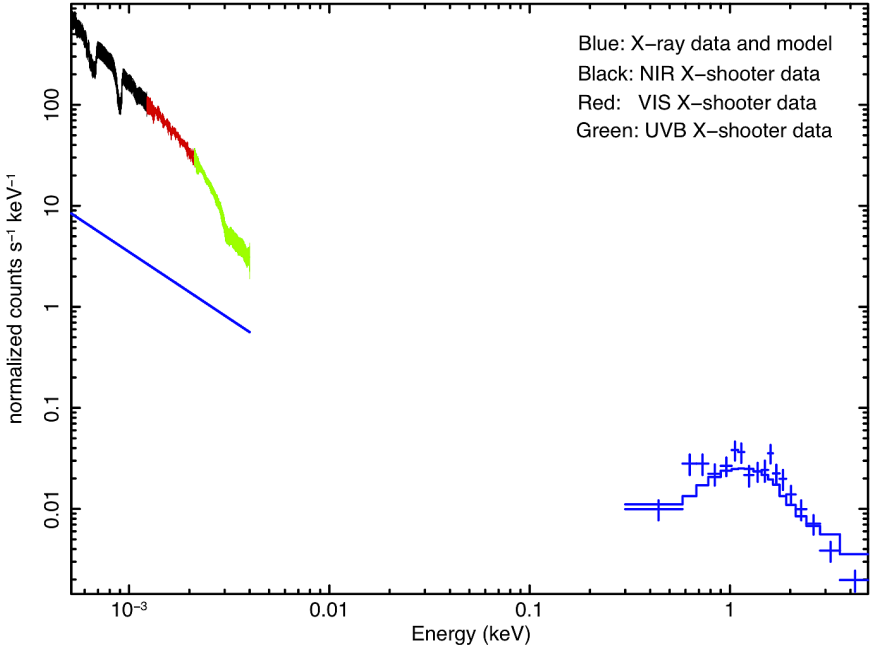}
\caption{Folded data and best fit band model (restricted to $\alpha<-2/3$). The X-shooter data have been binned for illustrative purposes.}
\label{fig:band}
\end{figure}

We find that this model provides a reasonably good fit (reduced $\chi^2=2.7$), and as seen in Fig.~\ref{fig:band}, the spectrum falls off fast enough, so that no tail would be visible in the X-shooter data. We conclude that it is possible that the XRT data are dominated by flaring. If this is the case, then the intrinsic afterglow flux at X-ray energies must be lower than those observed by the XRT. That would mean that the intrinsic afterglow power law, and hence the extinction curve, must be even flatter. The extinction curve shown in Fig.~\ref{fig:curves} can then be considered an absolute upper limit in steepness.



\subsection{A differential dust destruction by the GRB?}\label{sec:dest}

The X-ray and UV radiation from the GRB can alter the circumburst environment, destroying dust grains. Models predict that the evaporation of dust from the radiation is biased towards smaller grains \citep{2000ApJ...537..796W,2003ApJ...585..775P}. This has been put forward as a possible explanation of flat extinction curves towards GRBs \citep[e.g.][]{2003ApJ...585..638S}. However, the model distance from the GRB is quite small, we would only expect to see an effect out to about $\sim20$\,pc. For the line-of-sight towards GRB\,121024A, we have determined the distance between the GRB and the absorbing gas to be $\sim600$\,pc and $>3.5$\,kpc for the two components respectively, see F15. Furthermore, we observe vibrationally excited molecular hydrogen lines from the ground-state of H$_2$ in the gas cloud furthest away from the GRB, but no H$_2^*$ lines. This is unlikely to have been the case, had the UV radiation been strong enough at this distance to destroy dust grains \citep{2002ApJ...569..780D}.

Differential dust destruction is also an implausible explanation for the other reported cases of grey dust towards GRBs, as the GRB-to-cloud distances determined in the literature are usually several hundreds of parsecs or more \citep[e.g.][]{2007A&A...468...83V,2013A&A...549A..22V,2008A&A...491..189F,2009A&A...506..661L}.

To sum up, the most convincing theory to explain both the SED and the abundance pattern is grey dust in the vicinity of the burst. It is improbable that this dust component is created by the GRB blast, so it is likely intrinsic to the GRB birthplace.

\section{GRB extinction curve samples}\label{sec:sample}

Having determined that there is very likely grey dust in the line-of-sight towards GRB\,121024A, we now try to quantify how pervasive this type of dust is towards GRBs in general.

The physical properties of dust within GRB hosts have been the subject of several recent papers presenting samples of GRB extinction curves from SED fitting. The largest sample (to date) of extinction curves outside the local group is presented in \cite{zafar11}: 42 GRB lines-of-sight. They find that approximately two thirds of the sample are consistent with an SMC-type extinction, while about one fourth (11/41) is consistent with no dust extinction (while $>7\,\%$ have a well-defined bump at 2175\,\AA). Of the 11 bursts with no measured extinction, 4 have SED fits with a cooling-break frequency at an energy just under the X-ray data cut-off, while 3 more have a break frequency between the optical and X-ray data-regions. These lines-of-sight are candidates for containing grey dust, as there exists no constraints on the break, nor other indications of the amount of dust in the host. The remaining four are fit with a single power law, but the X-ray extrapolation leaves room for varying quantities of grey dust for all but GRB\,080319B. Similarly several of the SEDs that are well fit with a small amount of SMC-type dust show no apparent reddening in the optical data, and have X-ray data that are either not good enough to constrain a fit, or for which a flat extinction curve would fit just as well (most notably GRB\,060729). Overall, up to about 30\% of the \cite{zafar11} sample could potentially have a flatter extinction law. We sum up the grey dust candidates in Table~\ref{tab:candidates}. 

\begin{table}
\centering
\caption{Grey dust candidates in the \cite{zafar11} sample}
\renewcommand*{\arraystretch}{1.3}
\begin{tabular}{@{} l c l @{}}
\hline\hline
GRB				&	Break?			&	Comment									\\ \hline
050824			&	Yes				&	X-rays fit well with no break, but grey dust			\\
060512			&	Yes				&	Break could be moved						\\
060614			&	Yes				&	X-ray fit badly constrained					\\
060729			&	No				&	Grey dust improves fit						\\
060906			&	Yes				&	Break right below X-ray energies				\\
060927			&	No				&	Poor data									\\
061021			&	Yes				&	Break right below X-ray energies				\\
061110A			&	No				&	X-ray fit badly constrained					\\
070110			&	Yes				&	Break right below X-ray energies				\\
071031			&	Yes				&	Break could be moved						\\
071112C			&	Yes				&	Break could be moved						\\
080707			&	Yes				&	Break right below X-ray energies				\\
080916A			&	Yes				&	Break right below X-ray energies				\\ \hline
\end{tabular}
\label{tab:candidates}
\end{table}

\cite{schady12} study a sample of 17 which only contains lines-of-sight to GRB hosts with $A_V<1$. They set an upper limit of 12\% on flat host extinction curves. We note though that they have excluded from their sample bursts with no apparent extinction ($\sim16$\%), which could potentially have contained grey dust, so the actual upper limit might be as high as $\sim25\%$ (the line-of-sight to GRB\,121024A would have been concluded to have no extinction, if no alternative dust estimate existed). Complementary to this sample, \cite{2011A&A...534A.108K} present a small sample of 8 GRB SEDs and hosts with $A_V>1$ (overlapping with the \cite{zafar11} sample). As these are chosen to be dusty lines-of-sight, the optical data do appear reddened compared to an extrapolation from the X-rays. At high visual extinction they find that the SMC is no longer typically the best fit of the three local curves. Both LMC and MW give comparable or better fit, with half the sample having a significant 2175\,\AA \ bump.

For the bursts in Table~\ref{tab:candidates}, we went through the literature to look for any potential additional information on the amount of dust. Several of the bursts do have an observed afterglow spectrum, predominantly observed with VLT/FORS1+2 (FOcal Reducer and low dispersion Spectrographs), see \cite{2009ApJS..185..526F}. They did detect absorption lines, but at the resolution of FORS (R\,$=440$--$2140$), we cannot set usable limits on the depletion. The only burst we could find with well resolved absorption lines in the spectrum is GRB\,071031, which was observed with UVES (Ultraviolet and Visual Echelle Spectrograph, mounted on the VLT). The dust depletion as recorded by \cite{2013A&A...560A..88D} for this burst is consistent with a low amount of dust. However, they do find substantial dust depletion for both GRB\,080330, for which \cite{2011A&A...526A..30G} report $A_{\text{V}}\sim0.1$, and GRB\,990123 for which \cite{2007ApJ...661..787S} find $E(B-V)<0.03$\,mag.

We conclude that while we do not have enough information to determine the fraction of GRB lines-of-sight with significant amounts of grey dust, given the growing amount of reported cases, it is not unlikely that up to $\sim25\%$ may have some contribution to their extinction from grey dust.  

\section{Grey Dust}\label{grey}

\subsection{The origin of grey dust}
Before discussing the formation of grey dust, we first examine the exact dust composition in more detail. To determine the contribution of extinction from different dust grain sizes to the extinction curve of GRB\,121024A we have calculated synthetic extinction curves.

The dust extinction was calculated using DDSCATv7.1 \citep{2010arXiv1002.1505D}, for grain sizes 0.01 to 0.5\,$\mu$m, in steps of 0.001\,$\mu$m, in the wavelength range $0.1$-- $2$\,$\mu$m. DDSCAT is a Fortran code for calculating scattering and absorption of light by irregular particles by replacing the considered grain by a cubic array of point dipoles. For this study we used 17256 dipoles for each grain size. For all calculations we assumed the grains to be homogeneous isotropic silicate spheres. The silicate optical properties used were taken from \cite{2003ApJ...598.1017D}, representing a fictitious ''interstellar silicate'' with an imaginary index of refraction chosen specifically to reproduce the observed interstellar MW extinction curve. 

To calculate the A$_\lambda$/A$_v$ for our synthetic extinction curve, we used the Beer-Lambert law in
the form $\rho(Q_{\text{ext}}/a) = \sigma_{\text{eff}}$, with $\rho$ being the grain density, $a$ the grain radius, $Q_{\text{ext}}$ 
the extinction and $\sigma_{\text{eff}}$ the effective cross section. From $\tau_{\lambda} = A_{\lambda} (0.4/\text{log}(e))$ and 
$\tau_{\nu} = A_{\nu} (0.4/\text{log}(e))$ it is possible to derive $A_{\lambda}/A_{\nu}$ as
\begin{equation}
\tau_{\lambda} = l \int_{a_{\text{min}}}^{a_{\text{max}}} n(a) \sigma_{\lambda}(a) da
\end{equation}
using the MNR \citep{1977ApJ...217..425M} power-law distribution,
\begin{equation}
n(a) \propto a^{-\alpha}
\end{equation}
with $\alpha = 3.4$ and grains ranging in size between $a_{\text{min}}$ and $a_{\text{max}}$. 

\begin{figure}[ht]
\resizebox{\hsize}{!}{\includegraphics[bb=80 410 554 754,width=1.0\columnwidth]{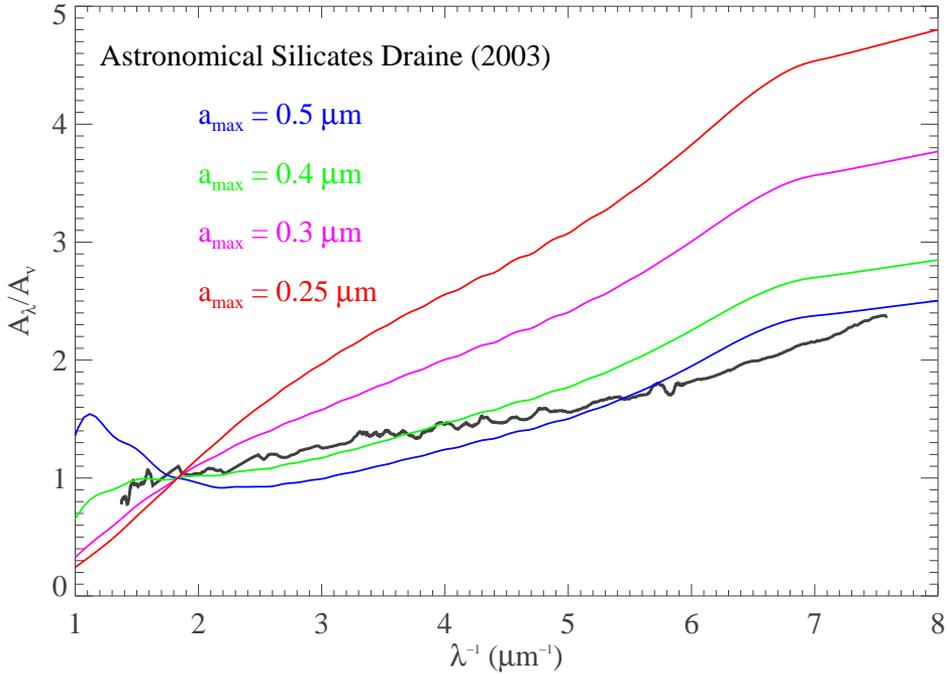}}
 \caption{The extinction curve of GRB\,121024A (black, using the broken power law solution) along with four different calculated synthetic extinction curves based on astronomical silicates \citep{2003ApJ...598.1017D} and assuming an MNR \citep{1977ApJ...217..425M} power-law distribution with $\alpha = 3.4$. The minimum grain size is $a_{\text{min}} = 0.01$\,$\mu$m while $a_{\text{max}}$ has been varied between 0.25 and 0.5\,$\mu$m.}
 \label{fig:syntext}
\end{figure}

Four synthetic extinction curves are shown in Fig.\,\ref{fig:syntext} compared to the observed extinction curve of
GRB\,121024A. For the fitting $a_{\text{min}}$ was fixed to $0.01$\,$\mu$m while $a_{\text{max}}$ was varied. As can be
seen from the figure the grain size distribution needs an $a_{\text{max}}$ larger than $0.25$\,$\mu$m indicating that rather large grains are needed, compared to the MW.  

\cite{2014Natur.511..326G} interpret the spectral evolution of SN 2010jl as evidence for evolution of dust production in the circum-stellar medium. They fit different dust models to the extinction curve they infer under this interpretation and find that to model the extinction their dust grain size distribution must also extend to sizes significantly larger than those observed in the MW ($\gtrsim0.25\,\mu$m). They suggest that the dust may be produced through a two-stage process, with early dust formation in a cool, dense shell, followed by accelerated dust formation in the ejected material, where the large grains, predominantly, survive reverse shock interactions, as shown in simulations \citep{2010ApJ...715.1575S}. The efficiency of the differential destruction depends on the ambient medium, so that a large ambient density leads to a flat extinction curve \citep{2008MNRAS.384.1725H}.
\cite{2015A&A...575A..95S} model the dust production in Type II-P SNe, finding that large density variations in the ejecta, in the form of gas 'clumps', increase the grain size. Since long GRBs are found in star-forming regions, we expect core-collapse SNe in the same area, which may for some reason have had particular clumpy ejecta, or a large surrounding density (consistent with the large observed hydrogen column density) which effectively skewed the dust composition towards larger grains. Another possibility is preferential dust destruction by a strong UV radiation field from young massive stars in the star-forming region.

Alternatively, the large dust grains could be a result of grain growth through coagulation. Grain growth is naturally more efficient in high-density environments \citep[see e.g.][]{2008ApJ...684.1228S}, with dust in the cores of dense molecular clouds being observed to consist of $\mu$m-sized grains through so-called 'coreshine' detections \citep{2013MNRAS.434L..70H}. We observe a large hydrogen column density, with a significant molecular content in the line-of-sight towards GRB\,121024A, consistent with the conditions needed for grain growth.



\subsection{Ubiquity of grey dust}
As discussed in Section~\ref{sec:dest}, the flatness of the extinction curve for GRB\,121024A and other bursts is unlikely to be directly related with the bursts themselves. If the presence of the grey dust is caused by e.g. a high UV field from young, massive stars, or from peculiar SNe, it is reasonable to assume that we can extrapolate the discussion so far to star-forming areas in general, not just in GRB hosts. The same is true if the larger grain sizes are caused by grain growth through coagulation in the ISM, in which case the flat extinction may be linked to a high-density environment, such as the DLAs probed by GRBs. DLAs the majority of the neutral gas at high $z$, which is then turned into stars over time. GRBs have been suggested as tracers of star-formation, though they are possibly biased towards low-metallicity environments. This metallicity threshold is unlikely to be significantly below solar though \citep[see e.g.][]{2015A&A...579A.126S,2015arXiv150402479P}, in which case GRBs unbiasedly trace star-formation at $z\gtrsim2$. It is hence possible that the existence of grey dust in up to $25$\% of lines-of-sight towards GRBs can be extrapolated to star-forming areas in general at these redshifts. 


\subsection{Consequences of applying the wrong extinction curve}\label{sec:wrong}
To illustrate the danger of assuming a local extinction law, we again use the example of GRB\,121024A. In the X-shooter spectrum of this burst we detect 6 emission lines from the GRB host. These lines are extinction corrected using the Balmer decrement, assuming the attenuation law of \cite{2000ApJ...533..682C}. Given the lack of reddening we find a small correction of $E(B-V)=0.04\pm0.09$\,mag, i.e. consistent with no correction within the \cite{2000ApJ...533..682C} law. If we instead assume that the dust properties we observe for the GRB line-of-sight are representative for the star-forming regions of the entire host galaxy, then we may perform the flux extinction correction using the curve plotted in Fig.~\ref{fig:curves} (although there is a difference between attenuation and extinction, but dust scattering into the line-of-sight will only produce an even greyer curve, so the effects found may be considered lower limits). We now find flux correction factors ranging from 2.1 for [\mbox{N\,{\sc ii}}] (the line at the highest wavelength) to 2.8 for [\mbox{O\,{\sc ii}}] (the shortest wavelength). This means that rather than a SFR of $42\pm11$\,M$_\odot$\,yr$^{-1}$ calculated from the H$\alpha$ using the \cite{1998ARA&A..36..189K} relation, the actual SFR would be $96\pm27$\,M$_\odot$\,yr$^{-1}$ (with an estimated error on the flux correction factor of $\pm0.1$). A caveat is that the extinction law for the line-of-sight to GRB\,121024A may not represent the general dust distribution in star-forming regions of the host. We also note that since the emission line flux is the integrated value for the entire host, the least dust obscured regions are likely to contribute the most. Nonetheless, we have not found any evidence that this dust distribution is connected to the presence of the GRB, so there is a strong possibility that we are underestimating the SFR to some degree. 

The Balmer decrement is a measure of the spectral reddening between the H$\alpha$ and the H$\beta$ lines. Often either one of the \cite{2000ApJ...533..682C} or \cite{pei} attenuation/extinction curves are assumed without much consequence, as they are fairly similar at these wavelengths, see Fig.~\ref{fig:curves} (the rest-frame wavelength of H$\alpha$  is $6562.80$\,\AA \ = $1.52\,\mu$m$^{-1}$, while for H$\beta$ it is $4861.33$\,\AA \ = $2.06\,\mu$m$^{-1}$). However, if the intrinsic dust distribution is similar to that in the line-of-sight to GRB\,121024A, then it makes a significant difference which curve is used.

The current picture of the star-formation history (SFH) of the Universe consists of a rise in the star formation rate density (SFRD) until approximately $z\sim3$, at which point the star-formation starts to slow down and peak at $z\sim1.5$--$2$, after which it falls off exponentially until $z=0$ \citep[for a review see e.g.][]{2014ARA&A..52..415M}. The SFRD is well mapped out to a redshift $z\sim1$. At larger look-back times, particularly at $z\gtrsim2.5$, the full data set becomes limited, and we often have to rely solely on rest-frame UV data alone. The UV luminosity of galaxies is dominated by massive young stars, and is hence used to trace the star formation. It is, however, degenerate with stellar-population age and dust extinction, which can severely affect the reported SFR \citep[see e.g.][]{2001ApJ...559..620P}. The latter is determined from fitting synthetic stellar populations to photometry, or, if only a narrow wavelength range is available as is often the case, from the UV slope $\beta$ through the IRX-$\beta$ relation \citep[IRX: infrared excess, see e.g.][]{1999ApJ...521...64M,2011ApJ...726L...7O}. This relation is known to be relatively tight in the local Universe, following the \cite{2000ApJ...533..682C} attenuation curve \citep[e.g.][]{2013ApJ...762..125N}, but even here it depends on galaxy classification \citep[ultraluminous infrared galaxies, for instance, do not follow the same relation as UV bright galaxies, see e.g.][]{2002ApJ...568..651G}. At higher redshifts the limited amount of data makes definite conclusions difficult, see e.g. \cite{2008ApJ...689..883C}, \cite{2011ApJ...726L...7O} and \cite{2012ApJ...744..154R}. At the highest redshifts ($\gtrsim6$), only upper limits have been set on the IRX, e.g. \cite{2015A&A...574A..19S}. Recently, \cite{2015arXiv151001514W} determined $\beta$ values for $z\sim10$ galaxies, reporting that while no dust extinction was inferred from the \cite{1999ApJ...521...64M} relation, hydro-dynamical simulations of galaxy formation result in a significant extinction, even at these redshifts.

Inconsistencies between UV and IR SFRs are well known. \cite{2014A&A...566A..19C} find that UV determined SFRs for Lyman break galaxies at $z\sim3$ are generally 2--4 times smaller than the equivalent from other SFR indicators. They attribute this to calibrations being based on solar metallicity, while the $z=3$ galaxies have significantly lower metallicity. Even SFRs determined from H$\alpha$ through the \cite{1998ARA&A..36..189K} relation can be severely affected by missing highly-attenuated sources, see e.g. \cite{2015arXiv150602670O} who compared their H$\alpha$-emitters (HAEs) with \emph{Herschel} detections for 9 PACS/SPIRE-detected HAEs finding a significant difference in SFRs.

As we have demonstrated in this paper, up to $\sim20$--$30\%$ of GRB hosts could potentially contain significant amounts of grey dust. If we extrapolate this result to star-forming galaxies at high redshift, then it becomes apparent, that using the UV slope alone to constrain the dust extinction is not sufficient, as we could end up significantly underestimating the SFR. This could potentially have an effect on the SFH, since, while IR data broadly agrees with the trends observed, the peak in SFRD around $z\sim1.5$ -- $2$ appears less sharp \citep[e.g.][]{2013A&A...554A..70B}, and the slope after $z\sim3$ is less constrained.

To illustrate the potential effect of grey dust, we again use the extinction curve from GRB\,121024A. If we introduce grey dust as an extra component to the extinction then $A_{\text{UV}}^{\text{new}}=A_{\text{UV}}^{\text{old}}+A_{\text{UV}}^{\text{grey}}$. If the SFRD from the UV is given as: \[\text{SFRD}(z) = k\times\text{LD}_{\text{UV}}(z)\times10^{0.4A_{\text{UV}}(z)}\]  where $k$ is the conversion factor between SFR and the UV luminosity, e.g. from \cite{1998ARA&A..36..189K} and LD$_{\text{UV}}$ is the UV luminosity density, then the correction factor to the SFRD is: \[\text{SFRD}_{\text{new}}=10^{0.4A_{\text{grey}}}\times\text{SFRD}_{\text{old}}\] at a given $z$.

For the extinction curve of GRB\,121024A, $A_{\text{UV}}\sim2.1$\,mag. If we now assume that, more realistically, 15\% of the UV luminosity we observe is attenuated by this extra component of grey dust, then the observation should be corrected with: $A_{\text{grey}}=0.15\times2.1=0.32$, i.e. \[\frac{\text{SFRD}^{\text{new}}}{\text{SFRD}^{\text{old}}}=10^{0.4\times0.32}=1.3\] So, at a given $z$ and at this level of grey dust contribution, the UV determined SFRD reported in the literature is $\sim30\%$ lower than the actual value. If we assume that the fraction of grey dust is different at different redshifts, then this correction could potentially change the exact shape of the SFH as probed in the UV. If for instance grey dust was more prominent before $z\sim1.5$--$2$, then the correction factor could make the peak in star formation with redshift flatter, or move it to a higher redshift, as hinted at by radio and IR observations \citep[e.g.][]{2011ApJ...730...61K,2013A&A...554A..70B}, and consistent with UV observations, once the uncertainty in dust obscuration is considered \citep[e.g.][]{2015arXiv150705629P}.


\section{Conclusions}
We have found clear evidence for grey dust in the line-of-sight to GRB\,121024A. Several other such components have previously been reported, and a significant ($\sim25$\,\%) fraction of GRB hosts are potential candidates for containing grey dust. We find that the dust composition fits well with MW-type dust, but with a significantly larger maximum grain size. We propose that grey dust may be present in a significant fraction of galaxies that share properties with GRB hosts. Since these galaxies are star forming, this could potentially have an impact on the SFH of the Universe, as SFRs could be underestimated with as much as a factor of $\sim2$--$3$.

While it is not our intention here to claim that a high fraction of high-$z$ galaxies are dominated by grey dust, we do recommend exercising care when applying local extinction laws, as a small, but significant, number of physical properties such as SFRs, might be underestimated.   

\section*{Acknowledgements}
MF acknowledges support from the University of Iceland Research fund. The
research leading to these results has received funding from the European
Research Council under the European Union's Seventh Framework Program
(FP7/2007-2013)/ERC Grant agreement no.  EGGS-278202. This work made use of data supplied by the UK Swift Science Data Centre at the University of Leicester.

\cleardoublepage  
\chapter{Summary and Prospects}\label{sec:summary_thesis}
In this thesis, numerous aspects of the physics of GRBs and their surrounding environment has been examined. The work presented gives a thorough overview of how much information it is possible to extract from GRB afterglow spectroscopy, ranging from examining the burst mechanism itself from the soft X-ray afterglow (Chapters~\ref{chap:spec} and~\ref{sec:paper_thermal}), to infer physical conditions of the burst host galaxy and birth place from optical/NIR spectroscopy (Chapters~\ref{chapter:host} and~\ref{sec:paper_121024A}), and lastly to answer larger astrophysical questions about dust in the non-local Universe through a combination of the two (X-ray and optical afterglow, see Chapters~\ref{sec:dust} and~\ref{chap:ext}).

The first part of the thesis is motivated by the fairly recent consensus in the research field that the combination of power laws previously used to fit the GRB spectrum gives an inadequate description of time resolved spectra and the temporal evolution observed. Instead, models including a strong blackbody component became favoured. This has let us to propose a new model for the thermal emission observed in the soft X-ray GRB afterglow, linking this component to the prompt thermal emission. We have studied the ubiquitousness of this component, by searching for it in a sample of the brightest bursts observed with the XRT. Finding that the whole sample could potentially have a thermal component (although not always statistically preferred over the power laws), we proceeded to compare the blackbody parameters with those observed in the prompt phase. We conclude that our model of the soft X-ray thermal component as late-time photospheric emission from the GRB jet is a well motivated and self-consistent theory.

There is still large uncertainty regarding the exact physics of GRB emission. Debate is still ongoing with regards to the contribution from (mainly) synchrotron versus photospheric radiation. In the couple of years since our paper was published, models of photospheric emission from a jet moving at relativistic speed has been improved, and there is now evidence to suggest that although most of the GRB emission is indeed photospheric, it is not in the form of a blackbody at these extreme conditions (F. Ryde, private communication). If the data is seen to support this, our theory will have to be revisited. One way to distinguish between synchrotron and relativistic-thermal radiation, is by measuring the polarisation of the emission. Instruments are currently not ideally designed to measure this polarisation, but new instrumentation may be funded in the future.

The second part of this thesis is a presentation of a GRB host galaxy study. The optical/NIR spectrum taken of GRB\,121024A with the X-shooter spectrograph is very rich in features, and highlights the use of GRBs as cosmic lighthouses. From this spectrum, we were able to study several separate components of the gas along the line-of-sight to the burst, by isolating velocity components in the absorption lines, showing lines from both low- and high-ionisation states, fine-structure lines and molecular hydrogen. We also observed several nebular lines in emission from the host galaxy, tracing the star-forming regions of the host. From these lines we could determine that the host is highly star forming and, for the first time in a GRB host, we got a metallicity estimate from both absorption and emission lines. We found that the two methods are in agreement, which, given that the absorption and emission lines trace different regions of the galaxy, was not necessarily expected to be the case. With GRB afterglow samples such as the X-shooter sample, hopefully several such metallicity comparisons will be possible in the future, so that we might make a statistical comparison of the metallicity in the different regions of host galaxies. These cases are rare though (however, see the recent GRB\,150915A), and are further complicated by problems in calibrating the strong-line diagnostics used to get the metallicity estimate from emission line fluxes.

By including X-ray data from the XRT, we were also able to fit the spectral energy distribution from NIR to X-rays, to constrain the dust extinction along the line-of-sight. During this analysis, we discovered a discrepancy between the dust extinction inferred from depletion analysis and SED fitting. In order to reconcile the two, a very flat extinction curve is needed, indicating the presence of grey dust in the line-of-sight to the GRB.

This grey dust is the subject of the third part of the thesis, where we examine the possible physics behind its presence. We find that the grey dust is likely dust with larger grain sizes than observed on average in the Milky Way, and conclude that it is implausible to be the influence of the GRB itself creating this dust. We also examine a sample of GRB extinction curves, to determine how common the presence of grey dust could be in GRB birth environments. We find that $\sim25\%$ of lines-of-sight towards GRBs are candidates for containing grey dust, a result which if extrapolated to star-forming regions in general, and choosing a more realistic value of $15\%$, would mean that we are potentially underestimating the UV-determined SFRD at a given redshift with as much as $30\%$.

Since grey dust is difficult to detect, we do not know how common this component might be outside the local Universe. One way to possibly constrain this, is by conducting large-scale surveys in the rest-frame infrared, but this will be observationally expensive at high redshifts, and hence unlikely to be prioritised in the near future. One possibility is to use the MIR camera onboard the planned \emph{James Webb Space Telescope}, and then use scaling relations to determine the total IR luminosity, see e.g. \cite{2013A&A...558A.136M}. Another option is to use SN type Ia observations, as the intrinsic luminosity can be obtained from the light curve, and hence the total extinction can be determined. Observations of heavily obscured SNe currently indicate the opposite of grey dust, i.e. a very low selective extinction, see for example \cite{2007AJ....133...58K} and \cite{2015MNRAS.453.3300A}. It has been suggested though, that this low value could be associated with other processes besides extinction \citep{2008A&A...487...19N}. At high redshifts, multiple-imaged quasar systems can be used to determine the total extinction. By comparing different lines-of-sight, the differential extinction can be determined, see e.g. \cite{2006ApJS..166..443E}. The \emph{Euclid} satellite \citep{2011arXiv1110.3193L} should help increase the sample size of these studies. With the X-shooter sample, we will be able to analyse depletion patterns for several bursts, to determine the visual extinction, as well as constraining the allowed extinction curves from SED fitting of bursts with an overlap in X-ray and optical light curves, and hence may constrain the fraction of GRB hosts that could contain grey dust.

As demonstrated in this thesis, observations of GRBs provide a versatile tool, enabling the study of many aspects of astronomy and astrophysics. Although we are closer to understanding the mechanisms behind the bursts and afterglows, theories are still being refined and compared, as more data streams in. Improved insight into the physical processes leading to a GRB, will help determine the possible progenitors, which in turn will help us understand the different biases in birth environment, so that we may confidently use GRBs as probes of e.g. star formation across the Universe. Inversely, by studying the burst environments (and possible biases) directly through observations of the afterglow or host, we can put constraints on the allowed GRB models. In the near future, important work will be done at both ends; studies of the burst emission through \emph{Fermi} and \emph{Swift} observations will attempt to determine the dominant radiative mechanism, while analysis of data from e.g. the X-shooter GRB program will shed light on the environment and host properties, e.g. examining thresholds in metallicity and preferences regarding dust and molecular content in the vicinity of the bursts. Through these studies, GRBs have the potential to help solve some of the Universe' big mysteries.

\cleardoublepage

\bibliography{mybib}

\begin{thebibliography}{455}
\expandafter\ifx\csname natexlab\endcsname\relax\def\natexlab#1{#1}\fi

\bibitem[{{Abadie} {et~al.}(2012){Abadie}, {Abbott}, {Abbott}, {Abbott},
  {Abernathy}, {Accadia}, {Acernese}, {Adams}, {Adhikari}, {Affeldt}, \&
  et~al.}]{2012ApJ...760...12A}
{Abadie}, J., {Abbott}, B.~P., {Abbott}, R., {et~al.} 2012, \apj, 760, 12

\bibitem[{{Abel} {et~al.}(2002){Abel}, {Bryan}, \&
  {Norman}}]{2002Sci...295...93A}
{Abel}, T., {Bryan}, G.~L., \& {Norman}, M.~L. 2002, Science, 295, 93

\bibitem[{{Aguirre}(1999)}]{1999ApJ...525..583A}
{Aguirre}, A. 1999, \apj, 525, 583

\bibitem[{{Amanullah} {et~al.}(2015){Amanullah}, {Johansson}, {Goobar},
  {Ferretti}, {Papadogiannakis}, {Petrushevska}, {Brown}, {Cao}, {Contreras},
  {Dahle}, {Elias-Rosa}, {Fynbo}, {Gorosabel}, {Guaita}, {Hangard}, {Howell},
  {Hsiao}, {Kankare}, {Kasliwal}, {Leloudas}, {Lundqvist}, {Mattila}, {Nugent},
  {Phillips}, {Sandberg}, {Stanishev}, {Sullivan}, {Taddia}, {{\"O}stlin},
  {Asadi}, {Herrero-Illana}, {Jensen}, {Karhunen}, {Lazarevic}, {Varenius},
  {Santos}, {Sridhar}, {Wallstr{\"o}m}, \& {Wiegert}}]{2015MNRAS.453.3300A}
{Amanullah}, R., {Johansson}, J., {Goobar}, A., {et~al.} 2015, \mnras, 453,
  3300

\bibitem[{{Arabsalmani} {et~al.}(2015){Arabsalmani}, {M{\o}ller}, {Fynbo},
  {Christensen}, {Freudling}, {Savaglio}, \& {Zafar}}]{arabsalmani}
{Arabsalmani}, M., {M{\o}ller}, P., {Fynbo}, J.~P.~U., {et~al.} 2015, \mnras,
  446, 990

\bibitem[{{Asplund} {et~al.}(2009){Asplund}, {Grevesse}, {Sauval}, \&
  {Scott}}]{asplund09}
{Asplund}, M., {Grevesse}, N., {Sauval}, A.~J., \& {Scott}, P. 2009, \araa, 47,
  481

\bibitem[{{Axelsson} {et~al.}(2012){Axelsson}, {Baldini}, {Barbiellini},
  {Baring}, {Bellazzini}, {Bregeon}, {Brigida}, {Bruel}, {Buehler},
  {Caliandro}, {Cameron}, {Caraveo}, {Cecchi}, {Chaves}, {Chekhtman}, {Chiang},
  {Claus}, {Conrad}, {Cutini}, {D'Ammando}, {de Palma}, {Dermer}, {Silva},
  {Drell}, {Favuzzi}, {Fegan}, {Ferrara}, {Focke}, {Fukazawa}, {Fusco},
  {Gargano}, {Gasparrini}, {Gehrels}, {Germani}, {Giglietto}, {Giroletti},
  {Godfrey}, {Guiriec}, {Hadasch}, {Hanabata}, {Hayashida}, {Hou}, {Iyyani},
  {Jackson}, {Kocevski}, {Kuss}, {Larsson}, {Larsson}, {Longo}, {Loparco},
  {Lundman}, {Mazziotta}, {McEnery}, {Mizuno}, {Monzani}, {Moretti},
  {Morselli}, {Murgia}, {Nuss}, {Nymark}, {Ohno}, {Omodei}, {Pesce-Rollins},
  {Piron}, {Pivato}, {Racusin}, {Rain{\`o}}, {Razzano}, {Razzaque}, {Reimer},
  {Roth}, {Ryde}, {Sanchez}, {Sgr{\`o}}, {Siskind}, {Spandre}, {Spinelli},
  {Stamatikos}, {Tibaldo}, {Tinivella}, {Usher}, {Vandenbroucke}, {Vasileiou},
  {Vianello}, {Vitale}, {Waite}, {Winer}, {Wood}, {Burgess}, {Bhat},
  {Bissaldi}, {Briggs}, {Connaughton}, {Fishman}, {Fitzpatrick}, {Foley},
  {Gruber}, {Kippen}, {Kouveliotou}, {Jenke}, {McBreen}, {McGlynn}, {Meegan},
  {Paciesas}, {Pelassa}, {Preece}, {Tierney}, {von Kienlin}, {Wilson-Hodge},
  {Xiong}, \& {Pe'er}}]{2012ApJ...757L..31A}
{Axelsson}, M., {Baldini}, L., {Barbiellini}, G., {et~al.} 2012, \apjl, 757,
  L31

\bibitem[{{Balashev} {et~al.}(2015){Balashev}, {Noterdaeme}, {Klimenko},
  {Petitjean}, {Srianand}, {Ledoux}, {Ivanchik}, \&
  {Varshalovich}}]{balashev15}
{Balashev}, S.~A., {Noterdaeme}, P., {Klimenko}, V.~V., {et~al.} 2015, \aap,
  575, L8

\bibitem[{{Band} {et~al.}(1993){Band}, {Matteson}, {Ford}, {Schaefer},
  {Palmer}, {Teegarden}, {Cline}, {Briggs}, {Paciesas}, {Pendleton}, {Fishman},
  {Kouveliotou}, {Meegan}, {Wilson}, \& {Lestrade}}]{1993ApJ...413..281B}
{Band}, D., {Matteson}, J., {Ford}, L., {et~al.} 1993, \apj, 413, 281

\bibitem[{{Baret} {et~al.}(2011){Baret}, {Bartos}, {Bouhou}, {Corsi}, {Palma},
  {Donzaud}, {Elewyck}, {Finley}, {Jones}, {Kouchner}, {M{\'a}rka},
  {M{\'a}rka}, {Moscoso}, {Chassande-Mottin}, {Papa}, {Pradier}, {Raffai},
  {Rollins}, \& {Sutton}}]{2011APh....35....1B}
{Baret}, B., {Bartos}, I., {Bouhou}, B., {et~al.} 2011, Astroparticle Physics,
  35, 1

\bibitem[{{Baring} \& {Braby}(2004)}]{2004ApJ...613..460B}
{Baring}, M.~G. \& {Braby}, M.~L. 2004, \apj, 613, 460

\bibitem[{{Barthelmy} {et~al.}(2005){Barthelmy}, {Barbier}, {Cummings},
  {Fenimore}, {Gehrels}, {Hullinger}, {Krimm}, {Markwardt}, {Palmer},
  {Parsons}, {Sato}, {Suzuki}, {Takahashi}, {Tashiro}, \&
  {Tueller}}]{Barthelmy}
{Barthelmy}, S.~D., {Barbier}, L.~M., {Cummings}, J.~R., {et~al.} 2005, \ssr,
  120, 143

\bibitem[{{Becker} \& {D\"oring}(1935)}]{becker35}
{Becker}, R. \& {D\"oring}, W. 1935, Annalen der Physik, 416, 719

\bibitem[{{Becker} {et~al.}(2001){Becker}, {Fan}, {White}, {Strauss},
  {Narayanan}, {Lupton}, {Gunn}, {Annis}, {Bahcall}, {Brinkmann}, {Connolly},
  {Csabai}, {Czarapata}, {Doi}, {Heckman}, {Hennessy}, {Ivezi{\'c}}, {Knapp},
  {Lamb}, {McKay}, {Munn}, {Nash}, {Nichol}, {Pier}, {Richards}, {Schneider},
  {Stoughton}, {Szalay}, {Thakar}, \& {York}}]{2001AJ....122.2850B}
{Becker}, R.~H., {Fan}, X., {White}, R.~L., {et~al.} 2001, \aj, 122, 2850

\bibitem[{{Belczynski} {et~al.}(2006){Belczynski}, {Perna}, {Bulik},
  {Kalogera}, {Ivanova}, \& {Lamb}}]{2006ApJ...648.1110B}
{Belczynski}, K., {Perna}, R., {Bulik}, T., {et~al.} 2006, \apj, 648, 1110

\bibitem[{{Berg} {et~al.}(2015){Berg}, {Ellison}, {Prochaska}, {Venn}, \&
  {Dessauges-Zavadsky}}]{2015MNRAS.452.4326B}
{Berg}, T.~A.~M., {Ellison}, S.~L., {Prochaska}, J.~X., {Venn}, K.~A., \&
  {Dessauges-Zavadsky}, M. 2015, \mnras, 452, 4326

\bibitem[{{Berger}(2014)}]{2014ARA&A..52...43B}
{Berger}, E. 2014, \araa, 52, 43

\bibitem[{{Berger} {et~al.}(2008){Berger}, {Foley}, {Simcoe}, \&
  {Irwin}}]{2008GCN..8434....1B}
{Berger}, E., {Foley}, R., {Simcoe}, R., \& {Irwin}, J. 2008, GRB Coordinates
  Network, 8434, 1

\bibitem[{{Berger} {et~al.}(2013){Berger}, {Fong}, \&
  {Chornock}}]{2013ApJ...774L..23B}
{Berger}, E., {Fong}, W., \& {Chornock}, R. 2013, \apjl, 774, L23

\bibitem[{{Berger} \& {Fox}(2009)}]{2009GCN..9156....1B}
{Berger}, E. \& {Fox}, D.~B. 2009, GRB Coordinates Network, 9156, 1

\bibitem[{{Bertoldi} {et~al.}(2003){Bertoldi}, {Carilli}, {Cox}, {Fan},
  {Strauss}, {Beelen}, {Omont}, \& {Zylka}}]{2003A&A...406L..55B}
{Bertoldi}, F., {Carilli}, C.~L., {Cox}, P., {et~al.} 2003, \aap, 406, L55

\bibitem[{{Bhat} {et~al.}(1992){Bhat}, {Fishman}, {Meegan}, {Wilson}, {Brock},
  \& {Paciesas}}]{1992Natur.359..217B}
{Bhat}, P.~N., {Fishman}, G.~J., {Meegan}, C.~A., {et~al.} 1992, \nat, 359, 217

\bibitem[{{Bhattacharya}(2001)}]{2001BASI...29..107B}
{Bhattacharya}, D. 2001, Bulletin of the Astronomical Society of India, 29, 107

\bibitem[{{Biscaro} \& {Cherchneff}(2014)}]{2014A&A...564A..25B}
{Biscaro}, C. \& {Cherchneff}, I. 2014, \aap, 564, A25

\bibitem[{{Bloom} {et~al.}(1999){Bloom}, {Kulkarni}, {Djorgovski},
  {Eichelberger}, {C{\^o}t{\'e}}, {Blakeslee}, {Odewahn}, {Harrison}, {Frail},
  {Filippenko}, {Leonard}, {Riess}, {Spinrad}, {Stern}, {Bunker}, {Dey},
  {Grossan}, {Perlmutter}, {Knop}, {Hook}, \& {Feroci}}]{1999Natur.401..453B}
{Bloom}, J.~S., {Kulkarni}, S.~R., {Djorgovski}, S.~G., {et~al.} 1999, \nat,
  401, 453

\bibitem[{{Bloom} {et~al.}(2006){Bloom}, {Prochaska}, {Pooley}, {Blake},
  {Foley}, {Jha}, {Ramirez-Ruiz}, {Granot}, {Filippenko}, {Sigurdsson},
  {Barth}, {Chen}, {Cooper}, {Falco}, {Gal}, {Gerke}, {Gladders}, {Greene},
  {Hennanwi}, {Ho}, {Hurley}, {Koester}, {Li}, {Lubin}, {Newman}, {Perley},
  {Squires}, \& {Wood-Vasey}}]{2006ApJ...638..354B}
{Bloom}, J.~S., {Prochaska}, J.~X., {Pooley}, D., {et~al.} 2006, \apj, 638, 354

\bibitem[{{Blustin} {et~al.}(2006){Blustin}, {Band}, {Barthelmy}, {Boyd},
  {Capalbi}, {Holland}, {Marshall}, {Mason}, {Perri}, {Poole}, {Roming},
  {Rosen}, {Schady}, {Still}, {Zhang}, {Angelini}, {Barbier}, {Beardmore},
  {Breeveld}, {Burrows}, {Cummings}, {Cannizzo}, {Campana}, {Chester},
  {Chincarini}, {Cominsky}, {Cucchiara}, {de Pasquale}, {Fenimore}, {Gehrels},
  {Giommi}, {Goad}, {Gronwall}, {Grupe}, {Hill}, {Hinshaw}, {Hunsberger},
  {Hurley}, {Ivanushkina}, {Kennea}, {Krimm}, {Kumar}, {Landsman}, {La Parola},
  {Markwardt}, {McGowan}, {M{\'e}sz{\'a}ros}, {Mineo}, {Moretti}, {Morgan},
  {Nousek}, {O'Brien}, {Osborne}, {Page}, {Page}, {Palmer}, {Parsons},
  {Rhoads}, {Romano}, {Sakamoto}, {Sato}, {Tagliaferri}, {Tueller}, {Wells}, \&
  {White}}]{2006ApJ...637..901B}
{Blustin}, A.~J., {Band}, D., {Barthelmy}, S., {et~al.} 2006, \apj, 637, 901

\bibitem[{{Bohlin} {et~al.}(1978){Bohlin}, {Savage}, \&
  {Drake}}]{1978ApJ...224..132B}
{Bohlin}, R.~C., {Savage}, B.~D., \& {Drake}, J.~F. 1978, \apj, 224, 132

\bibitem[{{Bornancini} {et~al.}(2004){Bornancini}, {Mart{\'{\i}}nez}, {Lambas},
  {Le Floc'h}, {Mirabel}, \& {Minniti}}]{2004ApJ...614...84B}
{Bornancini}, C.~G., {Mart{\'{\i}}nez}, H.~J., {Lambas}, D.~G., {et~al.} 2004,
  \apj, 614, 84

\bibitem[{{Bowen} {et~al.}(2005){Bowen}, {Jenkins}, {Pettini}, \&
  {Tripp}}]{bowen05}
{Bowen}, D.~V., {Jenkins}, E.~B., {Pettini}, M., \& {Tripp}, T.~M. 2005, \apj,
  635, 880

\bibitem[{{Bromm} \& {Loeb}(2003)}]{2003Natur.425..812B}
{Bromm}, V. \& {Loeb}, A. 2003, \nat, 425, 812

\bibitem[{{Bruzual}(2007)}]{2007ASPC..374..303B}
{Bruzual}, G. 2007, in Astronomical Society of the Pacific Conference Series,
  Vol. 374, From Stars to Galaxies: Building the Pieces to Build Up the
  Universe, ed. A.~{Vallenari}, R.~{Tantalo}, L.~{Portinari}, \& A.~{Moretti},
  303

\bibitem[{{Bruzual} \& {Charlot}(2003)}]{BC03}
{Bruzual}, G. \& {Charlot}, S. 2003, \mnras, 344, 1000

\bibitem[{{Bucciantini} {et~al.}(2009){Bucciantini}, {Quataert}, {Metzger},
  {Thompson}, {Arons}, \& {Del Zanna}}]{2009MNRAS.396.2038B}
{Bucciantini}, N., {Quataert}, E., {Metzger}, B.~D., {et~al.} 2009, \mnras,
  396, 2038

\bibitem[{{Bufano} {et~al.}(2012){Bufano}, {Pian}, {Sollerman}, {Benetti},
  {Pignata}, {Valenti}, {Covino}, {D'Avanzo}, {Malesani}, {Cappellaro}, {Della
  Valle}, {Fynbo}, {Hjorth}, {Mazzali}, {Reichart}, {Starling}, {Turatto},
  {Vergani}, {Wiersema}, {Amati}, {Bersier}, {Campana}, {Cano},
  {Castro-Tirado}, {Chincarini}, {D'Elia}, {de Ugarte Postigo}, {Deng},
  {Ferrero}, {Filippenko}, {Goldoni}, {Gorosabel}, {Greiner}, {Hammer},
  {Jakobsson}, {Kaper}, {Kawabata}, {Klose}, {Levan}, {Maeda}, {Masetti},
  {Milvang-Jensen}, {Mirabel}, {M{\o}ller}, {Nomoto}, {Palazzi}, {Piranomonte},
  {Salvaterra}, {Stratta}, {Tagliaferri}, {Tanaka}, {Tanvir}, \&
  {Wijers}}]{2012ApJ...753...67B}
{Bufano}, F., {Pian}, E., {Sollerman}, J., {et~al.} 2012, \apj, 753, 67

\bibitem[{{Burgarella} {et~al.}(2013){Burgarella}, {Buat}, {Gruppioni},
  {Cucciati}, {Heinis}, {Berta}, {B{\'e}thermin}, {Bock}, {Cooray}, {Dunlop},
  {Farrah}, {Franceschini}, {Le Floc'h}, {Lutz}, {Magnelli}, {Nordon},
  {Oliver}, {Page}, {Popesso}, {Pozzi}, {Riguccini}, {Vaccari}, \&
  {Viero}}]{2013A&A...554A..70B}
{Burgarella}, D., {Buat}, V., {Gruppioni}, C., {et~al.} 2013, \aap, 554, A70

\bibitem[{{Burlon} {et~al.}(2008){Burlon}, {Ghirlanda}, {Ghisellini},
  {Lazzati}, {Nava}, {Nardini}, \& {Celotti}}]{2008ApJ...685L..19B}
{Burlon}, D., {Ghirlanda}, G., {Ghisellini}, G., {et~al.} 2008, \apjl, 685, L19

\bibitem[{{Burrows} {et~al.}(2007){Burrows}, {Falcone}, {Chincarini}, {Morris},
  {Romano}, {Hill}, {Godet}, {Moretti}, {Krimm}, {Osborne}, {Racusin},
  {Mangano}, {Page}, {Perri}, {Stroh}, \& {Swift XRT
  Team}}]{2007RSPTA.365.1213B}
{Burrows}, D.~N., {Falcone}, A., {Chincarini}, G., {et~al.} 2007, Royal Society
  of London Philosophical Transactions Series A, 365, 1213

\bibitem[{{Burrows} {et~al.}(2012){Burrows}, {Fox}, {Palmer}, {Romano},
  {Mangano}, {La Parola}, {Falcone}, \& {Roming}}]{2012MSAIS..21...59B}
{Burrows}, D.~N., {Fox}, D., {Palmer}, D., {et~al.} 2012, Memorie della Societa
  Astronomica Italiana Supplementi, 21, 59

\bibitem[{{Burrows} {et~al.}(2006){Burrows}, {Grupe}, {Capalbi}, {Panaitescu},
  {Patel}, {Kouveliotou}, {Zhang}, {M{\'e}sz{\'a}ros}, {Chincarini}, {Gehrels},
  \& {Wijers}}]{2006ApJ...653..468B}
{Burrows}, D.~N., {Grupe}, D., {Capalbi}, M., {et~al.} 2006, \apj, 653, 468

\bibitem[{{Butler}(2007)}]{2007ApJ...656.1001B}
{Butler}, N.~R. 2007, \apj, 656, 1001

\bibitem[{{Cabrera Lavers} {et~al.}(2011){Cabrera Lavers}, {de Ugarte Postigo},
  {Castro-Tirado}, {Gorosabel}, {Thoene}, \& {Dominguez}}]{2011GCN..12234...1C}
{Cabrera Lavers}, A., {de Ugarte Postigo}, A., {Castro-Tirado}, A.~J., {et~al.}
  2011, GRB Coordinates Network, 12234, 1

\bibitem[{{Calzetti} {et~al.}(2000){Calzetti}, {Armus}, {Bohlin}, {Kinney},
  {Koornneef}, \& {Storchi-Bergmann}}]{2000ApJ...533..682C}
{Calzetti}, D., {Armus}, L., {Bohlin}, R.~C., {et~al.} 2000, \apj, 533, 682

\bibitem[{{Campana} {et~al.}(2006{\natexlab{a}}){Campana}, {Mangano},
  {Blustin}, {Brown}, {Burrows}, {Chincarini}, {Cummings}, {Cusumano}, {Della
  Valle}, {Malesani}, {M{\'e}sz{\'a}ros}, {Nousek}, {Page}, {Sakamoto},
  {Waxman}, {Zhang}, {Dai}, {Gehrels}, {Immler}, {Marshall}, {Mason},
  {Moretti}, {O'Brien}, {Osborne}, {Page}, {Romano}, {Roming}, {Tagliaferri},
  {Cominsky}, {Giommi}, {Godet}, {Kennea}, {Krimm}, {Angelini}, {Barthelmy},
  {Boyd}, {Palmer}, {Wells}, \& {White}}]{2006Natur.442.1008C}
{Campana}, S., {Mangano}, V., {Blustin}, A.~J., {et~al.} 2006{\natexlab{a}},
  \nat, 442, 1008

\bibitem[{{Campana} {et~al.}(2006{\natexlab{b}}){Campana}, {Tagliaferri},
  {Lazzati}, {Chincarini}, {Covino}, {Page}, {Romano}, {Moretti}, {Cusumano},
  {Mangano}, {Mineo}, {La Parola}, {Giommi}, {Perri}, {Capalbi}, {Zhang},
  {Barthelmy}, {Cummings}, {Sakamoto}, {Burrows}, {Kennea}, {Nousek},
  {Osborne}, {O'Brien}, {Godet}, \& {Gehrels}}]{2006A&A...454..113C}
{Campana}, S., {Tagliaferri}, G., {Lazzati}, D., {et~al.} 2006{\natexlab{b}},
  \aap, 454, 113

\bibitem[{{Cano}(2013)}]{cano13}
{Cano}, Z. 2013, \mnras, 434, 1098

\bibitem[{{Cano} {et~al.}(2011){Cano}, {Bersier}, {Guidorzi}, {Margutti},
  {Svensson}, {Kobayashi}, {Melandri}, {Wiersema}, {Pozanenko}, {van der
  Horst}, {Pooley}, {Fernandez-Soto}, {Castro-Tirado}, {Postigo}, {Im},
  {Kamble}, {Sahu}, {Alonso-Lorite}, {Anupama}, {Bibby}, {Burgdorf}, {Clay},
  {Curran}, {Fatkhullin}, {Fruchter}, {Garnavich}, {Gomboc}, {Gorosabel},
  {Graham}, {Gurugubelli}, {Haislip}, {Huang}, {Huxor}, {Ibrahimov}, {Jeon},
  {Jeon}, {Ivarsen}, {Kasen}, {Klunko}, {Kouveliotou}, {Lacluyze}, {Levan},
  {Loznikov}, {Mazzali}, {Moskvitin}, {Mottram}, {Mundell}, {Nugent},
  {Nysewander}, {O'Brien}, {Park}, {Peris}, {Pian}, {Reichart}, {Rhoads},
  {Rol}, {Rumyantsev}, {Scowcroft}, {Shakhovskoy}, {Small}, {Smith}, {Sokolov},
  {Starling}, {Steele}, {Strom}, {Tanvir}, {Tsapras}, {Urata}, {Vaduvescu},
  {Volnova}, {Volvach}, {Wijers}, {Woosley}, \& {Young}}]{2011MNRAS.413..669C}
{Cano}, Z., {Bersier}, D., {Guidorzi}, C., {et~al.} 2011, \mnras, 413, 669

\bibitem[{{Carilli} {et~al.}(2008){Carilli}, {Lee}, {Capak}, {Schinnerer},
  {Lee}, {McCraken}, {Yun}, {Scoville}, {Smol{\v c}i{\'c}}, {Giavalisco},
  {Datta}, {Taniguchi}, \& {Urry}}]{2008ApJ...689..883C}
{Carilli}, C.~L., {Lee}, N., {Capak}, P., {et~al.} 2008, \apj, 689, 883

\bibitem[{{Castellano} {et~al.}(2014){Castellano}, {Sommariva}, {Fontana},
  {Pentericci}, {Santini}, {Grazian}, {Amorin}, {Donley}, {Dunlop}, {Ferguson},
  {Fiore}, {Galametz}, {Giallongo}, {Guo}, {Huang}, {Koekemoer}, {Maiolino},
  {McLure}, {Paris}, {Schaerer}, {Troncoso}, \&
  {Vanzella}}]{2014A&A...566A..19C}
{Castellano}, M., {Sommariva}, V., {Fontana}, A., {et~al.} 2014, \aap, 566, A19

\bibitem[{{Castro Cer{\'o}n} {et~al.}(2010){Castro Cer{\'o}n},
  {Micha{\l}owski}, {Hjorth}, {Malesani}, {Gorosabel}, {Watson}, {Fynbo}, \&
  {Morales Calder{\'o}n}}]{ceron10}
{Castro Cer{\'o}n}, J.~M., {Micha{\l}owski}, M.~J., {Hjorth}, J., {et~al.}
  2010, \apj, 721, 1919

\bibitem[{{Castro-Tirado} {et~al.}(2007){Castro-Tirado}, {Bremer}, {McBreen},
  {Gorosabel}, {Guziy}, {Fakthullin}, {Sokolov}, {Gonz{\'a}lez Delgado},
  {Bihain}, {Pandey}, {Jel{\'{\i}}nek}, {de Ugarte Postigo}, {Misra}, {Sagar},
  {Bama}, {Kamble}, {Anupama}, {Licandro}, {P{\'e}rez-Ram{\'{\i}}rez},
  {Bhattacharya}, {Aceituno}, \& {Neri}}]{2007A&A...475..101C}
{Castro-Tirado}, A.~J., {Bremer}, M., {McBreen}, S., {et~al.} 2007, \aap, 475,
  101

\bibitem[{{Castro-Tirado} \& {Gorosabel}(1999)}]{1999A&AS..138..449C}
{Castro-Tirado}, A.~J. \& {Gorosabel}, J. 1999, \aaps, 138, 449

\bibitem[{{Cenarro} {et~al.}(2003){Cenarro}, {Gorgas}, {Vazdekis}, {Cardiel},
  \& {Peletier}}]{2003MNRAS.339L..12C}
{Cenarro}, A.~J., {Gorgas}, J., {Vazdekis}, A., {Cardiel}, N., \& {Peletier},
  R.~F. 2003, \mnras, 339, L12

\bibitem[{{Cenko} {et~al.}(2011){Cenko}, {Hora}, \&
  {Bloom}}]{2011GCN..11638...1C}
{Cenko}, S.~B., {Hora}, J.~L., \& {Bloom}, J.~S. 2011, GRB Coordinates Network,
  11638, 1

\bibitem[{{Cenko} {et~al.}(2013){Cenko}, {Kulkarni}, {Horesh}, {Corsi}, {Fox},
  {Carpenter}, {Frail}, {Nugent}, {Perley}, {Gruber}, {Gal-Yam}, {Groot},
  {Hallinan}, {Ofek}, {Rau}, {MacLeod}, {Miller}, {Bloom}, {Filippenko},
  {Kasliwal}, {Law}, {Morgan}, {Polishook}, {Poznanski}, {Quimby}, {Sesar},
  {Shen}, {Silverman}, \& {Sternberg}}]{2013ApJ...769..130C}
{Cenko}, S.~B., {Kulkarni}, S.~R., {Horesh}, A., {et~al.} 2013, \apj, 769, 130

\bibitem[{{Cenko} {et~al.}(2009){Cenko}, {Perley}, {Junkkarinen}, {Burbidge},
  {Diego}, \& {Miller}}]{2009GCN..9518....1C}
{Cenko}, S.~B., {Perley}, D.~A., {Junkkarinen}, V., {et~al.} 2009, GRB
  Coordinates Network, 9518, 1

\bibitem[{{Cenko} {et~al.}(2015){Cenko}, {Urban}, {Perley}, {Horesh}, {Corsi},
  {Fox}, {Cao}, {Kasliwal}, {Lien}, {Arcavi}, {Bloom}, {Butler}, {Cucchiara},
  {de Diego}, {Filippenko}, {Gal-Yam}, {Gehrels}, {Georgiev}, {Jes{\'u}s
  Gonz{\'a}lez}, {Graham}, {Greiner}, {Kann}, {Klein}, {Knust}, {Kulkarni},
  {Kutyrev}, {Laher}, {Lee}, {Nugent}, {Prochaska}, {Ramirez-Ruiz}, {Richer},
  {Rubin}, {Urata}, {Varela}, {Watson}, \& {Wozniak}}]{2015ApJ...803L..24C}
{Cenko}, S.~B., {Urban}, A.~L., {Perley}, D.~A., {et~al.} 2015, \apjl, 803, L24

\bibitem[{{Cepa} {et~al.}(2000){Cepa}, {Aguiar}, {Escalera},
  {Gonzalez-Serrano}, {Joven-Alvarez}, {Peraza}, {Rasilla}, {Rodriguez-Ramos},
  {Gonzalez}, {Cobos Duenas}, {Sanchez}, {Tejada}, {Bland-Hawthorn},
  {Militello}, \& {Rosa}}]{cepa00}
{Cepa}, J., {Aguiar}, M., {Escalera}, V.~G., {et~al.} 2000, in Society of
  Photo-Optical Instrumentation Engineers (SPIE) Conference Series, Vol. 4008,
  Optical and IR Telescope Instrumentation and Detectors, ed. M.~{Iye} \& A.~F.
  {Moorwood}, 623--631

\bibitem[{{Chabrier}(2003)}]{chabrier03}
{Chabrier}, G. 2003, \pasp, 115, 763

\bibitem[{{Chandler} \& {Sargent}(1997)}]{1997ASPC..122...25C}
{Chandler}, C.~J. \& {Sargent}, A.~I. 1997, in Astronomical Society of the
  Pacific Conference Series, Vol. 122, From Stardust to Planetesimals, ed.
  Y.~J. {Pendleton}, 25

\bibitem[{{Chandra} \& {Frail}(2012)}]{2012ApJ...746..156C}
{Chandra}, P. \& {Frail}, D.~A. 2012, \apj, 746, 156

\bibitem[{{Chen}(2012)}]{hw}
{Chen}, H.-W. 2012, \mnras, 419, 3039

\bibitem[{{Cherchneff}(2010)}]{kinetic}
{Cherchneff}, I. 2010, in Astronomical Society of the Pacific Conference
  Series, Vol. 425, Astronomical Society of the Pacific Conference Series, ed.
  C.~{Leitherer}, P.~{Bennett}, P.~{Morris}, \& J.~{van Loon}, 237

\bibitem[{{Chincarini} {et~al.}(2010){Chincarini}, {Mao}, {Margutti},
  {Bernardini}, {Guidorzi}, {Pasotti}, {Giannios}, {Della Valle}, {Moretti},
  {Romano}, {D'Avanzo}, {Cusumano}, \& {Giommi}}]{2010MNRAS.406.2113C}
{Chincarini}, G., {Mao}, J., {Margutti}, R., {et~al.} 2010, \mnras, 406, 2113

\bibitem[{{Chisari} {et~al.}(2010){Chisari}, {Tissera}, \&
  {Pellizza}}]{2010MNRAS.408..647C}
{Chisari}, N.~E., {Tissera}, P.~B., \& {Pellizza}, L.~J. 2010, \mnras, 408, 647

\bibitem[{{Chornock} {et~al.}(2010){Chornock}, {Berger}, {Levesque},
  {Soderberg}, {Foley}, {Fox}, {Frebel}, {Simon}, {Bochanski}, {Challis},
  {Kirshner}, {Podsiadlowski}, {Roth}, {Rutledge}, {Schmidt}, {Sheppard}, \&
  {Simcoe}}]{2010arXiv1004.2262C}
{Chornock}, R., {Berger}, E., {Levesque}, E.~M., {et~al.} 2010, ArXiv e-prints

\bibitem[{{Chornock} {et~al.}(2009){Chornock}, {Perley}, {Cenko}, \&
  {Bloom}}]{2009GCN..9243....1C}
{Chornock}, R., {Perley}, D.~A., {Cenko}, S.~B., \& {Bloom}, J.~S. 2009, GRB
  Coordinates Network, 9243, 1

\bibitem[{{Christensen} {et~al.}(2012){Christensen}, {Laursen}, {Richard},
  {Hjorth}, {Milvang-Jensen}, {Dessauges-Zavadsky}, {Limousin}, {Grillo}, \&
  {Ebeling}}]{christensen12}
{Christensen}, L., {Laursen}, P., {Richard}, J., {et~al.} 2012, \mnras, 427,
  1973

\bibitem[{{Christensen} {et~al.}(2014){Christensen}, {M{\o}ller}, {Fynbo}, \&
  {Zafar}}]{christensen14}
{Christensen}, L., {M{\o}ller}, P., {Fynbo}, J.~P.~U., \& {Zafar}, T. 2014,
  \mnras, 445, 225

\bibitem[{{Coburn} \& {Boggs}(2003)}]{2003Natur.423..415C}
{Coburn}, W. \& {Boggs}, S.~E. 2003, \nat, 423, 415

\bibitem[{{Cohen} {et~al.}(1997){Cohen}, {Katz}, {Piran}, {Sari}, {Preece}, \&
  {Band}}]{1997ApJ...488..330C}
{Cohen}, E., {Katz}, J.~I., {Piran}, T., {et~al.} 1997, \apj, 488, 330

\bibitem[{{Compi{\`e}gne} {et~al.}(2011){Compi{\`e}gne}, {Verstraete}, {Jones},
  {Bernard}, {Boulanger}, {Flagey}, {Le Bourlot}, {Paradis}, \&
  {Ysard}}]{2011A&A...525A.103C}
{Compi{\`e}gne}, M., {Verstraete}, L., {Jones}, A., {et~al.} 2011, \aap, 525,
  A103

\bibitem[{{Conroy} {et~al.}(2009){Conroy}, {Gunn}, \&
  {White}}]{2009ApJ...699..486C}
{Conroy}, C., {Gunn}, J.~E., \& {White}, M. 2009, \apj, 699, 486

\bibitem[{{Cordier} {et~al.}(2014){Cordier}, {Wei}, {Claret}, {Han}, {Klotz},
  {Lachaud}, {Wu}, {Wang}, {Xin}, {Atteia}, {Basa}, {Deng}, {Dong}, {Godet},
  {Goldwurm}, {Gotz}, {Osborne}, {Qiu}, {Schanne}, {Wu}, {Zhang}, \&
  {Zhang}}]{2014styd.confE...5C}
{Cordier}, B., {Wei}, J., {Claret}, A., {et~al.} 2014, in Proceedings of Swift:
  10 Years of Discovery (SWIFT 10), held 2-5 December 2014 at La Sapienza
  University, Rome, Italy., 5

\bibitem[{{Costa} {et~al.}(1997){Costa}, {Frontera}, {Heise}, {Feroci}, {in't
  Zand}, {Fiore}, {Cinti}, {Dal Fiume}, {Nicastro}, {Orlandini}, {Palazzi},
  {Rapisarda\#}, {Zavattini}, {Jager}, {Parmar}, {Owens}, {Molendi},
  {Cusumano}, {Maccarone}, {Giarrusso}, {Coletta}, {Antonelli}, {Giommi},
  {Muller}, {Piro}, \& {Butler}}]{1997Natur.387..783C}
{Costa}, E., {Frontera}, F., {Heise}, J., {et~al.} 1997, \nat, 387, 783

\bibitem[{{Covino} {et~al.}(2013){Covino}, {Melandri}, {Salvaterra}, {Campana},
  {Vergani}, {Bernardini}, {D'Avanzo}, {D'Elia}, {Fugazza}, {Ghirlanda},
  {Ghisellini}, {Gomboc}, {Jin}, {Kr{\"u}hler}, {Malesani}, {Nava},
  {Sbarufatti}, \& {Tagliaferri}}]{2013MNRAS.432.1231C}
{Covino}, S., {Melandri}, A., {Salvaterra}, R., {et~al.} 2013, \mnras, 432,
  1231

\bibitem[{{Crider} {et~al.}(1998){Crider}, {Liang}, \&
  {Preece}}]{1998AIPC..428..359C}
{Crider}, A., {Liang}, E.~P., \& {Preece}, R.~D. 1998, in American Institute of
  Physics Conference Series, Vol. 428, Gamma-Ray Bursts, 4th Hunstville
  Symposium, ed. C.~A. {Meegan}, R.~D. {Preece}, \& T.~M. {Koshut}, 359--363

\bibitem[{{Crowther}(2007)}]{2007ARA&A..45..177C}
{Crowther}, P.~A. 2007, \araa, 45, 177

\bibitem[{{Cucchiara} {et~al.}(2015{\natexlab{a}}){Cucchiara}, {Fumagalli},
  {Rafelski}, {Kocevski}, {Prochaska}, {Cooke}, \& {Becker}}]{cucchiara14}
{Cucchiara}, A., {Fumagalli}, M., {Rafelski}, M., {et~al.} 2015{\natexlab{a}},
  \apj, 804, 51

\bibitem[{{Cucchiara} {et~al.}(2015{\natexlab{b}}){Cucchiara}, {Veres},
  {Corsi}, {Cenko}, {Perley}, {Marshall}, {Pagani}, {Toy}, {Capone}, {Frail},
  {Horesh}, {Modjaz}, {Butler}, {Littlejohns}, {Watson}, {Kutyrev}, {Lee},
  {Richer}, {Klein}, {Fox}, {Prochaska}, {Bloom}, {Troja}, {Ramirez-Ruiz}, {de
  Diego}, {Georgiev}, {Gonzalez}, {Roman-Zuniga}, {Gehrels}, \&
  {Moseley}}]{2015arXiv151000996C}
{Cucchiara}, A., {Veres}, P., {Corsi}, A., {et~al.} 2015{\natexlab{b}}, ArXiv
  e-prints

\bibitem[{{Daigne} {et~al.}(2011){Daigne}, {Bo{\v s}njak}, \&
  {Dubus}}]{2011A&A...526A.110D}
{Daigne}, F., {Bo{\v s}njak}, {\v Z}., \& {Dubus}, G. 2011, \aap, 526, A110

\bibitem[{{Dale} {et~al.}(2007){Dale}, {Gil de Paz}, {Gordon}, {Hanson},
  {Armus}, {Bendo}, {Bianchi}, {Block}, {Boissier}, {Boselli}, {Buckalew},
  {Buat}, {Burgarella}, {Calzetti}, {Cannon}, {Engelbracht}, {Helou},
  {Hollenbach}, {Jarrett}, {Kennicutt}, {Leitherer}, {Li}, {Madore}, {Martin},
  {Meyer}, {Murphy}, {Regan}, {Roussel}, {Smith}, {Sosey}, {Thilker}, \&
  {Walter}}]{2007ApJ...655..863D}
{Dale}, D.~A., {Gil de Paz}, A., {Gordon}, K.~D., {et~al.} 2007, \apj, 655, 863

\bibitem[{{D'Avanzo} {et~al.}(2008){D'Avanzo}, {D'Elia}, \&
  {Covino}}]{2008GCN..8350....1D}
{D'Avanzo}, P., {D'Elia}, V., \& {Covino}, S. 2008, GRB Coordinates Network,
  8350, 1

\bibitem[{{De Cia} {et~al.}(2011){De Cia}, {Jakobsson}, {Bj{\"o}rnsson},
  {Vreeswijk}, {Dhillon}, {Marsh}, {Chapman}, {Fynbo}, {Ledoux}, {Littlefair},
  {Malesani}, {Schulze}, {Smette}, {Zafar}, \& {Gudmundsson}}]{annalisa11}
{De Cia}, A., {Jakobsson}, P., {Bj{\"o}rnsson}, G., {et~al.} 2011, \mnras, 412,
  2229

\bibitem[{{De Cia} {et~al.}(2012){De Cia}, {Ledoux}, {Fox}, {Vreeswijk},
  {Smette}, {Petitjean}, {Bj{\"o}rnsson}, {Fynbo}, {Hjorth}, \&
  {Jakobsson}}]{2012A&A...545A..64D}
{De Cia}, A., {Ledoux}, C., {Fox}, A.~J., {et~al.} 2012, \aap, 545, A64

\bibitem[{{De Cia} {et~al.}(2013){De Cia}, {Ledoux}, {Savaglio}, {Schady}, \&
  {Vreeswijk}}]{2013A&A...560A..88D}
{De Cia}, A., {Ledoux}, C., {Savaglio}, S., {Schady}, P., \& {Vreeswijk}, P.~M.
  2013, \aap, 560, A88

\bibitem[{{De Marchi} \& {Panagia}(2014)}]{2014MNRAS.445...93D}
{De Marchi}, G. \& {Panagia}, N. 2014, \mnras, 445, 93

\bibitem[{{de Ugarte Postigo} {et~al.}(2009){de Ugarte Postigo}, {Gorosabel},
  {Malesani}, {Fynbo}, \& {Levan}}]{2009GCN..9383....1D}
{de Ugarte Postigo}, A., {Gorosabel}, J., {Malesani}, D., {Fynbo}, J.~P.~U., \&
  {Levan}, A.~J. 2009, GRB Coordinates Network, 9383, 1

\bibitem[{{de Ugarte Postigo} {et~al.}(2011){de Ugarte Postigo}, {Horv{\'a}th},
  {Veres}, {Bagoly}, {Kann}, {Th{\"o}ne}, {Balazs}, {D'Avanzo}, {Aloy},
  {Foley}, {Campana}, {Mao}, {Jakobsson}, {Covino}, {Fynbo}, {Gorosabel},
  {Castro-Tirado}, {Amati}, \& {Nardini}}]{2011A&A...525A.109D}
{de Ugarte Postigo}, A., {Horv{\'a}th}, I., {Veres}, P., {et~al.} 2011, \aap,
  525, A109

\bibitem[{{de Ugarte Postigo} {et~al.}(2012){de Ugarte Postigo}, {Lundgren},
  {Mart{\'{\i}}n}, {Garcia-Appadoo}, {de Gregorio Monsalvo}, {Peck},
  {Micha{\l}owski}, {Th{\"o}ne}, {Campana}, {Gorosabel}, {Tanvir}, {Wiersema},
  {Castro-Tirado}, {Schulze}, {De Breuck}, {Petitpas}, {Hjorth}, {Jakobsson},
  {Covino}, {Fynbo}, {Winters}, {Bremer}, {Levan}, {Llorente},
  {S{\'a}nchez-Ram{\'{\i}}rez}, {Tello}, \& {Salvaterra}}]{2012A&A...538A..44D}
{de Ugarte Postigo}, A., {Lundgren}, A., {Mart{\'{\i}}n}, S., {et~al.} 2012,
  \aap, 538, A44

\bibitem[{{D'Elia} {et~al.}(2014){D'Elia}, {Fynbo}, {Goldoni}, {Covino}, {de
  Ugarte Postigo}, {Ledoux}, {Calura}, {Gorosabel}, {Malesani}, {Matteucci},
  {S{\'a}nchez-Ram{\'{\i}}rez}, {Savaglio}, {Castro-Tirado}, {Hartoog},
  {Kaper}, {Mu{\~n}oz-Darias}, {Pian}, {Piranomonte}, {Tagliaferri}, {Tanvir},
  {Vergani}, {Watson}, \& {Xu}}]{delia14}
{D'Elia}, V., {Fynbo}, J.~P.~U., {Goldoni}, P., {et~al.} 2014, \aap, 564, A38

\bibitem[{{Della Valle} {et~al.}(2006){Della Valle}, {Chincarini}, {Panagia},
  {Tagliaferri}, {Malesani}, {Testa}, {Fugazza}, {Campana}, {Covino},
  {Mangano}, {Antonelli}, {D'Avanzo}, {Hurley}, {Mirabel}, {Pellizza},
  {Piranomonte}, \& {Stella}}]{2006Natur.444.1050D}
{Della Valle}, M., {Chincarini}, G., {Panagia}, N., {et~al.} 2006, \nat, 444,
  1050

\bibitem[{{Della Valle} {et~al.}(2003){Della Valle}, {Malesani}, {Benetti},
  {Testa}, {Hamuy}, {Antonelli}, {Chincarini}, {Cocozza}, {Covino}, {D'Avanzo},
  {Fugazza}, {Ghisellini}, {Gilmozzi}, {Lazzati}, {Mason}, {Mazzali}, \&
  {Stella}}]{2003A&A...406L..33D}
{Della Valle}, M., {Malesani}, D., {Benetti}, S., {et~al.} 2003, \aap, 406, L33

\bibitem[{{Dessart} {et~al.}(2012){Dessart}, {Hillier}, {Waldman}, {Livne}, \&
  {Blondin}}]{2012MNRAS.426L..76D}
{Dessart}, L., {Hillier}, D.~J., {Waldman}, R., {Livne}, E., \& {Blondin}, S.
  2012, \mnras, 426, L76

\bibitem[{{Dessauges-Zavadsky} {et~al.}(2006){Dessauges-Zavadsky}, {Prochaska},
  {D'Odorico}, {Calura}, \& {Matteucci}}]{2006A&A...445...93D}
{Dessauges-Zavadsky}, M., {Prochaska}, J.~X., {D'Odorico}, S., {Calura}, F., \&
  {Matteucci}, F. 2006, \aap, 445, 93

\bibitem[{{Donn} \& {Nuth}(1985)}]{1985ApJ...288..187D}
{Donn}, B. \& {Nuth}, J.~A. 1985, \apj, 288, 187

\bibitem[{{Draine}(1989)}]{1989IAUS..135..313D}
{Draine}, B. 1989, in IAU Symposium, Vol. 135, Interstellar Dust, ed. L.~J.
  {Allamandola} \& A.~G.~G.~M. {Tielens}, 313

\bibitem[{{Draine}(2000)}]{draine00}
{Draine}, B.~T. 2000, \apj, 532, 273

\bibitem[{{Draine}(2003{\natexlab{a}})}]{2003ARA&A..41..241D}
{Draine}, B.~T. 2003{\natexlab{a}}, \araa, 41, 241

\bibitem[{{Draine}(2003{\natexlab{b}})}]{2003ApJ...598.1017D}
{Draine}, B.~T. 2003{\natexlab{b}}, \apj, 598, 1017

\bibitem[{{Draine}(2011)}]{2011piim.book.....D}
{Draine}, B.~T. 2011, {Physics of the Interstellar and Intergalactic Medium}

\bibitem[{{Draine} \& {Flatau}(2010)}]{2010arXiv1002.1505D}
{Draine}, B.~T. \& {Flatau}, P.~J. 2010, ArXiv e-prints

\bibitem[{{Draine} \& {Hao}(2002)}]{2002ApJ...569..780D}
{Draine}, B.~T. \& {Hao}, L. 2002, \apj, 569, 780

\bibitem[{{Draine} \& {Malhotra}(1993)}]{1993ApJ...414..632D}
{Draine}, B.~T. \& {Malhotra}, S. 1993, \apj, 414, 632

\bibitem[{{El{\'{\i}}asd{\'o}ttir} {et~al.}(2009){El{\'{\i}}asd{\'o}ttir},
  {Fynbo}, {Hjorth}, {Ledoux}, {Watson}, {Andersen}, {Malesani}, {Vreeswijk},
  {Prochaska}, {Sollerman}, \& {Jaunsen}}]{2009ApJ...697.1725E}
{El{\'{\i}}asd{\'o}ttir}, {\'A}., {Fynbo}, J.~P.~U., {Hjorth}, J., {et~al.}
  2009, \apj, 697, 1725

\bibitem[{{El{\'{\i}}asd{\'o}ttir} {et~al.}(2006){El{\'{\i}}asd{\'o}ttir},
  {Hjorth}, {Toft}, {Burud}, \& {Paraficz}}]{2006ApJS..166..443E}
{El{\'{\i}}asd{\'o}ttir}, {\'A}., {Hjorth}, J., {Toft}, S., {Burud}, I., \&
  {Paraficz}, D. 2006, \apjs, 166, 443

\bibitem[{{Erickson}(1957)}]{1957ApJ...126..480E}
{Erickson}, W.~C. 1957, \apj, 126, 480

\bibitem[{{Esteban} {et~al.}(2004){Esteban}, {Peimbert}, {Garc{\'{\i}}a-Rojas},
  {Ruiz}, {Peimbert}, \& {Rodr{\'{\i}}guez}}]{esteban04}
{Esteban}, C., {Peimbert}, M., {Garc{\'{\i}}a-Rojas}, J., {et~al.} 2004,
  \mnras, 355, 229

\bibitem[{{Evans} {et~al.}(2007){Evans}, {Beardmore}, {Page}, {Tyler},
  {Osborne}, {Goad}, {O'Brien}, {Vetere}, {Racusin}, {Morris}, {Burrows},
  {Capalbi}, {Perri}, {Gehrels}, \& {Romano}}]{2007A&A...469..379E}
{Evans}, P.~A., {Beardmore}, A.~P., {Page}, K.~L., {et~al.} 2007, \aap, 469,
  379

\bibitem[{{Evans} {et~al.}(2014){Evans}, {Willingale}, {Osborne}, {O'Brien},
  {Tanvir}, {Frederiks}, {Pal'shin}, {Svinkin}, {Lien}, {Cummings}, {Xiong},
  {Zhang}, {G{\"o}tz}, {Savchenko}, {Negoro}, {Nakahira}, {Suzuki}, {Wiersema},
  {Starling}, {Castro-Tirado}, {Beardmore}, {S{\'a}nchez-Ram{\'{\i}}rez},
  {Gorosabel}, {Jeong}, {Kennea}, {Burrows}, \&
  {Gehrels}}]{2014MNRAS.444..250E}
{Evans}, P.~A., {Willingale}, R., {Osborne}, J.~P., {et~al.} 2014, \mnras, 444,
  250

\bibitem[{{Fan} {et~al.}(2006){Fan}, {Piran}, \& {Xu}}]{2006JCAP...09..013F}
{Fan}, Y.-Z., {Piran}, T., \& {Xu}, D. 2006, \jcap, 9, 13

\bibitem[{{Fan} {et~al.}(2013){Fan}, {Tam}, {Zhang}, {Liang}, {He}, {Zhou},
  {Yang}, {Jin}, \& {Wei}}]{2013ApJ...776...95F}
{Fan}, Y.-Z., {Tam}, P.~H.~T., {Zhang}, F.-W., {et~al.} 2013, \apj, 776, 95

\bibitem[{{Ferrara} \& {Dettmar}(1994)}]{1994ApJ...427..155F}
{Ferrara}, A. \& {Dettmar}, R.-J. 1994, \apj, 427, 155

\bibitem[{{Ferrara} {et~al.}(1991){Ferrara}, {Ferrini}, {Barsella}, \&
  {Franco}}]{1991ApJ...381..137F}
{Ferrara}, A., {Ferrini}, F., {Barsella}, B., \& {Franco}, J. 1991, \apj, 381,
  137

\bibitem[{{Filgas} {et~al.}(2012){Filgas}, {Greiner}, {Schady}, {de Ugarte
  Postigo}, {Oates}, {Nardini}, {Kr{\"u}hler}, {Panaitescu}, {Kann}, {Klose},
  {Afonso}, {Allen}, {Castro-Tirado}, {Christie}, {Dong}, {Elliott}, {Natusch},
  {Nicuesa Guelbenzu}, {Olivares E.}, {Rau}, {Rossi}, {Sudilovsky}, \&
  {Yock}}]{2012A&A...546A.101F}
{Filgas}, R., {Greiner}, J., {Schady}, P., {et~al.} 2012, \aap, 546, A101

\bibitem[{{Finkbeiner}(2004)}]{2004ApJ...614..186F}
{Finkbeiner}, D.~P. 2004, \apj, 614, 186

\bibitem[{{Fitzpatrick} \& {Massa}(1986)}]{1986ApJ...307..286F}
{Fitzpatrick}, E.~L. \& {Massa}, D. 1986, \apj, 307, 286

\bibitem[{{Fitzpatrick} \& {Massa}(1990)}]{1990ApJS...72..163F}
{Fitzpatrick}, E.~L. \& {Massa}, D. 1990, \apjs, 72, 163

\bibitem[{{Fitzpatrick} \& {Massa}(2005)}]{FM}
{Fitzpatrick}, E.~L. \& {Massa}, D. 2005, \aj, 130, 1127

\bibitem[{{Fitzpatrick} \& {Massa}(2007)}]{2007ApJ...663..320F}
{Fitzpatrick}, E.~L. \& {Massa}, D. 2007, \apj, 663, 320

\bibitem[{{Foley} {et~al.}(2006){Foley}, {Watson}, {Gorosabel}, {Fynbo},
  {Sollerman}, {McGlynn}, {McBreen}, \& {Hjorth}}]{2006A&A...447..891F}
{Foley}, S., {Watson}, D., {Gorosabel}, J., {et~al.} 2006, \aap, 447, 891

\bibitem[{{F{\"o}rster Schreiber} {et~al.}(2009){F{\"o}rster Schreiber},
  {Genzel}, {Bouch{\'e}}, {Cresci}, {Davies}, {Buschkamp}, {Shapiro},
  {Tacconi}, {Hicks}, {Genel}, {Shapley}, {Erb}, {Steidel}, {Lutz},
  {Eisenhauer}, {Gillessen}, {Sternberg}, {Renzini}, {Cimatti}, {Daddi},
  {Kurk}, {Lilly}, {Kong}, {Lehnert}, {Nesvadba}, {Verma}, {McCracken},
  {Arimoto}, {Mignoli}, \& {Onodera}}]{sins}
{F{\"o}rster Schreiber}, N.~M., {Genzel}, R., {Bouch{\'e}}, N., {et~al.} 2009,
  \apj, 706, 1364

\bibitem[{{Fox} {et~al.}(2008){Fox}, {Ledoux}, {Vreeswijk}, {Smette}, \&
  {Jaunsen}}]{2008A&A...491..189F}
{Fox}, A.~J., {Ledoux}, C., {Vreeswijk}, P.~M., {Smette}, A., \& {Jaunsen},
  A.~O. 2008, \aap, 491, 189

\bibitem[{{Fox} {et~al.}(2005){Fox}, {Frail}, {Price}, {Kulkarni}, {Berger},
  {Piran}, {Soderberg}, {Cenko}, {Cameron}, {Gal-Yam}, {Kasliwal}, {Moon},
  {Harrison}, {Nakar}, {Schmidt}, {Penprase}, {Chevalier}, {Kumar}, {Roth},
  {Watson}, {Lee}, {Shectman}, {Phillips}, {Roth}, {McCarthy}, {Rauch},
  {Cowie}, {Peterson}, {Rich}, {Kawai}, {Aoki}, {Kosugi}, {Totani}, {Park},
  {MacFadyen}, \& {Hurley}}]{2005Natur.437..845F}
{Fox}, D.~B., {Frail}, D.~A., {Price}, P.~A., {et~al.} 2005, \nat, 437, 845

\bibitem[{{Frail} {et~al.}(1997){Frail}, {Kulkarni}, {Nicastro}, {Feroci}, \&
  {Taylor}}]{1997Natur.389..261F}
{Frail}, D.~A., {Kulkarni}, S.~R., {Nicastro}, L., {Feroci}, M., \& {Taylor},
  G.~B. 1997, \nat, 389, 261

\bibitem[{{Friis} {et~al.}(2015){Friis}, {De Cia}, {Kr{\"u}hler}, {Fynbo},
  {Ledoux}, {Vreeswijk}, {Watson}, {Malesani}, {Gorosabel}, {Starling},
  {Jakobsson}, {Varela}, {Wiersema}, {Drachmann}, {Trotter}, {Th{\"o}ne}, {de
  Ugarte Postigo}, {D'Elia}, {Elliott}, {Maturi}, {Goldoni}, {Greiner},
  {Haislip}, {Kaper}, {Knust}, {LaCluyze}, {Milvang-Jensen}, {Reichart},
  {Schulze}, {Sudilovsky}, {Tanvir}, \& {Vergani}}]{2015MNRAS.451.4686F}
{Friis}, M., {De Cia}, A., {Kr{\"u}hler}, T., {et~al.} 2015, \mnras, 451, 167

\bibitem[{{Fruchter} {et~al.}(2006){Fruchter}, {Levan}, {Strolger},
  {Vreeswijk}, {Thorsett}, {Bersier}, {Burud}, {Castro Cer{\'o}n},
  {Castro-Tirado}, {Conselice}, {Dahlen}, {Ferguson}, {Fynbo}, {Garnavich},
  {Gibbons}, {Gorosabel}, {Gull}, {Hjorth}, {Holland}, {Kouveliotou}, {Levay},
  {Livio}, {Metzger}, {Nugent}, {Petro}, {Pian}, {Rhoads}, {Riess}, {Sahu},
  {Smette}, {Tanvir}, {Wijers}, \& {Woosley}}]{2006Natur.441..463F}
{Fruchter}, A.~S., {Levan}, A.~J., {Strolger}, L., {et~al.} 2006, \nat, 441,
  463

\bibitem[{{Fryer} {et~al.}(2006){Fryer}, {Rockefeller}, \&
  {Young}}]{2006ApJ...647.1269F}
{Fryer}, C.~L., {Rockefeller}, G., \& {Young}, P.~A. 2006, \apj, 647, 1269

\bibitem[{{Fryer} \& {Woosley}(1998)}]{1998ApJ...502L...9F}
{Fryer}, C.~L. \& {Woosley}, S.~E. 1998, \apjl, 502, L9

\bibitem[{{Fynbo} {et~al.}(2013){Fynbo}, {Geier}, {Christensen}, {Gallazzi},
  {Krogager}, {Kr{\"u}hler}, {Ledoux}, {Maund}, {M{\o}ller}, {Noterdaeme},
  {Rivera-Thorsen}, \& {Vestergaard}}]{fynbo13}
{Fynbo}, J.~P.~U., {Geier}, S.~J., {Christensen}, L., {et~al.} 2013, \mnras,
  436, 361

\bibitem[{{Fynbo} {et~al.}(2009){Fynbo}, {Jakobsson}, {Prochaska}, {Malesani},
  {Ledoux}, {de Ugarte Postigo}, {Nardini}, {Vreeswijk}, {Wiersema}, {Hjorth},
  {Sollerman}, {Chen}, {Th{\"o}ne}, {Bj{\"o}rnsson}, {Bloom}, {Castro-Tirado},
  {Christensen}, {De Cia}, {Fruchter}, {Gorosabel}, {Graham}, {Jaunsen},
  {Jensen}, {Kann}, {Kouveliotou}, {Levan}, {Maund}, {Masetti},
  {Milvang-Jensen}, {Palazzi}, {Perley}, {Pian}, {Rol}, {Schady}, {Starling},
  {Tanvir}, {Watson}, {Xu}, {Augusteijn}, {Grundahl}, {Telting}, \&
  {Quirion}}]{2009ApJS..185..526F}
{Fynbo}, J.~P.~U., {Jakobsson}, P., {Prochaska}, J.~X., {et~al.} 2009, \apjs,
  185, 526

\bibitem[{{Fynbo} {et~al.}(2014){Fynbo}, {Kr{\"u}hler}, {Leighly}, {Ledoux},
  {Vreeswijk}, {Schulze}, {Noterdaeme}, {Watson}, {Wijers}, {Bolmer}, {Cano},
  {Christensen}, {Covino}, {D'Elia}, {Flores}, {Friis}, {Goldoni}, {Greiner},
  {Hammer}, {Hjorth}, {Jakobsson}, {Japelj}, {Kaper}, {Klose}, {Knust},
  {Leloudas}, {Levan}, {Malesani}, {Milvang-Jensen}, {M{\o}ller}, {Nicuesa
  Guelbenzu}, {Oates}, {Pian}, {Schady}, {Sparre}, {Tagliaferri}, {Tanvir},
  {Th{\"o}ne}, {de Ugarte Postigo}, {Vergani}, {Wiersema}, {Xu}, \&
  {Zafar}}]{2014A&A...572A..12F}
{Fynbo}, J.~P.~U., {Kr{\"u}hler}, T., {Leighly}, K., {et~al.} 2014, \aap, 572,
  A12

\bibitem[{{Fynbo} {et~al.}(2011{\natexlab{a}}){Fynbo}, {Ledoux}, {Noterdaeme},
  {Christensen}, {M{\o}ller}, {Durgapal}, {Goldoni}, {Kaper}, {Krogager},
  {Laursen}, {Maund}, {Milvang-Jensen}, {Okoshi}, {Rasmussen}, {Thorsen},
  {Toft}, \& {Zafar}}]{fynbo11}
{Fynbo}, J.~P.~U., {Ledoux}, C., {Noterdaeme}, P., {et~al.} 2011{\natexlab{a}},
  \mnras, 413, 2481

\bibitem[{{Fynbo} {et~al.}(2011{\natexlab{b}}){Fynbo}, {Ledoux}, {Noterdaeme},
  {Christensen}, {M{\o}ller}, {Durgapal}, {Goldoni}, {Kaper}, {Krogager},
  {Laursen}, {Maund}, {Milvang-Jensen}, {Okoshi}, {Rasmussen}, {Thorsen},
  {Toft}, \& {Zafar}}]{2011MNRAS.413.2481F}
{Fynbo}, J.~P.~U., {Ledoux}, C., {Noterdaeme}, P., {et~al.} 2011{\natexlab{b}},
  \mnras, 413, 2481

\bibitem[{{Fynbo} {et~al.}(2002){Fynbo}, {M{\"o}ller}, {Thomsen}, {Hjorth},
  {Gorosabel}, {Andersen}, {Egholm}, {Holland}, {Jensen}, {Pedersen}, \&
  {Weidinger}}]{2002A&A...388..425F}
{Fynbo}, J.~P.~U., {M{\"o}ller}, P., {Thomsen}, B., {et~al.} 2002, \aap, 388,
  425

\bibitem[{{Fynbo} {et~al.}(2004){Fynbo}, {Sollerman}, {Hjorth}, {Grundahl},
  {Gorosabel}, {Weidinger}, {M{\o}ller}, {Jensen}, {Vreeswijk}, {Fransson},
  {Ramirez-Ruiz}, {Jakobsson}, {J{\o}rgensen}, {Vinter}, {Andersen}, {Castro
  Cer{\'o}n}, {Castro-Tirado}, {Fruchter}, {Greiner}, {Kouveliotou}, {Levan},
  {Klose}, {Masetti}, {Pedersen}, {Palazzi}, {Pian}, {Rhoads}, {Rol},
  {Sekiguchi}, {Tanvir}, {Tristram}, {de Ugarte Postigo}, {Wijers}, \& {van den
  Heuvel}}]{2004ApJ...609..962F}
{Fynbo}, J.~P.~U., {Sollerman}, J., {Hjorth}, J., {et~al.} 2004, \apj, 609, 962

\bibitem[{{Fynbo} {et~al.}(2006){Fynbo}, {Watson}, {Th{\"o}ne}, {Sollerman},
  {Bloom}, {Davis}, {Hjorth}, {Jakobsson}, {J{\o}rgensen}, {Graham},
  {Fruchter}, {Bersier}, {Kewley}, {Cassan}, {Castro Cer{\'o}n}, {Foley},
  {Gorosabel}, {Hinse}, {Horne}, {Jensen}, {Klose}, {Kocevski}, {Marquette},
  {Perley}, {Ramirez-Ruiz}, {Stritzinger}, {Vreeswijk}, {Wijers}, {Woller},
  {Xu}, \& {Zub}}]{2006Natur.444.1047F}
{Fynbo}, J.~P.~U., {Watson}, D., {Th{\"o}ne}, C.~C., {et~al.} 2006, \nat, 444,
  1047

\bibitem[{{Fynbo} {et~al.}(2001){Fynbo}, {Gorosabel}, {Dall}, {Hjorth},
  {Pedersen}, {Andersen}, {M{\o}ller}, {Holland}, {Smail}, {Kobayashi}, {Rol},
  {Vreeswijk}, {Burud}, {Jensen}, {Thomsen}, {Henden}, {Vrba}, {Canzian},
  {Castro Cer{\'o}n}, {Castro-Tirado}, {Cline}, {Goto}, {Greiner}, {Hanski},
  {Hurley}, {Lund}, {Pursimo}, {{\O}stensen}, {Solheim}, {Tanvir}, \&
  {Terada}}]{2001A&A...373..796F}
{Fynbo}, J.~U., {Gorosabel}, J., {Dall}, T.~H., {et~al.} 2001, \aap, 373, 796

\bibitem[{{Galama} {et~al.}(1998){Galama}, {Vreeswijk}, {van Paradijs},
  {Kouveliotou}, {Augusteijn}, {B{\"o}hnhardt}, {Brewer}, {Doublier},
  {Gonzalez}, {Leibundgut}, {Lidman}, {Hainaut}, {Patat}, {Heise}, {in't Zand},
  {Hurley}, {Groot}, {Strom}, {Mazzali}, {Iwamoto}, {Nomoto}, {Umeda},
  {Nakamura}, {Young}, {Suzuki}, {Shigeyama}, {Koshut}, {Kippen}, {Robinson},
  {de Wildt}, {Wijers}, {Tanvir}, {Greiner}, {Pian}, {Palazzi}, {Frontera},
  {Masetti}, {Nicastro}, {Feroci}, {Costa}, {Piro}, {Peterson}, {Tinney},
  {Boyle}, {Cannon}, {Stathakis}, {Sadler}, {Begam}, \&
  {Ianna}}]{1998Natur.395..670G}
{Galama}, T.~J., {Vreeswijk}, P.~M., {van Paradijs}, J., {et~al.} 1998, \nat,
  395, 670

\bibitem[{{Gall} {et~al.}(2011){Gall}, {Hjorth}, \& {Andersen}}]{GHA}
{Gall}, C., {Hjorth}, J., \& {Andersen}, A.~C. 2011, \aapr, 19, 43

\bibitem[{{Gall} {et~al.}(2014){Gall}, {Hjorth}, {Watson}, {Dwek}, {Maund},
  {Fox}, {Leloudas}, {Malesani}, \& {Day-Jones}}]{2014Natur.511..326G}
{Gall}, C., {Hjorth}, J., {Watson}, D., {et~al.} 2014, \nat, 511, 326

\bibitem[{{Gao} {et~al.}(2009){Gao}, {Jiang}, \& {Li}}]{li08}
{Gao}, J., {Jiang}, B.~W., \& {Li}, A. 2009, \apj, 707, 89

\bibitem[{{Gehrels} {et~al.}(2004){Gehrels}, {Chincarini}, {Giommi}, {Mason},
  {Nousek}, {Wells}, {White}, {Barthelmy}, {Burrows}, {Cominsky}, {Hurley},
  {Marshall}, {M{\'e}sz{\'a}ros}, {Roming}, {Angelini}, {Barbier}, {Belloni},
  {Campana}, {Caraveo}, {Chester}, {Citterio}, {Cline}, {Cropper}, {Cummings},
  {Dean}, {Feigelson}, {Fenimore}, {Frail}, {Fruchter}, {Garmire}, {Gendreau},
  {Ghisellini}, {Greiner}, {Hill}, {Hunsberger}, {Krimm}, {Kulkarni}, {Kumar},
  {Lebrun}, {Lloyd-Ronning}, {Markwardt}, {Mattson}, {Mushotzky}, {Norris},
  {Osborne}, {Paczynski}, {Palmer}, {Park}, {Parsons}, {Paul}, {Rees},
  {Reynolds}, {Rhoads}, {Sasseen}, {Schaefer}, {Short}, {Smale}, {Smith},
  {Stella}, {Tagliaferri}, {Takahashi}, {Tashiro}, {Townsley}, {Tueller},
  {Turner}, {Vietri}, {Voges}, {Ward}, {Willingale}, {Zerbi}, \&
  {Zhang}}]{swift}
{Gehrels}, N., {Chincarini}, G., {Giommi}, P., {et~al.} 2004, \apj, 611, 1005

\bibitem[{{Gehrels} {et~al.}(1994){Gehrels}, {Chipman}, \&
  {Kniffen}}]{1994ApJS...92..351G}
{Gehrels}, N., {Chipman}, E., \& {Kniffen}, D. 1994, \apjs, 92, 351

\bibitem[{{Gehrels} \& {M{\'e}sz{\'a}ros}(2012)}]{2012Sci...337..932G}
{Gehrels}, N. \& {M{\'e}sz{\'a}ros}, P. 2012, Science, 337, 932

\bibitem[{{Gehrels} {et~al.}(2002){Gehrels}, {Piro}, \&
  {Leonard}}]{2002SciAm.287f..52G}
{Gehrels}, N., {Piro}, L., \& {Leonard}, P.~J.~T. 2002, Scientific American,
  287, 52

\bibitem[{{Gehrels} {et~al.}(2005){Gehrels}, {Sarazin}, {O'Brien}, {Zhang},
  {Barbier}, {Barthelmy}, {Blustin}, {Burrows}, {Cannizzo}, {Cummings}, {Goad},
  {Holland}, {Hurkett}, {Kennea}, {Levan}, {Markwardt}, {Mason}, {Meszaros},
  {Page}, {Palmer}, {Rol}, {Sakamoto}, {Willingale}, {Angelini}, {Beardmore},
  {Boyd}, {Breeveld}, {Campana}, {Chester}, {Chincarini}, {Cominsky},
  {Cusumano}, {de Pasquale}, {Fenimore}, {Giommi}, {Gronwall}, {Grupe}, {Hill},
  {Hinshaw}, {Hjorth}, {Hullinger}, {Hurley}, {Klose}, {Kobayashi},
  {Kouveliotou}, {Krimm}, {Mangano}, {Marshall}, {McGowan}, {Moretti},
  {Mushotzky}, {Nakazawa}, {Norris}, {Nousek}, {Osborne}, {Page}, {Parsons},
  {Patel}, {Perri}, {Poole}, {Romano}, {Roming}, {Rosen}, {Sato}, {Schady},
  {Smale}, {Sollerman}, {Starling}, {Still}, {Suzuki}, {Tagliaferri},
  {Takahashi}, {Tashiro}, {Tueller}, {Wells}, {White}, \&
  {Wijers}}]{2005Natur.437..851G}
{Gehrels}, N., {Sarazin}, C.~L., {O'Brien}, P.~T., {et~al.} 2005, \nat, 437,
  851

\bibitem[{{Gendre} {et~al.}(2013){Gendre}, {Stratta}, {Atteia}, {Basa},
  {Bo{\"e}r}, {Coward}, {Cutini}, {D'Elia}, {Howell}, {Klotz}, \&
  {Piro}}]{2013ApJ...766...30G}
{Gendre}, B., {Stratta}, G., {Atteia}, J.~L., {et~al.} 2013, \apj, 766, 30

\bibitem[{{Genet} {et~al.}(2007){Genet}, {Daigne}, \&
  {Mochkovitch}}]{2007MNRAS.381..732G}
{Genet}, F., {Daigne}, F., \& {Mochkovitch}, R. 2007, \mnras, 381, 732

\bibitem[{{Ghirlanda} {et~al.}(2011){Ghirlanda}, {Ghisellini}, \&
  {Nava}}]{2011MNRAS.418L.109G}
{Ghirlanda}, G., {Ghisellini}, G., \& {Nava}, L. 2011, \mnras, 418, L109

\bibitem[{{Ghirlanda} {et~al.}(2009){Ghirlanda}, {Nava}, {Ghisellini},
  {Celotti}, \& {Firmani}}]{2009A&A...496..585G}
{Ghirlanda}, G., {Nava}, L., {Ghisellini}, G., {Celotti}, A., \& {Firmani}, C.
  2009, \aap, 496, 585

\bibitem[{{Ghisellini} {et~al.}(2007{\natexlab{a}}){Ghisellini}, {Ghirlanda},
  \& {Tavecchio}}]{2007MNRAS.382L..77G}
{Ghisellini}, G., {Ghirlanda}, G., \& {Tavecchio}, F. 2007{\natexlab{a}},
  \mnras, 382, L77

\bibitem[{{Ghisellini} {et~al.}(2007{\natexlab{b}}){Ghisellini}, {Ghirlanda},
  \& {Tavecchio}}]{2007MNRAS.375L..36G}
{Ghisellini}, G., {Ghirlanda}, G., \& {Tavecchio}, F. 2007{\natexlab{b}},
  \mnras, 375, L36

\bibitem[{{Goldader} {et~al.}(2002){Goldader}, {Meurer}, {Heckman}, {Seibert},
  {Sanders}, {Calzetti}, \& {Steidel}}]{2002ApJ...568..651G}
{Goldader}, J.~D., {Meurer}, G., {Heckman}, T.~M., {et~al.} 2002, \apj, 568,
  651

\bibitem[{{Goldoni} {et~al.}(2006){Goldoni}, {Royer}, {Fran{\c c}ois},
  {Horrobin}, {Blanc}, {Vernet}, {Modigliani}, \& {Larsen}}]{pipeline}
{Goldoni}, P., {Royer}, F., {Fran{\c c}ois}, P., {et~al.} 2006, in Society of
  Photo-Optical Instrumentation Engineers (SPIE) Conference Series, Vol. 6269,
  Society of Photo-Optical Instrumentation Engineers (SPIE) Conference Series,
  2

\bibitem[{{Golenetskii} {et~al.}(2014){Golenetskii}, {Aptekar}, {Frederiks},
  {Pal'Shin}, {Oleynik}, {Ulanov}, {Svinkin}, {Tsvetkova}, {Lysenko}, \&
  {Cline}}]{2014GCN..17108...1G}
{Golenetskii}, S., {Aptekar}, R., {Frederiks}, D., {et~al.} 2014, GRB
  Coordinates Network, 17108, 1

\bibitem[{{Golenetskii} {et~al.}(2015){Golenetskii}, {Aptekar}, {Frederiks},
  {Pal'Shin}, {Oleynik}, {Ulanov}, {Svinkin}, {Tsvetkova}, {Lysenko}, \&
  {Cline}}]{2015GCN..17357...1G}
{Golenetskii}, S., {Aptekar}, R., {Frederiks}, D., {et~al.} 2015, GRB
  Coordinates Network, 17357, 1

\bibitem[{{Gompertz} {et~al.}(2014){Gompertz}, {O'Brien}, \&
  {Wynn}}]{2014MNRAS.438..240G}
{Gompertz}, B.~P., {O'Brien}, P.~T., \& {Wynn}, G.~A. 2014, \mnras, 438, 240

\bibitem[{{Gorosabel} {et~al.}(2003){Gorosabel}, {Christensen}, {Hjorth},
  {Fynbo}, {Pedersen}, {Jensen}, {Andersen}, {Lund}, {Jaunsen}, {Castro
  Cer{\'o}n}, {Castro-Tirado}, {Fruchter}, {Greiner}, {Pian}, {Vreeswijk},
  {Burud}, {Frontera}, {Kaper}, {Klose}, {Kouveliotou}, {Masetti}, {Palazzi},
  {Rhoads}, {Rol}, {Salamanca}, {Tanvir}, {Wijers}, \& {van den
  Heuvel}}]{2003A&A...400..127G}
{Gorosabel}, J., {Christensen}, L., {Hjorth}, J., {et~al.} 2003, \aap, 400, 127

\bibitem[{{Gorosabel} {et~al.}(2010){Gorosabel}, {de Ugarte Postigo},
  {Castro-Tirado}, {Agudo}, {Jel{\'{\i}}nek}, {Leon}, {Augusteijn}, {Fynbo},
  {Hjorth}, {Micha{\l}owski}, {Xu}, {Ferrero}, {Kann}, {Klose}, {Rossi},
  {Madrid}, {Llorente}, {Bremer}, \& {Winters}}]{2010A&A...522A..14G}
{Gorosabel}, J., {de Ugarte Postigo}, A., {Castro-Tirado}, A.~J., {et~al.}
  2010, \aap, 522, A14

\bibitem[{{Gorosabel} {et~al.}(2004){Gorosabel}, {Lund}, {Brandt},
  {Westergaard}, \& {Castro Cer{\'o}n}}]{2004A&A...427...87G}
{Gorosabel}, J., {Lund}, N., {Brandt}, S., {Westergaard}, N.~J., \& {Castro
  Cer{\'o}n}, J.~M. 2004, \aap, 427, 87

\bibitem[{{G{\"o}tz} {et~al.}(2013){G{\"o}tz}, {Covino}, {Fern{\'a}ndez-Soto},
  {Laurent}, \& {Bo{\v s}njak}}]{2013MNRAS.431.3550G}
{G{\"o}tz}, D., {Covino}, S., {Fern{\'a}ndez-Soto}, A., {Laurent}, P., \&
  {Bo{\v s}njak}, {\v Z}. 2013, \mnras, 431, 3550

\bibitem[{{Graham} \& {Fruchter}(2012)}]{2012IAUS..279..232G}
{Graham}, J.~F. \& {Fruchter}, A.~S. 2012, in IAU Symposium, Vol. 279, Death of
  Massive Stars: Supernovae and Gamma-Ray Bursts, 232--236

\bibitem[{{Granot}(2006)}]{2006NCimB.121.1073G}
{Granot}, J. 2006, Nuovo Cimento B Serie, 121, 1073

\bibitem[{{Greiner} {et~al.}(2008){Greiner}, {Bornemann}, {Clemens}, {Deuter},
  {Hasinger}, {Honsberg}, {Huber}, {Huber}, {Krauss}, {Kr{\"u}hler},
  {K{\"u}pc{\"u} Yolda{\c s}}, {Mayer-Hasselwander}, {Mican}, {Primak},
  {Schrey}, {Steiner}, {Szokoly}, {Th{\"o}ne}, {Yolda{\c s}}, {Klose}, {Laux},
  \& {Winkler}}]{grond2}
{Greiner}, J., {Bornemann}, W., {Clemens}, C., {et~al.} 2008, \pasp, 120, 405

\bibitem[{{Greiner} {et~al.}(2007){Greiner}, {Bornemann}, {Clemens}, {Deuter},
  {Hasinger}, {Honsberg}, {Huber}, {Huber}, {Krauss}, {Kr{\"u}hler},
  {K{\"u}pc{\"u} Yoldas}, {Mayer-Hasselwander}, {Mican}, {Primak}, {Schrey},
  {Steiner}, {Szokoly}, {Th{\"o}ne}, {Yoldas}, {Klose}, {Laux}, \&
  {Winkler}}]{grond1}
{Greiner}, J., {Bornemann}, W., {Clemens}, C., {et~al.} 2007, The Messenger,
  130, 12

\bibitem[{{Greiner} {et~al.}(2011){Greiner}, {Kr{\"u}hler}, {Klose}, {Afonso},
  {Clemens}, {Filgas}, {Hartmann}, {K{\"u}pc{\"u} Yolda{\c s}}, {Nardini},
  {Olivares E.}, {Rau}, {Rossi}, {Schady}, \& {Updike}}]{2011A&A...526A..30G}
{Greiner}, J., {Kr{\"u}hler}, T., {Klose}, S., {et~al.} 2011, \aap, 526, A30

\bibitem[{{Greiner} {et~al.}(2015){Greiner}, {Mazzali}, {Kann}, {Kr{\"u}hler},
  {Pian}, {Prentice}, {Olivares E.}, {Rossi}, {Klose}, {Taubenberger}, {Knust},
  {Afonso}, {Ashall}, {Bolmer}, {Delvaux}, {Diehl}, {Elliott}, {Filgas},
  {Fynbo}, {Graham}, {Guelbenzu}, {Kobayashi}, {Leloudas}, {Savaglio},
  {Schady}, {Schmidl}, {Schweyer}, {Sudilovsky}, {Tanga}, {Updike}, {van
  Eerten}, \& {Varela}}]{2015Natur.523..189G}
{Greiner}, J., {Mazzali}, P.~A., {Kann}, D.~A., {et~al.} 2015, \nat, 523, 189

\bibitem[{{Grindlay}(2010)}]{2010AIPC.1279..212G}
{Grindlay}, J.~E. 2010, in American Institute of Physics Conference Series,
  Vol. 1279, American Institute of Physics Conference Series, ed. N.~{Kawai} \&
  S.~{Nagataki}, 212--219

\bibitem[{{Gruber} {et~al.}(2014){Gruber}, {Goldstein}, {Weller von Ahlefeld},
  {Narayana Bhat}, {Bissaldi}, {Briggs}, {Byrne}, {Cleveland}, {Connaughton},
  {Diehl}, {Fishman}, {Fitzpatrick}, {Foley}, {Gibby}, {Giles}, {Greiner},
  {Guiriec}, {van der Horst}, {von Kienlin}, {Kouveliotou}, {Layden}, {Lin},
  {Meegan}, {McGlynn}, {Paciesas}, {Pelassa}, {Preece}, {Rau}, {Wilson-Hodge},
  {Xiong}, {Younes}, \& {Yu}}]{2014ApJS..211...12G}
{Gruber}, D., {Goldstein}, A., {Weller von Ahlefeld}, V., {et~al.} 2014, \apjs,
  211, 12

\bibitem[{{Grupe} {et~al.}(2006){Grupe}, {Burrows}, {Patel}, {Kouveliotou},
  {Zhang}, {M{\'e}sz{\'a}ros}, {Wijers}, \& {Gehrels}}]{2006ApJ...653..462G}
{Grupe}, D., {Burrows}, D.~N., {Patel}, S.~K., {et~al.} 2006, \apj, 653, 462

\bibitem[{{Habergham} {et~al.}(2010){Habergham}, {Anderson}, \&
  {James}}]{2010ApJ...717..342H}
{Habergham}, S.~M., {Anderson}, J.~P., \& {James}, P.~A. 2010, \apj, 717, 342

\bibitem[{{Hao} \& {Yuan}(2013)}]{2013ApJ...772...42H}
{Hao}, J.-M. \& {Yuan}, Y.-F. 2013, \apj, 772, 42

\bibitem[{{Harrison} {et~al.}(1999){Harrison}, {Bloom}, {Frail}, {Sari},
  {Kulkarni}, {Djorgovski}, {Axelrod}, {Mould}, {Schmidt}, {Wieringa}, {Wark},
  {Subrahmanyan}, {McConnell}, {McCarthy}, {Schaefer}, {McMahon}, {Markze},
  {Firth}, {Soffitta}, \& {Amati}}]{1999ApJ...523L.121H}
{Harrison}, F.~A., {Bloom}, J.~S., {Frail}, D.~A., {et~al.} 1999, \apjl, 523,
  L121

\bibitem[{{Harry} \& {LIGO Scientific
  Collaboration}(2010)}]{2010CQGra..27h4006H}
{Harry}, G.~M. \& {LIGO Scientific Collaboration}. 2010, Classical and Quantum
  Gravity, 27, 084006

\bibitem[{{Heesen} {et~al.}(2014){Heesen}, {Brinks}, {Leroy}, {Heald}, {Braun},
  {Bigiel}, \& {Beck}}]{2014AJ....147..103H}
{Heesen}, V., {Brinks}, E., {Leroy}, A.~K., {et~al.} 2014, \aj, 147, 103

\bibitem[{{Hirashita} \& {Li}(2013)}]{2013MNRAS.434L..70H}
{Hirashita}, H. \& {Li}, Z.-Y. 2013, \mnras, 434, L70

\bibitem[{{Hirashita} {et~al.}(2008){Hirashita}, {Nozawa}, {Takeuchi}, \&
  {Kozasa}}]{2008MNRAS.384.1725H}
{Hirashita}, H., {Nozawa}, T., {Takeuchi}, T.~T., \& {Kozasa}, T. 2008, \mnras,
  384, 1725

\bibitem[{{Hjorth} \& {Bloom}(2012)}]{2012grbu.book..169H}
{Hjorth}, J. \& {Bloom}, J.~S. 2012, {The Gamma-Ray Burst - Supernova
  Connection}, 169--190

\bibitem[{{Hjorth} {et~al.}(2012){Hjorth}, {Malesani}, {Jakobsson}, {Jaunsen},
  {Fynbo}, {Gorosabel}, {Kr{\"u}hler}, {Levan}, {Micha{\l}owski},
  {Milvang-Jensen}, {M{\o}ller}, {Schulze}, {Tanvir}, \&
  {Watson}}]{2012ApJ...756..187H}
{Hjorth}, J., {Malesani}, D., {Jakobsson}, P., {et~al.} 2012, \apj, 756, 187

\bibitem[{{Hjorth} {et~al.}(2005{\natexlab{a}}){Hjorth}, {Sollerman},
  {Gorosabel}, {Granot}, {Klose}, {Kouveliotou}, {Melinder}, {Ramirez-Ruiz},
  {Starling}, {Thomsen}, {Andersen}, {Fynbo}, {Jensen}, {Vreeswijk}, {Castro
  Cer{\'o}n}, {Jakobsson}, {Levan}, {Pedersen}, {Rhoads}, {Tanvir}, {Watson},
  \& {Wijers}}]{2005ApJ...630L.117H}
{Hjorth}, J., {Sollerman}, J., {Gorosabel}, J., {et~al.} 2005{\natexlab{a}},
  \apjl, 630, L117

\bibitem[{{Hjorth} {et~al.}(2003){Hjorth}, {Sollerman}, {M{\o}ller}, {Fynbo},
  {Woosley}, {Kouveliotou}, {Tanvir}, {Greiner}, {Andersen}, {Castro-Tirado},
  {Castro Cer{\'o}n}, {Fruchter}, {Gorosabel}, {Jakobsson}, {Kaper}, {Klose},
  {Masetti}, {Pedersen}, {Pedersen}, {Pian}, {Palazzi}, {Rhoads}, {Rol}, {van
  den Heuvel}, {Vreeswijk}, {Watson}, \& {Wijers}}]{2003Natur.423..847H}
{Hjorth}, J., {Sollerman}, J., {M{\o}ller}, P., {et~al.} 2003, \nat, 423, 847

\bibitem[{{Hjorth} {et~al.}(2005{\natexlab{b}}){Hjorth}, {Watson}, {Fynbo},
  {Price}, {Jensen}, {J{\o}rgensen}, {Kubas}, {Gorosabel}, {Jakobsson},
  {Sollerman}, {Pedersen}, \& {Kouveliotou}}]{2005Natur.437..859H}
{Hjorth}, J., {Watson}, D., {Fynbo}, J.~P.~U., {et~al.} 2005{\natexlab{b}},
  \nat, 437, 859

\bibitem[{{Houck} \& {Denicola}(2000)}]{isis}
{Houck}, J.~C. \& {Denicola}, L.~A. 2000, in Astronomical Society of the
  Pacific Conference Series, Vol. 216, Astronomical Data Analysis Software and
  Systems IX, ed. N.~{Manset}, C.~{Veillet}, \& D.~{Crabtree}, 591

\bibitem[{{Hui} \& {Haiman}(2003)}]{2003ApJ...596....9H}
{Hui}, L. \& {Haiman}, Z. 2003, \apj, 596, 9

\bibitem[{{Hunt} {et~al.}(2014){Hunt}, {Palazzi}, {Micha{\l}owski}, {Rossi},
  {Savaglio}, {Basa}, {Berta}, {Bianchi}, {Covino}, {D'Elia}, {Ferrero},
  {G{\"o}tz}, {Greiner}, {Klose}, {Le Borgne}, {Le Floc'h}, {Pian},
  {Piranomonte}, {Schady}, \& {Vergani}}]{hunt14}
{Hunt}, L.~K., {Palazzi}, E., {Micha{\l}owski}, M.~J., {et~al.} 2014, \aap,
  565, A112

\bibitem[{{Hunter} {et~al.}(1993){Hunter}, {Hawley}, \&
  {Gallagher}}]{1993AJ....106.1797H}
{Hunter}, D.~A., {Hawley}, W.~N., \& {Gallagher}, III, J.~S. 1993, \aj, 106,
  1797

\bibitem[{{Jakobsson} {et~al.}(2005){Jakobsson}, {Bj{\"o}rnsson}, {Fynbo},
  {J{\'o}hannesson}, {Hjorth}, {Thomsen}, {M{\o}ller}, {Watson}, {Jensen},
  {{\"O}stlin}, {Gorosabel}, \& {Gudmundsson}}]{2005MNRAS.362..245J}
{Jakobsson}, P., {Bj{\"o}rnsson}, G., {Fynbo}, J.~P.~U., {et~al.} 2005, \mnras,
  362, 245

\bibitem[{{Jakobsson} {et~al.}(2006){Jakobsson}, {Fynbo}, {Ledoux},
  {Vreeswijk}, {Kann}, {Hjorth}, {Priddey}, {Tanvir}, {Reichart}, {Gorosabel},
  {Klose}, {Watson}, {Sollerman}, {Fruchter}, {de Ugarte Postigo}, {Wiersema},
  {Bj{\"o}rnsson}, {Chapman}, {Th{\"o}ne}, {Pedersen}, \&
  {Jensen}}]{Jakobsson2006}
{Jakobsson}, P., {Fynbo}, J.~P.~U., {Ledoux}, C., {et~al.} 2006, \aap, 460, L13

\bibitem[{{Jakobsson} {et~al.}(2004){Jakobsson}, {Hjorth}, {Fynbo}, {Watson},
  {Pedersen}, {Bj{\"o}rnsson}, \& {Gorosabel}}]{2004ApJ...617L..21J}
{Jakobsson}, P., {Hjorth}, J., {Fynbo}, J.~P.~U., {et~al.} 2004, \apjl, 617,
  L21

\bibitem[{{Jakobsson} {et~al.}(2012{\natexlab{a}}){Jakobsson}, {Hjorth},
  {Malesani}, {Chapman}, {Fynbo}, {Tanvir}, {Milvang-Jensen}, {Vreeswijk},
  {Letawe}, \& {Starling}}]{2012ApJ...752...62J}
{Jakobsson}, P., {Hjorth}, J., {Malesani}, D., {et~al.} 2012{\natexlab{a}},
  \apj, 752, 62

\bibitem[{{Jakobsson} {et~al.}(2012{\natexlab{b}}){Jakobsson}, {Hjorth},
  {Malesani}, {Fynbo}, {Kr{\"u}hler}, {Milvang-Jensen}, \&
  {Tanvir}}]{2012IAUS..279..187J}
{Jakobsson}, P., {Hjorth}, J., {Malesani}, D., {et~al.} 2012{\natexlab{b}}, in
  IAU Symposium, Vol. 279, Death of Massive Stars: Supernovae and Gamma-Ray
  Bursts, 187--190

\bibitem[{{Jaunsen} {et~al.}(2008){Jaunsen}, {Rol}, {Watson}, {Malesani},
  {Fynbo}, {Milvang-Jensen}, {Hjorth}, {Vreeswijk}, {Ovaldsen}, {Wiersema},
  {Tanvir}, {Gorosabel}, {Levan}, {Schirmer}, \&
  {Castro-Tirado}}]{2008ApJ...681..453J}
{Jaunsen}, A.~O., {Rol}, E., {Watson}, D.~J., {et~al.} 2008, \apj, 681, 453

\bibitem[{{Jenkins}(2009)}]{jenkins09}
{Jenkins}, E.~B. 2009, \apj, 700, 1299

\bibitem[{{Jensen} {et~al.}(2001){Jensen}, {Fynbo}, {Gorosabel}, {Hjorth},
  {Holland}, {M{\"o}ller}, {Thomsen}, {Bj{\"o}rnsson}, {Pedersen}, {Burud},
  {Henden}, {Tanvir}, {Davis}, {Vreeswijk}, {Rol}, {Hurley}, {Cline},
  {Trombka}, {McClanahan}, {Starr}, {Goldsten}, {Castro-Tirado}, {Greiner},
  {Bailer-Jones}, {K{\"u}mmel}, \& {Mundt}}]{2001A&A...370..909J}
{Jensen}, B.~L., {Fynbo}, J.~U., {Gorosabel}, J., {et~al.} 2001, \aap, 370, 909

\bibitem[{{Jing} {et~al.}(2011){Jing}, {He}, {Brucato}, {De Sio}, {Tozzetti},
  \& {Vidali}}]{2011ApJ...741L...9J}
{Jing}, D., {He}, J., {Brucato}, J., {et~al.} 2011, \apjl, 741, L9

\bibitem[{{Jorgenson} \& {Wolfe}(2014)}]{JW14}
{Jorgenson}, R.~A. \& {Wolfe}, A.~M. 2014, \apj, 785, 16

\bibitem[{{Kalberla} {et~al.}(2005){Kalberla}, {Burton}, {Hartmann}, {Arnal},
  {Bajaja}, {Morras}, \& {P{\"o}ppel}}]{2005A&A...440..775K}
{Kalberla}, P.~M.~W., {Burton}, W.~B., {Hartmann}, D., {et~al.} 2005, \aap,
  440, 775

\bibitem[{{Kaneko} {et~al.}(2006){Kaneko}, {Preece}, {Briggs}, {Paciesas},
  {Meegan}, \& {Band}}]{2006ApJS..166..298K}
{Kaneko}, Y., {Preece}, R.~D., {Briggs}, M.~S., {et~al.} 2006, \apjs, 166, 298

\bibitem[{{Kann} {et~al.}(2006){Kann}, {Klose}, \& {Zeh}}]{2006ApJ...641..993K}
{Kann}, D.~A., {Klose}, S., \& {Zeh}, A. 2006, \apj, 641, 993

\bibitem[{{Karim} {et~al.}(2011){Karim}, {Schinnerer},
  {Mart{\'{\i}}nez-Sansigre}, {Sargent}, {van der Wel}, {Rix}, {Ilbert},
  {Smol{\v c}i{\'c}}, {Carilli}, {Pannella}, {Koekemoer}, {Bell}, \&
  {Salvato}}]{2011ApJ...730...61K}
{Karim}, A., {Schinnerer}, E., {Mart{\'{\i}}nez-Sansigre}, A., {et~al.} 2011,
  \apj, 730, 61

\bibitem[{{Katz}(1994)}]{1994ApJ...422..248K}
{Katz}, J.~I. 1994, \apj, 422, 248

\bibitem[{{Katz} \& {Piran}(1997)}]{1997ApJ...490..772K}
{Katz}, J.~I. \& {Piran}, T. 1997, \apj, 490, 772

\bibitem[{{Kelly} {et~al.}(2008){Kelly}, {Kirshner}, \&
  {Pahre}}]{2008ApJ...687.1201K}
{Kelly}, P.~L., {Kirshner}, R.~P., \& {Pahre}, M. 2008, \apj, 687, 1201

\bibitem[{{Kennicutt}(1998)}]{1998ARA&A..36..189K}
{Kennicutt}, Jr., R.~C. 1998, \araa, 36, 189

\bibitem[{{Kerr} {et~al.}(1998){Kerr}, {Hibbins}, {Fossey}, {Miles}, \&
  {Sarre}}]{1998ApJ...495..941K}
{Kerr}, T.~H., {Hibbins}, R.~E., {Fossey}, S.~J., {Miles}, J.~R., \& {Sarre},
  P.~J. 1998, \apj, 495, 941

\bibitem[{{Kewley} \& {Ellison}(2008)}]{KE}
{Kewley}, L.~J. \& {Ellison}, S.~L. 2008, \apj, 681, 1183

\bibitem[{{Khare} {et~al.}(2012){Khare}, {vanden Berk}, {York}, {Lundgren}, \&
  {Kulkarni}}]{2012MNRAS.419.1028K}
{Khare}, P., {vanden Berk}, D., {York}, D.~G., {Lundgren}, B., \& {Kulkarni},
  V.~P. 2012, \mnras, 419, 1028

\bibitem[{{Kistler} {et~al.}(2008){Kistler}, {Y{\"u}ksel}, {Beacom}, \&
  {Stanek}}]{2008ApJ...673L.119K}
{Kistler}, M.~D., {Y{\"u}ksel}, H., {Beacom}, J.~F., \& {Stanek}, K.~Z. 2008,
  \apjl, 673, L119

\bibitem[{{Klebesadel} {et~al.}(1973){Klebesadel}, {Strong}, \&
  {Olson}}]{1973ApJ...182L..85K}
{Klebesadel}, R.~W., {Strong}, I.~B., \& {Olson}, R.~A. 1973, \apjl, 182, L85

\bibitem[{{Kogut} {et~al.}(2003){Kogut}, {Spergel}, {Barnes}, {Bennett},
  {Halpern}, {Hinshaw}, {Jarosik}, {Limon}, {Meyer}, {Page}, {Tucker},
  {Wollack}, \& {Wright}}]{2003ApJS..148..161K}
{Kogut}, A., {Spergel}, D.~N., {Barnes}, C., {et~al.} 2003, \apjs, 148, 161

\bibitem[{{Komatsu} {et~al.}(2011){Komatsu}, {Smith}, {Dunkley}, {Bennett},
  {Gold}, {Hinshaw}, {Jarosik}, {Larson}, {Nolta}, {Page}, {Spergel},
  {Halpern}, {Hill}, {Kogut}, {Limon}, {Meyer}, {Odegard}, {Tucker}, {Weiland},
  {Wollack}, \& {Wright}}]{2011ApJS..192...18K}
{Komatsu}, E., {Smith}, K.~M., {Dunkley}, J., {et~al.} 2011, \apjs, 192, 18

\bibitem[{{Kopa{\v c}} {et~al.}(2015){Kopa{\v c}}, {Mundell}, {Kobayashi},
  {Virgili}, {Harrison}, {Japelj}, {Guidorzi}, {Melandri}, \&
  {Gomboc}}]{2015ApJ...806..179K}
{Kopa{\v c}}, D., {Mundell}, C.~G., {Kobayashi}, S., {et~al.} 2015, \apj, 806,
  179

\bibitem[{{Kouveliotou} {et~al.}(1993){Kouveliotou}, {Meegan}, {Fishman},
  {Bhat}, {Briggs}, {Koshut}, {Paciesas}, \& {Pendleton}}]{1993ApJ...413L.101K}
{Kouveliotou}, C., {Meegan}, C.~A., {Fishman}, G.~J., {et~al.} 1993, \apjl,
  413, L101

\bibitem[{{Krisciunas} {et~al.}(2007){Krisciunas}, {Garnavich}, {Stanishev},
  {Suntzeff}, {Prieto}, {Espinoza}, {Gonzalez}, {Salvo}, {Elias de la Rosa},
  {Smartt}, {Maund}, \& {Kudritzki}}]{2007AJ....133...58K}
{Krisciunas}, K., {Garnavich}, P.~M., {Stanishev}, V., {et~al.} 2007, \aj, 133,
  58

\bibitem[{{Krogager} {et~al.}(2013){Krogager}, {Fynbo}, {Ledoux},
  {Christensen}, {Gallazzi}, {Laursen}, {M{\o}ller}, {Noterdaeme},
  {P{\'e}roux}, {Pettini}, \& {Vestergaard}}]{krogager13}
{Krogager}, J.-K., {Fynbo}, J.~P.~U., {Ledoux}, C., {et~al.} 2013, \mnras, 433,
  3091

\bibitem[{{Krogager} {et~al.}(2015){Krogager}, {Fynbo}, {Noterdaeme}, {Zafar},
  {M{\o}ller}, {Ledoux}, {Kr{\"u}hler}, \& {Stockton}}]{2015arXiv151004695K}
{Krogager}, J.-K., {Fynbo}, J.~P.~U., {Noterdaeme}, P., {et~al.} 2015, ArXiv
  e-prints

\bibitem[{{Kroupa}(2001)}]{2001MNRAS.322..231K}
{Kroupa}, P. 2001, \mnras, 322, 231

\bibitem[{{Kr{\"u}hler} {et~al.}(2012){Kr{\"u}hler}, {Fynbo}, {Geier},
  {Hjorth}, {Malesani}, {Milvang-Jensen}, {Levan}, {Sparre}, {Watson}, \&
  {Zafar}}]{thomas2}
{Kr{\"u}hler}, T., {Fynbo}, J.~P.~U., {Geier}, S., {et~al.} 2012, \aap, 546, A8

\bibitem[{{Kr{\"u}hler} {et~al.}(2011){Kr{\"u}hler}, {Greiner}, {Schady},
  {Savaglio}, {Afonso}, {Clemens}, {Elliott}, {Filgas}, {Gruber}, {Kann},
  {Klose}, {K{\"u}pc{\"u}-Yolda{\c s}}, {McBreen}, {Olivares}, {Pierini},
  {Rau}, {Rossi}, {Nardini}, {Nicuesa Guelbenzu}, {Sudilovsky}, \&
  {Updike}}]{2011A&A...534A.108K}
{Kr{\"u}hler}, T., {Greiner}, J., {Schady}, P., {et~al.} 2011, \aap, 534, A108

\bibitem[{{Kr{\"u}hler} {et~al.}(2008){Kr{\"u}hler}, {K{\"u}pc{\"u} Yolda{\c
  s}}, {Greiner}, {Clemens}, {McBreen}, {Primak}, {Savaglio}, {Yolda{\c s}},
  {Szokoly}, \& {Klose}}]{thomas08}
{Kr{\"u}hler}, T., {K{\"u}pc{\"u} Yolda{\c s}}, A., {Greiner}, J., {et~al.}
  2008, \apj, 685, 376

\bibitem[{{Kr{\"u}hler} {et~al.}(2013){Kr{\"u}hler}, {Ledoux}, {Fynbo},
  {Vreeswijk}, {Schmidl}, {Malesani}, {Christensen}, {De Cia}, {Hjorth},
  {Jakobsson}, {Kann}, {Kaper}, {Vergani}, {Afonso}, {Covino}, {de Ugarte
  Postigo}, {D'Elia}, {Filgas}, {Goldoni}, {Greiner}, {Hartoog},
  {Milvang-Jensen}, {Nardini}, {Piranomonte}, {Rossi},
  {S{\'a}nchez-Ram{\'{\i}}rez}, {Schady}, {Schulze}, {Sudilovsky}, {Tanvir},
  {Tagliaferri}, {Watson}, {Wiersema}, {Wijers}, \& {Xu}}]{thomas1}
{Kr{\"u}hler}, T., {Ledoux}, C., {Fynbo}, J.~P.~U., {et~al.} 2013, \aap, 557,
  A18

\bibitem[{{Kudritzki} {et~al.}(2012){Kudritzki}, {Urbaneja}, {Gazak},
  {Bresolin}, {Przybilla}, {Gieren}, \& {Pietrzy{\'n}ski}}]{Kudritzki}
{Kudritzki}, R.-P., {Urbaneja}, M.~A., {Gazak}, Z., {et~al.} 2012, \apj, 747,
  15

\bibitem[{{Lamb} \& {Reichart}(2000)}]{2000ApJ...536....1L}
{Lamb}, D.~Q. \& {Reichart}, D.~E. 2000, \apj, 536, 1

\bibitem[{{Larmor}(1897)}]{larmor}
{Larmor}, J. 1897, Philosophical Transactions of the Royal Society, 190, 205

\bibitem[{{Larsson} {et~al.}(2011){Larsson}, {Ryde}, {Lundman}, {McGlynn},
  {Larsson}, {Ohno}, \& {Yamaoka}}]{2011MNRAS.414.2642L}
{Larsson}, J., {Ryde}, F., {Lundman}, C., {et~al.} 2011, \mnras, 414, 2642

\bibitem[{{Laureijs} {et~al.}(2011){Laureijs}, {Amiaux}, {Arduini},
  {Augu{\`e}res}, {Brinchmann}, {Cole}, {Cropper}, {Dabin}, {Duvet}, {Ealet},
  \& et~al.}]{2011arXiv1110.3193L}
{Laureijs}, R., {Amiaux}, J., {Arduini}, S., {et~al.} 2011, ArXiv e-prints

\bibitem[{{Lazzati} {et~al.}(2013){Lazzati}, {Morsony}, {Margutti}, \&
  {Begelman}}]{2013arXiv1301.3920L}
{Lazzati}, D., {Morsony}, B.~J., {Margutti}, R., \& {Begelman}, M.~C. 2013,
  ArXiv e-prints

\bibitem[{{Le Floc'h} {et~al.}(2003){Le Floc'h}, {Duc}, {Mirabel}, {Sanders},
  {Bosch}, {Diaz}, {Donzelli}, {Rodrigues}, {Courvoisier}, {Greiner},
  {Mereghetti}, {Melnick}, {Maza}, \& {Minniti}}]{2003A&A...400..499L}
{Le Floc'h}, E., {Duc}, P.-A., {Mirabel}, I.~F., {et~al.} 2003, \aap, 400, 499

\bibitem[{{Ledoux} {et~al.}(2002){Ledoux}, {Bergeron}, \&
  {Petitjean}}]{2002A&A...385..802L}
{Ledoux}, C., {Bergeron}, J., \& {Petitjean}, P. 2002, \aap, 385, 802

\bibitem[{{Ledoux} {et~al.}(2006){Ledoux}, {Petitjean}, {Fynbo}, {M{\o}ller},
  \& {Srianand}}]{ledoux06}
{Ledoux}, C., {Petitjean}, P., {Fynbo}, J.~P.~U., {M{\o}ller}, P., \&
  {Srianand}, R. 2006, \aap, 457, 71

\bibitem[{{Ledoux} {et~al.}(2003){Ledoux}, {Petitjean}, \&
  {Srianand}}]{ledoux03}
{Ledoux}, C., {Petitjean}, P., \& {Srianand}, R. 2003, \mnras, 346, 209

\bibitem[{{Ledoux} {et~al.}(2009){Ledoux}, {Vreeswijk}, {Smette}, {Fox},
  {Petitjean}, {Ellison}, {Fynbo}, \& {Savaglio}}]{2009A&A...506..661L}
{Ledoux}, C., {Vreeswijk}, P.~M., {Smette}, A., {et~al.} 2009, \aap, 506, 661

\bibitem[{{Lee} {et~al.}(2013){Lee}, {Hwang}, \& {Ko}}]{2013ApJ...774...62L}
{Lee}, J.~C., {Hwang}, H.~S., \& {Ko}, J. 2013, \apj, 774, 62

\bibitem[{{Lef{\`e}vre} {et~al.}(2014){Lef{\`e}vre}, {Pagani}, {Juvela},
  {Paladini}, {Lallement}, {Marshall}, {Andersen}, {Bacmann}, {McGehee},
  {Montier}, {Noriega-Crespo}, {Pelkonen}, {Ristorcelli}, \&
  {Steinacker}}]{2014A&A...572A..20L}
{Lef{\`e}vre}, C., {Pagani}, L., {Juvela}, M., {et~al.} 2014, \aap, 572, A20

\bibitem[{{Levan}(2015)}]{2015arXiv150603960L}
{Levan}, A.~J. 2015, ArXiv e-prints

\bibitem[{{Levan} {et~al.}(2014){Levan}, {Tanvir}, {Starling}, {Wiersema},
  {Page}, {Perley}, {Schulze}, {Wynn}, {Chornock}, {Hjorth}, {Cenko},
  {Fruchter}, {O'Brien}, {Brown}, {Tunnicliffe}, {Malesani}, {Jakobsson},
  {Watson}, {Berger}, {Bersier}, {Cobb}, {Covino}, {Cucchiara}, {de Ugarte
  Postigo}, {Fox}, {Gal-Yam}, {Goldoni}, {Gorosabel}, {Kaper}, {Kr{\"u}hler},
  {Karjalainen}, {Osborne}, {Pian}, {S{\'a}nchez-Ram{\'{\i}}rez}, {Schmidt},
  {Skillen}, {Tagliaferri}, {Th{\"o}ne}, {Vaduvescu}, {Wijers}, \&
  {Zauderer}}]{2014ApJ...781...13L}
{Levan}, A.~J., {Tanvir}, N.~R., {Starling}, R.~L.~C., {et~al.} 2014, \apj,
  781, 13

\bibitem[{{Li} {et~al.}(2008){Li}, {Liang}, {Kann}, {Wei}, {Klose}, \&
  {Wang}}]{2008ApJ...685.1046L}
{Li}, A., {Liang}, S.~L., {Kann}, D.~A., {et~al.} 2008, \apj, 685, 1046

\bibitem[{{Li} {et~al.}(2014){Li}, {de Grijs}, \& {Deng}}]{2014Natur.516..367L}
{Li}, C., {de Grijs}, R., \& {Deng}, L. 2014, \nat, 516, 367

\bibitem[{{Li}(2007)}]{2007MNRAS.375..240L}
{Li}, L.-X. 2007, \mnras, 375, 240

\bibitem[{{Liang} {et~al.}(2010){Liang}, {Yi}, {Zhang}, {L{\"u}}, {Zhang}, \&
  {Zhang}}]{2010ApJ...725.2209L}
{Liang}, E.-W., {Yi}, S.-X., {Zhang}, J., {et~al.} 2010, \apj, 725, 2209

\bibitem[{{Lloyd} \& {Petrosian}(2000)}]{2000ApJ...543..722L}
{Lloyd}, N.~M. \& {Petrosian}, V. 2000, \apj, 543, 722

\bibitem[{{Lodders} {et~al.}(2009){Lodders}, {Palme}, \& {Gail}}]{lodders09}
{Lodders}, K., {Palme}, H., \& {Gail}, H.-P. 2009, Landolt B{\"o}rnstein, 44

\bibitem[{{L{\"u}} {et~al.}(2015){L{\"u}}, {Zhang}, {Lei}, {Li}, \&
  {Lasky}}]{2015ApJ...805...89L}
{L{\"u}}, H.-J., {Zhang}, B., {Lei}, W.-H., {Li}, Y., \& {Lasky}, P.~D. 2015,
  \apj, 805, 89

\bibitem[{{Lyutikov} \& {Blandford}(2003)}]{2003astro.ph.12347L}
{Lyutikov}, M. \& {Blandford}, R. 2003, ArXiv Astrophysics e-prints

\bibitem[{{Madau} \& {Dickinson}(2014)}]{2014ARA&A..52..415M}
{Madau}, P. \& {Dickinson}, M. 2014, \araa, 52, 415

\bibitem[{{Madau} {et~al.}(1996){Madau}, {Ferguson}, {Dickinson}, {Giavalisco},
  {Steidel}, \& {Fruchter}}]{1996MNRAS.283.1388M}
{Madau}, P., {Ferguson}, H.~C., {Dickinson}, M.~E., {et~al.} 1996, \mnras, 283,
  1388

\bibitem[{{Magdis} {et~al.}(2013){Magdis}, {Rigopoulou}, {Helou}, {Farrah},
  {Hurley}, {Alonso-Herrero}, {Bock}, {Burgarella}, {Chapman}, {Charmandaris},
  {Cooray}, {Dai}, {Dale}, {Elbaz}, {Feltre}, {Hatziminaoglou}, {Huang},
  {Morrison}, {Oliver}, {Page}, {Scott}, \& {Shi}}]{2013A&A...558A.136M}
{Magdis}, G.~E., {Rigopoulou}, D., {Helou}, G., {et~al.} 2013, \aap, 558, A136

\bibitem[{{Mangano} {et~al.}(2007){Mangano}, {Holland}, {Malesani}, {Troja},
  {Chincarini}, {Zhang}, {La Parola}, {Brown}, {Burrows}, {Campana}, {Capalbi},
  {Cusumano}, {Della Valle}, {Gehrels}, {Giommi}, {Grupe}, {Guidorzi}, {Mineo},
  {Moretti}, {Osborne}, {Pandey}, {Perri}, {Romano}, {Roming}, \&
  {Tagliaferri}}]{2007A&A...470..105M}
{Mangano}, V., {Holland}, S.~T., {Malesani}, D., {et~al.} 2007, \aap, 470, 105

\bibitem[{{Mannucci} {et~al.}(2010){Mannucci}, {Cresci}, {Maiolino}, {Marconi},
  \& {Gnerucci}}]{mannucci10}
{Mannucci}, F., {Cresci}, G., {Maiolino}, R., {Marconi}, A., \& {Gnerucci}, A.
  2010, \mnras, 408, 2115

\bibitem[{{Mannucci} {et~al.}(2011){Mannucci}, {Salvaterra}, \&
  {Campisi}}]{2011MNRAS.414.1263M}
{Mannucci}, F., {Salvaterra}, R., \& {Campisi}, M.~A. 2011, \mnras, 414, 1263

\bibitem[{{Marggraf} {et~al.}(2004){Marggraf}, {Bluhm}, \& {de
  Boer}}]{2004A&A...416..251M}
{Marggraf}, O., {Bluhm}, H., \& {de Boer}, K.~S. 2004, \aap, 416, 251

\bibitem[{{Margutti} {et~al.}(2011){Margutti}, {Chincarini}, {Granot},
  {Guidorzi}, {Berger}, {Bernardini}, {Gehrels}, {Soderberg}, {Stamatikos}, \&
  {Zaninoni}}]{2011MNRAS.417.2144M}
{Margutti}, R., {Chincarini}, G., {Granot}, J., {et~al.} 2011, \mnras, 417,
  2144

\bibitem[{{Massey} {et~al.}(2005){Massey}, {Plez}, {Levesque}, {Olsen},
  {Clayton}, \& {Josselin}}]{2005ApJ...634.1286M}
{Massey}, P., {Plez}, B., {Levesque}, E.~M., {et~al.} 2005, \apj, 634, 1286

\bibitem[{{Mathis} {et~al.}(1977){Mathis}, {Rumpl}, \&
  {Nordsieck}}]{1977ApJ...217..425M}
{Mathis}, J.~S., {Rumpl}, W., \& {Nordsieck}, K.~H. 1977, \apj, 217, 425

\bibitem[{{Mattsson} {et~al.}(2014){Mattsson}, {De Cia}, {Andersen}, \&
  {Zafar}}]{mattsson}
{Mattsson}, L., {De Cia}, A., {Andersen}, A.~C., \& {Zafar}, T. 2014, \mnras,
  440, 1562

\bibitem[{{Mazzali} {et~al.}(2008){Mazzali}, {Valenti}, {Della Valle},
  {Chincarini}, {Sauer}, {Benetti}, {Pian}, {Piran}, {D'Elia}, {Elias-Rosa},
  {Margutti}, {Pasotti}, {Antonelli}, {Bufano}, {Campana}, {Cappellaro},
  {Covino}, {D'Avanzo}, {Fiore}, {Fugazza}, {Gilmozzi}, {Hunter}, {Maguire},
  {Maiorano}, {Marziani}, {Masetti}, {Mirabel}, {Navasardyan}, {Nomoto},
  {Palazzi}, {Pastorello}, {Panagia}, {Pellizza}, {Sari}, {Smartt},
  {Tagliaferri}, {Tanaka}, {Taubenberger}, {Tominaga}, {Trundle}, \&
  {Turatto}}]{2008Sci...321.1185M}
{Mazzali}, P.~A., {Valenti}, S., {Della Valle}, M., {et~al.} 2008, Science,
  321, 1185

\bibitem[{{McGaugh}(1991)}]{mcgaugh91}
{McGaugh}, S.~S. 1991, \apj, 380, 140

\bibitem[{{M{\'e}sz{\'a}ros} \& {Rees}(2001)}]{2001ApJ...556L..37M}
{M{\'e}sz{\'a}ros}, P. \& {Rees}, M.~J. 2001, \apjl, 556, L37

\bibitem[{{Metzger} {et~al.}(2008){Metzger}, {Quataert}, \&
  {Thompson}}]{2008MNRAS.385.1455M}
{Metzger}, B.~D., {Quataert}, E., \& {Thompson}, T.~A. 2008, \mnras, 385, 1455

\bibitem[{{Meurer} {et~al.}(1999){Meurer}, {Heckman}, \&
  {Calzetti}}]{1999ApJ...521...64M}
{Meurer}, G.~R., {Heckman}, T.~M., \& {Calzetti}, D. 1999, \apj, 521, 64

\bibitem[{{Micha{\l}owski} {et~al.}(2009){Micha{\l}owski}, {Hjorth},
  {Malesani}, {Micha{\l}owski}, {Castro Cer{\'o}n}, {Reinfrank}, {Garrett},
  {Fynbo}, {Watson}, \& {J{\o}rgensen}}]{michal}
{Micha{\l}owski}, M.~J., {Hjorth}, J., {Malesani}, D., {et~al.} 2009, \apj,
  693, 347

\bibitem[{{Milvang-Jensen} {et~al.}(2010){Milvang-Jensen}, {Goldoni}, {Tanvir},
  {Wiersema}, {Malesani}, {de Ugarte Postigo}, {D'Elia}, {Vergani}, {Fynbo},
  {Kaper}, {Sollerman}, {Updike}, {Greiner}, \&
  {Kruehler}}]{2010GCN..10876...1M}
{Milvang-Jensen}, B., {Goldoni}, P., {Tanvir}, N.~R., {et~al.} 2010, GRB
  Coordinates Network, 10876, 1

\bibitem[{{Mirabal} \& {Halpern}(2006)}]{2006GCN..4792....1M}
{Mirabal}, N. \& {Halpern}, J.~P. 2006, GRB Coordinates Network, 4792, 1

\bibitem[{{Modjaz} {et~al.}(2008){Modjaz}, {Kewley}, {Kirshner}, {Stanek},
  {Challis}, {Garnavich}, {Greene}, {Kelly}, \& {Prieto}}]{2008AJ....135.1136M}
{Modjaz}, M., {Kewley}, L., {Kirshner}, R.~P., {et~al.} 2008, \aj, 135, 1136

\bibitem[{{M{\o}ller}(2000)}]{moller}
{M{\o}ller}, P. 2000, The Messenger, 99, 31

\bibitem[{{M{\o}ller} {et~al.}(2013){M{\o}ller}, {Fynbo}, {Ledoux}, \&
  {Nilsson}}]{moller13}
{M{\o}ller}, P., {Fynbo}, J.~P.~U., {Ledoux}, C., \& {Nilsson}, K.~K. 2013,
  \mnras, 430, 2680

\bibitem[{{Monnier} {et~al.}(2002){Monnier}, {Tuthill}, \&
  {Danchi}}]{2002ApJ...567L.137M}
{Monnier}, J.~D., {Tuthill}, P.~G., \& {Danchi}, W.~C. 2002, \apjl, 567, L137

\bibitem[{{Moretti} {et~al.}(2008){Moretti}, {Margutti}, {Pasotti},
  {Beardmore}, {Campana}, {Chincarini}, {Covino}, {Godet}, {Guidorzi},
  {Osborne}, {Romano}, \& {Tagliaferri}}]{2008A&A...478..409M}
{Moretti}, A., {Margutti}, R., {Pasotti}, F., {et~al.} 2008, \aap, 478, 409

\bibitem[{{Murray}(2011)}]{murray11}
{Murray}, N. 2011, \apj, 729, 133

\bibitem[{{Nakar}(2007)}]{2007PhR...442..166N}
{Nakar}, E. 2007, \physrep, 442, 166

\bibitem[{{Nath} {et~al.}(2008){Nath}, {Laskar}, \&
  {Shull}}]{2008ApJ...682.1055N}
{Nath}, B.~B., {Laskar}, T., \& {Shull}, J.~M. 2008, \apj, 682, 1055

\bibitem[{{Nobili} \& {Goobar}(2008)}]{2008A&A...487...19N}
{Nobili}, S. \& {Goobar}, A. 2008, \aap, 487, 19

\bibitem[{{Nordon} {et~al.}(2013){Nordon}, {Lutz}, {Saintonge}, {Berta},
  {Wuyts}, {F{\"o}rster Schreiber}, {Genzel}, {Magnelli}, {Poglitsch},
  {Popesso}, {Rosario}, {Sturm}, \& {Tacconi}}]{2013ApJ...762..125N}
{Nordon}, R., {Lutz}, D., {Saintonge}, A., {et~al.} 2013, \apj, 762, 125

\bibitem[{{Norris} \& {Bonnell}(2006)}]{2006ApJ...643..266N}
{Norris}, J.~P. \& {Bonnell}, J.~T. 2006, \apj, 643, 266

\bibitem[{{Noterdaeme} {et~al.}(2008){Noterdaeme}, {Ledoux}, {Petitjean}, \&
  {Srianand}}]{noterdaeme08}
{Noterdaeme}, P., {Ledoux}, C., {Petitjean}, P., \& {Srianand}, R. 2008, \aap,
  481, 327

\bibitem[{{Noterdaeme} {et~al.}(2012){Noterdaeme}, {Petitjean}, {Carithers},
  {P{\^a}ris}, {Font-Ribera}, {Bailey}, {Aubourg}, {Bizyaev}, {Ebelke},
  {Finley}, {Ge}, {Malanushenko}, {Malanushenko}, {Miralda-Escud{\'e}},
  {Myers}, {Oravetz}, {Pan}, {Pieri}, {Ross}, {Schneider}, {Simmons}, \&
  {York}}]{noterdaeme12}
{Noterdaeme}, P., {Petitjean}, P., {Carithers}, W.~C., {et~al.} 2012, \aap,
  547, L1

\bibitem[{{Nousek} {et~al.}(2006){Nousek}, {Kouveliotou}, {Grupe}, {Page},
  {Granot}, {Ramirez-Ruiz}, {Patel}, {Burrows}, {Mangano}, {Barthelmy},
  {Beardmore}, {Campana}, {Capalbi}, {Chincarini}, {Cusumano}, {Falcone},
  {Gehrels}, {Giommi}, {Goad}, {Godet}, {Hurkett}, {Kennea}, {Moretti},
  {O'Brien}, {Osborne}, {Romano}, {Tagliaferri}, \&
  {Wells}}]{2006ApJ...642..389N}
{Nousek}, J.~A., {Kouveliotou}, C., {Grupe}, D., {et~al.} 2006, \apj, 642, 389

\bibitem[{{Nysewander} {et~al.}(2009){Nysewander}, {Fruchter}, \&
  {Pe'er}}]{2009ApJ...701..824N}
{Nysewander}, M., {Fruchter}, A.~S., \& {Pe'er}, A. 2009, \apj, 701, 824

\bibitem[{{Oates} {et~al.}(2007){Oates}, {de Pasquale}, {Page}, {Blustin},
  {Zane}, {McGowan}, {Mason}, {Poole}, {Schady}, {Roming}, {Page}, {Falcone},
  \& {Gehrels}}]{2007MNRAS.380..270O}
{Oates}, S.~R., {de Pasquale}, M., {Page}, M.~J., {et~al.} 2007, \mnras, 380,
  270

\bibitem[{{O'Brien} {et~al.}(2006){O'Brien}, {Willingale}, {Osborne}, \&
  {Goad}}]{2006NJPh....8..121O}
{O'Brien}, P.~T., {Willingale}, R., {Osborne}, J.~P., \& {Goad}, M.~R. 2006,
  New Journal of Physics, 8, 121

\bibitem[{{Offner} {et~al.}(2014){Offner}, {Clark}, {Hennebelle}, {Bastian},
  {Bate}, {Hopkins}, {Moraux}, \& {Whitworth}}]{2014prpl.conf...53O}
{Offner}, S.~S.~R., {Clark}, P.~C., {Hennebelle}, P., {et~al.} 2014, Protostars
  and Planets VI, 53

\bibitem[{{O'Meara} {et~al.}(2010){O'Meara}, {Chen}, \&
  {Prochaska}}]{2010GCN..11089...1O}
{O'Meara}, J., {Chen}, H.-W., \& {Prochaska}, J.~X. 2010, GRB Coordinates
  Network, 11089, 1

\bibitem[{{Omont} {et~al.}(2003){Omont}, {Beelen}, {Bertoldi}, {Cox},
  {Carilli}, {Priddey}, {McMahon}, \& {Isaak}}]{2003A&A...398..857O}
{Omont}, A., {Beelen}, A., {Bertoldi}, F., {et~al.} 2003, \aap, 398, 857

\bibitem[{{Omont} {et~al.}(2001){Omont}, {Cox}, {Bertoldi}, {McMahon},
  {Carilli}, \& {Isaak}}]{2001A&A...374..371O}
{Omont}, A., {Cox}, P., {Bertoldi}, F., {et~al.} 2001, \aap, 374, 371

\bibitem[{{Omukai}(2000)}]{2000ApJ...534..809O}
{Omukai}, K. 2000, \apj, 534, 809

\bibitem[{{Osterbrock}(1989)}]{balmer}
{Osterbrock}, D.~E. 1989, {Astrophysics of gaseous nebulae and active galactic
  nuclei}

\bibitem[{{Oteo} {et~al.}(2015){Oteo}, {Sobral}, {Ivison}, {Smail}, {Best},
  {Cepa}, \& {P{\'e}rez-Garc{\'{\i}}a}}]{2015arXiv150602670O}
{Oteo}, I., {Sobral}, D., {Ivison}, R.~J., {et~al.} 2015, ArXiv e-prints

\bibitem[{{Overzier} {et~al.}(2011){Overzier}, {Heckman}, {Wang}, {Armus},
  {Buat}, {Howell}, {Meurer}, {Seibert}, {Siana}, {Basu-Zych}, {Charlot},
  {Gon{\c c}alves}, {Martin}, {Neill}, {Rich}, {Salim}, \&
  {Schiminovich}}]{2011ApJ...726L...7O}
{Overzier}, R.~A., {Heckman}, T.~M., {Wang}, J., {et~al.} 2011, \apjl, 726, L7

\bibitem[{{Page} {et~al.}(2011){Page}, {Starling}, {Fitzpatrick}, {Pandey},
  {Osborne}, {Schady}, {McBreen}, {Campana}, {Ukwatta}, {Pagani}, {Beardmore},
  \& {Evans}}]{2011MNRAS.416.2078P}
{Page}, K.~L., {Starling}, R.~L.~C., {Fitzpatrick}, G., {et~al.} 2011, \mnras,
  416, 2078

\bibitem[{{Panaitescu}(2007)}]{2007MNRAS.380..374P}
{Panaitescu}, A. 2007, \mnras, 380, 374

\bibitem[{{Panaitescu} \& {M{\'e}sz{\'a}ros}(2000)}]{2000ApJ...544L..17P}
{Panaitescu}, A. \& {M{\'e}sz{\'a}ros}, P. 2000, \apjl, 544, L17

\bibitem[{{Panaitescu} {et~al.}(2006){Panaitescu}, {M{\'e}sz{\'a}ros},
  {Burrows}, {Nousek}, {Gehrels}, {O'Brien}, \&
  {Willingale}}]{2006MNRAS.369.2059P}
{Panaitescu}, A., {M{\'e}sz{\'a}ros}, P., {Burrows}, D., {et~al.} 2006, \mnras,
  369, 2059

\bibitem[{{Papovich} {et~al.}(2001){Papovich}, {Dickinson}, \&
  {Ferguson}}]{2001ApJ...559..620P}
{Papovich}, C., {Dickinson}, M., \& {Ferguson}, H.~C. 2001, \apj, 559, 620

\bibitem[{{Parsa} {et~al.}(2015){Parsa}, {Dunlop}, {McLure}, \&
  {Mortlock}}]{2015arXiv150705629P}
{Parsa}, S., {Dunlop}, J.~S., {McLure}, R.~J., \& {Mortlock}, A. 2015, ArXiv
  e-prints

\bibitem[{{Pe'er}(2008)}]{2008ApJ...682..463P}
{Pe'er}, A. 2008, \apj, 682, 463

\bibitem[{{Pe'er} {et~al.}(2006){Pe'er}, {M{\'e}sz{\'a}ros}, \&
  {Rees}}]{2006ApJ...652..482P}
{Pe'er}, A., {M{\'e}sz{\'a}ros}, P., \& {Rees}, M.~J. 2006, \apj, 652, 482

\bibitem[{{Pe'er} {et~al.}(2007{\natexlab{a}}){Pe'er}, {M{\'e}sz{\'a}ros}, \&
  {Rees}}]{2007RSPTA.365.1171P}
{Pe'er}, A., {M{\'e}sz{\'a}ros}, P., \& {Rees}, M.~J. 2007{\natexlab{a}}, Royal
  Society of London Philosophical Transactions Series A, 365, 1171

\bibitem[{{Pe'er} \& {Ryde}(2010)}]{2010arXiv1003.2582P}
{Pe'er}, A. \& {Ryde}, F. 2010, ArXiv e-prints

\bibitem[{{Pe'er} {et~al.}(2007{\natexlab{b}}){Pe'er}, {Ryde}, {Wijers},
  {M{\'e}sz{\'a}ros}, \& {Rees}}]{2007ApJ...664L...1P}
{Pe'er}, A., {Ryde}, F., {Wijers}, R.~A.~M.~J., {M{\'e}sz{\'a}ros}, P., \&
  {Rees}, M.~J. 2007{\natexlab{b}}, \apjl, 664, L1

\bibitem[{{Pe'er} {et~al.}(2012){Pe'er}, {Zhang}, {Ryde}, {McGlynn}, {Zhang},
  {Preece}, \& {Kouveliotou}}]{2012MNRAS.420..468P}
{Pe'er}, A., {Zhang}, B.-B., {Ryde}, F., {et~al.} 2012, \mnras, 420, 468

\bibitem[{{Pei}(1992)}]{pei}
{Pei}, Y.~C. 1992, \apj, 395, 130

\bibitem[{{Pei} {et~al.}(1991){Pei}, {Fall}, \&
  {Bechtold}}]{1991ApJ...378....6P}
{Pei}, Y.~C., {Fall}, S.~M., \& {Bechtold}, J. 1991, \apj, 378, 6

\bibitem[{{Perley} {et~al.}(2008){Perley}, {Bloom}, {Butler}, {Pollack},
  {Holtzman}, {Blake}, {Kocevski}, {Vestrand}, {Li}, {Foley}, {Bellm}, {Chen},
  {Prochaska}, {Starr}, {Filippenko}, {Falco}, {Szentgyorgyi}, {Wren},
  {Wozniak}, {White}, \& {Pergande}}]{2008ApJ...672..449P}
{Perley}, D.~A., {Bloom}, J.~S., {Butler}, N.~R., {et~al.} 2008, \apj, 672, 449

\bibitem[{{Perley} {et~al.}(2013){Perley}, {Levan}, {Tanvir}, {Cenko}, {Bloom},
  {Hjorth}, {Kr{\"u}hler}, {Filippenko}, {Fruchter}, {Fynbo}, {Jakobsson},
  {Kalirai}, {Milvang-Jensen}, {Morgan}, {Prochaska}, \&
  {Silverman}}]{perley13}
{Perley}, D.~A., {Levan}, A.~J., {Tanvir}, N.~R., {et~al.} 2013, \apj, 778, 128

\bibitem[{{Perley} {et~al.}(2011){Perley}, {Morgan}, {Updike}, {Yuan},
  {Akerlof}, {Miller}, {Bloom}, {Cenko}, {Li}, {Filippenko}, {Prochaska},
  {Kann}, {Tanvir}, {Levan}, {Butler}, {Christian}, {Hartmann}, {Milne},
  {Rykoff}, {Rujopakarn}, {Wheeler}, \& {Williams}}]{2011AJ....141...36P}
{Perley}, D.~A., {Morgan}, A.~N., {Updike}, A., {et~al.} 2011, \aj, 141, 36

\bibitem[{{Perley} {et~al.}(2015{\natexlab{a}}){Perley}, {Perley}, {Hjorth},
  {Micha{\l}owski}, {Cenko}, {Jakobsson}, {Kr{\"u}hler}, {Levan}, {Malesani},
  \& {Tanvir}}]{2015ApJ...801..102P}
{Perley}, D.~A., {Perley}, R.~A., {Hjorth}, J., {et~al.} 2015{\natexlab{a}},
  \apj, 801, 102

\bibitem[{{Perley} {et~al.}(2015{\natexlab{b}}){Perley}, {Tanvir}, {Hjorth},
  {Laskar}, {Berger}, {Chary}, {de Ugarte Postigo}, {Fynbo}, {Kr{\"u}hler},
  {Levan}, {Micha{\l}owski}, \& {Schulze}}]{2015arXiv150402479P}
{Perley}, D.~A., {Tanvir}, N.~R., {Hjorth}, J., {et~al.} 2015{\natexlab{b}},
  ArXiv e-prints

\bibitem[{{Perna} {et~al.}(2003){Perna}, {Lazzati}, \&
  {Fiore}}]{2003ApJ...585..775P}
{Perna}, R., {Lazzati}, D., \& {Fiore}, F. 2003, \apj, 585, 775

\bibitem[{{P{\'e}roux} {et~al.}(2012){P{\'e}roux}, {Bouch{\'e}}, {Kulkarni},
  {York}, \& {Vladilo}}]{peroux}
{P{\'e}roux}, C., {Bouch{\'e}}, N., {Kulkarni}, V.~P., {York}, D.~G., \&
  {Vladilo}, G. 2012, \mnras, 419, 3060

\bibitem[{{Petrov} {et~al.}(2015){Petrov}, {Gahm}, {Djupvik}, {Babina},
  {Artemenko}, \& {Grankin}}]{2015A&A...577A..73P}
{Petrov}, P.~P., {Gahm}, G.~F., {Djupvik}, A.~A., {et~al.} 2015, \aap, 577, A73

\bibitem[{{Pettini} {et~al.}(1998){Pettini}, {Kellogg}, {Steidel}, {Dickinson},
  {Adelberger}, \& {Giavalisco}}]{1998ApJ...508..539P}
{Pettini}, M., {Kellogg}, M., {Steidel}, C.~C., {et~al.} 1998, \apj, 508, 539

\bibitem[{{Pettini} \& {Pagel}(2004)}]{PP}
{Pettini}, M. \& {Pagel}, B.~E.~J. 2004, \mnras, 348, L59

\bibitem[{{Pettini} {et~al.}(2002){Pettini}, {Rix}, {Steidel}, {Adelberger},
  {Hunt}, \& {Shapley}}]{pettini02}
{Pettini}, M., {Rix}, S.~A., {Steidel}, C.~C., {et~al.} 2002, \apj, 569, 742

\bibitem[{{Pettini} {et~al.}(2001){Pettini}, {Shapley}, {Steidel}, {Cuby},
  {Dickinson}, {Moorwood}, {Adelberger}, \& {Giavalisco}}]{2001ApJ...554..981P}
{Pettini}, M., {Shapley}, A.~E., {Steidel}, C.~C., {et~al.} 2001, \apj, 554,
  981

\bibitem[{{Pettini} {et~al.}(1994){Pettini}, {Smith}, {Hunstead}, \&
  {King}}]{pettini94}
{Pettini}, M., {Smith}, L.~J., {Hunstead}, R.~W., \& {King}, D.~L. 1994, \apj,
  426, 79

\bibitem[{{Pian} {et~al.}(2006){Pian}, {Mazzali}, {Masetti}, {Ferrero},
  {Klose}, {Palazzi}, {Ramirez-Ruiz}, {Woosley}, {Kouveliotou}, {Deng},
  {Filippenko}, {Foley}, {Fynbo}, {Kann}, {Li}, {Hjorth}, {Nomoto}, {Patat},
  {Sauer}, {Sollerman}, {Vreeswijk}, {Guenther}, {Levan}, {O'Brien}, {Tanvir},
  {Wijers}, {Dumas}, {Hainaut}, {Wong}, {Baade}, {Wang}, {Amati}, {Cappellaro},
  {Castro-Tirado}, {Ellison}, {Frontera}, {Fruchter}, {Greiner}, {Kawabata},
  {Ledoux}, {Maeda}, {M{\o}ller}, {Nicastro}, {Rol}, \&
  {Starling}}]{2006Natur.442.1011P}
{Pian}, E., {Mazzali}, P.~A., {Masetti}, N., {et~al.} 2006, \nat, 442, 1011

\bibitem[{{Pilyugin}(2001)}]{2001A&A...374..412P}
{Pilyugin}, L.~S. 2001, \aap, 374, 412

\bibitem[{{Piran}(1999)}]{1999PhR...314..575P}
{Piran}, T. 1999, \physrep, 314, 575

\bibitem[{{Piran}(2004)}]{2004RvMP...76.1143P}
{Piran}, T. 2004, Reviews of Modern Physics, 76, 1143

\bibitem[{{Piro} {et~al.}(2014){Piro}, {Troja}, {Gendre}, {Ghisellini},
  {Ricci}, {Bannister}, {Fiore}, {Kidd}, {Piranomonte}, \&
  {Wieringa}}]{2014ApJ...790L..15P}
{Piro}, L., {Troja}, E., {Gendre}, B., {et~al.} 2014, \apjl, 790, L15

\bibitem[{{Pirronello} {et~al.}(2004){Pirronello}, {Manic{\'o}}, {Roser}, \&
  {Vidali}}]{2004ASPC..309..529P}
{Pirronello}, V., {Manic{\'o}}, G., {Roser}, J., \& {Vidali}, G. 2004, in
  Astronomical Society of the Pacific Conference Series, Vol. 309, Astrophysics
  of Dust, ed. A.~N. {Witt}, G.~C. {Clayton}, \& B.~T. {Draine}, 529

\bibitem[{{Planck Collaboration}(2015)}]{2015arXiv150201589P}
{Planck Collaboration}. 2015, ArXiv e-prints

\bibitem[{{Preece} {et~al.}(1998{\natexlab{a}}){Preece}, {Briggs}, {Mallozzi},
  {Pendleton}, {Paciesas}, \& {Band}}]{1998ApJ...506L..23P}
{Preece}, R.~D., {Briggs}, M.~S., {Mallozzi}, R.~S., {et~al.}
  1998{\natexlab{a}}, \apjl, 506, L23

\bibitem[{{Preece} {et~al.}(1998{\natexlab{b}}){Preece}, {Pendleton}, {Briggs},
  {Mallozzi}, {Paciesas}, {Band}, {Matteson}, \&
  {Meegan}}]{1998ApJ...496..849P}
{Preece}, R.~D., {Pendleton}, G.~N., {Briggs}, M.~S., {et~al.}
  1998{\natexlab{b}}, \apj, 496, 849

\bibitem[{{Price}(2006)}]{2006GCN..5104....1P}
{Price}, P.~A. 2006, GRB Coordinates Network, 5104, 1

\bibitem[{{Prochaska} {et~al.}(2006{\natexlab{a}}){Prochaska}, {Bloom}, {Chen},
  {Foley}, {Perley}, {Ramirez-Ruiz}, {Granot}, {Lee}, {Pooley}, {Alatalo},
  {Hurley}, {Cooper}, {Dupree}, {Gerke}, {Hansen}, {Kalirai}, {Newman}, {Rich},
  {Richer}, {Stanford}, {Stern}, \& {van Breugel}}]{2006ApJ...642..989P}
{Prochaska}, J.~X., {Bloom}, J.~S., {Chen}, H.-W., {et~al.} 2006{\natexlab{a}},
  \apj, 642, 989

\bibitem[{{Prochaska} {et~al.}(2006{\natexlab{b}}){Prochaska}, {Chen}, \&
  {Bloom}}]{prochaska06}
{Prochaska}, J.~X., {Chen}, H.-W., \& {Bloom}, J.~S. 2006{\natexlab{b}}, \apj,
  648, 95

\bibitem[{{Prochaska} {et~al.}(2007){Prochaska}, {Chen}, {Dessauges-Zavadsky},
  \& {Bloom}}]{prochaska07}
{Prochaska}, J.~X., {Chen}, H.-W., {Dessauges-Zavadsky}, M., \& {Bloom}, J.~S.
  2007, \apj, 666, 267

\bibitem[{{Prochaska} {et~al.}(2009){Prochaska}, {Sheffer}, {Perley}, {Bloom},
  {Lopez}, {Dessauges-Zavadsky}, {Chen}, {Filippenko}, {Ganeshalingam}, {Li},
  {Miller}, \& {Starr}}]{2009ApJ...691L..27P}
{Prochaska}, J.~X., {Sheffer}, Y., {Perley}, D.~A., {et~al.} 2009, \apjl, 691,
  L27

\bibitem[{{Prochaska} \& {Wolfe}(2002)}]{2002ApJ...566...68P}
{Prochaska}, J.~X. \& {Wolfe}, A.~M. 2002, \apj, 566, 68

\bibitem[{{Prochter} {et~al.}(2006){Prochter}, {Prochaska}, {Chen}, {Bloom},
  {Dessauges-Zavadsky}, {Foley}, {Lopez}, {Pettini}, {Dupree}, \&
  {Guhathakurta}}]{2006ApJ...648L..93P}
{Prochter}, G.~E., {Prochaska}, J.~X., {Chen}, H.-W., {et~al.} 2006, \apjl,
  648, L93

\bibitem[{{Racusin} {et~al.}(2008){Racusin}, {Karpov}, {Sokolowski}, {Granot},
  {Wu}, {Pal'Shin}, {Covino}, {van der Horst}, {Oates}, {Schady}, {Smith},
  {Cummings}, {Starling}, {Piotrowski}, {Zhang}, {Evans}, {Holland}, {Malek},
  {Page}, {Vetere}, {Margutti}, {Guidorzi}, {Kamble}, {Curran}, {Beardmore},
  {Kouveliotou}, {Mankiewicz}, {Melandri}, {O'Brien}, {Page}, {Piran},
  {Tanvir}, {Wrochna}, {Aptekar}, {Barthelmy}, {Bartolini}, {Beskin}, {Bondar},
  {Bremer}, {Campana}, {Castro-Tirado}, {Cucchiara}, {Cwiok}, {D'Avanzo},
  {D'Elia}, {Della Valle}, {de Ugarte Postigo}, {Dominik}, {Falcone}, {Fiore},
  {Fox}, {Frederiks}, {Fruchter}, {Fugazza}, {Garrett}, {Gehrels},
  {Golenetskii}, {Gomboc}, {Gorosabel}, {Greco}, {Guarnieri}, {Immler},
  {Jelinek}, {Kasprowicz}, {La Parola}, {Levan}, {Mangano}, {Mazets},
  {Molinari}, {Moretti}, {Nawrocki}, {Oleynik}, {Osborne}, {Pagani}, {Pandey},
  {Paragi}, {Perri}, {Piccioni}, {Ramirez-Ruiz}, {Roming}, {Steele}, {Strom},
  {Testa}, {Tosti}, {Ulanov}, {Wiersema}, {Wijers}, {Winters}, {Zarnecki},
  {Zerbi}, {M{\'e}sz{\'a}ros}, {Chincarini}, \&
  {Burrows}}]{2008Natur.455..183R}
{Racusin}, J.~L., {Karpov}, S.~V., {Sokolowski}, M., {et~al.} 2008, \nat, 455,
  183

\bibitem[{{Rafelski} {et~al.}(2012){Rafelski}, {Wolfe}, {Prochaska},
  {Neeleman}, \& {Mendez}}]{2012ApJ...755...89R}
{Rafelski}, M., {Wolfe}, A.~M., {Prochaska}, J.~X., {Neeleman}, M., \&
  {Mendez}, A.~J. 2012, \apj, 755, 89

\bibitem[{{Reddy} {et~al.}(2012){Reddy}, {Dickinson}, {Elbaz}, {Morrison},
  {Giavalisco}, {Ivison}, {Papovich}, {Scott}, {Buat}, {Burgarella},
  {Charmandaris}, {Daddi}, {Magdis}, {Murphy}, {Altieri}, {Aussel},
  {Dannerbauer}, {Dasyra}, {Hwang}, {Kartaltepe}, {Leiton}, {Magnelli}, \&
  {Popesso}}]{2012ApJ...744..154R}
{Reddy}, N., {Dickinson}, M., {Elbaz}, D., {et~al.} 2012, \apj, 744, 154

\bibitem[{{Rees} \& {Meszaros}(1992)}]{1992MNRAS.258P..41R}
{Rees}, M.~J. \& {Meszaros}, P. 1992, \mnras, 258, 41P

\bibitem[{{Rees} \& {Meszaros}(1994)}]{1994ApJ...430L..93R}
{Rees}, M.~J. \& {Meszaros}, P. 1994, \apjl, 430, L93

\bibitem[{{Reichart} {et~al.}(2005){Reichart}, {Nysewander}, {Moran},
  {Bartelme}, {Bayliss}, {Foster}, {Clemens}, {Price}, {Evans}, {Salmonson},
  {Trammell}, {Carney}, {Keohane}, \& {Gotwals}}]{prompt}
{Reichart}, D., {Nysewander}, M., {Moran}, J., {et~al.} 2005, Nuovo Cimento C
  Geophysics Space Physics C, 28, 767

\bibitem[{{Rhoads}(1997)}]{1997ApJ...487L...1R}
{Rhoads}, J.~E. 1997, \apjl, 487, L1

\bibitem[{{Robertson} \& {Ellis}(2012)}]{2012ApJ...744...95R}
{Robertson}, B.~E. \& {Ellis}, R.~S. 2012, \apj, 744, 95

\bibitem[{{Rodigas} {et~al.}(2012){Rodigas}, {Hinz}, {Leisenring},
  {Vaitheeswaran}, {Skemer}, {Skrutskie}, {Su}, {Bailey}, {Schneider}, {Close},
  {Mannucci}, {Esposito}, {Arcidiacono}, {Pinna}, {Argomedo}, {Agapito},
  {Apai}, {Bono}, {Boutsia}, {Briguglio}, {Brusa}, {Busoni}, {Cresci},
  {Currie}, {Desidera}, {Eisner}, {Falomo}, {Fini}, {Follette}, {Fontana},
  {Garnavich}, {Gratton}, {Green}, {Guerra}, {Hill}, {Hoffmann}, {Jones},
  {Krejny}, {Kulesa}, {Males}, {Masciadri}, {Mesa}, {McCarthy}, {Meyer},
  {Miller}, {Nelson}, {Puglisi}, {Quiros-Pacheco}, {Riccardi}, {Sani},
  {Stefanini}, {Testa}, {Wilson}, {Woodward}, \&
  {Xompero}}]{2012ApJ...752...57R}
{Rodigas}, T.~J., {Hinz}, P.~M., {Leisenring}, J., {et~al.} 2012, \apj, 752, 57

\bibitem[{{Rol} {et~al.}(2005){Rol}, {Wijers}, {Kouveliotou}, {Kaper}, \&
  {Kaneko}}]{2005ApJ...624..868R}
{Rol}, E., {Wijers}, R.~A.~M.~J., {Kouveliotou}, C., {Kaper}, L., \& {Kaneko},
  Y. 2005, \apj, 624, 868

\bibitem[{{Roman-Duval} {et~al.}(2014){Roman-Duval}, {Gordon}, {Meixner},
  {Bot}, {Bolatto}, {Hughes}, {Wong}, {Babler}, {Bernard}, {Clayton}, {Fukui},
  {Galametz}, {Galliano}, {Glover}, {Hony}, {Israel}, {Jameson},
  {Lebouteiller}, {Lee}, {Li}, {Madden}, {Misselt}, {Montiel}, {Okumura},
  {Onishi}, {Panuzzo}, {Reach}, {Remy-Ruyer}, {Robitaille}, {Rubio}, {Sauvage},
  {Seale}, {Sewilo}, {Staveley-Smith}, \& {Zhukovska}}]{2014ApJ...797...86R}
{Roman-Duval}, J., {Gordon}, K.~D., {Meixner}, M., {et~al.} 2014, \apj, 797, 86

\bibitem[{{Romano} {et~al.}(2006){Romano}, {Campana}, {Chincarini}, {Cummings},
  {Cusumano}, {Holland}, {Mangano}, {Mineo}, {Page}, {Pal'Shin}, {Rol},
  {Sakamoto}, {Zhang}, {Aptekar}, {Barbier}, {Barthelmy}, {Beardmore}, {Boyd},
  {Burrows}, {Capalbi}, {Fenimore}, {Frederiks}, {Gehrels}, {Giommi}, {Goad},
  {Godet}, {Golenetskii}, {Guetta}, {Kennea}, {La Parola}, {Malesani},
  {Marshall}, {Moretti}, {Nousek}, {O'Brien}, {Osborne}, {Perri}, \&
  {Tagliaferri}}]{2006A&A...456..917R}
{Romano}, P., {Campana}, S., {Chincarini}, G., {et~al.} 2006, \aap, 456, 917

\bibitem[{{Romero} {et~al.}(2010){Romero}, {Reynoso}, \&
  {Christiansen}}]{2010A&A...524A...4R}
{Romero}, G.~E., {Reynoso}, M.~M., \& {Christiansen}, H.~R. 2010, \aap, 524, A4

\bibitem[{{Rowlinson} {et~al.}(2010){Rowlinson}, {Wiersema}, {Levan}, {Tanvir},
  {O'Brien}, {Rol}, {Hjorth}, {Th{\"o}ne}, {de Ugarte Postigo}, {Fynbo},
  {Jakobsson}, {Pagani}, \& {Stamatikos}}]{2010MNRAS.408..383R}
{Rowlinson}, A., {Wiersema}, K., {Levan}, A.~J., {et~al.} 2010, \mnras, 408,
  383

\bibitem[{{Rybicki} \& {Lightman}(1979)}]{Rybicki}
{Rybicki}, G.~B. \& {Lightman}, A.~P. 1979, {Radiative Processes in
  Astrophysics} (John Wiley \& Sons, Inc.)

\bibitem[{{Ryde}(2004)}]{2004ApJ...614..827R}
{Ryde}, F. 2004, \apj, 614, 827

\bibitem[{{Ryde}(2005)}]{2005ApJ...625L..95R}
{Ryde}, F. 2005, \apjl, 625, L95

\bibitem[{{Ryde} {et~al.}(2010){Ryde}, {Axelsson}, {Zhang}, {McGlynn}, {Pe'er},
  {Lundman}, {Larsson}, {Battelino}, {Zhang}, {Bissaldi}, {Bregeon}, {Briggs},
  {Chiang}, {de Palma}, {Guiriec}, {Larsson}, {Longo}, {McBreen}, {Omodei},
  {Petrosian}, {Preece}, \& {van der Horst}}]{2010ApJ...709L.172R}
{Ryde}, F., {Axelsson}, M., {Zhang}, B.~B., {et~al.} 2010, \apjl, 709, L172

\bibitem[{{Ryde} {et~al.}(2005){Ryde}, {Kocevski}, {Bagoly}, {Ryde}, \&
  {M{\'e}sz{\'a}ros}}]{2005A&A...432..105R}
{Ryde}, F., {Kocevski}, D., {Bagoly}, Z., {Ryde}, N., \& {M{\'e}sz{\'a}ros}, A.
  2005, \aap, 432, 105

\bibitem[{{Ryde} \& {Pe'er}(2009)}]{2009ApJ...702.1211R}
{Ryde}, F. \& {Pe'er}, A. 2009, \apj, 702, 1211

\bibitem[{{Rykoff} {et~al.}(2006){Rykoff}, {Mangano}, {Yost}, {Sari},
  {Aharonian}, {Akerlof}, {Ashley}, {Barthelmy}, {Burrows}, {Gehrels},
  {G{\"o}{\v g}{\"u}{\c s}}, {G{\"u}ver}, {Horns}, {K{\i}z{\i}lo{\v g}lu},
  {Krimm}, {McKay}, {{\"O}zel}, {Phillips}, {Quimby}, {Rowell}, {Rujopakarn},
  {Schaefer}, {Smith}, {Swan}, {Vestrand}, {Wheeler}, {Wren}, \&
  {Yuan}}]{2006ApJ...638L...5R}
{Rykoff}, E.~S., {Mangano}, V., {Yost}, S.~A., {et~al.} 2006, \apjl, 638, L5

\bibitem[{{Salim} {et~al.}(2007){Salim}, {Rich}, {Charlot}, {Brinchmann},
  {Johnson}, {Schiminovich}, {Seibert}, {Mallery}, {Heckman}, {Forster},
  {Friedman}, {Martin}, {Morrissey}, {Neff}, {Small}, {Wyder}, {Bianchi},
  {Donas}, {Lee}, {Madore}, {Milliard}, {Szalay}, {Welsh}, \&
  {Yi}}]{2007ApJS..173..267S}
{Salim}, S., {Rich}, R.~M., {Charlot}, S., {et~al.} 2007, \apjs, 173, 267

\bibitem[{{Salpeter}(1955)}]{1955ApJ...121..161S}
{Salpeter}, E.~E. 1955, \apj, 121, 161

\bibitem[{{Salvaterra} {et~al.}(2009){Salvaterra}, {Della Valle}, {Campana},
  {Chincarini}, {Covino}, {D'Avanzo}, {Fern{\'a}ndez-Soto}, {Guidorzi},
  {Mannucci}, {Margutti}, {Th{\"o}ne}, {Antonelli}, {Barthelmy}, {de Pasquale},
  {D'Elia}, {Fiore}, {Fugazza}, {Hunt}, {Maiorano}, {Marinoni}, {Marshall},
  {Molinari}, {Nousek}, {Pian}, {Racusin}, {Stella}, {Amati}, {Andreuzzi},
  {Cusumano}, {Fenimore}, {Ferrero}, {Giommi}, {Guetta}, {Holland}, {Hurley},
  {Israel}, {Mao}, {Markwardt}, {Masetti}, {Pagani}, {Palazzi}, {Palmer},
  {Piranomonte}, {Tagliaferri}, \& {Testa}}]{2009Natur.461.1258S}
{Salvaterra}, R., {Della Valle}, M., {Campana}, S., {et~al.} 2009, \nat, 461,
  1258

\bibitem[{{Sana} {et~al.}(2012){Sana}, {Lacour}, {Le Bouquin}, {de Koter},
  {Moni-Bidin}, {Muijres}, {Schnurr}, \& {Zinnecker}}]{2012ASPC..465..363S}
{Sana}, H., {Lacour}, S., {Le Bouquin}, J., {et~al.} 2012, in Astronomical
  Society of the Pacific Conference Series, Vol. 465, Proceedings of a
  Scientific Meeting in Honor of Anthony F. J. Moffat, ed. L.~{Drissen},
  C.~{Robert}, N.~{St-Louis}, \& A.~F.~J. {Moffat}, 363

\bibitem[{{Sarangi} \& {Cherchneff}(2015)}]{2015A&A...575A..95S}
{Sarangi}, A. \& {Cherchneff}, I. 2015, \aap, 575, A95

\bibitem[{{Sari}(1998)}]{sari98}
{Sari}, R. 1998, \apjl, 494, L49

\bibitem[{{Sari} \& {Piran}(1997)}]{1997ApJ...485..270S}
{Sari}, R. \& {Piran}, T. 1997, \apj, 485, 270

\bibitem[{{Sari} \& {Piran}(1999{\natexlab{a}})}]{1999ApJ...520..641S}
{Sari}, R. \& {Piran}, T. 1999{\natexlab{a}}, \apj, 520, 641

\bibitem[{{Sari} \& {Piran}(1999{\natexlab{b}})}]{1999A&AS..138..537S}
{Sari}, R. \& {Piran}, T. 1999{\natexlab{b}}, \aaps, 138, 537

\bibitem[{{Sari} {et~al.}(1998){Sari}, {Piran}, \&
  {Narayan}}]{1998ApJ...497L..17S}
{Sari}, R., {Piran}, T., \& {Narayan}, R. 1998, \apjl, 497, L17

\bibitem[{{Savage} \& {Sembach}(1996)}]{SS96}
{Savage}, B.~D. \& {Sembach}, K.~R. 1996, \araa, 34, 279

\bibitem[{{Savaglio}(2001)}]{savaglio01}
{Savaglio}, S. 2001, in IAU Symposium, Vol. 204, The Extragalactic Infrared
  Background and its Cosmological Implications, ed. M.~{Harwit} \& M.~G.
  {Hauser}, 307

\bibitem[{{Savaglio} \& {Fall}(2004)}]{2004ApJ...614..293S}
{Savaglio}, S. \& {Fall}, S.~M. 2004, \apj, 614, 293

\bibitem[{{Savaglio} {et~al.}(2003){Savaglio}, {Fall}, \&
  {Fiore}}]{2003ApJ...585..638S}
{Savaglio}, S., {Fall}, S.~M., \& {Fiore}, F. 2003, \apj, 585, 638

\bibitem[{{Savaglio} {et~al.}(2009){Savaglio}, {Glazebrook}, \& {Le
  Borgne}}]{2009ApJ...691..182S}
{Savaglio}, S., {Glazebrook}, K., \& {Le Borgne}, D. 2009, \apj, 691, 182

\bibitem[{{Savaglio} {et~al.}(2012){Savaglio}, {Rau}, {Greiner}, {Kr{\"u}hler},
  {McBreen}, {Hartmann}, {Updike}, {Filgas}, {Klose}, {Afonso}, {Clemens},
  {K{\"u}pc{\"u} Yolda{\c s}}, {Olivares E.}, {Sudilovsky}, \&
  {Szokoly}}]{savaglio12}
{Savaglio}, S., {Rau}, A., {Greiner}, J., {et~al.} 2012, \mnras, 420, 627

\bibitem[{{Schady} {et~al.}(2012){Schady}, {Dwelly}, {Page}, {Kr{\"u}hler},
  {Greiner}, {Oates}, {de Pasquale}, {Nardini}, {Roming}, {Rossi}, \&
  {Still}}]{schady12}
{Schady}, P., {Dwelly}, T., {Page}, M.~J., {et~al.} 2012, \aap, 537, A15

\bibitem[{{Schady} {et~al.}(2015){Schady}, {Kr{\"u}hler}, {Greiner}, {Graham},
  {Kann}, {Bolmer}, {Delvaux}, {Elliott}, {Klose}, {Knust}, {Nicuesa
  Guelbenzu}, {Rau}, {Rossi}, {Savaglio}, {Schmidl}, {Schweyer}, {Sudilovsky},
  {Tanga}, {Tanvir}, {Varela}, \& {Wiseman}}]{2015A&A...579A.126S}
{Schady}, P., {Kr{\"u}hler}, T., {Greiner}, J., {et~al.} 2015, \aap, 579, A126

\bibitem[{{Schaerer} {et~al.}(2015){Schaerer}, {Boone}, {Zamojski}, {Staguhn},
  {Dessauges-Zavadsky}, {Finkelstein}, \& {Combes}}]{2015A&A...574A..19S}
{Schaerer}, D., {Boone}, F., {Zamojski}, M., {et~al.} 2015, \aap, 574, A19

\bibitem[{{Schanne} {et~al.}(2010){Schanne}, {Paul}, {Wei}, {Zhang}, {Basa},
  {Atteia}, {Barret}, {Claret}, {Cordier}, {Daigne}, {Evans}, {Fraser},
  {Godet}, {G{\"o}tz}, {Mandrou}, \& {Osborne}}]{2010arXiv1005.5008S}
{Schanne}, S., {Paul}, J., {Wei}, J., {et~al.} 2010, ArXiv e-prints

\bibitem[{{Schlegel} {et~al.}(1998){Schlegel}, {Finkbeiner}, \&
  {Davis}}]{schlegel}
{Schlegel}, D.~J., {Finkbeiner}, D.~P., \& {Davis}, M. 1998, \apj, 500, 525

\bibitem[{{Schnee} {et~al.}(2008){Schnee}, {Li}, {Goodman}, \&
  {Sargent}}]{2008ApJ...684.1228S}
{Schnee}, S., {Li}, J., {Goodman}, A.~A., \& {Sargent}, A.~I. 2008, \apj, 684,
  1228

\bibitem[{{Schneider} {et~al.}(2012){Schneider}, {Omukai}, {Limongi},
  {Ferrara}, {Salvaterra}, {Chieffi}, \& {Bianchi}}]{2012MNRAS.423L..60S}
{Schneider}, R., {Omukai}, K., {Limongi}, M., {et~al.} 2012, \mnras, 423, L60

\bibitem[{{Schulze} {et~al.}(2014){Schulze}, {Malesani}, {Cucchiara}, {Tanvir},
  {Kr{\"u}hler}, {de Ugarte Postigo}, {Leloudas}, {Lyman}, {Bersier},
  {Wiersema}, {Perley}, {Schady}, {Gorosabel}, {Anderson}, {Castro-Tirado},
  {Cenko}, {De Cia}, {Ellerbroek}, {Fynbo}, {Greiner}, {Hjorth}, {Kann},
  {Kaper}, {Klose}, {Levan}, {Mart{\'{\i}}n}, {O'Brien}, {Page}, {Pignata},
  {Rapaport}, {S{\'a}nchez-Ram{\'{\i}}rez}, {Sollerman}, {Smith}, {Sparre},
  {Th{\"o}ne}, {Watson}, {Xu}, {Bauer}, {Bayliss}, {Bj{\"o}rnsson}, {Bremer},
  {Cano}, {Covino}, {D'Elia}, {Frail}, {Geier}, {Goldoni}, {Hartoog},
  {Jakobsson}, {Korhonen}, {Lee}, {Milvang-Jensen}, {Nardini}, {Nicuesa
  Guelbenzu}, {Oguri}, {Pandey}, {Petitpas}, {Rossi}, {Sandberg}, {Schmidl},
  {Tagliaferri}, {Tilanus}, {Winters}, {Wright}, \& {Wuyts}}]{schulze14}
{Schulze}, S., {Malesani}, D., {Cucchiara}, A., {et~al.} 2014, \aap, 566, A102

\bibitem[{{Sheffer} {et~al.}(2009){Sheffer}, {Prochaska}, {Draine}, {Perley},
  \& {Bloom}}]{sheffer09}
{Sheffer}, Y., {Prochaska}, J.~X., {Draine}, B.~T., {Perley}, D.~A., \&
  {Bloom}, J.~S. 2009, \apjl, 701, L63

\bibitem[{{Shustov} \& {Vibe}(1995)}]{1995AZh....72..650S}
{Shustov}, B.~M. \& {Vibe}, D.~Z. 1995, \azh, 72, 650

\bibitem[{{Silvia} {et~al.}(2010){Silvia}, {Smith}, \&
  {Shull}}]{2010ApJ...715.1575S}
{Silvia}, D.~W., {Smith}, B.~D., \& {Shull}, J.~M. 2010, \apj, 715, 1575

\bibitem[{{Smette} {et~al.}(2010){Smette}, {Sana}, \& {Horst}}]{smette}
{Smette}, A., {Sana}, H., \& {Horst}, H. 2010, Highlights of Astronomy, 15, 533

\bibitem[{{Smith} {et~al.}(2002){Smith}, {Tucker}, {Kent}, {Richmond},
  {Fukugita}, {Ichikawa}, {Ichikawa}, {Jorgensen}, {Uomoto}, {Gunn}, {Hamabe},
  {Watanabe}, {Tolea}, {Henden}, {Annis}, {Pier}, {McKay}, {Brinkmann}, {Chen},
  {Holtzman}, {Shimasaku}, \& {York}}]{smith02}
{Smith}, J.~A., {Tucker}, D.~L., {Kent}, S., {et~al.} 2002, \aj, 123, 2121

\bibitem[{{Soderberg} {et~al.}(2008){Soderberg}, {Berger}, {Page}, {Schady},
  {Parrent}, {Pooley}, {Wang}, {Ofek}, {Cucchiara}, {Rau}, {Waxman}, {Simon},
  {Bock}, {Milne}, {Page}, {Barentine}, {Barthelmy}, {Beardmore}, {Bietenholz},
  {Brown}, {Burrows}, {Burrows}, {Byrngelson}, {Cenko}, {Chandra}, {Cummings},
  {Fox}, {Gal-Yam}, {Gehrels}, {Immler}, {Kasliwal}, {Kong}, {Krimm},
  {Kulkarni}, {Maccarone}, {M{\'e}sz{\'a}ros}, {Nakar}, {O'Brien}, {Overzier},
  {de Pasquale}, {Racusin}, {Rea}, \& {York}}]{2008Natur.453..469S}
{Soderberg}, A.~M., {Berger}, E., {Page}, K.~L., {et~al.} 2008, \nat, 453, 469

\bibitem[{{Sollerman} {et~al.}(2007){Sollerman}, {Fynbo}, {Gorosabel},
  {Halpern}, {Hjorth}, {Jakobsson}, {Mirabal}, {Watson}, {Xu}, {Castro-Tirado},
  {F{\'e}ron}, {Jaunsen}, {Jel{\'{\i}}nek}, {Jensen}, {Kann}, {Ovaldsen},
  {Pozanenko}, {Stritzinger}, {Th{\"o}ne}, {de Ugarte Postigo}, {Guziy},
  {Ibrahimov}, {J{\"a}rvinen}, {Levan}, {Rumyantsev}, \&
  {Tanvir}}]{2007A&A...466..839S}
{Sollerman}, J., {Fynbo}, J.~P.~U., {Gorosabel}, J., {et~al.} 2007, \aap, 466,
  839

\bibitem[{{Sollerman} {et~al.}(2006){Sollerman}, {Jaunsen}, {Fynbo}, {Hjorth},
  {Jakobsson}, {Stritzinger}, {F{\'e}ron}, {Laursen}, {Ovaldsen}, {Selj},
  {Th{\"o}ne}, {Xu}, {Davis}, {Gorosabel}, {Watson}, {Duro}, {Ilyin}, {Jensen},
  {Lysfjord}, {Marquart}, {Nielsen}, {N{\"a}r{\"a}nen}, {Schwarz}, {Walch},
  {Wold}, \& {{\"O}stlin}}]{2006A&A...454..503S}
{Sollerman}, J., {Jaunsen}, A.~O., {Fynbo}, J.~P.~U., {et~al.} 2006, \aap, 454,
  503

\bibitem[{{Sparre} {et~al.}(2014){Sparre}, {Hartoog}, {Kr{\"u}hler}, {Fynbo},
  {Watson}, {Wiersema}, {D'Elia}, {Zafar}, {Afonso}, {Covino}, {de Ugarte
  Postigo}, {Flores}, {Goldoni}, {Greiner}, {Hjorth}, {Jakobsson}, {Kaper},
  {Klose}, {Levan}, {Malesani}, {Milvang-Jensen}, {Nardini}, {Piranomonte},
  {Sollerman}, {S{\'a}nchez-Ram{\'{\i}}rez}, {Schulze}, {Tanvir}, {Vergani}, \&
  {Wijers}}]{sparre13}
{Sparre}, M., {Hartoog}, O.~E., {Kr{\"u}hler}, T., {et~al.} 2014, \apj, 785,
  150

\bibitem[{{Sparre} {et~al.}(2011){Sparre}, {Sollerman}, {Fynbo}, {Malesani},
  {Goldoni}, {de Ugarte Postigo}, {Covino}, {D'Elia}, {Flores}, {Hammer},
  {Hjorth}, {Jakobsson}, {Kaper}, {Leloudas}, {Levan}, {Milvang-Jensen},
  {Schulze}, {Tagliaferri}, {Tanvir}, {Watson}, {Wiersema}, \&
  {Wijers}}]{sparre11}
{Sparre}, M., {Sollerman}, J., {Fynbo}, J.~P.~U., {et~al.} 2011, \apjl, 735,
  L24

\bibitem[{{Sparre} \& {Starling}(2012)}]{2012MNRAS.427.2965S}
{Sparre}, M. \& {Starling}, R.~L.~C. 2012, \mnras, 427, 2965

\bibitem[{{Springel} {et~al.}(2005){Springel}, {White}, {Jenkins}, {Frenk},
  {Yoshida}, {Gao}, {Navarro}, {Thacker}, {Croton}, {Helly}, {Peacock}, {Cole},
  {Thomas}, {Couchman}, {Evrard}, {Colberg}, \& {Pearce}}]{2005Natur.435..629S}
{Springel}, V., {White}, S.~D.~M., {Jenkins}, A., {et~al.} 2005, \nat, 435, 629

\bibitem[{{Stanek} {et~al.}(2003){Stanek}, {Matheson}, {Garnavich}, {Martini},
  {Berlind}, {Caldwell}, {Challis}, {Brown}, {Schild}, {Krisciunas}, {Calkins},
  {Lee}, {Hathi}, {Jansen}, {Windhorst}, {Echevarria}, {Eisenstein}, {Pindor},
  {Olszewski}, {Harding}, {Holland}, \& {Bersier}}]{2003ApJ...591L..17S}
{Stanek}, K.~Z., {Matheson}, T., {Garnavich}, P.~M., {et~al.} 2003, \apjl, 591,
  L17

\bibitem[{{Starling} {et~al.}(2012){Starling}, {Page}, {Pe'Er}, {Beardmore}, \&
  {Osborne}}]{2012MNRAS.427.2950S}
{Starling}, R.~L.~C., {Page}, K.~L., {Pe'Er}, A., {Beardmore}, A.~P., \&
  {Osborne}, J.~P. 2012, \mnras, 427, 2950

\bibitem[{{Starling} {et~al.}(2008){Starling}, {van der Horst}, {Rol},
  {Wijers}, {Kouveliotou}, {Wiersema}, {Curran}, \&
  {Weltevrede}}]{2008ApJ...672..433S}
{Starling}, R.~L.~C., {van der Horst}, A.~J., {Rol}, E., {et~al.} 2008, \apj,
  672, 433

\bibitem[{{Starling} {et~al.}(2011){Starling}, {Wiersema}, {Levan}, {Sakamoto},
  {Bersier}, {Goldoni}, {Oates}, {Rowlinson}, {Campana}, {Sollerman}, {Tanvir},
  {Malesani}, {Fynbo}, {Covino}, {D'Avanzo}, {O'Brien}, {Page}, {Osborne},
  {Vergani}, {Barthelmy}, {Burrows}, {Cano}, {Curran}, {de Pasquale}, {D'Elia},
  {Evans}, {Flores}, {Fruchter}, {Garnavich}, {Gehrels}, {Gorosabel}, {Hjorth},
  {Holland}, {van der Horst}, {Hurkett}, {Jakobsson}, {Kamble}, {Kouveliotou},
  {Kuin}, {Kaper}, {Mazzali}, {Nugent}, {Pian}, {Stamatikos}, {Th{\"o}ne}, \&
  {Woosley}}]{2011MNRAS.411.2792S}
{Starling}, R.~L.~C., {Wiersema}, K., {Levan}, A.~J., {et~al.} 2011, \mnras,
  411, 2792

\bibitem[{{Starling} {et~al.}(2007){Starling}, {Wijers}, {Wiersema}, {Rol},
  {Curran}, {Kouveliotou}, {van der Horst}, \&
  {Heemskerk}}]{2007ApJ...661..787S}
{Starling}, R.~L.~C., {Wijers}, R.~A.~M.~J., {Wiersema}, K., {et~al.} 2007,
  \apj, 661, 787

\bibitem[{{Stasi{\'n}ska}(2002)}]{2002RMxAC..12...62S}
{Stasi{\'n}ska}, G. 2002, in Revista Mexicana de Astronomia y Astrofisica
  Conference Series, Vol.~12, Revista Mexicana de Astronomia y Astrofisica
  Conference Series, ed. W.~J. {Henney}, J.~{Franco}, \& M.~{Martos}, 62--69

\bibitem[{{Stasi{\'n}ska}(2005)}]{2005A&A...434..507S}
{Stasi{\'n}ska}, G. 2005, \aap, 434, 507

\bibitem[{{Steidel} {et~al.}(1995){Steidel}, {Pettini}, \&
  {Hamilton}}]{1995AJ....110.2519S}
{Steidel}, C.~C., {Pettini}, M., \& {Hamilton}, D. 1995, \aj, 110, 2519

\bibitem[{{Stratta} {et~al.}(2005){Stratta}, {Perna}, {Lazzati}, {Fiore},
  {Antonelli}, \& {Conciatore}}]{2005A&A...441...83S}
{Stratta}, G., {Perna}, R., {Lazzati}, D., {et~al.} 2005, \aap, 441, 83

\bibitem[{{Sun} {et~al.}(2012){Sun}, {Liu}, {Gu}, \&
  {Lu}}]{2012ApJ...752...31S}
{Sun}, M.-Y., {Liu}, T., {Gu}, W.-M., \& {Lu}, J.-F. 2012, \apj, 752, 31

\bibitem[{{Suzuki} \& {Shigeyama}(2013)}]{2013ApJ...764L..12S}
{Suzuki}, A. \& {Shigeyama}, T. 2013, \apjl, 764, L12

\bibitem[{{Tanaka} {et~al.}(2009){Tanaka}, {Yamanaka}, {Maeda}, {Kawabata},
  {Hattori}, {Minezaki}, {Valenti}, {Della Valle}, {Sahu}, {Anupama},
  {Tominaga}, {Nomoto}, {Mazzali}, \& {Pian}}]{2009ApJ...700.1680T}
{Tanaka}, M., {Yamanaka}, M., {Maeda}, K., {et~al.} 2009, \apj, 700, 1680

\bibitem[{{Tanvir} {et~al.}(2009){Tanvir}, {Fox}, {Levan}, {Berger},
  {Wiersema}, {Fynbo}, {Cucchiara}, {Kr{\"u}hler}, {Gehrels}, {Bloom},
  {Greiner}, {Evans}, {Rol}, {Olivares}, {Hjorth}, {Jakobsson}, {Farihi},
  {Willingale}, {Starling}, {Cenko}, {Perley}, {Maund}, {Duke}, {Wijers},
  {Adamson}, {Allan}, {Bremer}, {Burrows}, {Castro-Tirado}, {Cavanagh}, {de
  Ugarte Postigo}, {Dopita}, {Fatkhullin}, {Fruchter}, {Foley}, {Gorosabel},
  {Kennea}, {Kerr}, {Klose}, {Krimm}, {Komarova}, {Kulkarni}, {Moskvitin},
  {Mundell}, {Naylor}, {Page}, {Penprase}, {Perri}, {Podsiadlowski}, {Roth},
  {Rutledge}, {Sakamoto}, {Schady}, {Schmidt}, {Soderberg}, {Sollerman},
  {Stephens}, {Stratta}, {Ukwatta}, {Watson}, {Westra}, {Wold}, \&
  {Wolf}}]{2009Natur.461.1254T}
{Tanvir}, N.~R., {Fox}, D.~B., {Levan}, A.~J., {et~al.} 2009, \nat, 461, 1254

\bibitem[{{Tanvir} {et~al.}(2013){Tanvir}, {Levan}, {Fruchter}, {Hjorth},
  {Hounsell}, {Wiersema}, \& {Tunnicliffe}}]{2013Natur.500..547T}
{Tanvir}, N.~R., {Levan}, A.~J., {Fruchter}, A.~S., {et~al.} 2013, \nat, 500,
  547

\bibitem[{{Tavani}(1996)}]{1996ApJ...466..768T}
{Tavani}, M. 1996, \apj, 466, 768

\bibitem[{{Th{\"o}ne} {et~al.}(2011){Th{\"o}ne}, {de Ugarte Postigo}, {Fryer},
  {Page}, {Gorosabel}, {Aloy}, {Perley}, {Kouveliotou}, {Janka}, {Mimica},
  {Racusin}, {Krimm}, {Cummings}, {Oates}, {Holland}, {Siegel}, {de Pasquale},
  {Sonbas}, {Im}, {Park}, {Kann}, {Guziy}, {Garc{\'{\i}}a}, {Llorente},
  {Bundy}, {Choi}, {Jeong}, {Korhonen}, {Kub{\`a}nek}, {Lim}, {Moskvitin},
  {Mu{\~n}oz-Darias}, {Pak}, \& {Parrish}}]{2011Natur.480...72T}
{Th{\"o}ne}, C.~C., {de Ugarte Postigo}, A., {Fryer}, C.~L., {et~al.} 2011,
  \nat, 480, 72

\bibitem[{{Tody}(1997)}]{tody97}
{Tody}, D. 1997, in Astronomical Society of the Pacific Conference Series, Vol.
  125, Astronomical Data Analysis Software and Systems VI, ed. G.~{Hunt} \&
  H.~{Payne}, 451

\bibitem[{{Treu} {et~al.}(2010){Treu}, {Auger}, {Koopmans}, {Gavazzi},
  {Marshall}, \& {Bolton}}]{2010ApJ...709.1195T}
{Treu}, T., {Auger}, M.~W., {Koopmans}, L.~V.~E., {et~al.} 2010, \apj, 709,
  1195

\bibitem[{{Treyer} {et~al.}(2007){Treyer}, {Schiminovich}, {Johnson},
  {Seibert}, {Wyder}, {Barlow}, {Conrow}, {Forster}, {Friedman}, {Martin},
  {Morrissey}, {Neff}, {Small}, {Bianchi}, {Donas}, {Heckman}, {Lee}, {Madore},
  {Milliard}, {Rich}, {Szalay}, {Welsh}, \& {Yi}}]{treyer07}
{Treyer}, M., {Schiminovich}, D., {Johnson}, B., {et~al.} 2007, \apjs, 173, 256

\bibitem[{{Tumlinson} {et~al.}(2007){Tumlinson}, {Prochaska}, {Chen},
  {Dessauges-Zavadsky}, \& {Bloom}}]{tumlinson07}
{Tumlinson}, J., {Prochaska}, J.~X., {Chen}, H.-W., {Dessauges-Zavadsky}, M.,
  \& {Bloom}, J.~S. 2007, \apj, 668, 667

\bibitem[{{Uhm} \& {Beloborodov}(2007)}]{2007ApJ...665L..93U}
{Uhm}, Z.~L. \& {Beloborodov}, A.~M. 2007, \apjl, 665, L93

\bibitem[{{Uhm} \& {Zhang}(2014)}]{2014NatPh..10..351U}
{Uhm}, Z.~L. \& {Zhang}, B. 2014, Nature Physics, 10, 351

\bibitem[{{van der Horst} {et~al.}(2011){van der Horst}, {Kamble}, {Paragi},
  {Sage}, {Pal}, {Taylor}, {Kouveliotou}, {Granot}, {Ramirez-Ruiz},
  {Ishwara-Chandra}, {Oosterloo}, {Wijers}, {Wiersema}, {Strom},
  {Bhattacharya}, {Rol}, {Starling}, {Curran}, \&
  {Garrett}}]{2011ApJ...726...99V}
{van der Horst}, A.~J., {Kamble}, A.~P., {Paragi}, Z., {et~al.} 2011, \apj,
  726, 99

\bibitem[{{van Eerten}(2014)}]{2014MNRAS.445.2414V}
{van Eerten}, H.~J. 2014, \mnras, 445, 2414

\bibitem[{{van Paradijs} {et~al.}(1997){van Paradijs}, {Groot}, {Galama},
  {Kouveliotou}, {Strom}, {Telting}, {Rutten}, {Fishman}, {Meegan}, {Pettini},
  {Tanvir}, {Bloom}, {Pedersen}, {N{\o}rdgaard-Nielsen}, {Linden-V{\o}rnle},
  {Melnick}, {van der Steene}, {Bremer}, {Naber}, {Heise}, {in't Zand},
  {Costa}, {Feroci}, {Piro}, {Frontera}, {Zavattini}, {Nicastro}, {Palazzi},
  {Bennett}, {Hanlon}, \& {Parmar}}]{1997Natur.386..686V}
{van Paradijs}, J., {Groot}, P.~J., {Galama}, T., {et~al.} 1997, \nat, 386, 686

\bibitem[{{Vernet} {et~al.}(2011){Vernet}, {Dekker}, {D'Odorico}, {Kaper},
  {Kjaergaard}, {Hammer}, {Randich}, {Zerbi}, {Groot}, {Hjorth}, {Guinouard},
  {Navarro}, {Adolfse}, {Albers}, {Amans}, {Andersen}, {Andersen}, {Binetruy},
  {Bristow}, {Castillo}, {Chemla}, {Christensen}, {Conconi}, {Conzelmann},
  {Dam}, {de Caprio}, {de Ugarte Postigo}, {Delabre}, {di Marcantonio},
  {Downing}, {Elswijk}, {Finger}, {Fischer}, {Flores}, {Fran{\c c}ois},
  {Goldoni}, {Guglielmi}, {Haigron}, {Hanenburg}, {Hendriks}, {Horrobin},
  {Horville}, {Jessen}, {Kerber}, {Kern}, {Kiekebusch}, {Kleszcz}, {Klougart},
  {Kragt}, {Larsen}, {Lizon}, {Lucuix}, {Mainieri}, {Manuputy}, {Martayan},
  {Mason}, {Mazzoleni}, {Michaelsen}, {Modigliani}, {Moehler}, {M{\o}ller},
  {Norup S{\o}rensen}, {N{\o}rregaard}, {P{\'e}roux}, {Patat}, {Pena}, {Pragt},
  {Reinero}, {Rigal}, {Riva}, {Roelfsema}, {Royer}, {Sacco}, {Santin},
  {Schoenmaker}, {Spano}, {Sweers}, {Ter Horst}, {Tintori}, {Tromp}, {van
  Dael}, {van der Vliet}, {Venema}, {Vidali}, {Vinther}, {Vola}, {Winters},
  {Wistisen}, {Wulterkens}, \& {Zacchei}}]{xshooter}
{Vernet}, J., {Dekker}, H., {D'Odorico}, S., {et~al.} 2011, \aap, 536, A105

\bibitem[{{Vestrand} {et~al.}(2005){Vestrand}, {Wozniak}, {Wren}, {Fenimore},
  {Sakamoto}, {White}, {Casperson}, {Davis}, {Evans}, {Galassi}, {McGowan},
  {Schier}, {Asa}, {Barthelmy}, {Cummings}, {Gehrels}, {Hullinger}, {Krimm},
  {Markwardt}, {McLean}, {Palmer}, {Parsons}, \&
  {Tueller}}]{2005Natur.435..178V}
{Vestrand}, W.~T., {Wozniak}, P.~R., {Wren}, J.~A., {et~al.} 2005, \nat, 435,
  178

\bibitem[{{Vestrand} {et~al.}(2014){Vestrand}, {Wren}, {Panaitescu}, {Wozniak},
  {Davis}, {Palmer}, {Vianello}, {Omodei}, {Xiong}, {Briggs}, {Elphick},
  {Paciesas}, \& {Rosing}}]{2014Sci...343...38V}
{Vestrand}, W.~T., {Wren}, J.~A., {Panaitescu}, A., {et~al.} 2014, Science,
  343, 38

\bibitem[{{Vestrand} {et~al.}(2006){Vestrand}, {Wren}, {Wozniak}, {Aptekar},
  {Golentskii}, {Pal'Shin}, {Sakamoto}, {White}, {Evans}, {Casperson}, \&
  {Fenimore}}]{2006Natur.442..172V}
{Vestrand}, W.~T., {Wren}, J.~A., {Wozniak}, P.~R., {et~al.} 2006, \nat, 442,
  172

\bibitem[{{Vink} {et~al.}(2001){Vink}, {de Koter}, \&
  {Lamers}}]{2001A&A...369..574V}
{Vink}, J.~S., {de Koter}, A., \& {Lamers}, H.~J.~G.~L.~M. 2001, \aap, 369, 574

\bibitem[{{Virgili} {et~al.}(2013){Virgili}, {Mundell}, {Pal'shin}, {Guidorzi},
  {Margutti}, {Melandri}, {Harrison}, {Kobayashi}, {Chornock}, {Henden},
  {Updike}, {Cenko}, {Tanvir}, {Steele}, {Cucchiara}, {Gomboc}, {Levan},
  {Cano}, {Mottram}, {Clay}, {Bersier}, {Kopa{\v c}}, {Japelj}, {Filippenko},
  {Li}, {Svinkin}, {Golenetskii}, {Hartmann}, {Milne}, {Williams}, {O'Brien},
  {Fox}, \& {Berger}}]{2013ApJ...778...54V}
{Virgili}, F.~J., {Mundell}, C.~G., {Pal'shin}, V., {et~al.} 2013, \apj, 778,
  54

\bibitem[{{Virgili} {et~al.}(2011){Virgili}, {Zhang}, {Nagamine}, \&
  {Choi}}]{2011MNRAS.417.3025V}
{Virgili}, F.~J., {Zhang}, B., {Nagamine}, K., \& {Choi}, J.-H. 2011, \mnras,
  417, 3025

\bibitem[{{Vreeswijk} {et~al.}(2004){Vreeswijk}, {Ellison}, {Ledoux}, {Wijers},
  {Fynbo}, {M{\o}ller}, {Henden}, {Hjorth}, {Masi}, {Rol}, {Jensen}, {Tanvir},
  {Levan}, {Castro Cer{\'o}n}, {Gorosabel}, {Castro-Tirado}, {Fruchter},
  {Kouveliotou}, {Burud}, {Rhoads}, {Masetti}, {Palazzi}, {Pian}, {Pedersen},
  {Kaper}, {Gilmore}, {Kilmartin}, {Buckle}, {Seigar}, {Hartmann}, {Lindsay},
  \& {van den Heuvel}}]{vreeswijk04}
{Vreeswijk}, P.~M., {Ellison}, S.~L., {Ledoux}, C., {et~al.} 2004, \aap, 419,
  927

\bibitem[{{Vreeswijk} {et~al.}(2013){Vreeswijk}, {Ledoux}, {Raassen}, {Smette},
  {De Cia}, {Wo{\'z}niak}, {Fox}, {Vestrand}, \&
  {Jakobsson}}]{2013A&A...549A..22V}
{Vreeswijk}, P.~M., {Ledoux}, C., {Raassen}, A.~J.~J., {et~al.} 2013, \aap,
  549, A22

\bibitem[{{Vreeswijk} {et~al.}(2007){Vreeswijk}, {Ledoux}, {Smette}, {Ellison},
  {Jaunsen}, {Andersen}, {Fruchter}, {Fynbo}, {Hjorth}, {Kaufer}, {M{\o}ller},
  {Petitjean}, {Savaglio}, \& {Wijers}}]{2007A&A...468...83V}
{Vreeswijk}, P.~M., {Ledoux}, C., {Smette}, A., {et~al.} 2007, \aap, 468, 83

\bibitem[{{Wainwright} {et~al.}(2007){Wainwright}, {Berger}, \&
  {Penprase}}]{2007ApJ...657..367W}
{Wainwright}, C., {Berger}, E., \& {Penprase}, B.~E. 2007, \apj, 657, 367

\bibitem[{{Wang} {et~al.}(2015){Wang}, {Zhang}, {Liang}, {Gao}, {Li}, {Deng},
  {Qin}, {Tang}, {Kann}, {Ryde}, \& {Kumar}}]{2015ApJS..219....9W}
{Wang}, X.-G., {Zhang}, B., {Liang}, E.-W., {et~al.} 2015, \apjs, 219, 9

\bibitem[{{Watson}(2011)}]{2011A&A...533A..16W}
{Watson}, D. 2011, \aap, 533, A16

\bibitem[{{Watson} {et~al.}(2004){Watson}, {Hjorth}, {Jakobsson}, {Pedersen},
  {Patel}, \& {Kouveliotou}}]{2004A&A...425L..33W}
{Watson}, D., {Hjorth}, J., {Jakobsson}, P., {et~al.} 2004, \aap, 425, L33

\bibitem[{{Waxman} \& {Draine}(2000)}]{2000ApJ...537..796W}
{Waxman}, E. \& {Draine}, B.~T. 2000, \apj, 537, 796

\bibitem[{{Waxman} {et~al.}(2007){Waxman}, {M{\'e}sz{\'a}ros}, \&
  {Campana}}]{2007ApJ...667..351W}
{Waxman}, E., {M{\'e}sz{\'a}ros}, P., \& {Campana}, S. 2007, \apj, 667, 351

\bibitem[{{Wei} \& {Lu}(2000)}]{2000A&A...360L..13W}
{Wei}, D.~M. \& {Lu}, T. 2000, \aap, 360, L13

\bibitem[{{Wesson} {et~al.}(2015){Wesson}, {Barlow}, {Matsuura}, \&
  {Ercolano}}]{2015MNRAS.446.2089W}
{Wesson}, R., {Barlow}, M.~J., {Matsuura}, M., \& {Ercolano}, B. 2015, \mnras,
  446, 2089

\bibitem[{{Wiersema}(2011)}]{2011MNRAS.414.2793W}
{Wiersema}, K. 2011, \mnras, 414, 2793

\bibitem[{{Wiersema} {et~al.}(2014){Wiersema}, {Covino}, {Toma}, {van der
  Horst}, {Varela}, {Min}, {Greiner}, {Starling}, {Tanvir}, {Wijers},
  {Campana}, {Curran}, {Fan}, {Fynbo}, {Gorosabel}, {Gomboc}, {G{\"o}tz},
  {Hjorth}, {Jin}, {Kobayashi}, {Kouveliotou}, {Mundell}, {O'Brien}, {Pian},
  {Rowlinson}, {Russell}, {Salvaterra}, {di Serego Alighieri}, {Tagliaferri},
  {Vergani}, {Elliott}, {Fari{\~n}a}, {Hartoog}, {Karjalainen}, {Klose},
  {Knust}, {Levan}, {Schady}, {Sudilovsky}, \& {Willingale}}]{wiersema}
{Wiersema}, K., {Covino}, S., {Toma}, K., {et~al.} 2014, \nat, 509, 201

\bibitem[{{Wiersema} {et~al.}(2010){Wiersema}, {D'Avanzo}, {Levan}, {Tanvir},
  {Malesani}, \& {Covino}}]{2010GCN..10525...1W}
{Wiersema}, K., {D'Avanzo}, P., {Levan}, A.~J., {et~al.} 2010, GRB Coordinates
  Network, 10525, 1

\bibitem[{{Wiersema} {et~al.}(2009){Wiersema}, {Levan}, {Kamble}, {Tanvir}, \&
  {Malesani}}]{GCN9673}
{Wiersema}, K., {Levan}, A., {Kamble}, A., {Tanvir}, N., \& {Malesani}, D.
  2009, GRB Coordinates Network, 9673

\bibitem[{{Wilkins} {et~al.}(2015){Wilkins}, {Bouwens}, {Oesch}, {Labbe},
  {Sargent}, {Caruana}, {Wardlow}, \& {Clay}}]{2015arXiv151001514W}
{Wilkins}, S.~M., {Bouwens}, R.~J., {Oesch}, P.~A., {et~al.} 2015, ArXiv
  e-prints

\bibitem[{{Willingale} {et~al.}(2013){Willingale}, {Starling}, {Beardmore},
  {Tanvir}, \& {O'Brien}}]{willingale13}
{Willingale}, R., {Starling}, R.~L.~C., {Beardmore}, A.~P., {Tanvir}, N.~R., \&
  {O'Brien}, P.~T. 2013, \mnras, 431, 394

\bibitem[{{Wilms} {et~al.}(2000){Wilms}, {Allen}, \& {McCray}}]{wilms}
{Wilms}, J., {Allen}, A., \& {McCray}, R. 2000, \apj, 542, 914

\bibitem[{{Witt} {et~al.}(1986){Witt}, {Bohlin}, \&
  {Stecher}}]{1986ApJ...305L..23W}
{Witt}, A.~N., {Bohlin}, R.~C., \& {Stecher}, T.~P. 1986, \apjl, 305, L23

\bibitem[{{Wolfe} {et~al.}(2005){Wolfe}, {Gawiser}, \&
  {Prochaska}}]{2005ARA&A..43..861W}
{Wolfe}, A.~M., {Gawiser}, E., \& {Prochaska}, J.~X. 2005, \araa, 43, 861

\bibitem[{{Woosley}(1993)}]{1993ApJ...405..273W}
{Woosley}, S.~E. 1993, \apj, 405, 273

\bibitem[{{Woosley} \& {Bloom}(2006)}]{2006ARA&A..44..507W}
{Woosley}, S.~E. \& {Bloom}, J.~S. 2006, \araa, 44, 507

\bibitem[{{Woosley} \& {Heger}(2006)}]{2006ApJ...637..914W}
{Woosley}, S.~E. \& {Heger}, A. 2006, \apj, 637, 914

\bibitem[{{Xu} {et~al.}(2013){Xu}, {de Ugarte Postigo}, {Leloudas},
  {Kr{\"u}hler}, {Cano}, {Hjorth}, {Malesani}, {Fynbo}, {Th{\"o}ne},
  {S{\'a}nchez-Ram{\'{\i}}rez}, {Schulze}, {Jakobsson}, {Kaper}, {Sollerman},
  {Watson}, {Cabrera-Lavers}, {Cao}, {Covino}, {Flores}, {Geier}, {Gorosabel},
  {Hu}, {Milvang-Jensen}, {Sparre}, {Xin}, {Zhang}, {Zheng}, \&
  {Zou}}]{2013ApJ...776...98X}
{Xu}, D., {de Ugarte Postigo}, A., {Leloudas}, G., {et~al.} 2013, \apj, 776, 98

\bibitem[{{Yang} {et~al.}(2015){Yang}, {Jin}, {Li}, {Covino}, {Zheng},
  {Hotokezaka}, {Fan}, {Piran}, \& {Wei}}]{2015NatCo...6E7323Y}
{Yang}, B., {Jin}, Z.-P., {Li}, X., {et~al.} 2015, Nature Communications, 6,
  7323

\bibitem[{{Yang} {et~al.}(2004){Yang}, {Chen}, \& {He}}]{2004A&A...414.1049Y}
{Yang}, X., {Chen}, P., \& {He}, J. 2004, \aap, 414, 1049

\bibitem[{{Yoon} \& {Langer}(2005)}]{2005A&A...443..643Y}
{Yoon}, S.-C. \& {Langer}, N. 2005, \aap, 443, 643

\bibitem[{{Zafar} \& {Watson}(2013)}]{zafar13}
{Zafar}, T. \& {Watson}, D. 2013, \aap, 560, A26

\bibitem[{{Zafar} {et~al.}(2012){Zafar}, {Watson}, {El{\'{\i}}asd{\'o}ttir},
  {Fynbo}, {Kr{\"u}hler}, {Schady}, {Leloudas}, {Jakobsson}, {Th{\"o}ne},
  {Perley}, {Morgan}, {Bloom}, \& {Greiner}}]{2012ApJ...753...82Z}
{Zafar}, T., {Watson}, D., {El{\'{\i}}asd{\'o}ttir}, {\'A}., {et~al.} 2012,
  \apj, 753, 82

\bibitem[{{Zafar} {et~al.}(2011){Zafar}, {Watson}, {Fynbo}, {Malesani},
  {Jakobsson}, \& {de Ugarte Postigo}}]{zafar11}
{Zafar}, T., {Watson}, D., {Fynbo}, J.~P.~U., {et~al.} 2011, \aap, 532, A143

\bibitem[{{Zeh} {et~al.}(2004){Zeh}, {Klose}, \&
  {Hartmann}}]{2004ApJ...609..952Z}
{Zeh}, A., {Klose}, S., \& {Hartmann}, D.~H. 2004, \apj, 609, 952

\bibitem[{{Zeh} {et~al.}(2006){Zeh}, {Klose}, \& {Kann}}]{2006ApJ...637..889Z}
{Zeh}, A., {Klose}, S., \& {Kann}, D.~A. 2006, \apj, 637, 889

\bibitem[{{Zhang} \& {Pe'er}(2009)}]{2009ApJ...700L..65Z}
{Zhang}, B. \& {Pe'er}, A. 2009, \apjl, 700, L65

\bibitem[{{Zhang} \& {Yan}(2011)}]{2011ApJ...726...90Z}
{Zhang}, B. \& {Yan}, H. 2011, \apj, 726, 90

\end{thebibliography}
\addcontentsline{toc}{chapter}{\sffamily\bfseries Bibliography}
\cleardoublepage
\end{document}